\documentclass[a4paper,11pt,openright,twoside]{report}
\usepackage{animate,amssymb,bm,braket,float,graphicx,enumitem,mathrsfs,mathtools,microtype,placeins,siunitx,subcaption}
\usepackage[toc,page]{appendix}

\usepackage[thicklines]{cancel}
\usepackage[bottom=3cm]{geometry}
\usepackage[hidelinks]{hyperref}
\usepackage{makecell}	
	
\usepackage[version=4]{mhchem}
\usepackage[utf8]{inputenc}
\usepackage[table]{xcolor}
\usepackage{pdfpages}	
\usepackage{cleveref}
\crefname{section}{§}{§§}
\Crefname{section}{§}{§§}

\usepackage[T1]{fontenc}
\usepackage{lmodern}

	\usepackage[backend=biber,style=numeric-comp,sorting=none,defernumbers=true]{biblatex}
	\DeclareCiteCommand{\citenum}{}{\printfield{labelnumber}}{}{}
	
	\makeatletter
	\let\blx@rerun@biber\relax
	\makeatother

	\newcommand*{\B}[1]{\relax\ifmmode\bm{#1}\else\textbf{#1}\fi}
	
	\newcommand*{\Cross}{\ensuremath{\B{\Red{\times}}}}

	\newcommand{\DOT}{\tikz\draw[blue!50!black,fill=blue!50!black] (0,0) circle (.5ex);}
	\newcommand{\oDOT}{\tikz\draw[blue!50!black,fill=none] (0,0) circle (.5ex);}
	\makeatletter
	\newcommand{\fns}[1]{\ensuremath{^{\@fnsymbol{#1}}}}
	\makeatother
	\newcommand*\circled[1]{\tikz[baseline=(char.base)]{
	            \node[shape=circle,draw,inner sep=1pt] (char) {#1};}}
	
	\DeclareMathOperator{\atanh}{atanh}

	\newcommand*{\Int}{\displaystyle\int}
	
	\newcommand*{\Pm}[1]{\begin{pmatrix}#1\end{pmatrix}}
	\newcommand*{\Sum}{\displaystyle\sum}
	
	\DeclarePairedDelimiter\floor{\lfloor}{\rfloor}

	\newcommand*{\ETh}{\ensuremath{E_\text{Th}}}
	\newcommand*{\Ha}{\ensuremath{\mathcal{H}}}
	
	\newcommand{\Le}{\ensuremath{\ell_\text{e}}}

	\newcommand{\vF}{\ensuremath{v_\text{F}}}
	\DeclareSIUnit\decade{dec}
	\DeclareSIUnit\sccm{sccm}

	\usepackage{footnote}
	\usepackage{etoolbox}
	\newcommand{\Quote}[2]{%
		\begin{flushright}\begin{minipage}[t]{0.45\textwidth}\itshape
		#1%
		\end{minipage}%
		\vspace*{0.5\baselineskip}
		
		-- #2\end{flushright}}
	\BeforeBeginEnvironment{minipage}{\savenotes}
	\AfterEndEnvironment{minipage}{\spewnotes}
	\BeforeBeginEnvironment{flushright}{\savenotes}
	\AfterEndEnvironment{flushright}{\spewnotes}

	\usepackage{tikz,pgf}
	\usetikzlibrary{calc,circuits.ee.IEC,circuits.superconducting,fit,shapes.misc,positioning}
	\usepackage[RPvoltages]{circuitikz}
	\ctikzset{bipoles/length=1.4cm,bipoles/thickness=1}
	\usetikzlibrary{arrows,decorations.markings,quantikz,fadings}
	\usepackage{pgfplots}
	\pgfplotsset{compat=1.17}

	\usetikzlibrary{patterns,tikzmark}
	\tikzset{
	    hatch distance/.store in=\hatchdistance,
	    hatch distance=10pt,
	    hatch thickness/.store in=\hatchthickness,
	    hatch thickness=2pt,
	    vertical distance/.store in=\verticaldistance,
	    vertical distance=10pt,
	    vertical thickness/.store in=\verticalthickness,
	    vertical thickness=2pt,
	DepositionA1/.pic = {
		\fill[Si] 			(0,0) -- (\FigX,0) -- (\FigX,10) -- (0,10) -- cycle;
		\fill[Pt] 			(0,10) -- (\FigX,10) -- (\FigX,20) -- (0,20) -- cycle;
		\fill[Pt!50!Si] 	(2.5,10) ellipse (2 and 1) (7.5,10) ellipse (2 and 1);
		},
	DepositionA2/.pic = {
		\fill[Si] 			(0,0) -- (\FigX,0) -- (\FigX,10) -- (0,10) -- cycle;
		\fill[Pt] 			(0,10) -- (\FigX,10) -- (\FigX,20) -- (0,20) -- cycle;
		\fill[Pt2Si] 		(0,10) -- (\FigX,10) -- (\FigX,12) -- (0,12) -- cycle;
		\fill[Pt!50!Si] 	(2.5,10) ellipse (2 and 1) (7.5,10) ellipse (2 and 1);
		},
	DepositionB1/.pic = {
		\fill[Si] 			(0,) -- (\FigX,) -- (\FigX,10) -- (0,10) -- cycle;
		\fill[Pt] 			(0,10) -- (\FigX,10) -- (\FigX,20) -- (0,20) -- cycle;
		\fill[Pt!50!Si] 	(0,9) -- (\FigX,9) -- (\FigX,11) -- (0,11) -- cycle;
		},
	DepositionB2/.pic = {
		\fill[Si] 			(0,) -- (\FigX,) -- (\FigX,10) -- (0,10) -- cycle;
		\fill[Pt] 			(0,10) -- (\FigX,10) -- (\FigX,20) -- (0,20) -- cycle;
		\fill[Pt!50!Si] 	(0,8) -- (\FigX,8) -- (\FigX,10) -- (0,10) -- cycle;
		\fill[Pt2Si] 		(0,10) -- (\FigX,10) -- (\FigX,12) -- (0,12) -- cycle;
		},
	DepositionC/.pic = {
		\fill[Si] 			(0,0) -- (\FigX,0) -- (\FigX,10) -- (0,10) -- cycle;
		\fill[Pt] 			(0,10) -- (\FigX,10) -- (\FigX,20) -- (0,20) -- cycle;
		\fill[bottom color=Si,top color=Pt] 	(0,5) -- (\FigX,5) -- (\FigX,15) -- (0,15) -- cycle;
		},
	RTA1A/.pic = {	
		\fill[Si] 			(0,0) -- (\FigX,0) -- (\FigX,10) -- (0,10) -- cycle;
		\fill[Pt] 			(0,10) -- (\FigX,10) -- (+\FigX,20) -- (0,20) -- cycle;
		\fill[Pt2Si] 		(0,7) -- (\FigX,7) -- (\FigX,14) -- (0,14) -- cycle;
		\fill[Pt!50!Si] 	(2.5,7) ellipse (2 and 1) (7.5,7) ellipse (2 and 1);
		},
	RTA1B/.pic = {
		\fill[Si] 			(0,0) -- (\FigX,0) -- (\FigX,10) -- (0,10) -- cycle;
		\fill[Pt] 			(0,10) -- (\FigX,10) -- (\FigX,20) -- (0,20) -- cycle;
		\fill[Pt!50!Si] 	(0,5) -- (\FigX,5) -- (\FigX,7) -- (0,7) -- cycle;
		\fill[Pt2Si] 		(0,7) -- (\FigX,7) -- (\FigX,14) -- (0,14) -- cycle;
		},
	SEA/.pic = {
		\fill[Si] 			(0,0) -- (\FigX,0) -- (\FigX,7) -- (0,7) -- cycle;
		\fill[Pt!50!Si] 	(2.5,7) ellipse (2 and 1) (7.5,7) ellipse (2 and 1);
		},
	SEB/.pic = {
		\fill[Si] 			(0,0) -- (\FigX,0) -- (\FigX,5) -- (0,5) -- cycle;
		\fill[Pt!50!Si] 	(0,5) -- (\FigX,5) -- (\FigX,7) -- (0,7) -- cycle;
		},
	Legend/.pic = {
		\fill[white] 	(-\SpaceX,20) -- (\GridX-\SpaceX,20) -- (\GridX-\SpaceX,1) -- (-\SpaceX,1) -- cycle;
		\fill[Pt] 				(-4,15) -- (2,15) -- (2,18) -- (-4,18) -- cycle;
		\node[anchor=west] at 	(4,16.5) {Pt};
		\fill[Pt2Si] 			(-4,11) -- (2,11) -- (2,14) -- (-4,14) -- cycle;
		\node[anchor=west] at 	(4,12.5) {Pt2Si};
		\fill[Pt!50!Si] 		(-4,7) -- (2,7) -- (2,10) -- (-4,10) -- cycle;
		\node[anchor=west] at 	(4,8.5) {PtSi};
		\fill[Si] 				(-4,3) -- (2,3) -- (2,6) -- (-4,6) -- cycle;
		\node[anchor=west] at 	(4,4.5) {Si};
		},
	LegendWithoutPt2Si/.pic = {
		\fill[white] 	(-\SpaceX,20) -- (\GridX-\SpaceX,20) -- (\GridX-\SpaceX,4) -- (-\SpaceX,4) -- cycle;
		\fill[Pt] 				(0,14) -- (4,14) -- (4,16) -- (0,16) -- cycle;
		\node[anchor=west] at 	(5,15) {Pt};
		\fill[Pt!50!Si] 		(0,11) -- (4,11) -- (4,13) -- (0,13) -- cycle;
		\node[anchor=west] at 	(5,12) {PtSi};
		\fill[Si] 				(0,8) -- (4,8) -- (4,10) -- (0,10) -- cycle;
		\node[anchor=west] at 	(5,9) {Si};
		},
	LegendPtSiSi/.pic = {
		\fill[white] 	(-\SpaceX,20) -- (\GridX-\SpaceX,20) -- (\GridX-\SpaceX,7) -- (-\SpaceX,7) -- cycle;
		\fill[Pt!50!Si] 		(0,14) -- (4,14) -- (4,16) -- (0,16) -- cycle;
		\node[anchor=west] at 	(5,15) {PtSi};
		\fill[Si] 				(0,11) -- (4,11) -- (4,13) -- (0,13) -- cycle;
		\node[anchor=west] at 	(5,12) {Si};
		}
	}
	\tikzset{gooddie/.style n args={2}{%
			rectangle,
			inner sep=0pt,
			fill=green,
			fill opacity=0.2,
			draw=none,
			fit={(#2-0.96,10.25-#1) (#2+0.04,9.25-#1)},
		}
	}
	\tikzset{baddie/.style n args={2}{%
			rectangle,
			inner sep=0pt,
			fill=red,
			fill opacity=0.2,
			draw=none,
			fit={(#2-0.96,10.25-#1) (#2+0.04,9.25-#1)},
		}
	}
		
	\tikzset{goodscribe/.style n args={2}{%
		    rectangle,
		    inner sep=0pt,
		    fill=green,
		    fill opacity=0.2,
		    draw=none,
		    fit={(#1-1+0.43,#2+2.3) (#1+0.43,#2+3.3)},
		    text=black,
		    text opacity=1
		}
	}
	\tikzset{badscribe/.style n args={2}{%
		    rectangle,
		    inner sep=0pt,
		    fill=red,
		    fill opacity=0.2,
		    draw=none,
		    fit={(#1-1+0.43,#2+2.3) (#1+0.43,#2+3.3)},
			text=black,
			text opacity=1
		}
	}
	\tikzset{scribetext/.style n args={2}{%
		    rectangle,
		    inner sep=0pt,
		    draw=none,
		    fit={(#1-1+0.43,#2+2.3) (#1+0.43,#2+3.3)},
		    text=white
		}
	}
	
	\newcommand{\goodscribe}[3]{%
		\node[goodscribe={#1}{#2}]{\\[-0.42em]\hspace*{0.25ex}\B{#3}};
		\node[scribetext={#1}{#2}]{\\[-0.55em]\B{#3}};
		}

	\tikzset{Tempgooddie/.style n args={2}{%
			rectangle,
			inner sep=0pt,
			fill=green,
			fill opacity=0.2,
			draw=none,
			fit={(#2-0.96,10.25-#1) (#2+0.04,9.25-#1)},
		}
	}
	\tikzset{Tempbaddie/.style n args={2}{%
			rectangle,
			inner sep=0pt,
			fill=red,
			fill opacity=0.2,
			draw=none,
			fit={(#2-0.96,10.25-#1) (#2+0.04,9.25-#1)},
		}
	}
		
	\tikzset{Tempgoodscribe/.style n args={2}{%
		    rectangle,
		    inner sep=0pt,
		    fill=green,
		    fill opacity=0.2,
		    draw=none,
		    fit={(7.5+0.5*\chipwidth-#2*\chipwidth/20+13*\chipwidth/20,15-0.5*\chiplength+#1*\chiplength/4-4*\chiplength/4) (7.5+0.5*\chipwidth-#2*\chipwidth/20+12*\chipwidth/20,15-0.5*\chiplength+#1*\chiplength/4-3*\chiplength/4)},
		    text=black,
		    text opacity=1
		}
	}
	\tikzset{Tempbadscribe/.style n args={2}{%
		    rectangle,
		    inner sep=0pt,
		    fill=red,
		    fill opacity=0.2,
		    draw=none,
		    fit={(7.5+0.5*\chipwidth-#2*\chipwidth/20+13*\chipwidth/20,15-0.5*\chiplength+#1*\chiplength/4-4*\chiplength/4) (7.5+0.5*\chipwidth-#2*\chipwidth/20+12*\chipwidth/20,15-0.5*\chiplength+#1*\chiplength/4-3*\chiplength/4)},
			text=black,
			text opacity=1
		}
	}
	\tikzset{Tempscribetext/.style n args={2}{%
		    rectangle,
		    inner sep=0pt,
		    draw=none,
		    fit={(7.5+0.5*\chipwidth-#2*\chipwidth/20+13*\chipwidth/20,15-0.5*\chiplength+#1*\chiplength/4-4*\chiplength/4) (7.5+0.5*\chipwidth-#2*\chipwidth/20+12*\chipwidth/20,15-0.5*\chiplength+#1*\chiplength/4-3*\chiplength/4)},
		    text=white
		}
	}
	
	\newcommand{\Tempgoodscribe}[3]{%
		\node[Tempgoodscribe={#1}{#2}]{};
		}

	\usepackage{xstring}
	\newcommand\IfStringInList[2]{\IfSubStr{,#2,}{,#1,}}
	\definecolor{PtSipresent}{RGB}{255,200,200}
	\definecolor{PtSiPtTSipresent}{RGB}{255,200,255}
	\definecolor{PtTSipresent}{RGB}{200,200,255}
	\makeatletter
	\pgfdeclarepatternformonly[\verticaldistance,\verticalthickness]{flexible horizontal}
	{\pgfqpoint{0pt}{0pt}}
	{\pgfqpoint{\verticaldistance}{\verticaldistance}}
	{\pgfpoint{\verticaldistance-1pt}{\verticaldistance-1pt}}%
	{
	    \pgfsetcolor{\tikz@pattern@color}
	    \pgfsetlinewidth{\verticalthickness}
	    \pgfpathmoveto{\pgfqpoint{0pt}{0pt}}
	    \pgfpathlineto{\pgfqpoint{\verticaldistance}{0pt}}
	    \pgfusepath{stroke}
	}
	\pgfdeclarepatternformonly[\hatchdistance,\hatchthickness]{flexible hatch}
	{\pgfqpoint{0pt}{0pt}}
	{\pgfqpoint{\hatchdistance}{\hatchdistance}}
	{\pgfpoint{\hatchdistance-1pt}{\hatchdistance-1pt}}%
	{
	    \pgfsetcolor{\tikz@pattern@color}
	    \pgfsetlinewidth{\hatchthickness}
	    \pgfpathmoveto{\pgfqpoint{0pt}{0pt}}
	    \pgfpathlineto{\pgfqpoint{\hatchdistance}{\hatchdistance}}
	    \pgfusepath{stroke}
	}
	\makeatother
	\definecolor{ArrowColor}{RGB}{195,195,180}
	\definecolor{Etch}{RGB}{255,255,255}
	\definecolor{Ge}{RGB}{102,255,102}
	\definecolor{Gold}{RGB}{255,236,175}
	\definecolor{Pt}{RGB}{255,80,60}		
	\definecolor{Pt2Si}{RGB}{200,70,255}		
	\definecolor{Si}{RGB}{180,180,190}
	\newcommand{\Sip}{[preaction={fill,Si},pattern=north west lines, pattern color=gray!20!white]}
	\definecolor{SiN}{RGB}{153,204,0}
	\definecolor{SiO2}{RGB}{255,204,102}
	\newcommand{\SiOdoped}{[pattern=dots, pattern color=SiO2]}
	\newcommand{\SiVseq}{[preaction={fill,Va},pattern=flexible horizontal, vertical distance=4pt, vertical thickness=1.5pt, pattern color=Si]}
	
	\definecolor{Va}{RGB}{255,52,52}
	\definecolor{VO}{RGB}{255,128,77}
	\definecolor{VSi2}{RGB}{205,137,144}
	\definecolor{V3Si}{RGB}{178,77,102}
	\definecolor{redbg}{RGB}{255,190,180}
	\definecolor{greenbg}{RGB}{210,255,180}
	\usepackage{listings}
	\usepackage{nicematrix}
	\newcommand{\gc}[1]{\cellcolor{gray!20!white}{#1}}

	\definecolor{Blue}{RGB}{0,0,220}
	\newcommand*{\Blue}[1]{\textcolor{Blue}{#1}}
	\newcommand*{\Bblue}[1]{\textcolor{Blue}{\B{#1}}}
	\definecolor{Gray}{RGB}{150,150,150}
	\definecolor{Green}{RGB}{0,160,0}
	
	\newcommand*{\Bgreen}[1]{\textcolor{Green}{\B{#1}}}
	\definecolor{Purple}{RGB}{220,0,220}
	
	\definecolor{Red}{RGB}{220,0,0}
	\newcommand*{\Red}[1]{\textcolor{Red}{#1}}
	\newcommand*{\Bred}[1]{\textcolor{Red}{\B{#1}}}
	
	\newcounter{nalg}[chapter] 
	\renewcommand{\thenalg}{\thechapter .\arabic{nalg}} 
	\DeclareCaptionLabelFormat{algocaption}{Algorithm \thenalg} 
	
	\lstnewenvironment{algorithm}[1][] 
	{   
	    \refstepcounter{nalg} 
	    \captionsetup{labelformat=algocaption,labelsep=colon} 
	    \lstset{ 
		mathescape=true,
		frame=tB,
		numbers=left, 
		numberstyle=\tiny,
		basicstyle=\scriptsize, 
		keywordstyle=\color{black}\bfseries,
		keywords={, begin, buy, datatype, else, end, for, foreach, function, if, in, input, output, return, solve, sort, while, } 
		numbers=left,
		xleftmargin=.04\textwidth,
		emph={exit,False,float,len,None,print,range,True},
		emphstyle={\color{codepurple}\bfseries},
		emph={[2]1,2,3,4,5,6,7,8,9,0,-,=,+,[,],(,),\{,\},:,*,!},
		emphstyle=[2]{\color{codered}\bfseries},
		#1 
	    }
	}
	{}

	\newcommand{\graybox}[1]{\edef\oldfboxsep{\the\fboxsep}%
	\fboxsep=15pt%
	\noindent\colorbox{gray!15!white}{%
	\parbox{\dimexpr\textwidth-2\fboxsep-2\fboxrule-\leftmargin\relax}{%
		\fboxsep=\oldfboxsep%
			#1%
		}}
	\fboxsep=\oldfboxsep\\}

	\let\plainappendixpage\appendixpage
	\makeatletter
	\renewcommand{\appendixpage}{%
		\begingroup
		\let\ps@plain\ps@empty
		\plainappendixpage
		\endgroup}
	\makeatother

	\graphicspath{{img/}}

\title{Silicide-based Josephson field effect transistors for superconducting qubits}
\author{Tom Doekle Vethaak}
\date{\today}

\begin{document}

\pagenumbering{gobble}
\includepdf{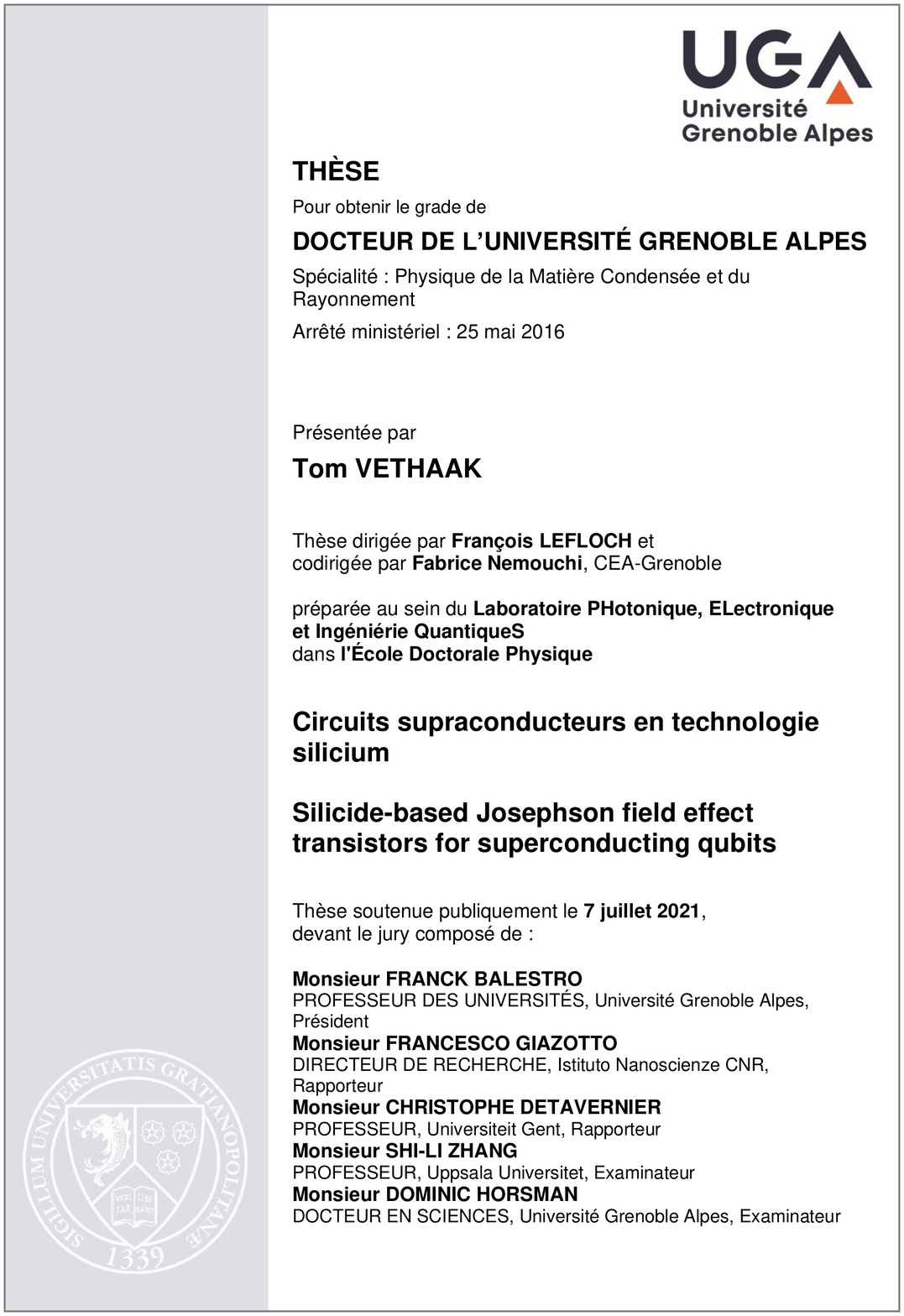}
\cleardoublepage

\chapter*{Abstract}

Scalability in the fabrication and operation of quantum computers is key to move beyond the NISQ era.
So far, superconducting transmon qubits based on aluminum Josephson tunnel junctions have demonstrated the most advanced results, though this technology is difficult to implement with large-scale facilities.
An alternative ``gatemon'' qubit has recently appeared, which uses hybrid superconducting/semiconducting (S/Sm) devices as gate-tuned Josephson junctions.
Current implementations of these use nanowires however, of which the large-scale fabrication has not yet matured either.
A scalable gatemon design could be made with CMOS Josephson Field-Effect Transistors as tunable weak link, where an ideal device has leads with a large superconducting gap that contact a short channel through high-transparency interfaces.
High transparency, or low contact resistance, is achieved in the microelectronics industry with silicides, of which some turn out to be superconducting.
The first part of the experimental work in this thesis covers material studies on two such materials: \ce{V3Si} and PtSi, which are interesting for their high $T_\text{c}$, and mature integration, respectively.
The second part covers experimental results on \SI{50}{\nano\meter} gate length PtSi transistors, where the transparency of the S/Sm interfaces is modulated by the gate voltage.
At low voltages, the transport shows no conductance at low energy, and well-defined features at the superconducting gap.
The barrier height at the S/Sm interface is reduced by increasing the gate voltage, until a zero-bias peak appears around zero drain voltage, which reveals the appearance of an Andreev current.
The successful gate modulation of Andreev current in a silicon-based transistor represents a step towards fully CMOS-integrated superconducting quantum computers.
\cleardoublepage

\chapter*{Résumé}

L'évolution de la fabrication et l'exploitation des ordinateurs quantiques est essentielle pour dépasser l'ère NISQ. Jusqu'à présent, les qubits transmon supraconducteurs basés sur des jonctions tunnel Josephson en aluminium ont donné les résultats les plus avancés, bien que cette technologie soit difficile à mettre en œuvre dans des installations à grande échelle. Un autre qubit "gatemon" est apparu récemment, qui utilise des dispositifs hybrides supraconducteurs/semiconducteurs (S/Sm) comme jonctions Josephson accordées par un contrôle électrostatique. Cependant, les implémentations actuelles de ces dispositifs utilisent des nanofils, dont la fabrication à grande échelle n'a pas encore atteint sa maturité. Une conception évolutive de gatemon pourrait être réalisée avec des transistors à effet de champ Josephson CMOS comme lien faible accordable, où un dispositif idéal est constitué de réservoirs supraconducteurs qui contactent un canal court par des interfaces à haute transparence. Une transparence élevée, ou une faible résistance de contact, est obtenue dans l'industrie de la microélectronique avec des siliciures, dont certains s'avèrent être supraconducteurs. La première partie du travail expérimental de cette thèse couvre les études de matériaux sur deux de ces matériaux : \ce{V3Si} et PtSi, qui sont intéressants pour leur $T_\text{c}$ élevé, et leur intégration mature. La deuxième partie couvre des résultats expérimentaux sur des transistors PtSi de longueur de grille \SI{50}{\nano\m}, où la transparence des interfaces S/Sm est modulée par la tension de grille. A basse tension, le transport ne montre aucune conductance à basse énergie, et des pics de cohérence bien définis au niveau du gap supraconducteur. La hauteur de la barrière à l'interface S/Sm est réduite en augmentant la tension de grille, jusqu'à ce qu'un pic de conductance apparaisse autour d’une tension de drain nulle, ce qui révèle l'apparition d'un courant d'Andreev. La modulation du courant d'Andreev par une grille dans un transistor à base de silicium, représente une étape vers des ordinateurs quantiques supraconducteurs entièrement intégrés en CMOS.
\cleardoublepage

\tableofcontents
\cleardoublepage
\pagenumbering{arabic}

\begin{refsection}
	\graphicspath{{img/intro/}}
	\chapter*{Introduction}

\Quote{``You have nothing to do but mention the quantum theory, and people will take your voice for the voice of science, and believe anything.''}{George Bernard Shaw}

In front of you are $10^7$ bits of information. 
That sounds like a lot, and it certainly took me a while to put together, but it isn't nearly enough to describe what is going on inside a chip like Sycamore at any point in time~\cite{arute2019quantum}, even after accounting for redundancies due to error rates and such~\cite{zhou2020limits}.
What's remarkable is that all this complexity is accounted for by only 53 physical qubits.
This exponential scaling of the Hilbert space with the number of qubits is the basis for the general excitement about the promise of quantum computing.
Possible applications range far beyond the familiar breaking of RSA cryptography, from optimization problems and quantum machine learning, to a complete overhaul of chemistry, materials science and biology by the simulation of nano-scale electronic interactions.
Broadening the range of computable problems is of course an exciting, rewarding and even economically sensible goal, and justifies the recent surge in activity.
On the other hand, the cautious skepticism towards claims of any imminent \emph{useful} advantage is equally justified.
But even if quantum computers only become competitive with supercomputer clusters on actual applications after our lifespans, or if there turn out to be practical limits to the scale at which coherence can be maintained, this field still has an additional appeal on a more basic level.

What gets me excited is the realization that at its most fundamental, existence itself is computation~\cite{feynman1982simulating,deutsch1985quantum,lloyd2002computational}, and the laws of physics don't deal with classical bits, they deal with qubits~\cite{deutsch1998fabric,aaronson2013quantum,dowling2013schrodinger}.
No matter what underlying structure there turns out to be to the universe, whether it contains supersymmetric siblings of every elementary particle, or if they are all excitations of strings vibrating in 10 or 11 dimensions~\cite{susskind2008black,hossenfelder2018lost}, in the end it is all described by \emph{information}; a vast register of qubits in a cosmic quantum computer.
The quantum nature of reality has been tested (and always verified) since Einstein's interpretation of the photoelectric effect, and improvements in the understanding of its principles has led to revolutions in science and technology alike~\cite{dowling2003quantum}.
But only recently have we begun to isolate, couple and probe individual degrees of freedom in physical qubits, such that we can now set, flip or measure the state of a single fundamental unit of information.
Now that we have broken it all down to this core component, we can start to build something of our own.
I do not much care what the appropriate level of skepticism is towards any predictions of near-term realizations of large-scale quantum computers, for to me \emph{this}, the gaining of control over qubits, is already a turning point in science.

This thesis explores a path towards large-scale quantum computers.
The central idea is to leverage the existing fabrication processes already in use for classical computer chips, and adapt them to mass-produce quantum computers made of superconducting circuits.
We will first discuss the broader context of quantum computing in chapter~\ref{sec:quantum_computing}, before delving into the details of designing a superconducting qubit in chapter~\ref{sec:cmos_gatemons}.
Here we will make the argument that, like with so many instances of technological innovation, the key is in the development of the right materials.
In our case, this means studying silicides, intermetallic compounds that are part silicon and part metal, that are traditionally used in the contacts to the semiconducting silicon channel of transistors.
Chapter~\ref{sec:silicides} then elaborates on experiments with two such silicides that superconduct, \ce{V3Si} and \ce{PtSi}.
This last material was used in the late 90's in a special kind of transistor where, unusually, the contact silicides extended to underneath the gate electrode.
As we will discuss in chapter~\ref{sec:jofets}, the particular combination of design parameters made it hard to switch this transistor off at room temperature, rendering it unsuitable for its original purpose.
Fortunately, these very same properties allowed for current to flow even when the device was cooled to a few tens of millikelvin, leading to the observation of induced superconductivity inside the silicon channel.
Although work remains to be done, we hope to convince you that this is a promising result that hints at the possibility of large-scale fabrication of silicon-based superconducting qubits.

\printbibliography

\chapter*{Acknowledgements}

\Quote{``Teamwork makes the dream work.''}{Scott Kelly}

De tous ceux que je dois énumérer, sans doute premier parmi ses pairs est Frederic Gustavo.
Fred, merci de m'avoir appris les bonnes manières de la salle blanche, et de m'avoir aidé à supporter la charge des expériences.
C'est un fait indiscutable que pas un seul des résultats discutés dans cette thèse n'a été obtenu sans ton aide.
Mais surtout, je veux te remercier pour les nombreuses heures pendant lesquelles nous avons parlé de tout \emph{sauf} de science.
Je suis sûr que je me souviendrai avec émotion de nos discussions et dîners longtemps après avoir oublié la plupart des résultats scientifiques dans ce manuscrit.

As with Fred, I would like to thank also François Lefloch in the first place for the human component that he brings to his research.
His concern for the well-being of those around him is of a level that is rare not just in scientific environments.
To name just one instance, I'm glad that I could take off an entire month when my grandmother was about to pass away, without any questioning of my commitment, nor any push to work remotely.
This freedom extended to the work itself, and has helped me enormously to gain the confidence to do research independently.
As co-supervisor, Fabrice ``the magician'' Nemouchi has been absolutely irreplaceable to help me navigate the LETI, and has taught by example the importance of all the soft skills involved in any collaborative effort.
The sense of humor that he shares with his taller alter ego Philippe Rodriquez has brought color to even the most mundane discussions.

I'm greatly indebted to Valérie Lapras, who has introduced me to all the principles of the cleanroom, and who has sacrificed many weeks of her time to cook up the process flows for nearly all the lots that were prepared.
For the materials part of this thesis I have relied heavily on Shi-Li Zhang, who became my mentor on all things related to silicides, and who together with Tomas Kubart took good care of me in Uppsala.
Last, I am grateful to Sylvan Brocard, who has been my soundboard and intellectual anchor throughout the past three years.
\end{refsection}

\begin{refsection}
	\graphicspath{{img/ch1/}}
	\chapter{\label{sec:quantum_computing}Quantum computing}

\section{Introduction}
	
	\Quote{``It is difficult to make predictions, especially about the future.''\footnote{\emph{``Det er vanskeligt at spaa, især naar det gælder Fremtiden.''}}}{Danish proverb}
	
	\noindent Quantum computers will change our world.
	
	To see why, first consider how we became aware of the power of \emph{classical} computation. Around the time that Alan Turing wrote his prescient paper \emph{``Computing machinery and intelligence''}~\cite{turing1950computing}, where he introduced the concept of machine intelligence, the general public at large had not yet had any interaction with an artificial computer\footnote{``Artificial'', to distinguish the machines from the profession.}.
	It then took 40 years from the invention of the integrated circuit~\cite{kilby1959miniaturized} to the explosion in software applications that shaped our modern world~\cite{schmidt2020keynote}, and only this last decade have we seen convincing implementations akin to Turing's vision.
	What Turing had touched on that allowed him to preconceive the future that we are only now embarking on, was a concept called \emph{universality}\footnote{While it is now known as a ``universal Turing machine'' due to Church, he originally named his invention more humbly a ``universal computing machine'' in the paper cited above. He starts section~6 with the phrase \emph{``It is possible to invent a single machine which can be used to compute any computable sequence''}, which is now known as the Church-Turing thesis. }~\cite{turing1937computable,turing1938computablecorrection}, the idea that there is a small set of operations that all logical processes can be reduced to\footnote{The original example of a universal machine was a tape reader with a set of internal states, that moves along a tape with bits of 0 and 1. It would move, read, write, or change its configuration depending on the combination of its current configuration and the value of the bit it was on.
	}.
	Since there was (is?) no reason to assume that human thought is not a logical process\footnote{At the time that Turing, Shannon and von Neumann were making these comparisons, McCulloch and Pitts' Boolean framework~\cite{mcculloch1943logical} was still current. Cobb details how we now know that the brain is not just a network of logical gates (signals are non-binary, bi-directional, and can be modulated by semi-global variables)~\cite{cobb2020idea}, but far as I understand computational complexity theory, there is still a chance that it can be simulated efficiently on a Turing machine.}, he applied this principle to reason that computing machines, once they would be fast enough, could emulate us as well.
	The impact of computation on the world was predicted not by extrapolating from contemporary capabilities or applications, but by taking a fundamental statement about computability to its logical conclusion.
	
	Now, how does this relate to the current situation around quantum computers, and what kind of fundamental statement can we make about them?
	After Turing had established that all naturally computable functions can be computed by a universal machine \emph{in principle}, the field started wondering whether such a machine could always do this \emph{efficiently}.
	And so Church and Turing's original thesis was extended, to state that for any problem, if it is possible to make a special dedicated machine tailored to that problem that can solve it efficiently, then a Turing machine can solve it efficiently as well (with at most polynomial overhead).
	This qualifier about efficiency makes an important difference in practical terms.
	For example, if a traveling salesperson wanted to minimize the distance she would have to travel to visit each of a set of cities, she would need an amount of time exponential in the number of cities when consulting a Turing machine\footnote{Though ``determining'' the solution is exponentially hard, checking it can be done in polynomial time, which means that its complexity is \emph{non-deterministically polynomial} (NP). What's more, once you have found the shortest route, you can use that solution to efficiently solve a range of other NP problems, such as finding the lowest-energy configuration of an Ising spin glass. This equivalence groups these problems together in what is called NP-complete.}.
	Given the extended thesis, she would concede that there is not much point in building a specialized tool either, since whatever speedup it would give, it could still never do better than exponential time\footnote{Since $e^x$ is an infinite series of all powers of $x$, division by any finite polynomial still returns an exponential.}.
	The extended thesis puts a cap on the potential of computers: a Turing machine is the most powerful thing that you could possibly build, and those problems that are out of its reach will remain forever so.
	But here's the crux: it turns out to be \emph{wrong}~\cite{bernstein1997quantum}.
	
	Turing's universal machine appears not to be universally efficient after all: there are problems that can be solved in polynomial time on a quantum computer, that are exponentially hard on a classical one.
	Though not proven rigorously before 1993~\cite{bernstein1993quantum}, the difficulty of simulating quantum systems and therefore the relative advantage of computers based on them was already realized by Feynman in 1982~\cite{feynman1982simulating}\footnote{The paper contains many profound insights, such as the idea that the correctness of a law depends on its computability, since the things that that law describes could themselves be used as the computer. Once you realize that quantum mechanics cannot be efficiently simulated by a classical computer, two options remain: either you assume the extended Church-Turing thesis, taking computability to mean \emph{computable by a Turing machine}, and conclude that our physical laws must be wrong, or you take quantum physics to be true and conclude that the extended thesis is wrong.
	Luckily the Bell test was done a decade earlier by Freedman and Clauser, helping him conclude the latter~\cite{freedman1972experimental}.
	He also posited the idea of a probabilistic computer, which are now known to be faster in some situations (e.g. they can test whether a number is prime in polynomial time~\cite{aaronson2013quantum}), giving rise to the complexity class BPP.
	It was relative to BPP that Bernstein and Vazirani proved quantum computers to be more powerful, BPP$\subseteq$BQP and BPP$\neq$BQP~\cite{bernstein1997quantum} (here BPP stands for Bounded-error Probabilistic Polynomial time).
	}.
	We do not yet exactly know, however, how many such problems there are.
	For example, whether the traveling salesperson mentioned above may one day find her shortest route by sending a query to a quantum server is unknown: no-one has yet proven whether NP-complete problems fall within BQP (Bounded-error Quantum Polynomial, loosely speaking the set of problems that can be solved in polynomial time on a quantum computer), or any other complexity class that takes advantage of quantum mechanics~\cite{aaronson2013quantum}.
	Neither do we know whether the exponential speedups of some of our best quantum algorithms are truly insurmountable by any classical means.
	
	Nobody has yet found proof that Shor's prime factorization algorithm~\cite{shor1994algorithms}, arguably the best-known example of the potential that quantum computers hold, cannot be equaled in speed by something run on a classical Turing machine.
	As far as we know, finding the prime factors of large numbers is exponentially hard on a classical computer.
	This is so hard that even if every particle in the entire universe were to be used as a logical component in an enormous computer, which would be left to calculate prime factors for the entire age of the universe, it could at most have factored only a million-bit number (less than a megabyte)~\cite{lloyd2002computational}\footnote{The hardness of this problem is the basis for RSA encryption~\cite{rivest1978method}.}.
	But this is only true if our best classical algorithms really are the best ones \emph{possible}.
	In July 2018, 18-year old Ewin Tang posted a preprint~\cite{tang2019quantum} describing a classical algorithm that could solve the problem of sampling sparse matrices in polylogarithmic time.
	This was a shock to the computational complexity community, who had considered the quantum algorithm for this problem~\cite{kerenidis2016quantum} a promising indication that quantum machine learning (one of three loosely defined branches of quantum algorithms, along optimization and simulation) could offer exponential speedups.
	To further illustrate the state of the field, not only are we unsure of BQP's relationship to NP, we cannot even say with certainty that P$\neq$NP!
	
	So here is where we are today: we do not yet really know what limits there are to the computational power of a classical computer, but we are sure that whatever it \emph{can} do, a quantum computer can do more.
	We already have many quantum algorithms for more or less useless problems that are certainly faster than anything that could run on a classical computer, and a healthy number of quantum algorithms for actually useful problems that are probably faster than classical alternatives.
	Though currently more or less limited to the simulation of quantum systems, finding prime factors and quantum walk graph traversal, large efforts are under way to add to the list of useful problems with certain, exponential quantum advantage.
	
\section{Quantum circuits and gates}
	
	Feynman ended the first half of his 1982 talk by saying \emph{``I therefore believe it's true that with a suitable class of quantum machines you could imitate any quantum system, including the physical world.''}, and asked \emph{``What, in other words, is the universal quantum simulator?''}
	This question was answered by David Deutsch three years later~\cite{deutsch1985quantum}, who gave a proof of existence of a universal quantum Turing machine\footnote{That quantum Turing machines can simulate each other efficiently was only proven in 1993~\cite{yao1993quantum}.}.
	In short, Deutsch showed that there exist initialization programs $\rho$ for each $L$-qubit state $\ket{\psi}$ such that $\rho\ket{\psi}=\ket{0_L}$.
	These programs together with their inverses, combined with the ability to multiply the ground state by a phase factor, can then perform arbitrary unitary transformations on any $L$-qubit state.
	Though this was an important result as it proved that we can construct a universal quantum Turing machine in principle, the proposed operations are not very practical.
	Much simpler and more convenient gate sets have since been developed, and we will see below that any universal set of single-qubit gates plus a single entangling two-qubit gate is sufficient.
	
	\subsection{Single-qubit gates}
	
		When discussing single-qubit gates, the word ``universal'' refers to the possibility of obtaining any (normalized) complex combination of $\ket{0}$ and $\ket{1}$ in a unitary manner\footnote{Unitarity implies reversibility, which means that information is always preserved. This fundamental principle of quantum mechanics led Don Page to conclude that black holes must preserve information, contradicting his PhD supervisors Stephen Hawking and Kip Thorne. It became a productive controversy, leading to the holographic principle~\cite{susskind2008black} (which in turn is one of the pillars of Erik Verlinde's entropic gravity theory, that does away with the need for dark matter~\cite{verlinde2011origin}), and fruitful links between general relativity and quantum information theory~\cite{sachdev1993gapless,kitaev2015simple}, two opposite ends of the field of Physics that famously appear incompatible (e.g., their estimates of the vacuum energy density are off by 121 orders of magnitude, predicting $10^{-9}$ and $10^{113}$~\si{\joule/\meter\cubed}, respectively~\cite{griffiths2005introduction}). To bring us full circle, these links can actually be studied with quantum computers!~\cite{brown2019quantum,hossenfelder2019analog}}.		
		
\begin{figure}
			\centering
			\begin{subfigure}[t]{0.33\textwidth}
				\centering
				\includegraphics[width=0.9\textwidth]{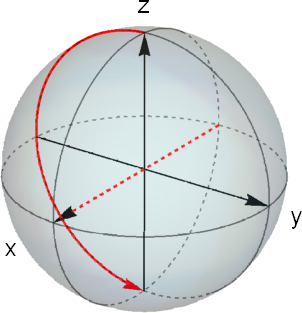}
				\caption{$\sigma_x\,\ket{0}=\ket{1}$}
			\end{subfigure}\begin{subfigure}[t]{0.33\textwidth}
				\centering
				\includegraphics[width=0.9\textwidth]{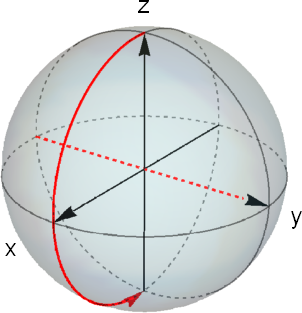}
				\caption{$\sigma_y\ket{0}=i\ket{1}$}
			\end{subfigure}\begin{subfigure}[t]{0.33\textwidth}
				\centering
				\includegraphics[width=0.9\textwidth]{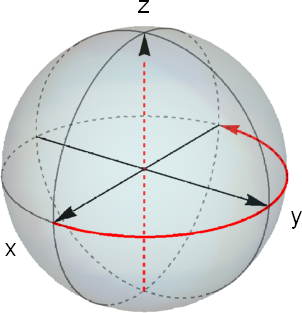}
				\caption{$\sigma_z\,\ket{+}=\ket{-}$}
			\end{subfigure}
			\caption{\label{fig:blochxyz}Each of the Pauli matrices is a $\pi$ rotation around the relevant axis. Since $\sigma_x$ turns $\ket{0}$ into $\ket{1}$ and vice versa, it is also known as the ``bit-flip operator''. Rotations around the $y$-axis do the same, except that they also introduce a phase shift. This distinction between $x$ and $y$ is of course just a matter of choice in the case of single qubits, where we can forget about the global phase. Once we entangle multiple qubits, however, the phase becomes relative and really does matter (hence the name of the $i$SWAP operation that we will meet in section~\ref{sec:two_qubit_gates}). The identity matrix, $\sigma_0$, preserves the state.}
		\end{figure}

		One might think that the familiar Pauli matrices,
		\begin{equation}\sigma_0=\Pm{1&0\\0&1},\quad\sigma_x=\Pm{0&1\\1&0},\quad\sigma_y=\Pm{0&-i\\i&0},\quad\sigma_z=\Pm{1&0\\0&-1},\end{equation}
		which form a complete basis for the $2\times2$ space of Hermitian matrices, should be able to rotate the qubit in any direction.
		In fact these matrices all cause rotations of the basis states by multiples of $\pi$, so you cannot, for example, use them to go from a $z$-eigenstate to an $x$-eigenstate.
		One popular way of visualizing this is the Bloch sphere, shown in Fig.~\ref{fig:blochxyz}, in which the ground state points up along the $z$-axis, the excited state points down, and the equator represents equal superpositions of these two with varying phases.
		
\begin{figure}
			\centering
			\begin{subfigure}[t]{0.33\textwidth}
				\centering
				\includegraphics[width=0.9\textwidth]{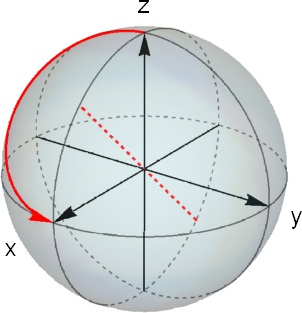}
				\caption{$H=\dfrac{1}{\sqrt{2}}\Pm{1&1\\1&-1}$}
			\end{subfigure}\begin{subfigure}[t]{0.33\textwidth}
				\centering
				\includegraphics[width=0.9\textwidth]{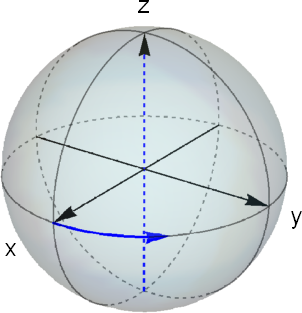}
				\caption{$T=\dfrac{1}{\sqrt{2}}\Pm{1&0\\0&e^{i\pi/4}}$}
			\end{subfigure}\begin{subfigure}[t]{0.33\textwidth}
				\centering
				\includegraphics[width=0.9\textwidth]{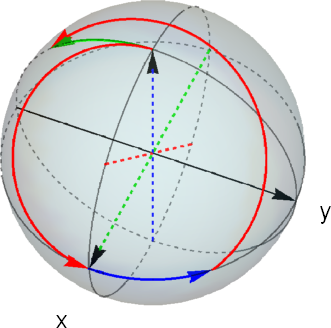}
				\caption{$X_{\pi/4}=HTH\phantom{\!\!\!\!\!\!\!\!\!\!\!\!\Pm{0\\0}}$.}
			\end{subfigure}
			\caption{\label{fig:blochhadamardt}\B{(a)} The Hadamard gate $H$ rotates by $\pi$ around a diagonal axis, allowing for the creation of a superposition. \B{(b)} The $T$ gate introduces a small rotation of only $\pi/4$ about the $z$-axis. \B{(c)} By combining the two, smaller rotations about other axes can be generated, which in turn can be combined to reach any point on the Bloch sphere.}
		\end{figure}

		To create a superposition, the Hadamard gate can be used,
		\begin{equation}H=\dfrac{1}{\sqrt{2}}\Pm{1&1\\1&-1}=\ket{+}\bra{0}+\ket{-}\bra{1},\end{equation}
		which is a rotation around a tilted axis that maps $\ket{0}$ to $\ket{+}=\sqrt{1/2}(\ket{0}+\ket{1})$ and $\ket{1}$ to $\ket{-}=\sqrt{1/2}(\ket{0}-\ket{1})$ (and vice versa), as illustrated in Fig.~\ref{fig:blochhadamardt}.
		Though superposition and entanglement are purely quantum-mechanical phenomena that cannot appear in classical systems, they still are not sufficient to give a computational advantage.
		Gottesman and Knill showed~\cite{gottesman1998heisenberg} that any combination of Pauli and Hadamard gates, together with an entangling operation, can be simulated efficiently (i.e. in polynomial time) on a probabilistic classical computer.		
		We therefore need to add one more gate, which gives a smaller rotation around the $z$-axis: the $T$ or $\pi/8$ gate,
		\begin{equation}T=\dfrac{1}{\sqrt{2}}\Pm{1&0\\0&e^{i\pi/4}}=e^{i\pi/8}\Pm{e^{-i\pi/8}&0\\0&e^{i\pi/8}}.\end{equation}
		Luckily, adding this final gate is enough to access any point on the Bloch sphere, and we do not need to include gates with infinitesimal rotations\footnote{These $T$ gates are so essential to quantum computers' advantage over classical computers, that some researchers estimate that they will make up around 90\% of the gates in a typical quantum algorithm. Since they are costly to implement, efforts are underway to bring this number down~\cite{heyfron2018efficient}.}.
		It was proven by Kitaev and Solovay~\cite{kitaev1997quantum,dawson2005solovay} that as long as we have a gate that can perform a rotation by $\pi/4$ around one of the three axes, in addition to the Hadamard, we can approximate any other gate to arbitrary precision $\epsilon$ in logarithmic time $\mathcal{O}(log(1/\epsilon))$.
		
		Though a qubit state can be prepared at any point on the Bloch sphere, such detail can only ever be known to us in a statistical sense\footnote{The state can also be \emph{inside} the Bloch sphere, which means that there is less information about its state. This happens for example when two qubits are entangled, at which point the individual states no longer exist. Once you learn something about one of the qubits, the other will purify and move towards the surface of the Bloch sphere.}.
		Every time a measurement is performed, all that we will get out is either a $\ket{0}$ or a $\ket{1}$ in the measurement basis, with the chance of either described by Born's rule.
		This rule is usually stated as follows: when given a state $\Psi$, the probability of measuring outcome $A$ is proportional to~\cite{griffiths2005introduction}
		\begin{equation}P(A)\propto|\braket{A|\Psi}|^2,\end{equation}
		i.e. the absolute square of $\Psi$'s amplitude over that state.
		But since physicists don't quite agree what it means to \emph{measure} something\footnote{Does the universe split into as many branches as there are basis states with nonzero amplitude~\cite{deutsch1998fabric}, does the wavefunction ``collapse'' to classical certainty (Copenhagen interpretation), do we become entangled with the measured object and remain forever in superposition~\cite{everett1957foundations,coleman1994sidney}, or should we just not be allowed to even ask this kind of question?}, I prefer to follow Hossenfelder~\cite{hossenfelder2020derivation} and phrase it just as the transition probability from state $\Psi_0$ to state $\Psi_1$:
		\begin{equation}P(\ket{\Psi_0}\rightarrow\ket{\Psi_1})\propto|\braket{\Psi_1|\Psi_0}|^2,\end{equation}	
		where $\Psi_1$ can then be some basis state aligned with a ``measurement outcome'' $A$ (e.g. $\ket{0}$ or $\ket{1}$), and leave it at that. This relation, at least, is free of controversy, and can even be derived from first principles~\cite{hossenfelder2020derivation}.

	\subsection{\label{sec:two_qubit_gates}Two-qubit gates}
		
		Algorithms with exponential speedups are the holy grail of quantum computing, hence the fame of Deutsch's black box problem~\cite{deutsch1985quantum}, Shor's algorithm for finding the prime factors of large integers~\cite{shor1994algorithms}, and Childs' procedure for quantum random walks on a graph~\cite{childs2003exponential} (each forms the basis of their own class of algorithms, where they are used as subroutines).
		This impressive advantage relative to classical computers has its origin in the amount of information that can be stored in a set of entangled qubits.
		To see this, consider first the four possible ways that we can combine two classical bits:
		\begin{equation}00,\quad 01,\quad 10,\quad 11.\end{equation}
		In general, though we can generate $2^n$ numbers with $n$ bits, we only need $n$ fundamental ``units'' of information~\cite{shannon1948mathematical} to describe their state.
		Two quantum bits can also be in these four configurations\footnote{This choice of basis vectors is common in quantum computation. For some platforms it may be useful to instead choose a basis where three of the states have a total spin of $1$ (the ``triplet''), and one with zero spin (the ``singlet''):
		\begin{equation}s=1:\;\left\{\begin{array}{l@{\quad}l}
			\ket{\uparrow\uparrow}		& (m=1)\\
			\dfrac{1}{\sqrt{2}}\left(\ket{\uparrow\downarrow}+\ket{\downarrow\uparrow}\right)			& (m=0)\\
			\ket{\downarrow\downarrow}	& (m=-1)\\
		\end{array}\right.,\quad s=0:\; \dfrac{1}{\sqrt{2}}\left(\ket{\uparrow\downarrow}-\ket{\downarrow\uparrow}\right)\quad(m=0),\end{equation}
		where $s$ is the amplitude of the spin, and $m$ its projection along the $z$-axis.
		In platform-agnostic notation, $\ket{0}=\ket{\uparrow}$ and $\ket{1}=\ket{\downarrow}$.},
		\begin{equation}\ket{00},\quad \ket{01},\quad \ket{10},\quad \ket{11},\end{equation}
		but what sets them apart from classical bits is that they can be in any linear combination or \emph{superposition} of them,
		\begin{equation}\label{eq:superposition}\ket{\Psi}=\alpha\ket{00}+\beta\ket{01}+\gamma\ket{10}+\delta\ket{11}.\end{equation}
		So now, to fully describe the state, two numbers are no longer enough: we cannot just use one number to express the state of the first qubit, and then a second one to indicate that of the other.
		Neither can we extract such individual information: if we were to measure the first bit and got $0$, and then the second one and got $0$ again, then all that we would know is that $\alpha\neq0$.
		Once two qubits become entangled, they give up their individual identity, and it no longer makes sense to talk about what state each of them is in.
		In turn, this loss of information is more than made up for by the exponential amount of it contained in the correlations between the two qubits.
		In general, we need $2^n-1$ complex numbers to describe an $n$-qubit state, where the $-1$ comes from a global phase that can be ignored\footnote{Of course the global phase of an entangled state \emph{exists}, but in order to estimate it with the quantum phase estimation algorithm (QPE, based on the QFT discussed in section~\ref{sec:qft_shor})~\cite{nielsen2010quantum}, you need to introduce an extra qubit that you will measure, and then you still don't know the global phase of \emph{the whole system} including that extra qubit! Hence there is no useful information in the global phase.}.
		
		To create these correlations, we need two-qubit entangling gates. The simplest of these is the controlled-NOT (CNOT for short), which we draw as follows\footnote{The $\oplus$ symbol is used in math for ``modulo plus'', which means that you add two numbers \emph{modulo} the base. So if you use base 10, then $5\oplus7=12\mod10=2$. In binary this is the same as the XOR (eXclusive OR) gate, and you can see in truth table~\eqref{eq:cnot_truth_table} that the output of the target qubit is just the XOR of the two input states.}.
		\begin{equation}
			\begin{quantikz}
				\ket{\psi_1}	& \ctrl{1}	& \qw\\
				\ket{\psi_2}	& \targ{}	& \qw
			\end{quantikz}
		\end{equation}
		The horizontal lines are the qubits, labeled $\ket{\psi_1}$ and $\ket{\psi_2}$, which are connected through the CNOT gate in such a way that the state of the first ``controls'' the second.
		If $\ket{\psi_1}=\ket{0}$, then nothing happens to $\ket{\psi_2}$, but if $\ket{\psi_1}=\ket{1}$, $\ket{\psi_2}$ will be flipped, which can be conveniently summarized in a truth table:
		\begin{equation}\label{eq:cnot_truth_table}\text{CNOT}:\quad\begin{tabular}{l@{\quad}l}
			input		& output\\\hline
			$\ket{00}$	& $\ket{00}$\\
			$\ket{01}$	& $\ket{01}$\\
			$\ket{10}$	& $\ket{11}$\\
			$\ket{11}$	& $\ket{10}$
		\end{tabular}\end{equation}
		This CNOT is universal on classical machines, so we already see that we can always simulate a classical Turing machine on a quantum one.
		This gate by itself is not enough to generate entanglement though, which has the requirement that the states cannot be separated.
		If we really do have as input one of the two-qubit basis states, e.g. $\ket{11}$, such that we end up with $\ket{10}$ after the operation, this state can still be written as
		\begin{equation}\ket{10}=\underbrace{\ket{1}}_{\psi_1}\otimes\underbrace{\ket{0}}_{\psi_2},\end{equation}
		which means that we can get information about the first qubit without learning anything about the second: we can still talk about the two states separately.
		To properly entangle the two states such that their individual identity disappears, we need to combine the CNOT with a gate that creates a superposition, such as the Hadamard introduced earlier (Fig.~\ref{fig:blochhadamardt}):
		\begin{equation}
			\begin{quantikz}
				\ket{\psi_1}=\ket{0}	& \gate{H} \slice{\Red{$\psi_1=\ket{+}$}}	& \ctrl{1}	& \qw\\
				\ket{\psi_2}=\ket{0}	& \qw 								& \targ{}		& \qw
			\end{quantikz}.
		\end{equation}
		Here we start with both qubits in the ground state $\ket{0}$, rotate the first by $\pi$ around an axis tilted by $\pi/4$ from the $z$-direction (see Fig.~\ref{fig:blochhadamardt}) to create the superposition $\ket{\psi_1}=\sqrt{1/2}\,(\ket{0}+\ket{1})=\ket{+}$ in the top line, and \emph{then} apply the CNOT.
		Since the input state right before this last operation is now a superposition of $\ket{00}$ and $\ket{10}$ (which is still separable), this finally gives us the inseparable state
		\begin{equation}\ket{\psi}=\dfrac{1}{\sqrt{2}}\,\big(\ket{01}+\ket{10}\big)\neq\ket{\psi_1}\otimes\ket{\psi_2},\end{equation}
		which cannot be rewritten as a single product of two single-qubit states.
		
		We can make other two-qubit gates using the same idea of controlling one with the other, by placing other single-qubit gates on the target timeline, e.g.
		\begin{equation}
			\begin{quantikz}
				\ket{\psi_1}	& \ctrl{1}	& \qw\\
				\ket{\psi_2}	& \gate{Y}	& \qw
			\end{quantikz}	\quad,\quad
			\begin{quantikz}
				\ket{\psi_1}	& \ctrl{1}	& \qw\\
				\ket{\psi_2}	& \gate{Z}	& \qw
			\end{quantikz}	\quad,\quad
			\begin{quantikz}
				\ket{\psi_1}	& \ctrl{1}	& \qw\\
				\ket{\psi_2}	& \gate{H}	& \qw
			\end{quantikz}	\quad\text{, etc\dots}
		\end{equation}
		The matrix form of these gates is easy to find when the first qubit is the control.
		In this case the matrix simply has the identity in the top-left, and the gate in the bottom-right:
		\begin{equation}\begin{quantikz}
			\qw	& \ctrl{1}	& \qw\\
			\qw	& \targ{}	& \qw
		\end{quantikz}\;=\;
		\begin{pNiceMatrix}[create-large-nodes,margin,last-row,nullify-dots]
			1&0&0&0\\
			0&1&0&0\\
			0&0&0&1\\
			0&0&1&0\\
			 & &\multicolumn{2}{c}{\sigma_x}\\
		\CodeAfter
		\tikz \draw[rounded corners,red,thick] ($(3-3.north west)+(-0.1,0.1)$) rectangle ($(4-4.south east)+(0.1,-0.1)$);
		\end{pNiceMatrix},\qquad
		\begin{quantikz}
			\qw	& \ctrl{1}	& \qw\\
			\qw	& \gate{Y}	& \qw
		\end{quantikz}\;=\;
		\begin{pNiceMatrix}[create-large-nodes,margin,last-row,nullify-dots]
			1&0&0&0\\
			0&1&0&0\\
			0&0&0&i\\
			0&0&-i&0\\
	 		 & &\multicolumn{2}{c}{\sigma_y}\\
		\CodeAfter
		\tikz \draw[rounded corners,red,thick] ($(3-3.north west)+(-0.2,0.1)$) rectangle ($(4-4.south east)+(0.1,-0.1)$);
		\end{pNiceMatrix}.\end{equation}
		When the second qubit is used as control, matrices can be found using truth tables like the one in eq.~\eqref{eq:cnot_truth_table}, 
		\begin{equation}\label{eq:cy21_truth_table}
		\begin{quantikz}
			\qw	& \gate{Y}	& \qw\\
			\qw	& \ctrl{-1}	& \qw
		\end{quantikz}\quad=\quad\begin{tabular}{l@{\quad}r}
			input		& output\\\hline
			$\ket{00}$	& $\ket{00}$\\
			$\ket{01}$	& $-i\ket{11}$\\
			$\ket{10}$	& $\ket{10}$\\
			$\ket{11}$	& $i\ket{01}$
		\end{tabular}\quad=\quad
		\begin{pNiceMatrix}[create-large-nodes,margin,last-row,nullify-dots]
			1&0&0&0\\
			0&0&0&-i\\
			0&0&1&0\\
			0&i&0&0\\
	 		 &\multicolumn{3}{c}{\sigma_y^T}\\
		\CodeAfter
		\tikz \draw[rounded corners,red,thick] ($(2-2.north west)+(-0.1,0.1)$) rectangle ($(2-2.south east)+(0.1,-0.1)$);
		\tikz \draw[rounded corners,red,thick] ($(2-4.north west)+(-0.2,0.1)$) rectangle ($(2-4.south east)+(0.1,-0.1)$);
		\tikz \draw[rounded corners,red,thick] ($(4-2.north west)+(-0.1,0.1)$) rectangle ($(4-2.south east)+(0.1,-0.1)$);
		\tikz \draw[rounded corners,red,thick] ($(4-4.north west)+(-0.1,0.1)$) rectangle ($(4-4.south east)+(0.1,-0.1)$);
		\end{pNiceMatrix},\end{equation}
		where in general you'll find that you need to switch the middle rows and columns and transpose the target matrix\footnote{The logic here is that you could just list the basis states in a different order, such that the first qubit becomes the second and vice versa.}.		
		Luckily we don't need all of these, since it turns out that the combination of single-qubit $H$ and $T$, and the two-qubit CNOT (known as the ``Clifford+T'' gate set) is universal~\cite{barenco1995elementary}.
		
		However, since this is a thesis about superconducting qubits, there is one more important gate that we need to discuss, that is especially relevant to transmon qubits: the $i$SWAP,
		\begin{equation}
			\begin{quantikz}
				\qw&\gate[wires=2][2cm]{i\text{SWAP}}	& \qw\\
				\qw& 									& \qw
			\end{quantikz}\quad=\quad\Pm{1&0&0&0\\0&0&-i&0\\0&-i&0&0\\0&0&0&1}.
		\end{equation}
		As will be discussed in more detail in section~\ref{sec:superconducting_qubits}, in the transmon limit of large Josephson coupling and small charging energy (large capacitance), $E_\text{J}\gg E_\text{C}$, little energy is associated with the number of charges on either side of the Josephson junction, and the qubit states are instead mostly defined by the phase fluctuations across the junction (though they are not phase states~\cite{koch2007charge}!).
		This means that (in the limit where the capacitance \emph{between} the qubits is much smaller than the capacitances of the qubits themselves) a capacitive coupling between two transmons is \emph{transverse} (orthogonal) to the quantization axis of the system~\cite{krantz2019quantum}, and the off-diagonal interaction causes an exchange of energy.
		The simplest capacitive interaction occurs when the two qubits have their resonance frequencies tuned to the same energy, such that there is no energy difference between $\ket{01}$ and $\ket{10}$, and the two-qubit spectrum hybridizes at a so-called avoided crossing~\cite{majer2007coupling}.
		At this point an excitation in one qubit can be transferred to the other without having to either release energy to, or absorb energy from the environment, and a lossless ``swap'' can be performed.
		
		The matrix form of this interaction can be derived by first recognizing that there are two terms: a lowering operator on the excited qubit times a raising operator on the other, for each the two relevant input states $\ket{01}$ and $\ket{10}$ that can be swapped~\cite{jaynes1963comparison},
		\begin{equation}H_\text{transv. int.}=g\,\Big(\!\!\!\!\!\!\underbrace{\sigma_+^{(1)}}_{\Pm{0&0\\1&0}\phantom{mn}}\!\!\!\!\!\!\!\otimes\!\!\!\!\!\!\!\underbrace{\sigma_-^{(2)}}_{\phantom{mn}\Pm{0&1\\0&0}}\!\!\!\!\!\!+\;\sigma_-^{(1)}\otimes\sigma_+^{(2)}\Big)=
		g\begin{pNiceMatrix}[create-large-nodes,first-row,nullify-dots]
			\\
			0&0&0&0\\
			0&0&1&0\\
			0&1&0&0\\
			0&0&0&0
		\CodeAfter
			\tikz \draw[rounded corners,red,thick] ($(2-3.north west)+(-0.1,0.1)$) rectangle ($(2-3.south east)+(0.1,-0.1)$);
			\tikz \draw[red] ($(2-3.north east)+(0.1,0.1)$) -- +(0.2,0.5) node[anchor=south]{$\ket{10}\rightarrow\ket{01}$};
			\tikz \draw[rounded corners,blue!80!black,thick] ($(3-2.north west)+(-0.1,0.1)$) rectangle ($(3-2.south east)+(0.1,-0.1)$);
			\tikz \draw[blue!80!black] ($(3-2.south west)+(-0.1,-0.1)$) -- +(-0.2,-0.6) node[anchor=north]{$\ket{01}\rightarrow\ket{10}$};
		\end{pNiceMatrix}.\end{equation}
		This then gives rise to the unitary evolution~\cite{blais2004cavity}
		\begin{equation}U_\text{transv. int.}=e^{iHt}=\Pm{1&0&0&0\\0&\cos(gt)&-i\sin(gt)&0\\0&-i\sin(gt)&\cos(gt)&0\\0&0&0&1},\end{equation}
		which causes the complex amplitudes associated with $\ket{01}$ and $\ket{10}$ to oscillate out of phase with a period\footnote{To see why we get $1$s in the top-left and bottom-right corners of this $4\times4$ unitary time evolution matrix, while the Hamiltonian matrix itself only has nonzero entries in the central $2\times2$ block, expand the exponential $\exp(iHt)$:
		\begin{equation*}\exp\Pm{\ddots}=\Sum_{n=0}^\infty\dfrac{1}{n!}\,\Pm{\ddots}^n=I+\Pm{\ddots}+\dfrac{1}{2}\Pm{\ddots}^2+\dots,\end{equation*}
		where the identity gives us those corner $1$s, while the higher-order terms will only change the central block. We then get sines and cosines by summing the odd (off-diagonal) and even (diagonal) matrix products, respectively, keeping the diagonal real since it only has even factors of the imaginary $i$.} of $\tau=2\pi/g$.
		One might expect a swap to occur after half a period, but that would only flip the signs of each component.
		Instead, when the interaction is turned on for just a \emph{quarter} oscillation, we get
		\begin{equation}U_\text{transv. int.}\left(t=\dfrac{\pi}{2g}\right)=\Pm{1&0&0&0\\0&0&-i&0\\0&-i&0&0\\0&0&0&1},\end{equation}
		revealing the $i$ in the name ``$i$SWAP''. Coupling the two qubits for an even shorter amount of time of only an eighth of a period, will perform half a swap or $\sqrt{i\text{SWAP}}$, turning either of the swappable two-qubit basis states into entangled pairs,
		\begin{equation}\sqrt{i\text{SWAP}}\ket{01}=\dfrac{1}{\sqrt{2}}\,\big(\ket{01}-i\ket{10}\big),\quad \sqrt{i\text{SWAP}}\ket{10}=\dfrac{1}{\sqrt{2}}\,\big(-i\ket{01}+\ket{10}\big).\end{equation}
		This gate is thus a very efficient ``native'' or ``primitive'' means of creating large multi-qubit entangled states in circuits of tunable-frequency transmons, able to generate $2^n$-qubit entanglement after a circuit depth\footnote{In qubit-speak, the ``volume'' is the square of the number of qubits that can effectively be used on your physical chip (or $2$ to the power of that number if you work at IonQ~\cite{ionq2020q2b20}), the ``size'' of a logical circuit is the total number of gates, and the ``depth'' is the maximum number of sequential operations on a single qubit in terms of the gates native to the physical system.} $n$.
		
		Though such entanglement will allow us to access an exponentially large Hilbert space, unfortunately this does not mean that we can also get exponentially much information out of the computation.
		The Holevo bound simply states that since we will always have to perform a measurement at the end, all we can ever expect is an output of as many classical bits as we use qubits~\cite{holevo1973bounds}.
		This means that to have a speedup, interference of all these exponentially many quantum states needs to be orchestrated between the initialization and readout steps~\cite{deutsch1985quantum}.
		It is this design of interference patterns, in a way that depends on the specifics of the problem at hand, that we are concerned with when developing quantum algorithms.
		
	\section{Algorithms and computational complexity}

		\Quote{``Shut up and calculate!''}{David Mermin}
		\Quote{``OK, but what?''}{Scott Aaronson}
		
		Consciously or not, we use algorithms every day, and have in fact been using them for much longer than we have had digital computers.
		An algorithm is any set of logical operations that solves a problem, and can be as mundane as the following thought process that you may go through when choosing a pack of toilet paper in the supermarket:
		
		\noindent\begin{minipage}{\textwidth}\begin{algorithm}[gobble=8,tabsize=4,caption={Choose a product in the supermarket.},label={alg:tp},mathescape]
		PriceWeight 	$\leftarrow$ -2
		QualityWeight 	$\leftarrow$ 1
		foreach brand in tpbrands:
			brand.score $\leftarrow$ PriceWeight * brand.Price
			brand.score += QualityWeight * brand.PerceivedQuality
		tpbrands 	$\leftarrow$ sort(tpbrands,score)
		buy(tpbrands[0])
		\end{algorithm}\end{minipage}
		
		\noindent More complex routines like baking a cake or writing a thesis can in principle also be reduced to similar sets of operations, as prescribed by the Church-Turing thesis discussed earlier~\cite{turing1950computing}.
		
		One important property of an algorithm is its complexity in terms of time, energy, memory and other resources that it requires.
		These are in turn linked to the complexity of the problem that it is designed to solve, and are bound by the physical characteristics of the hardware it is running on.
		The extended Church-Turing thesis states that any efficiently computable function is also efficiently computable on a universal Turing machine, which means that there is at most a polynomial overhead when an algorithm is compiled to a form that a Turing machine can interpret.
		The following sections will pick apart how the existence of quantum algorithms have shattered this bedrock of classical computer science.

		\subsection{Types of quantum algorithms}
		
			There are different ways that we can group quantum algorithms that have an advantage relative to classical ones.
			One popular way is to sort them into three broad groups by the type of problem that they solve~\cite{babbush2020q2b20}.
			\begin{itemize}[leftmargin=*]
				\item[] \B{Optimization:} This can generally be rephrased as minimizing a cost function, such as the one shown in algorithm~\ref{alg:tp}.
				\begin{itemize}
					\item[] \emph{Problems:} finding the ground state of an Ising Hamiltonian, compiling an investment portfolio or picking locations for vaccination centres.
					\item[] \emph{Algorithms:} Quantum annealing (QA), quantum approximate optimization algorithm (QAOA), and combinations thereof~\cite{brady2020optimal}.
				\end{itemize}
				\item[] \B{Quantum machine learning (QML)}: Reducible to sampling probability distributions, as problems can be reformulated to finding the probability that a function accepts (i.e. returns $1$ or \texttt{TRUE}) over many degrees of freedom.
				\begin{itemize}
					\item[] \emph{Problems:} analyzing data from quantum circuits, possibly using quantum circuits to analyze classical data (image processing, unsupervised learning). Not yet clear what kind of problems can have quantum speedups~\cite{babbush2020q2b20}.
					\item[] \emph{Algorithms:} quantum neural networks, Gaussian boson sampling with gradient descent~\cite{banchi2020training}, quantum circuit Born machines (QCBM)~\cite{coyle2020born}, variational quantum algorithms (VQA)~\cite{cerezo2020variational}.
				\end{itemize}				
				\item[] \B{Simulation:} Quantum systems are intrinsically hard to simulate with classical machines~\cite{feynman1982simulating,laughlin2000cover}; calculating the energy spectrum of a single large atom takes more computing power than all supercomputers in the world combined can deliver in any reasonable time, while all it takes on a quantum computer is a few hundred qubits~\cite{lloyd1996universal,kassal2008polynomial,dowling2013schrodinger}.
				\begin{itemize}
					\item[] \emph{Problems:} quantum chemistry, condensed matter physics, calculating the ground state or the dynamics of any quantum system.
					\item[] \emph{Algorithms:} Hamiltonian simulation~\cite{lloyd1996universal,georgescu2014quantum}, variational quantum eigensolver (VQE)~\cite{peruzzo2014variational}.
				\end{itemize}
			\end{itemize}
			At the time of this writing, the first category may seem to be off to a head start, as commercial superconducting quantum annealing chips are running optimization problems with thousands of qubits~\cite{teplukhin2020electronic}.
			This is in part because it is perceived to have direct applications to lucrative disciplines~\cite{economist2020quantum}, leading analysts to predict that it will continue to be an important part of the end-user market~\cite{sorensen2020q2b20}.
			It should be noted however, that known quantum optimization algorithms give at most a polynomial speedup relative to classical alternatives, with practical examples typically solving problems only quadratically faster.

			In these discussions, we generally only care about the \emph{asymptotic} behavior of an algorithm; the way that the demand on some resource scales with the input size of the problem or accuracy of the output without caring about constant prefactors.
			We can describe the complexity of an algorithm with at least one of the following:
			\begin{itemize}
				\item $\B{\Omega}$: the lower bound, best-case complexity. See e.g. section~\ref{sec:deutsch_problem}.
				\item $\B{\mathcal{O}}$ (``big-oh''): the upper bound, worst-case complexity. See sections~\ref{sec:qft_shor} and~\ref{sec:qa_qaoa}.
				\item $\B{\Theta}$: both a lower and upper bound: the best and worst case scenarios scale in the same way. See section~\ref{sec:grover}.
			\end{itemize}
			
			Figuring out the \emph{type} of speedup that can be achieved is essential, since it is the way that the computation time scales with the size of the problem that give quantum computers their edge.
			Performing a single two-qubit operation takes on the order of $10^3$ to $10^6$ times longer than a simple switch of a transistor, even without taking into account the overhead introduced by error correction\footnote{It is estimated that error correction requires anywhere from a 13 to a $10^4$ physical to logical qubit ratio, depending on the two-qubit gate fidelity~\cite{fowler2012surface}. Ion traps have the advantage of higher fidelity, which also means that they can run larger circuits before even needing error correction at all, but a single gate operation takes around a thousand times longer than on a superconducting system~\cite{linke2017experimental,gambetta2017building}. By the way, the existence of error correction~\cite{feynman1985quantum} is guaranteed by the linearity of quantum mechanics, preventing errors from cascading into noise~\cite{vazirani1998power,aaronson2013quantum}.}.
			Fully fault-tolerant two-qubit operations may even be up to $10^7$ times slower than classical two-bit operations, which means that a quadratic speedup with $n_\text{Q}=\sqrt{n_\text{C}}$ operations and a fault-tolerant gate time of $\tau_\text{Q}=10^{-2}\,\si{\second}$ still gives cross-over times on the order of
			\begin{equation}t=\tau_\text{Q}\,\dfrac{\tau_\text{Q}}{\tau_\text{C}}=10^{-2}\,\si{\second}\times 10^7=10^{5}\,\si{\second}\approx\text{1 day}.\end{equation}
			More comprehensive estimates that also take into account classical parallelization and typical circuit implementations range from hundreds of days ($10^3$ classical cores) to thousands of years ($10^6$ cores)~\cite{babbush2020focus}.
			Improving the order of the polynomial speedup to cubic or quartic can bring these cross-over times down to mere hours or minutes.
			Exponential speedups such as those seen for finding the prime factors of large numbers~\cite{shor1994algorithms} or traversing graphs~\cite{childs2003exponential} would give a more fundamental quantum advantage.
			
			It may therefore be more productive to have a bottom-up classification of quantum algorithms, looking first at the kinds of asymptotic speedups that we know of for existing problems, focusing on those that are better than quadratic, and then finding real-world examples that they can be applied to.
			In this spirit, the following sections will briefly review the black-box problems of Deutsch and Simon, the period-finding Quantum Fourier Transform (QFT), the unsorted database search by Grover, and the quantum annealing and quantum approximate optimization algorithms.
		
		\subsection{\label{sec:deutsch_problem}Deutsch's problem}
			
			Deutsch's problem deals with what we call an ``oracle'': some black-box function that we take to exist, and that we want to find information about.
			Though this specific problem was clearly contrived for the express purpose of demonstrating a quantum advantage, oracles are actually found in many real-life applications.
			Historically, Deutsch's problem as it was posed in 1985~\cite{deutsch1985quantum} was the first to show a concrete example of a situation (besides simulation) in which a quantum computer would give a speedup.
			Given a function $f(x)$ with $x,\,f(x)\in\{0,1\}$, the problem is to find out whether $f(0)\oplus f(1)=0$ or $f(0)\oplus f(1)=1$.
			In the first case, the function is said to be ``constant'' since it gives the same output no matter the input, while in the second the function is ``balanced''.
			To illustrate, consider the four possible functions.
			\begin{equation}\label{eq:constant_balanced_table}\begin{tabular}{c@{\quad}c@{\quad}c}
				$f(0)$	& $f(1)$	& $f(0)\oplus f(1)$\\\hline
				0		& 0			& 0\\
				\gc{0}	& \gc{1}	& \gc{1}\\
				\gc{1}	& \gc{0}	& \gc{1}\\
				1		& 1			& 0
			\end{tabular}\quad\begin{tabular}{l}
				\\
				constant\\
				balanced\\
				balanced\\
				constant
			\end{tabular}\end{equation}
			Any classical algorithm deciding whether the function is balanced will need to query $f$ twice: once to extract $f(0)$, and then again for $f(1)$.
			Deutsch imagined a Turing-like machine where a quantum program $\pi(f,a,b)$ applies a function $f$ to slot $a$ and stores the result in slot $b$, thus performing the calculation\footnote{The storage in slot $b$ needs to be additive to ensure unitarity.}
			\begin{equation}\ket{\pi(f,2,3),i,j}\;\mapsto\;\ket{\pi(f,2,3),i,j\oplus f(i)}.\end{equation}
			If a superposition of $\ket{0}$ and $\ket{1}$ is placed in register $i$ while register $j$ is initialized in the ground state, then this program evaluates to
			\begin{equation}\dfrac{1}{\sqrt{2}}\,\Sum_{i\in\{0,1\}}\ket{\pi(f,2,3),i,0}\;\mapsto\;\dfrac{1}{\sqrt{2}}\,\Sum_{i\in\{0,1\}}\ket{\pi(f,2,3),i,f(i)}.\end{equation}
			We can then perform a simultaneous measurement on the last two qubits in the four-eigenstate basis
			\begin{equation}\left\{\begin{array}{l@{\;}c@{\;}l}
				\ket{\text{zero}}	& \equiv	& \dfrac{1}{2}\Big(\ket{00}-\ket{01}+\ket{10}-\ket{11}\Big),\\[1.4ex]
				\ket{\text{one}}	& \equiv	& \dfrac{1}{2}\Big(\ket{00}-\ket{01}-\ket{10}+\ket{11}\Big),\\[1.4ex]
				\ket{\text{fail}}	& \equiv	& \dfrac{1}{2}\Big(\ket{00}+\ket{01}+\ket{10}+\ket{11}\Big),\\[1.4ex]
				\ket{\text{error}}	& \equiv	& \dfrac{1}{2}\Big(\ket{00}+\ket{01}-\ket{10}-\ket{11}\Big),
			\end{array}\right.\end{equation}
			where the amplitudes for the outcomes of ``constant'' functions represented by the first and last rows in eq.~\eqref{eq:constant_balanced_table} add up in the eigenstates $\ket{\text{zero}}$ and $\ket{\text{fail}}$, while they cancel out in $\ket{\text{one}}$ and $\ket{\text{error}}$.
			Similarly, the ``balanced`` outcomes add up in $\ket{\text{one}}$ and $\ket{\text{fail}}$, while they cancel out in $\ket{\text{zero}}$ and $\ket{\text{error}}$.
			Constant functions thus have a 50\% chance of outputting $\ket{\text{zero}}$, while balanced functions have a 50\% chance of outputting $\ket{\text{one}}$, and both also have a 50\% chance of $\ket{\text{fail}}$, while $\ket{\text{error}}$ should  never be observed.
			This means that if we are allowed only a single query of $f$, Deutsch's algorithm has a 50\% chance of answering the question of whether it is constant, while a classical algorithm would have 0\% chance of solving it, giving it a complexity\footnote{In this case, no asymptote can be calculated, since the problem only exists for a single $n$.} $\Omega(1/2)$.
			
			The algorithm can be improved to have a 100\% chance of solving the problem by placing also the target qubit $j$ in superposition~\cite{cleve1998quantum}.
			This is most easily visualized using our familiar quantum circuit notation~\cite{deutsch1989quantum,nielsen2010quantum}:
			\begin{equation}\label{eq:cleve_algorithm}
				\left.\begin{quantikz}
					\ket{0}\;	& \gate{H} 	& \gate[wires=2][2cm]{U_f}
											\gateinput{$i$}\gateoutput{$i$}	& \gate{H}	&  \;\ket{f(0)\oplus f(1)}\qw\\
					\ket{1}\;	& \gate{H}	& \gateinput{$j$}\gateoutput{$j\oplus f(i)$}					& \qw		& \;\dfrac{1}{\sqrt{2}}\Big(\ket{0}-\ket{1}\Big)\qw
				\end{quantikz}\right\}\times\text{global phase.}
			\end{equation}
			Here the unitary $U_f$ is the circuit version of the program $\pi$ described above, and performs a rotation in the two-qubit Hilbert space.
			The perhaps counter-intuitive result that the solution is output on the top channel $i$, while the function $f(i)$ is added to the bottom qubit, is due to a phenomenon called ``phase kickback'', where the global phase is effectively \emph{kicked back} to the control qubit ($i$ in this case).
			This is seen more clearly when writing down the $U_f$ operation explicitly:
			\begin{equation}\begin{array}{r@{\;}c@{\;}l}
				\dfrac{U_f}{2}\Big[\big(\ket{0}+\ket{1}\big)\otimes\big(\ket{0}-\ket{1}\big)\Big]	& =	& \big(\ket{0}+\ket{1}\big)\otimes\dfrac{e^{i\pi (f(0)+f(1))}}{2}\big(\ket{0}-\ket{1}\big)\\\\
				& =	& \dfrac{e^{i\pi (f(0)+f(1))}}{2}\big(\ket{0}+\ket{1}\big)\otimes\big(\ket{0}-\ket{1}\big)\\\\
				& =	& \left\{\begin{array}{l@{\;}l}
					(i)&\big(\ket{0}+\ket{1}\big)\otimes\big(\ket{0}-\ket{1}\big)/2,\\\\
					(ii)&\big(\ket{0}-\ket{1}\big)\otimes\big(\ket{0}-\ket{1}\big)/2,\end{array}\right.
			\end{array}\end{equation}
			where options $(i)$ and $(ii)$ are:
			\begin{equation}\begin{array}{l@{\quad}l@{\quad}l}
				(i)		& f(0)\oplus f(1) = 0	& \text{(constant)},\\\\
				(ii)	& f(0)\oplus f(1) = 1	& \text{(balanced)}.
			\end{array}\end{equation}
			The final Hadamard on the first qubit ensures that we can measure the output in the $z$-basis.
			
			While the above version of the problem with only a single control qubit is important for having established that quantum computers can be faster in principle, the $n$-qubit version proves that even exponential speedups can be achieved~\cite{deutsch1992rapid,bernstein1993quantum,cleve1998quantum}.
			In this more general problem, the input for function $f$ is no longer just $0$ or $1$, but an $n$-bit string of $0$s and $1$s, and $f$ is said to be balanced iff the single-bit output is $0$ for exactly half the inputs.
			Since there are $2^n$ possible bit strings, this problem would require $2^n/2=2^{n-1}$ queries of the oracle in a classical setting.
			A quantum computer could solve this with a single query and by measuring only $n$ qubits, using a modified version of eq.~\eqref{eq:cleve_algorithm}~\cite{cleve1998quantum,norlen2020quantum,nielsen2010quantum}:
			\begin{equation}\label{eq:deutsch_jozsa}
				\begin{quantikz}
					\ket{0}^{\otimes n}\;	& \gate{H}\qwbundle[alternate]{}
					& \gate[wires=2][2cm]{U_f}\qwbundle[alternate]{}
											\gateinput{$i$}\gateoutput{$i$}	& \gate{H}\qwbundle[alternate]{}	&  \;2^{-n}\!\!\!\!\!\!\Sum_{x,y\in\{0,1\}^n}\!\!\!\!\!e^{i\pi x\cdot z+f(x)}\ket{y}\qwbundle[alternate]{}\\
					\ket{1}\;				& \gate{H}	
					& \gateinput{$j$}\gateoutput{$j\oplus f(i)$}					& \qw		& \;\dfrac{1}{\sqrt{2}}\big(\ket{0}-\ket{1}\big)\qw
				\end{quantikz}
			\end{equation}
			In this larger circuit, if the function is balanced, then there is at least one bit in the input string for which flipping it changes the outcome,
			\begin{equation}\exists m\in\{0,\dots,n-1\}:\quad f(\ket{\dots 0_m\dots})\oplus f(\ket{\dots 1_m\dots})=1.\end{equation}
			For example, if all odd inputs $x$ give $f(x)=0$ while all even ones give $f(x)=1$, such that the function only depends on the last qubit in the register ($m=n-1$), then that qubit will get a phase kickback.
			In general, any balanced function will necessarily have at least one (and possibly all) of the control qubits flip to $\ket{1}$ in eq.~\eqref{eq:deutsch_jozsa}, and finding a single nonzero entry is sufficient to determine that the function is balanced, giving us $\Omega(1)$ and $\mathcal{O}(n)$.
			In contrast, there will be no phase kickback at all to any of the $n$ control qubits if the function is constant, and the output will be $\ket{0}^{\otimes n}$.
			It takes at most one query of the oracle and $n$ measurements to analyze the function, an exponential speedup relative to the $2^n$ queries and measurements on a classical machine.
			
			This algorithm is an example of ``quantum parallelism'', and Deutsch famously remarked that the speedup due to it implies that Everett's multiverse interpretation of quantum mechanics\footnote{Everett never called it that, but used the name ``relative state interpretation'', insisting that the observed outcome of a measurement necessarily depends on the final state of the observer, just like the velocity of an object can only be measured relative to a rest frame.}~\cite{everett1957foundations} must be correct, for else~\cite{deutsch1985quantum,deutsch1998fabric}, \emph{where was it computed?}
			The term has unfortunately also given rise to one of quantum computer scientists' pet peeves in popular communications: that quantum computers somehow allow for ``massive parallel computation'' of any classically parallelizable circuit, simply performing all calculations in parallel in the exponentially large Hilbert space.
			Parallel operations can indeed be performed on all basis states in a superposition, but the idea of unbridled parallelism falls apart at the actual measurement of the output: while the information stored inside the register can in the case of maximum entanglement reach $2^n$ bits, only $n$ bits of classical information can ever be read out.
			This limit is known as the Holevo bound~\cite{holevo1973bounds}, and suggests a requirement for achieving a speedup: interference between the possible outcomes must be orchestrated, such that the information about the quadratic or exponential number of parallel computations can be read out from only a linear number of qubits.
			
			Deutsch's efforts inspired Simon to construct a similar problem~\cite{simon1994power,simon1997power}, where instead of a single bit of information (whether a function is balanced or not), an entire bit string can be extracted by using $n$ target qubits on the bottom line of eq.~\eqref{eq:deutsch_jozsa}.
			It was upon seeing this result that Shor realized a method for finding the prime factors of integers~\cite{shor1994algorithms}, which we will discuss next.
			
		\subsection{\label{sec:qft_shor}The Quantum Fourier Transform and Shor's algorithm}
			
			Any periodic function can be expressed as a (possibly infinite) series of sines and cosines, written most compactly using Euler's identity:
			\begin{equation}f(x)=\Sum_{n=-\infty}^\infty c_n\,e^{i n \pi x / L},\qquad 
			c_n=\dfrac{1}{2L}\Int_{-L}^L f(x)\,e^{-i n \pi x / L}.\end{equation}
			As an extension of this idea to non-periodic functions by letting $L\rightarrow \infty$, the Fourier transform takes as input such a function $f(x)$ and outputs its conjugate $\hat{f}(w)$, a continuous equivalent of the coefficients\footnote{Perhaps you prefer to normalize these transforms asymmetrically by placing a factor $1/2\pi$ only in front of the inverse Fourier transform, but then I suppose you also follow the mostly minuses convention of the metric tensor.} $c_n$~\cite{kreyszig2006advanced}:
			\begin{equation}\label{eq:fourier_transform}\begin{array}{r@{\;}c@{\;}c@{\;}c@{\;}l}
				\mathscr{F}\big(f(x)\big)				& =	& \hat{f}(w)	& =	& \dfrac{1}{\sqrt{2\pi}}\Int_{-\infty}^\infty f(x)\,e^{i w x}\,dx,\\\\
				\mathscr{F}^{-1}\big(\hat{f}(w)\big)	& =	& f(x)			& =	& \dfrac{1}{\sqrt{2\pi}}\Int_{-\infty}^\infty \hat{f}(w)\,e^{-i w x}\,dw.
			\end{array}\end{equation}
			The relevance of the Fourier transform to physics is perhaps clearest from the concept of conjugate variables, the transformation between which in the form of series expansions not only provides access to mathematical techniques~\cite{adams2010calculus,kreyszig2006advanced}, but also gives more direct insights from quantum mechanical uncertainty and commutation relations~\cite{griffiths2008introduction}.
			The importance to classical physics was stressed by Noether~\cite{noether1918invariante}, who pointed out that the symmetry of a system under differentiation to any variable implies a conservation of its conjugate (e.g. time invariance implies energy conservation, translational invariance implies momentum conservation etc).
			The transform also plays a role in computer science, where its application to signal processing is ubiquitous, and the bit-wise product representation that we will see below is a prime example of the divide and conquer strategy to solving recursive problems.
			Finding a faster way of performing this operation could thus have consequences for a broad range of studies and applications.
			
			Analytical methods to perform the Fourier transform are often unavailable, and we have to resort to numerics.
			In this case we replace the infinite integrals in eq.~\eqref{eq:fourier_transform} by finite sums,
			\begin{equation}\label{eq:dft}\begin{array}{r@{\;}c@{\;}l@{\qquad}r@{\;}c@{\;}l}
				\mathscr{F}(\B{f})				& =	& \B{\hat{f}},	&   
				\hat{f}_k	& =	& \dfrac{1}{\sqrt{N}}\Sum_{j=0}^{N-1}f_j\,e^{i k x_j},\\\\
				\mathscr{F}^{-1}(\B{\hat{f}})	& =	& \B{f},		&
				f_j			& =	& \dfrac{1}{\sqrt{N}}\Sum_{k=0}^{N-1}f_k\,e^{-i k x_j},
			\end{array}\end{equation}
			where $\B{f}$ and $\B{\hat{f}}$ are vectors of size $N$, and $x_j$ are $N$ points at which the function $f$ is evaluated (often equally spaced at $x_j=2\pi j/N$).
			Calculating all $N$ components of the transformed vector thus amounts to computing the multiplication by an $N\times N$ matrix, which naively takes $\mathcal{O}(N^2)$ logical operations, or $\mathcal{O}(N\log N)$ when the matrix is recursively divided (divide and conquer).
			
			Now imagine that we store the input vector $\B{f}$ of size $N$ in a $\log_2 N=n$ qubit string using the exponential scaling of the storage capacity as discussed around eq.~\eqref{eq:superposition}.
			When the $j$th component of the vector $\B{\hat{f}}$ in eq.~\eqref{eq:dft} is calculated, we don't store it in the $j$th register as we would classically, but assign it instead as an amplitude to the $j$th term in the $n$-qubit superposition.
			For example, if we would have $2$ qubits, we could store a four-component input vector $\B{f}$ like so:
			\begin{equation}\label{eq:two_qubit_basis}\B{f}=f_0\underbrace{\ket{00}}_{j=0}+f_1\underbrace{\ket{01}}_{j=1}+f_2\underbrace{\ket{10}}_{j=2}+f_3\underbrace{\ket{11}}_{j=3}.\end{equation}
			We can then write the definitions of the discrete Fourier transform in eq.~\eqref{eq:dft} in the appropriate braket notation~\cite{shor1994algorithms},
			\begin{equation}\B{\hat{f}}=\Sum_{k=0}^{N-1}\hat{f}_k\ket{k}=\dfrac{1}{\sqrt{N}}\Sum_{k=0}^{N-1}\Sum_{j=0}^{N-1}f_j\,e^{i k x_j}\ket{k},\end{equation}
			where each of the basis states $\ket{j}=\ket{j_0\cdots j_{N-1}}$, such as the four in eq.~\eqref{eq:two_qubit_basis}, contributes
			\begin{equation}\mathscr{F}\big(\ket{j}\big)=\dfrac{1}{\sqrt{N}}\Sum_{k=0}^{N-1}\,e^{i k x_j}\ket{k}=\dfrac{1}{\sqrt{N}}\Sum_{k=0}^{N-1}\,e^{2\pi i j k/N}\ket{k}.\end{equation}
			Here we assumed in the last step that the function is restricted to\footnote{That would be $[0,2\pi[$ in French notation.} $[0,2\pi)$ and that $x_j$ represents it optimally.
			Taking again the 2-qubit basis of eq.~\eqref{eq:two_qubit_basis} as example, this would result in
			\begin{equation}\label{eq:f_example_2_qubits}\begin{array}{r@{\;}c@{\;}l}
				\mathscr{F}\big(\ket{j_0j_1}\big)	& =	& \dfrac{1}{\sqrt{4}}\Sum_{k=0}^{3}\,e^{2\pi i [j_0j_1] k/4}\ket{k}\\\\
				& =	& \dfrac{1}{2}\Big(\ket{00}+e^{\pi i [j_0j_1]/2}\ket{01}+e^{\pi i [j_0j_1]}\ket{10}+e^{\pi i [j_0j_1] 3/2}\ket{11}\Big)\\\\
				& =	& \dfrac{1}{2}\Big(\ket{0}+e^{i\pi [\cancel{j_1}j_2]}\ket{1}\Big)\otimes\Big(\ket{0}+e^{i\pi [j_0j_1]/2}\ket{1}\Big),
			\end{array}\end{equation}
			where $[j_0j_1]$ is the binary number represented by the state $\ket{j}$, and we can drop any full rotations by $2\pi$ (hence the $[\cancel{j_1}j_2]=[j_1j_2]\bmod2$).
			In general, for an $n$-qubit register we find~\cite{griffiths1996semiclassical},
			\begin{equation}\label{eq:f_general_n_qubits}\hat{f}_k=2^{-n/2}\big(\ket{0}+e^{2\pi i 0.j_{n-1}}\ket{1}\big)\otimes\cdots\otimes\big(\ket{0}+e^{2\pi i 0.j_0\cdots j_{n-1}}\ket{1}\big),\end{equation}
			where $0.j_i\cdots j_n=(j_i\cdots j_{n-1})/2^n$.
			
			This expansion into product form is not unique to the qubit register, and can also be done with exponentially more classical bits, where the components of the vector $\B{f}$ are assigned to the $N=2^n$ bits themselves rather than the $2^n$ basis states:
			\begin{equation}\begin{array}{l@{\qquad}r@{\;}c@{\;}l@{\;}l@{\;}l@{\;}l@{\;}c@{\;}l}
				\text{Qubits:}	& \B{f}	& =	& 
			f_0\underbrace{\ket{00}}_{j=0}	& +f_1\underbrace{\ket{01}}_{j=1}	& +f_2\underbrace{\ket{10}}_{j=2}	& +f_3\underbrace{\ket{11}}_{j=3},& \\\\
				\text{Bits:}	& \B{f}	& =	& 
				f_0\underbrace{[0001]}_{j=0}	& +f_1\underbrace{[0010]}_{j=1}	& +f_2\underbrace{[0100]}_{j=2}	& +f_3\underbrace{[1000]}_{j=3}	& =[f_3f_2f_1f_0].
			\end{array}\end{equation}
			The exponential factors that we saw in eqs.~\eqref{eq:f_example_2_qubits} and~\eqref{eq:f_general_n_qubits} are now associated with whether we have the odd or even component of successive divisions of the problem,
			\begin{equation}\begin{array}{r@{\;}c@{\;}l}
				\mathscr{F}\big([j]\big)	& =	& \dfrac{1}{\sqrt{4}}\Sum_{k=0}^{3}\,e^{2\pi i j k/4}\ket{k}\\
				& =	& \dfrac{1}{2}\Big(\underbrace{[0001]}_{j=0}+
				\;\Blue{\underbrace{e^{\pi i j/2}}_{\Blue{j\text{ odd}}}}\,
				\underbrace{[0010]}_{j=1}+
				\Red{\overbrace{e^{\pi i j}}^{\Red{\floor{j/2}\text{ odd}}}}\!\!\!\!\!\underbrace{[0100]}_{j=2}+\;\Blue{\underbrace{e^{\pi i j/2}}_{\Blue{j\text{ odd}}}}\!\!\Red{\overbrace{e^{\pi i j}}^{\Red{\floor{j/2}\text{ odd}}}}\!\!\!\!\!\underbrace{[1000]}_{j=3}\Big).
			\end{array}\end{equation}
			This gives the idea of the divide and conquer approach to calculating the fast Fourier transform (FFT), where the problem is recursively divided into $\log_2 N=n$ sub-problems.
			Even though this is an exponential speedup in the calculation of the prefactors, the classical form still requires the evaluation of all $N=2^n$ terms, leading to a total complexity of $\mathcal{O}(n2^n)=\mathcal{O}(N\log N)$.
			
			The circuit for the quantum Fourier transform can be read almost directly from the product representation in eq.~\eqref{eq:f_general_n_qubits}.
			Each of the qubits is on the equator of the Bloch sphere, rotated by some angle:
			\begin{equation}\dfrac{1}{\sqrt{2}}\Big(\ket{0}+e^{i\Bred{\phi}}\ket{1}\Big)\quad=\quad\raisebox{-10.4ex}{\includegraphics[width=0.27\textwidth]{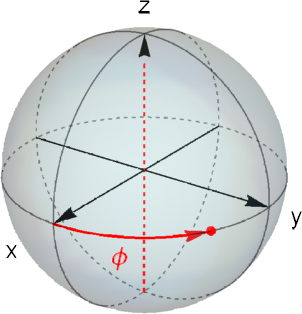}}.\end{equation}
			This angle, $2\pi i j_l/2^m$, is a rotation $2^{-m}$ times around the $z$-axis of the Bloch sphere, controlled by qubit $\ket{j_l}$:
			\begin{equation}\dfrac{1}{\sqrt{2}}\Big(\ket{0}+e^{2\pi i\,\scriptstyle{\overbrace{0.0\cdots}^{\displaystyle m\text{ zeros}} }j_l}\ket{1}\Big)\otimes\ket{j_l}\;=\;\begin{quantikz}[align equals at=1.35,row sep={1cm,between origins},column sep={1.3cm,between origins}]
				\sqrt{1/2}(\ket{0}+\ket{1})\;	& \qw	& \gate{\!\sqrt[2^m]{Z}\,}	& \qw\\
				\ket{j_l}\;	& \qw	&  \ctrl{-1}			& \qw
			\end{quantikz},\end{equation}
			where $\sqrt[2^m]{Z}=R_m$.
			This suggests a simple way of constructing the final state in eq.~\eqref{eq:f_general_n_qubits}, shown in pseudocode in algorithm~\ref{alg:qft}.
			
			\noindent\begin{minipage}{\textwidth}\begin{algorithm}[gobble=12,tabsize=4,caption={Quantum Fourier Transform (QFT).},label={alg:qft},mathescape]
			foreach Qubit in Register:
				rotate Qubit basis states to the equator
				foreach ControlQubit in Register[Qubit.index+1::]:
					target Qubit for a rotation
					rotate by 2\^(ControlQubit.index - Qubit.index)
			\end{algorithm}\end{minipage}			
			We can now write the QFT in circuit form: 
			\begin{equation}\begin{quantikz}[column sep=0.3cm,row sep={0.9cm,between origins}]
				\ket{j_0}\;\qw	& \gate{H}	& \gate{\!\sqrt[2]{Z}\,}	& \ \ldots\ \qw	& \gate{\!\sqrt[2^n]{Z}\,}	& \qw		& \ \ldots\ \qw	&
				\qw							& \ \ldots\ \qw	&  \qw			& \;\ket{k_n}\qw\\
				\ket{j_1}\;\qw	& \qw		& \ctrl{-1}					& \ \ldots\ \qw	& \qw						& \gate{H}	& \ \ldots\ \qw	&
				\gate{\!\sqrt[2^{n-1}]{Z}\,}& \ \ldots\ \qw	&  \qw			& \;\ket{k_{n-1}}\qw\\
				\vdots&&&&&&&&&&\vdots\\
				\ket{j_n}\;\qw	& \qw		& \qw						& \ \ldots\ \qw	& \ctrl{-3}					& \qw		& \ \ldots\ \qw	&
				\ctrl{-2}					& \ \ldots\ \qw	& \gate{H}		&  \;\ket{k_0}\qw
			\end{quantikz}.\end{equation}
			All in all, since each qubit line is targeted by at most $n-1$ other qubits, this only takes $\mathcal{O}(n^2)$ gates, exponentially fewer than the $\mathcal{O}(n2^n)$ required for the classical algorithm.
			
			As we saw above, the quantum Fourier transform allows us to efficiently find the periodicities in a string of numbers.
			Shor discovered that period finding can in turn, with many steps in between, be used to calculate with high probability the prime factors of an integer~\cite{shor1994algorithms}, something exponentially hard classically\footnote{The best classical algorithm is still the 2200-year old ``sieve of Eratosthenes''.}.
			There really isn't room here to discuss all these steps\footnote{This is a thesis about silicon transmons after all.}, which can be found in popular form in Ref.~\cite{aaronson2007Shor} and in even more popular form on pp.~122--125 of Ref.~\cite{dowling2013schrodinger}\footnote{An anecdote oft told by the late Jonathan Dowling gives some color to the historical fact that quantum computing as a field of research used to be almost entirely funded by the American security agencies. See p.~121 of his 2013 book~\cite{dowling2013schrodinger} or this webcomic: \url{https://www.smbc-comics.com/comic/jonathan-dowling}.}.
			The best-known direct implication of an efficient algorithm for prime factoring is the insecurity of RSA encryption~\cite{rivest1978method}, which relies on the hardness of this one-way problem (calculating $n=pq$ is easy given primes $p$ and $q$, but finding those factors given $n$ is hard).
			Though research on ``post-quantum cryptography'' is under way~\cite{schmidt2020keynote}, the hardness of predicting whether a given NP problem is unsolvable on a quantum computer (the relation of BQP to NP is currently unknown) means that the reliability of any non-quantum method of encryption is questionable~\cite{aaronson2020q2b20,dowling2013schrodinger}.
			
			By demonstrating an exponential speedup for a common classical problem, Shor's algorithm provided strong evidence that quantum computers could give significant speedups in areas of practical use.
			The QFT and the closely related phase estimation and order finding algorithms are all used to efficiently find global properties of sets of numbers, something known to be hard classically.
			They are therefore starting points in the search for quantum solutions to classically hard problems, giving hope of further exponential speedups.
			
		\subsection{\label{sec:grover}Grover's algorithm}
				
			Often times a technical problem can be reduced to finding an element in an unstructured list.
			You may have a set of possible answers, only some of which are actual solutions, and you know how to recognize those elements once you see them~\cite{supremecourt1964jacobellis}.
			If you know how to recognize the solution, then you could devise some function $f(x)$ that gives a phase shift iff $x$ is a solution, and leaves the element unchanged if it isn't.
			Grover found that if you use this function $f(x)$ as an oracle, you can extract the solutions from the list in time proportional to only the square root of the size of that list\footnote{Unfortunately, for black-box problems like these, Grover's quadratic speedup is the best you can ever get, even on quantum computers~\cite{bennett1997strengths}.}.
			This is a quadratic speedup relative to classical algorithms, where unstructured searches necessarily need to check a number of elements linear in the size of the list (on average, you'll find all solutions half-way through if there's one, two-thirds through if there are two, etc).
			
			Grover's algorithm uses two unitary operators, one of which is the operator mentioned above, which we can write as
			\begin{equation}\begin{quantikz}
				U_\text{oracle}\ket{\psi}=e^{i\pi f(\psi)}\ket{\psi}\quad:\quad\ket{\psi}\;\qwbundle[alternate]{}	& \gate{U_f}\qwbundle[alternate]{}	&  \;e^{i\pi f(\psi)}\ket{\psi}\qwbundle[alternate]{}
			\end{quantikz}.\end{equation}
			The other is a ``diffuser'' unitary, which rotates the multi-qubit state around the uniform superposition of all states.
			This rotation can be understood as follows.
			Recall that we can construct the Pauli matrices from the cross product of their positive eigenvectors\footnote{Or the negative ones, using $1-2\ket{-z}\bra{-z}$ etc.},
			\begin{equation}\begin{array}{r@{\;}c@{\;}r@{\quad}r@{\;}c@{\;}l@{\;}c@{\;}l}
				\ket{x}	& =	& \dfrac{1}{\sqrt{2}}\Pm{1\\1},& \sigma_x	& =	& \Pm{0&1\\1&0}	& =	& 2\ket{x}\bra{x}-1,\\\\
				\ket{y}	& =	& \dfrac{1}{\sqrt{2}}\Pm{1\\i},& \sigma_y	& =	& \Pm{0&-i\\i&0}	& =	& 2\ket{y}\bra{y}-1,\\\\
				\ket{z}	& =	& \Pm{1\\0},& \sigma_z	& =	& \Pm{1&0\\0&-1}	& =	& 2\ket{z}\bra{z}-1,
			\end{array}\end{equation}
			where $\sigma_x$ flips bits that are pointing along the $y$ and $z$-axes, $\sigma_y$ flips those along the $x$ and $z$-axes, etc.
			We can now construct a unitary $\sigma_\text{superposition}$ that flips multi-qubit states that point along the uniform superposition over all states:
			\begin{equation}\ket{s}=\ket{\text{superposition}}=\dfrac{1}{\sqrt{N}}\Pm{1\\\vdots\\1},\quad \sigma_s=2\ket{s}\bra{s}-1.\end{equation}
			This operation will rotate the qubit register in Hilbert space in such a way that its projection onto the uniform superposition is conserved, but any components orthogonal to it are flipped.
			As we will see below, this ``diffuses'' the wave function over the basis states by rectifying the phases relative to some tilted axis.
			
			\begin{figure}
				\centering
				\begin{tikzpicture}
					\node[anchor=center,inner sep=0] (image) at (0,0) {\includegraphics[width=0.4\textwidth]{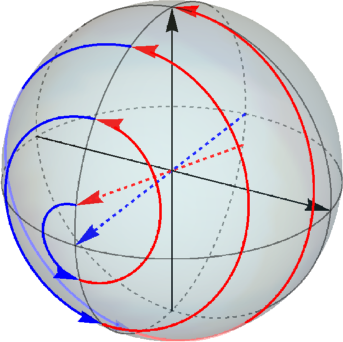}};
					\node[color=black] at (0,3.3) {$\ket{\text{solution}}$};
					\node[color=red] at (-4.4,-1) {$\ket{\text{superposition}}$};
					\node[color=blue] at (-4.7,-2) {$\ket{\text{superposition}}-\ket{\text{solution}}$};
				\end{tikzpicture}
				\caption{\label{fig:grover_bloch}Grover's algorithm visualized: The Bloch sphere is a three-dimensional slice through Hilbert space, with the $z$-axis the superposition of solution states, the \Blue{$x$-axis} the states orthogonal to those, and the angle around the \Blue{$x$-axis} representing the global phase. The \Blue{blue} rotations are multiplications by Grover's oracle, while the \Red{red} rotations are diffuser operations.}
			\end{figure}
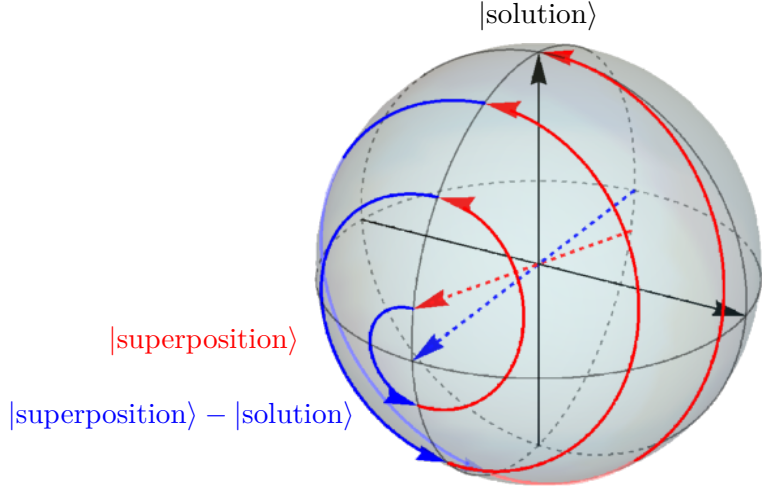
			
			In Fig.~\ref{fig:grover_bloch} a kind of Bloch sphere is shown, where instead of the usual $\ket{0}$ and $\ket{1}$ along the $z$-axis and their superpositions with different phases on the sphere at some angle in between, the $z$-axis now represents the superposition of all solution states.
			Orthogonal to the solutions are all the non-solutions, the wrong answers to the problem, indicated in blue.
			At some angle between the two (a small angle if most states are not solutions), we can find the vector that represents the uniform superposition over all states in the two sets.
			You can now see that if the register is initialized in the uniform distribution, applying the oracle $U_\text{oracle}$ will change only the phases of the solution states, effectively flipping its direction on the solution axis.
			Since the non-solution and uniform superposition axes are a bit off, applying the diffuser by rotating around the superposition then brings us closer to the solution state than we were before.
			After repeating the two operations on the order of $\mathcal{O}(\sqrt{N})$ times (the prefactor depends on the relative share of solution states), we will have a high probability of finding the solution when we perform a measurement.
			\begin{equation}\begin{quantikz}
				\ket{0}^{\otimes n}\;	& \qwbundle{n}		& \gate{H^{\otimes n}}\qw	& \gate{U_f}\gategroup[1,steps=2,style={dashed,rounded corners,fill=gray!20, inner xsep=3pt, outer sep=3pt},background]{$\times\,\mathcal{O}(\sqrt{N})$}\qw	& \gate{U_s}\qw	& \meter\qw	&  \;\ket{\text{solution}}\qw
			\end{quantikz}\end{equation}
			\noindent\begin{minipage}{\textwidth}\begin{algorithm}[gobble=12,tabsize=4,caption={Grover's algorithm.},label={alg:grover},mathescape]
			Apply the Hadamard to each qubit  // Create uniform superposition
			foreach AmplificationStep in 0::sqrt(Register.length):
				Apply the oracle 			  // Flips phase of solution qubit
				Apply the diffuser			  // Rotates around uniform superposition
			Measure the state				  // Extract solution
			\end{algorithm}\end{minipage}
			
		\subsection{\label{sec:qa_qaoa}Quantum annealing (QA) and the quantum approximate optimization algorithm (QAOA)}
			
			Quantum annealing is different from all the algorithms that we have discussed above in that it is not circuit based, i.e. it is not performed on quantum Turing machines, but on dedicated systems.
			The name is aptly chosen, as we can compare the process directly to the thermal annealing that we will discuss later on in chapter~\ref{sec:silicides}.
			When a material, such as the silicides that we study for Josephson field effect transistors, is first formed, it is not yet in its lowest-energy state.
			The Gibbs energy may be lowered significantly by aligning the crystal orientations in neighboring grains and removing the boundaries between them, but reaching that lower-energy state requires us to cross an energy barrier associated with all the chemical bonds that need to be broken.
			By annealing the material, sufficient heat is supplied to overcome this barrier, and the atoms are stimulated to move into configurations that are energetically favorable, lowering the Gibbs energy and leading to superior material properties such as higher superconducting critical temperatures.
			
			In the problems solved by quantum annealing, we are faced with a similar issue: we want to find out what the ground state of a system is, but if we just immediately apply its Hamiltonian to a set of qubits, the resulting state is rarely in the global minimum.
			Instead, the annealing approach is to start out from a state that we know how to initialize, for example with all qubits in $\ket{0}$,
			\begin{equation}\Ha_\text{init}=\sigma_z^{n}.\end{equation}
			The trick is then to heat up the system, such that the wave function spreads out from this initial state, while slowly changing the applied Hamiltonian to the one we want to find the ground state of,
			\begin{equation}\Ha(t)=u(t)\Ha_\text{init}+(1-u(t))\Ha_\text{problem},\end{equation}
			after which the system is cooled down again.
			This can of course be done classically as well\footnote{The classical analog is called ``simulated annealing'', and though it is sometimes used as a reference point by D-wave~\cite{condello2020q2b20}, it is not the fastest classical algorithm for finding approximate solutions to this problem~\cite{isakov2015optimised}.}, but the advantage of QA is that the escape from local minima is sped up by quantum tunneling.
			Current research focuses on what the ideal crossover function $u(t)$ is for specific problems~\cite{brady2020optimal,gorshkov2020q2b20}.
			An especially interesting problem Hamiltonian is the Ising spin glass~\cite{king2015benchmarking},
			\begin{equation}\Ha_\text{Ising}=\Sum_{i<j}J_{ij}\sigma_z^{(i)}\sigma_z^{(j)},\end{equation}
			where finding solutions for $J_{ij}>0$ (get as many neighbors as possible to have opposite spin) is in NP-complete.
			This class of problems is of special relevance since it contains the hardest problems in NP, a solution to any of which can be converted to a solution to any other NP-complete problem in polynomial time.
			This means, for example, that if you can find an approximate ground state of the Ising model, you can find good solutions to the traveling salesperson problem.
			It is currently unclear what kind of speedup quantum annealing can offer~\cite{zintchenko2015recent}, but it is believed that it is at most polynomial, likely quadratic~\cite{babbush2020q2b20,childs2020q2b20}.
			
			Quantum approximate optimization algorithms (QAOA) are a circuit-based equivalent to quantum annealing, where instead of slowly shifting from one Hamiltonian to another, the two Hamiltonians are applied sequentially for varying amounts of time~\cite{bouland2020q2b20,gorshkov2020q2b20}. 
			Again one starts out in an accessible ground state, but then instead of only slowly turning on the solution Hamiltonian, it is immediately applied for a relatively long stretch of time, ensuring that the system ends up at least close to some local minimum.
			Instead of heating up the system, the initialization Hamiltonian is then applied again to shake things up, and these two steps are repeated for a number of times.
			Perhaps an intuitive analogy is filling a cup with flour: after you tap a full cup on the counter a few times, it condenses until it is maybe only half full\footnote{Always measure the weight of your ingredients, never the volume!}.
			The level to which the cup is filled is our cost function, the precise locations of all the wheat granules is the approximate solution to the problem, and the number of times we need to tap is the computational complexity.
			
			One important result is that our final error, which we define as the difference between the energy of the final measured state and that of the actual ground state, goes down quadratically,
			\begin{equation}\text{Error}_\text{QAOA} = \braket{\psi_\text{final}|\Ha_\text{Ising}|\psi_\text{final}}-\braket{\psi_\text{0}|\Ha_\text{Ising}|\psi_\text{0}}\sim \dfrac{1}{n_\text{steps}^2}.\end{equation}
			In other words, if you want $n$ significant digits, the time complexity is $\mathcal{O}(\exp(n/2))$, while the optimal performance of classical computers is unknown~\cite{bouland2020q2b20,bravyi2021classical}.
			The catch is that you only get this optimal error scaling iff you use the optimal timing of applying the two Hamiltonians, which actually depends on the details of the problem~\cite{gorshkov2020q2b20}.
			It was also shown recently that QAOA even offers no quantum speedup at all if there are limits to the circuit depth (a real concern for NISQ devices)~\cite{bravyi2021classical}.
	
	\section{Physical implementations of quantum computers}

		\Quote{``Any physical experiment can be regarded as a computation, and any computation is a physical experiment.''}{David Deutsch}
		
		The brief descriptions below serve only to illustrate the variety of platforms that quantum circuits can be implemented on, and to give a rough idea of how a quantum system can be turned into a qubit.
		For a more exhaustive, yet accessible overview, see e.g. pp.~153--171 of Ref.~\cite{dowling2013schrodinger}.
		
		\subsection{Ion traps}
			
			The physical implementation of quantum computing arguably started in ion traps, for which a two-qubit CNOT gate was proposed in late 1994 by Cirac and Zoller~\cite{cirac1995quantum}.
			This gate was quickly demonstrated by Wineland and Monroe~\cite{monroe1995demonstration}, who had been working with ion traps for a long time in the context of atomic clocks.
			The development of theoretical understanding of gate operations in this physical system is of direct relevance to all other ``platforms'' (two of which will be discussed below), since \emph{``if the math is the same, the physics must be the same!''}~\cite{dowling2013schrodinger}.
			The cavity QED developed for gates based on ion-photon interactions~\cite{berman1994cavity,pellizzari1995decoherence,sleator1995realizable} later formed the basis for circuit QED~\cite{devoret1995quantum}, which is the framework for superconducting qubits.
			
			Ions are perhaps the most ``natural'' of the platforms, as the qubit states are encoded into the energy levels of a single electron in a half-occupied orbital of an ionized atom.
			These atoms are suspended by a rapidly oscillating electromagnetic field in ultra-high vacuum, and are operated on with laser beams so precise that they can be controlled almost down to individual photons~\cite{blinov2004observation}.
			The fact that electromagnetic trapping is even possible at all is perhaps counter-intuitive, as most freshman courses in electrodynamics will have you prove Earnshaw's theorem~\cite[p.~115]{griffiths2008introduction}:
			\begin{center}
				\begin{minipage}[C]{0.9\textwidth}
				\emph{``A charged particle cannot be held in a stable equilibrium by electrostatic forces alone.''}
				\end{minipage}
			\end{center}
			This is a direct consequence of the fact that the electric potential at any point in space is the spatial average of the potential at any fixed radius around it (Laplace's equation, valid as long as there are no charges inside te sphere).
			The clever way around this that got Wolfgang Paul the Nobel prize in Physics in 1989 is to create an electric field that has a local minimum along one axis that simultaneously is a local maximum along another, and then rotate these axes~\cite{paul1990electromagnetic}.
			
			Next we need to define the qubit levels in these trapped particles.
			The Pauli exclusion principle tells us that we cannot have multiple fermions like electrons occupying the same state. While the orbital quantum numbers uniquely define the orbits that electrons can be in, the spin degree of freedom ($\ket{\uparrow}$ and $\ket{\downarrow}$) still allows us to have two electrons in the same ``place''. If we now make an ion by removing one of the outer electrons from an atom with an even atomic number (e.g. \ce{_{12}Mg}, \ce{_{20}Ca}, \ce{_{48}Cd} or IonQ's \ce{_{70}Yb}), we are sure that the remaining electron is alone in its orbit.
			Once this ion is trapped inside the oscillating field, we can encode the qubit $\ket{0}$ and $\ket{1}$ states in its lone outer electron being excited or not.
			If the qubit were to decay from an excited $\ket{1}$ state, a photon would be emitted, which means that a superposition $\sqrt{1/2}(\ket{0}+\ket{1})$ is shared with the electromagnetic cavity both containing a photon and not.
			Lasers can then be used both to excite the ions and to bring them back to the ground state.
			When the same laser beam acts on two ions, a photon can be in a superposition of having been absorbed by the one or the other, effectively entangling the two qubits.
			
			This platform is especially well placed for integration with quantum communication systems, which as a long-distance technology necessarily operate with optical-range photons\footnote{At lower energies the thermal noise would be too strong, at higher frequencies you will have trouble building lenses.}~\cite{dowling2020schrodinger}.
			While the photons carried by the resonators in superconducting circuits are on the order of a few GHz (times the Planck constant)~\cite{martinis2020saving}, the energy levels in trapped ions are separated by energies on the order of hundreds of THz~\cite{bock2018high}.
			To quote Dowling and Milburn, \emph{``at optical frequencies, the world is very cold ($hf\gg k_\text{B}T$)''}~\cite{dowling2003quantum}.
			As a result, while superconducting circuits may need supercooled data links at ultra-high vacuum to couple distant chips~\cite{magnard2020microwave}, ion traps can be linked with simple room-temperature, ambient-pressure photonic interconnects~\cite{bock2018high,ionq2020q2b20,valivarthi2020teleportation}.
		
		\subsection{Trapped spins in semiconductors}
			
			While atoms may be the ``natural'' candidate for qubits, as each of them is identical to --- even indistinguishable from --- any other of the same isotope, we are limited in our choice of parameters by the number of stable elements\footnote{There are around 120 elements out there, only about 80 of which are stable. The number of options goes down further if we restrict ourselves to those with even atomic number, level spacings in the frequency range of existing lasers, non-toxicity, etc.}.
			This motivates the design of ``artificial atoms'', quantum systems where the energy levels can be engineered, such as quantum dots defined by electrostatic gates in semiconductors, or superconducting circuits.
			Just like in ordinary atoms, qubit states can be defined by the occupied orbital or the spin of an electron on the dot, or even the total number of electrons residing there~\cite{feynman1985quantum,leon2020coherent}.
			
			Besides their tunability, the interest in quantum dots defined in semiconductors is motivated by the already present technological prowess in the fabrication of micro-electronics in these materials.
			This is important not just for the manufacturing of the qubits themselves.
			Fault-tolerant quantum computers will require real-time error correction, performed in part by classical logic and control circuits, which in a scalable quantum computer should be implemented as close to the qubits as possible~\cite{annunziata2020q2b20}.
			The possibility of on-chip integration of classical CMOS circuitry~\cite{le2020low} minimizes time delays, removes any losses of signal due to the coupling of long microwave lines to electromagnetic radiation from the environment, and reduces the hardware requirements on dilution refrigerators and room-temperature instruments.
			
			In most physical implementations pursued today, the qubit degree of freedom is encoded in the spin orientation of an electron trapped on a quantum dot defined in semiconducting material by a gate or set of gates~\cite{chatterjee2020semiconductor}.
			Microwave signals sent through a gate that is capacitively coupled to the dot can be used to perform operations on a single qubit, as well as to read out its state.
			Electrons on nearby dots can be brought closer together to create entanglement by changing the electrostatic environment, while they can be coupled over longer distances with microwave resonators.
		
		\subsection{\label{sec:superconducting_qubits}Superconducting qubits}
			
			The focus of this thesis is a different kind of artificial atom, one where the energy levels are defined by the quantized resonance frequencies of an LC resonator.
			
			In its most common form, a superconducting qubit consists of a capacitively shunted inductor, where the inductance is provided by a superconducting weak link.
			The oscillations in the resonator can be understood as a constant tension between on the one hand the desire of the charges on opposite sides of a capacitor to even out by flowing around through the inductor, and on the other hand the resistance of that inductor against any change in the amount of current that flows through it.
			By the time the capacitor is empty, the inductor will make sure the current keeps flowing, until an equal but opposite charge has built up again.
			While macroscopic circuits of this kind would slowly lose energy through heating and inductive coupling to the environment, the oscillations on microscopic superconducting circuits are protected by the same principle of energy level quantization that prevents the electrons in real atoms from spiraling into the core.
			It is these quantized levels that we use as our qubit states.
			
			There are multiple degrees of freedom in these systems that can all be used as quantization axes for the qubit states, such as the number of Cooper pairs~\cite{cottet2002implementation} or the superconducting phase~\cite{steffen2006measurement} on one side of the capacitor, a combination of the two~\cite{vion2002manipulating}, or even the flux through the split Josephson junction that makes up the inductor~\cite{orlando1999superconducting}.
			We will focus on designs where the qubit states are a superposition of different numbers of Cooper pairs having collected on either side of the capacitor~\cite{koch2007charge}.
			In this limit relatively little energy is associated with a Cooper pair crossing the junction, $E_\text{J}\gg E_\text{C}$, and the energy levels are primarily defined by the Josephson operator.
			This operator couples charges states with different numbers of Cooper pairs~\cite{cottet2002implementation},
			\begin{equation}\Ha_\text{J}=-\dfrac{E_\text{J}}{2}\Sum_{n\in Z}\Big(\ket{n}\bra{n+1}+\ket{n+1}\bra{n}\Big)=
			-\dfrac{E_\text{J}}{2}\Pm{%
			\ddots\;\;	& 		&		& 		& \emptyset	\\
						& 0		& 1		&		&			\\
						& 1		& 0		& 1		& 			\\
						&		& 1		& 0		& 			\\
			\emptyset	&		&		& 		& \;\;\ddots},\end{equation}
			such that any eigenstate is a symmetric superposition of Cooper pair occupation numbers around some average $\braket{n}$ defined by the electrostatic environment of the capacitor.
			The larger the capacitance, the broader the spread of the qubit eigenstates over the number of Cooper pairs.
			In general the circuit is designed to have a ratio $E_\text{J}/E_\text{C}\approx 50$, such that the qubit levels become almost entirely insensitive to fluctuations in occupation while still maintaining some anharmonicity~\cite{koch2007charge}.
			
			The state of a superconducting qubit can be operated on by sending calibrated microwave pulses.
			Often the energy levels themselves can be changed as well by varying the Josephson energy $E_\text{J}$.
			In traditional designs, this energy level modulation is often done by splitting the Josephson junction into a superconducting quantum interference device (SQUID), and then tuning the external magnetic flux imposed on its surface with a current-carrying flux line nearby.
			Currents will be induced in the SQUID to bring the total flux that passes through it to an integer multiple of the flux quantum $\Phi_0=h/2e$, affecting in turn the junction's effective coupling.
			This way the energy levels of nearby qubits can be brought close together, allowing for lossless energy exchange between them, and entangling their states with the $i$SWAP gate discussed in section~\ref{sec:two_qubit_gates}.
			Some designs omit this individual tunability to improve coherence and reduce the number of cables, relying entirely on microwave signals to perform both single and two-qubit gate operations~\cite{gambetta2017building,krantz2019quantum}.
			Once a circuit has been executed, the state of a qubit can be read out by interrogating the same resonator that was used to operate on it, which will give a phase shift to a passing readout pulse that depends on the state of the qubit.
			
			Quantum computers of this sort have had some widely publicized early successes, achieving quantum simulations~\cite{bernien2017probing} and computations~\cite{arute2019quantum} that outperform classical machines.
			For truly large-scale quantum computers however, many improvements still need to be made in error rates, fabrication cost, size of control hardware etc.
			In the next chapter, we will discuss how designing qubits around a Josephson junction made from a CMOS field effect transistor can bring us closer to realizing these ambitions.

\printbibliography

\end{refsection}

\begin{refsection}
	\graphicspath{{img/ch2/}}
	\setcounter{chapter}{1}
\chapter{\label{sec:cmos_gatemons}Complementary Metal Oxide Silicon gatemons}

\section{Introduction}

	We saw in the previous chapter that superconducting qubits are built around Josephson junctions, which are weak links between two superconducting leads.
	Typically these are made of insulating layers of aluminum oxide between superconducting strips of pure aluminum, and often come in parallel pairs that form superconducting quantum interference devices (SQUIDs).
	These SQUIDs are what allow for the tunability in tunable transmons: their coupling strength is highly sensitive to the out-of-plane magnetic field, something that can easily and precisely be set with a simple current carrying wire nearby~\cite{cottet2002implementation}.
	The clear drawbacks to this design are that twice as many junctions are needed, a relatively large amount of current on the order of milli-Amps\footnote{For SQUIDs of about $10^2$ square micron large~\cite{chen2014qubit}, you need fields on the order of $B\approx\Phi_0/A\approx10\si{\micro\tesla}$. If the flux line is $\sqrt{A}=\SI{10}{\micro\meter}$ away, then Amp\`ere's law tells us that we need currents of around
	\begin{equation}I_\text{flux line}=\dfrac{2\pi \Phi_0 A}{\mu_0\sqrt{A}}=\dfrac{\pi h\sqrt{A}}{\mu_0 e}\approx \SI{1}{\milli\ampere}.\end{equation}
	You could bring this number down by making the SQUIDs bigger, but then you'll end up with antenna arrays instead of qubits.} linear in the number of qubits has to be supplied, fields intended for one qubit need to be compensated for in others\footnote{The compounding interactions of many nearby components need not necessarily be intractable; similar concerns about crosstalk in classical circuits were ultimately resolved by simple design rules~\cite{mead1980introduction,conway2012reminiscences,dowling2013schrodinger}.
	But this does of course not mean that exploring alternatives is not a good idea!}, and any extra degree of freedom in your Hamiltonian also acts as an extra channel for noise.
	Some of these concerns have led to the adoption of fixed-frequency transmons~\cite{mckay2016universal,gambetta2017building}, though this is only one of many viable approaches.
	The idea pursued in this thesis is to stick to the tunable transmon design for all its advantages, but to replace the cumbersome SQUID by a single gate-modulated Josephson junction~\cite{larsen2015semiconductor,de2015realization}.
	
	Before getting into the details of the experimental work on Josephson Field Effect transistors, we will review qubit Hamiltonians in section~\ref{sec:quantum_harmonic_oscillator}, some of the principles of superconducting junctions in section~\ref{sec:superconducting_transport}, and some basics of transistors in section~\ref{sec:transistors}.

\section{\label{sec:quantum_harmonic_oscillator}The quantum harmonic oscillator}
	
	\begin{figure}
		\begin{subfigure}[t]{0.45\textwidth}
			\includegraphics[width=\textwidth]{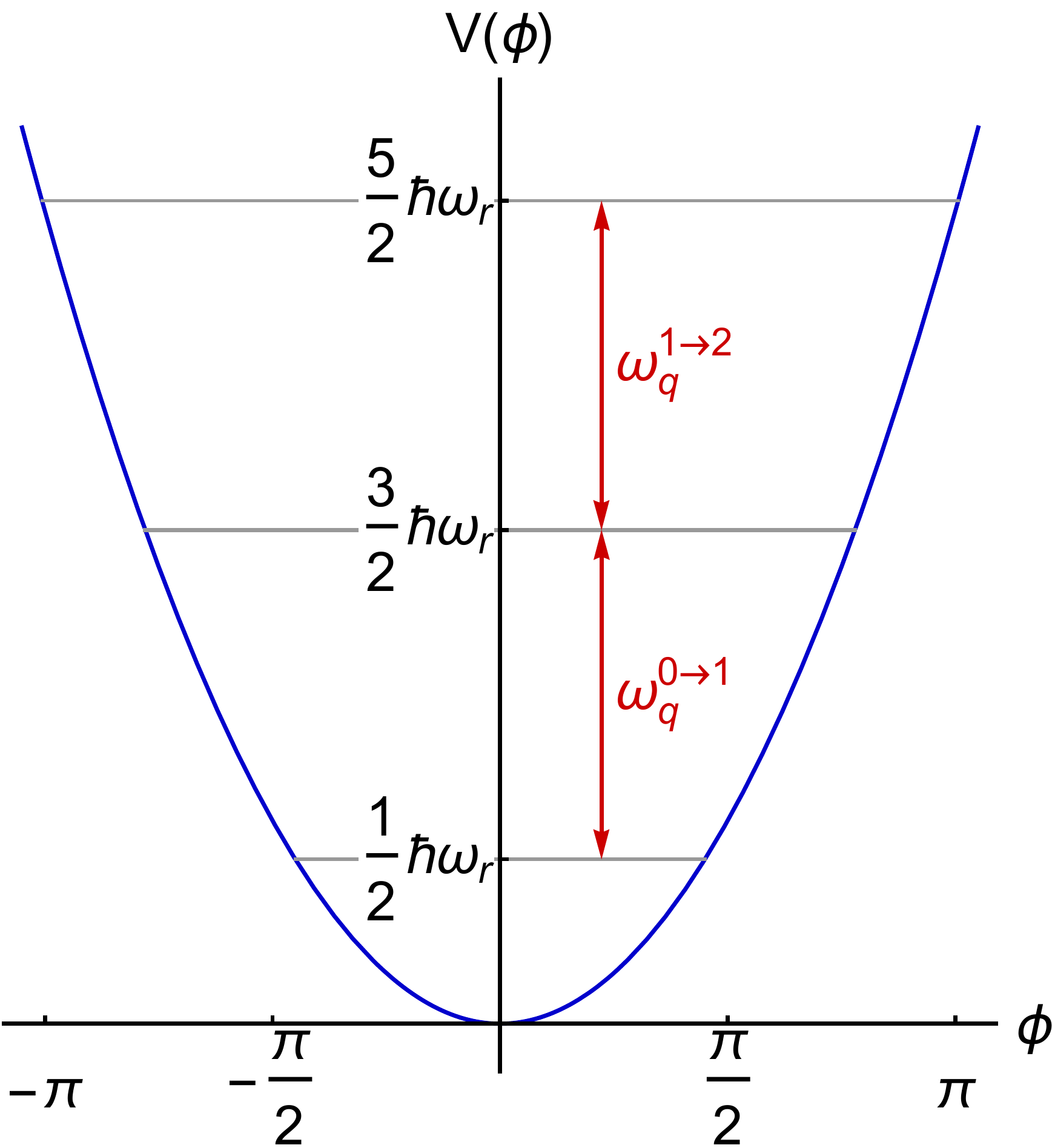}
		\end{subfigure}\hfill\begin{subfigure}[t]{0.45\textwidth}
			\includegraphics[width=\textwidth]{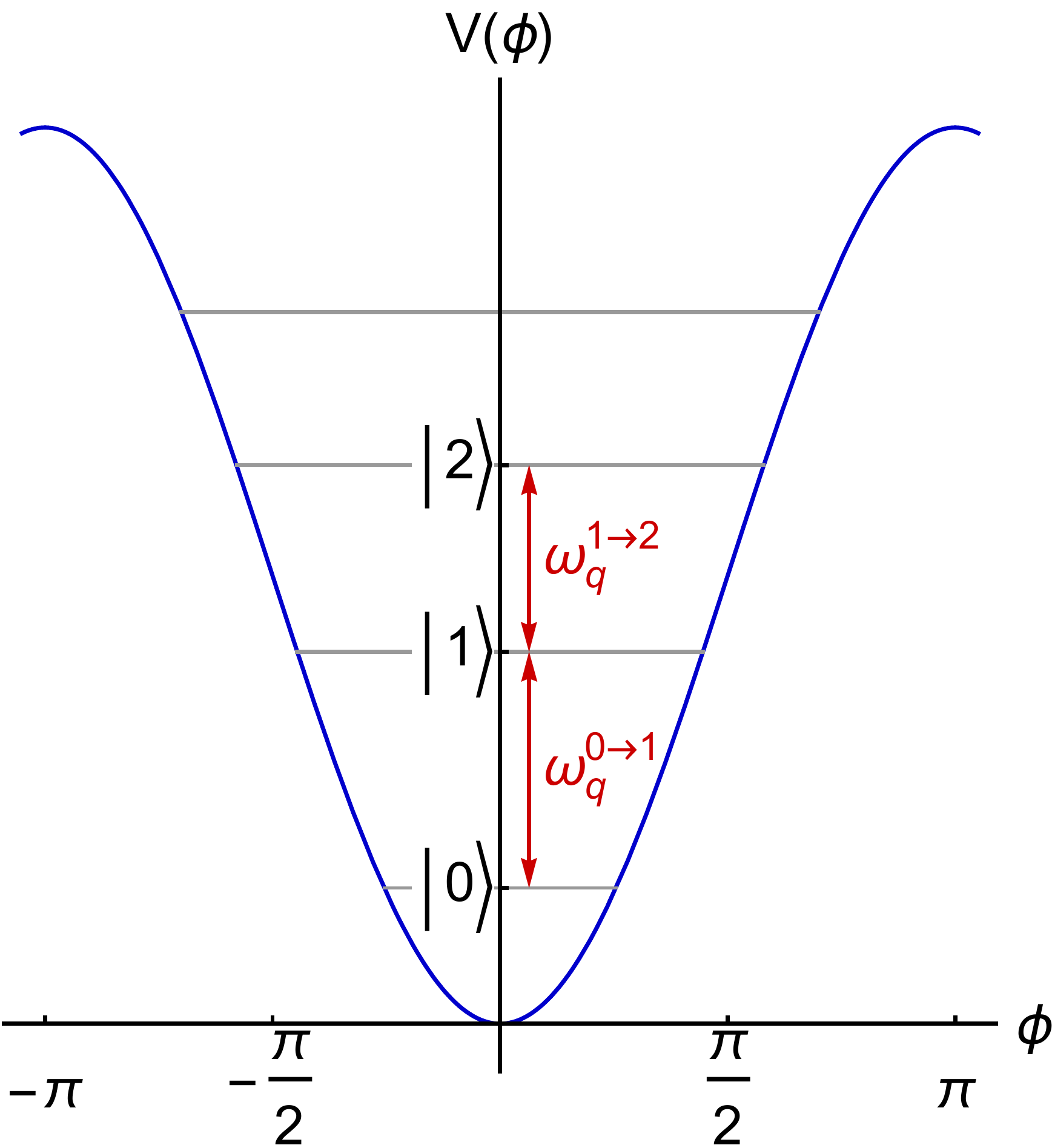}
		\end{subfigure}
		\caption{\label{fig:harmonic_potential}\B{(Left)} The quadratic potential energy of the quantum harmonic oscillator. On the $\phi$-axis we have some variable, which can be anything from the angle of a pendulum or the displacement of a mass on a spring, to the phase across a Josephson junction. The $y$-axis is the energy, such that the gray lines represent levels where it is constant: as the system oscillates around $\phi=0$, kinetic and potential energy are constantly interchanged. \B{(Right)} The almost-quadratic potential of an \emph{an}harmonic oscillator. Here the level spacings are different, and we can isolate the bottom transition as our qubit.}
	\end{figure}
	
	The only levels that we care about for quantum computational applications are the two with the lowest energy, which means that we always talk about local minima in the energy spectrum.
	And it turns out that at a local minimum, almost any potential can be well-approximated by a quadratic function, as can be seen by expanding some generic potential $V(\phi)$,
	\begin{equation}V(\phi)=V(\phi_0)+\partial_\phi V(\phi_0)(\phi-\phi_0)+\dfrac{1}{2}\partial^2_\phi V(\phi_0)(\phi-\phi_0)^2+\cdots\end{equation}
	We can forget about the first term as constant energies don't affect our system's dynamics, and we can ignore the second term since by definition of the ``local minimum'' it is strictly zero.
	As long as $\phi-\phi_0\ll1$ in the appropriate units, the higher-order terms $(\phi-\phi_0)^{n>2}$ will be negligible, and so all we are left with is the quadratic potential.
	This is what we call the harmonic oscillator, and for most practical purposes it is a good description of qubit systems.
	
	\subsection[Qubit from it]{Qubit from It}\label{sec:qubit_from_it}
	
		In principle, any system with quantized energy levels could be used as a quantum bit, which is great, because if you look closely enough, \emph{any} physical system exhibits quantum effects.
		Technically speaking however, for those bits to be qubits in any meaningful sense of the word, we also need them to to have the following properties.
		\begin{enumerate}
			\item The system needs to be isolated well enough that it stays coherent much longer than it takes to perform a gate operation.
			\item We need the spacings between the levels to be different, so that we can isolate a single transition between two levels as our degree of freedom.
			\item We have to be able to perform the universal gate set, i.e. perform $\pi/4$ and $\pi/2$ rotations about different axes on a single qubit, and couple multiple qubits to each other.
		\end{enumerate}
		
		We can break down how these conditions are all satisfied by transmons, the superconducting qubit type that we are aiming to build.
		The first of these conditions implies at the very least that the level spacing be greater than the temperature,
		\begin{equation}\hbar\omega_\text{q}\gg k_\text{B}T,\end{equation}
		for otherwise we would lose the coherence in the thermal noise.
		To give an idea, a typical superconducting qubit has an angular frequency of about 3 to \SI{5}{\giga\hertz}~\cite{krantz2019quantum}, while they are cooled to below $\SI{20}{\milli\kelvin}\,\hat{\approx}\,\SI{400}{\mega\hertz}$.
		Apart from heat, there are a host of other ways that a superconducting qubit can couple to its environment.
		Charge qubits are most sensitive to fluctuations in charge on elements that it is capacitively coupled to, flux qubits easily couple to modes of the electromagnetic field that are available in inductive elements, and phase qubits decay in the presence of current fluctuations~\cite{martinis1985energy}.
		Transmons, whose energy levels are hybridized charge states~\cite{koch2007charge}, mostly decay due to dielectric losses to parasitic capacitances~\cite{martinis2005decoherence} and quasiparticle poisoning~\cite{catelani2011relaxation,sun2012measurements}.
		
		The second condition limits how much we can achieve in the coherence time given by the first.
		Sending a signal to a many-level system in the ground state would first partially populate the $\ket{1}$ level, but instead of neatly oscillating back to $\ket{0}$, it would then spread into a superposition that also includes $\ket{2}$, $\ket{3}$, etc.
		We therefore need an \emph{anharmonicity}\footnote{Called so as it quantifies the degree to which the Hamiltonian is \emph{not} like a harmonic oscillator, the perfectly quadratic potential that gives rise to equally spaced levels.} $\alpha$,
		\begin{equation}\alpha=(\omega_{\ket{2}}-\omega_{\ket{1}})-(\omega_{\ket{1}}-\omega_{\ket{0}}),\end{equation}
		that is large enough that microwave signals will not excite level $\ket{2}$.
		Fourier transformation then tell us that any signal pulse shorter than some time $\tau$ will have energy components on the order of $\hbar/\tau$, so if you want to have e.g. \SI{100}{\nano\second} pulses, you need $\alpha/\hbar>\SI{10}{\mega\hertz}$.
		Obtaining a good ratio of gate operation to coherence time thus requires a large $\alpha$.	
		In practice, $\alpha\approx100$--\SI{300}{\mega\hertz} in transmons~\cite{koch2007charge}, so you could theoretically go down to pulse sequences with waveform features on the order of tens of nanoseconds.
		
		The first part of the third condition, performing at least two orthogonal single-qubit gates, means that we should be able to couple to our transmon in such a way that we rotate its state about multiple axes.
		The fact that we can address different axes becomes clear when we consider the time dependence of any general state~\cite{griffiths2005introduction},
		\begin{equation}\Psi(\B{r},t)=\psi(\B{r})e^{iEt/\hbar},\end{equation}
		which directly implies that a qubit satisfies
		\begin{equation}\ket{\Psi}(t)=\alpha(t)\ket{0}+\beta(t)\ket{1}=e^{i(E_1-E_0)t/\hbar}\left(\tilde{\alpha}\ket{0}+\beta(t)\ket{1}\right),\end{equation}
		so that it always rotates around the quantization axis at a rate $\omega_\text{q}=(E_1-E_0)/\hbar$.
		This means that we need to add a rotating component to any gate pulse that we send, but also that simply waiting for a time $1/4\omega$ will let you switch axes from $x$ to $y$.
		It is perhaps less obvious that these are the axes that microwave signals act on through capacitive coupling, but this will become clear after we have discussed the transmon Hamiltonian in section~\ref{sec:transmons}.
		Two-qubit transmon gates can be performed in two different ways~\cite{gambetta2017building}: either by temporarily tuning nearby gatemons to similar frequencies (see the description of the $i$SWAP gate in section~\ref{sec:two_qubit_gates}), or by permanently coupling the qubits such that the level separation of one depends on the state of the other, causing them to ``control'' each other's rotation when excited by microwave pulses~\cite{paraoanu2006microwave}.
		
	\subsection{\label{sec:transmons}The transmon Hamiltonian}

		\begin{figure}
			\centering
			\begin{circuitikz}[circuit ee IEC,x=1.75cm,y=1.75cm,very thick]
				\fill[rounded corners=1cm,gray!15!white] (0,0) rectangle (2,2.5);
				\draw (0.5,0.5) to [L] (0.5,2);
				\node[anchor=west] at (0.75,0.75) {$E_\text{L}=\dfrac{\Phi^2}{2L}$};

				\fill[rounded corners=1cm,gray!15!white] (2.25,0) rectangle (4.25,2.5);
				\draw (2.75,0.5) to [C] (2.75,2);
				\node[anchor=west] at (3,0.75) {$E_\text{C}=\dfrac{Q^2}{2C}$};

				\fill[rounded corners=1cm,gray!15!white] (4.5,0) rectangle (8.25,2.5);
				\draw (5,0.5) to [L] (5,2) -- (6.5,2) to [C] (6.5,0.5) -- (5,0.5);
				\node[anchor=west] at (6.75,0.75) {$E_\text{LC}=\dfrac{\hbar n_{\tilde{\gamma}}}{\sqrt{LC}}$};
			\end{circuitikz}
			\caption{\label{fig:lc_oscillator}\B{(Left)} A classical inductor. \B{(Middle)} A capacitor. \B{(Right)} A simple LC circuit with a resonance frequency $\omega=1/\sqrt{LC}$.}
		\end{figure}
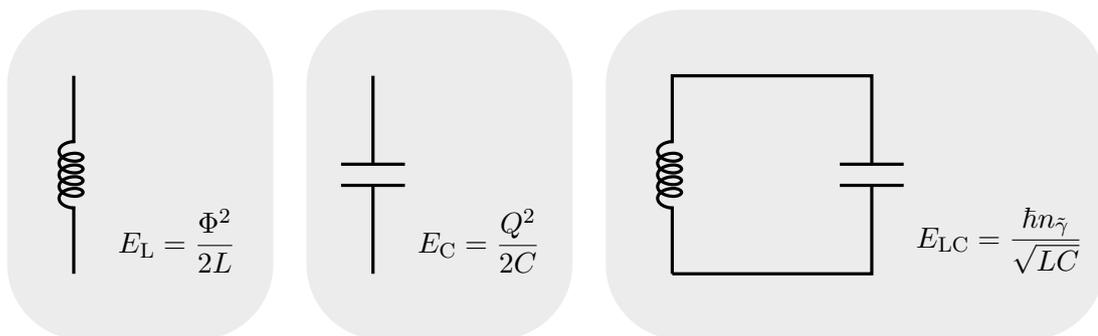
		
		The transmon is an anharmonic LC resonator consisting of a Josephson junction (L) in parallel with a capacitor (C).
		Classically, the resonance of an LC circuit can be understood as follows.
		Energy is stored in the electric field between the capacitor's plates as a voltage is built up across them, 
		\begin{equation}\label{eq:charging_energy}E_\text{C}=\dfrac{1}{2}\,CV^2=\dfrac{1}{2C}\,Q^2,\end{equation}
		while the inductor stores energy in a magnetic field as current flows through it,
		\begin{equation}\label{eq:inductive_energy}E_\text{L}=\dfrac{1}{2}LI^2=\dfrac{1}{2L}\,\Phi^2.\end{equation}
		This can only be a static situation if both energies are zero, as a fixed nonzero current would lead to a constantly growing charge on the capacitor, and a nonzero charge can only be fixed if the inductor has infinite inductance.
		The current and charge will therefore oscillate out of phase by an angle $\pi/2$: one will be maximum when the other has maximum derivative (and is zero), and vice versa.
		There is another way that we can state this phase shift, which is through their impedances $Z_\text{C}$ and $Z_\text{L}$, which express their resistance to forming a current:
		\begin{equation}Z=\dfrac{V}{I}:\qquad Z_\text{R}=R,\qquad Z_\text{C}=\dfrac{1}{i\omega C},\qquad Z_\text{L}=i\omega L.\end{equation}
		Here the phase shift of each element's current response to an applied voltage is expressed by the imaginary $i$, which represents a quarter turn ($\pi/2$) in the complex plane.
		Since the impedances of the capacitor and inductor have phase shifts in opposite directions, they are in total out of phase by $\pi$, as are their respective energies.
		The total impedance of the two elements then adds up to
		\begin{equation}\dfrac{1}{Z_\text{LC}}=\dfrac{1}{Z_\text{C}}+\dfrac{1}{Z_\text{L}}=\dfrac{i\omega L}{1-\omega^2 L C},\end{equation}
		and is minimal when
		\begin{equation}\omega_\text{LC}=1/\sqrt{LC},\end{equation}
		which means that the system will resist the least when it oscillates at this rate, giving us the resonance frequency.
		
		When proportional\footnote{The virial theorem states that a perfectly harmonic oscillator has on average equal amounts of energy stored in the kinetic and potential energies, $\braket{T}=\braket{V}$, or in our case $\braket{Q^2/2C}=\braket{\Phi^2/2L}$. Since pure undressed photons satisfy the relation $|\B{B}|^2=\mu_0\epsilon_0|\B{E}|^2$, this requires that $LC=\mu_0\epsilon_0A^2/d^2$, with $A$ the area of the inductor and $d$ the spacing of the capacitor.}, these out-of-phase oscillations of the electric and magnetic fields are nothing else than photons\footnote{This is best appreciated in the lumped-element approximation, where the wavelength is much longer than the circuit dimensions, clearly valid for $\lambda=2\pi c/\SI{4}{\giga\hertz}\approx\SI{50}{\centi\meter}$.
		In practice, the plasmon oscillations that define the qubit levels are never engineered to be pure photon, flux, or charge states, but some dressed combination of these.
		This way, the quantization axis can be chosen such that the levels are more isolated from the environment than the individual components it is made from, as has been demonstrated e.g. by dressing spin states with microwave photons~\cite{miao2020universal}.}~\cite{schuster2007resolving,houck2007generating}, and their number $n_\gamma$ will be the quantization axis of the harmonic oscillator,
		\begin{equation}\Ha=\hbar\omega_\text{LC}\left(a^\dagger a+\dfrac{1}{2}\right)=\hbar\omega_\text{LC}\left(n_\gamma+\dfrac{1}{2}\right).\end{equation}
		This photon number is but one of many quantization axes that can be chosen as qubit degree of freedom, and in general we will design the circuit to have energy levels that depend unevenly on electric and magnetic oscillations, to reduce sensitivity to particular kinds of noise.
		To better understand this general case, we go back to the Hamiltonian terms that we found earlier in equations~\eqref{eq:charging_energy} and~\eqref{eq:inductive_energy} for the capacitor and inductor separately, which combine to
		\begin{equation}\Ha=\dfrac{1}{2C}\,Q^2+\dfrac{1}{2L}\,\Phi^2=4E_\text{C}\,n^2+\dfrac{1}{2}E_\text{L}\,\phi^2,\end{equation}
		where $Q$ and $\Phi$ are rewritten in countable units of the number of Cooper pairs and flux quanta~\cite{devoret1995quantum},
		\begin{equation}n=\dfrac{Q}{2e},\qquad \phi=\dfrac{2\pi\Phi}{\Phi_0}=\dfrac{2e \Phi}{\hbar}.\end{equation}
		Here $\phi$ represents the ``gauge invariant phase difference'' across the inductor, or equivalently, the phase difference between the macroscopic wave functions of the superconducting condensates on opposite sides of a Josephson junction, and $n$ is the number of Cooper pairs built up at some node (in the case that one side of the loop is grounded, we choose the other as our node).
		These observables have their own operators that can be expressed in terms of the photon creation and annihilation operators~\cite{krantz2019quantum,rasmussen2021superconducting},
		\begin{equation}\hat{\phi}=\sqrt{\dfrac{4E_\text{C}}{\hbar\omega}}(a+a^\dagger)=\phi_\text{zpf}(a+a^\dagger),\qquad \hat{n}=\sqrt{\dfrac{\hbar\omega}{16E_\text{C}}}(a-a^\dagger)=n_\text{zpf}\,i(a-a^\dagger).\end{equation}
		Since $\phi$ and $n$ are each other's conjugate, they satisfy the appropriate commutation relation
		\begin{equation}[\hat{\Phi},\hat{Q}]=i\hbar,\qquad [\hat{\phi},\hat{n}]=i.\end{equation}
		An analogy can be drawn to the position and momentum operators in the traditional derivation~\cite{griffiths2005introduction},
		\begin{equation}\hat{x}=\sqrt{\dfrac{\hbar}{2m\omega}}(a^\dagger+a),\qquad \hat{p}=i\sqrt{\dfrac{\hbar m \omega}{2}}(a^\dagger-a),\end{equation}
		where the ``mass'' or inertia is now proportional to the capacitance, $m\,\hat{=}\,C/4e^2$.
		In this new formulation in terms of charge and flux, the excitations of the circuit are separated by the geometric mean of their energies,
		\begin{equation}\hbar\omega=\hbar/\sqrt{LC}=\sqrt{8E_\text{L}E_\text{C}},\end{equation}
		where flux and charge can each appear as approximate quantization axis for a qubit system.
		
		It is important to note that the Hamiltonians described above have exactly equal level spacing between any two consecutive numbers of excitations: $\Delta E=\hbar/\sqrt{LC}$ for each level that we go up.
		As we saw in section~\ref{sec:qubit_from_it}, we need to introduce some degree of \emph{anharmonicity} $\alpha=\omega^{1\rightarrow2}-\omega^{0\rightarrow1}$, which we can achieve by replacing the inductor by a Josephson junction with energy
		\begin{equation}E_\text{J}=\dfrac{I_\text{c}\Phi_0}{2\pi}=\dfrac{I_\text{c,0}\Phi_0}{2\pi}\,\cos(\phi).\end{equation}
		The energy contributed by this junction is proportional to the coupling strength between its two superconducting condensates, which in turn depends on their phase difference $\phi$.
		It is this sinusoidal dependence on $\phi$ that gives the Hamiltonian its non-quadratic anharmonicity,
		\begin{equation}\begin{array}{r@{\;}c@{\;}l}
			\Ha_\text{LC} 		& =	& 4E_\text{C}\,n^2+\dfrac{1}{2}E_\text{L}\,\Bred{\phi^2}\\
								& \Bgreen{\downarrow}	& \\
			\Ha_\text{transmon}	& =	& 4E_\text{C}\,n^2-\dfrac{I_\text{c,0}\Phi_0}{2\pi}\hspace{-1em}\underbrace{\Bgreen{\cos(\phi)}}_{\Bred{1-}\dfrac{\phi^2}{2!}+\dfrac{\phi^4}{4!}+\cdots},
		\end{array}\end{equation}
		where we discard the first term in the expansion since it is constant. 
		For small $\phi$, which is our usual operating point\footnote{For $\phi$ to be a reasonable quantum variable, the Josephson coupling needs to be stronger than other energy terms, e.g.~\cite{clark1980feasibility},
		\begin{equation}E_\text{J}>k_\text{B}T\quad\Rightarrow\quad I_\text{c,0}>\dfrac{2\pi k_\text{B}T}{\Phi_0},\end{equation}
		which for fridge temperatures of around $T=\SI{50}{\milli\kelvin}$ means that the critical current needs to be larger than \SI{2.1}{\nano\ampere}.
		Typical transmons~\cite{krantz2019quantum} with $\omega$=3--\SI{6}{\giga\hertz} and charging energies on the order of 100--\SI{300}{\mega\hertz} ($C\approx$0.1--\SI{0.4}{\pico\farad}) will need critical currents between and 1.2 and \SI{14}{\nano\ampere}.} (the spread in $n$ is much larger in transmons), higher orders in the cosine's Taylor expansion can be discarded, and we find an added term $\propto\phi^4$, which turns out to shift the second level by an amount $\alpha=-E_\text{C}$.
		See Fig.~\ref{fig:harmonic_potential} for a visual representation.
		While circuits with substantial $E_\text{C}$ have historical importance~\cite{vion2002manipulating,cottet2002implementation}, in what follows we will focus on designs with strong capacitive coupling and larger Josephson energies, such that $E_\text{J}\gtrsim50E_\text{C}$~\cite{koch2007charge}.
		
	\subsection{\label{sec:gatemons}Gatemons}

		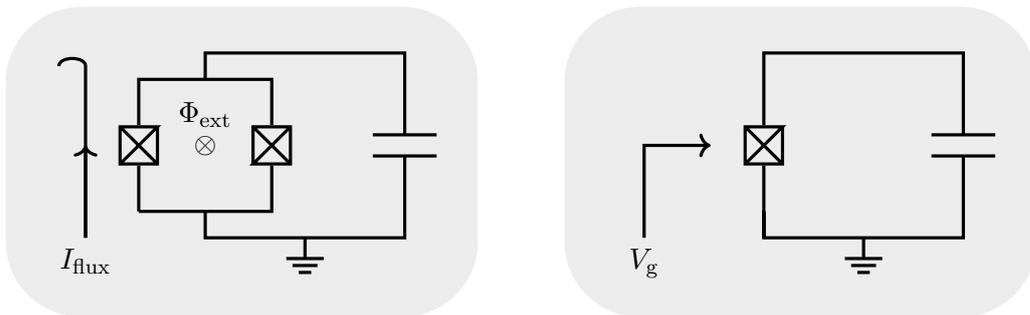
\begin{figure}
			\centering
			\begin{subfigure}[t]{0.5\textwidth}
				\centering
				\begin{circuitikz}[circuit ee IEC,x=1.75cm,y=1.75cm,very thick]
					\fill[rounded corners=1cm,gray!15!white] (-1,-0.8) rectangle (2.55,1.55);
					\draw (0,0) to [JJ cap] (0,1) -- (1,1) to [JJ cap] (1,0) -- (0,0);
					\node at (0.5,0.5) {$\otimes$};
					\draw (0.5,1) -- (0.5,1.2) -- (2,1.2) to [C] (2,-0.2) -- (0.5,-0.2) -- (0.5,0);
					\draw (1.25,-0.2) -- (1.25,-0.35) node[ground={pos=1.4},rotate=-90]{};
					\node[anchor=south,yshift=0.1cm] at (0.5,0.5) {$\Phi_\text{ext}$};
					\draw[->] (-0.4,-0.2) -- (-0.4,0.5) node[pos=0,anchor=north]{$I_\text{flux}$};
					\draw (-0.4,-0.2) -- (-0.4,1.1) to[out=90,in=90] (-0.6,1.1);
				\end{circuitikz}
			\end{subfigure}\begin{subfigure}[t]{0.5\textwidth}
				\centering
				\begin{circuitikz}[circuit ee IEC,x=1.75cm,y=1.75cm,very thick]
					\fill[rounded corners=1cm,gray!15!white] (-1,-0.8) rectangle (2.55,1.55);
					\draw (0.5,-0.2) to [JJ cap] (0.5,1.2) -- (2,1.2) to [C] (2,-0.2) -- (0.5,-0.2) -- (0.5,0);
					\draw (1.25,-0.2) -- (1.25,-0.35) node[ground={pos=1.4},rotate=-90]{};
					\draw[->] (-0.4,-0.2) -- (-0.4,0.5) -- (0.1,0.5);
					\node[anchor=north] at (-0.4,-0.2) {$V_\text{g}$};
				\end{circuitikz}
			\end{subfigure}
			\caption{\label{fig:tunable_transmon}\B{(Left)} The traditional flux-tunable transmon design~\cite{koch2007charge}, where the current $I_\text{flux}$ in a nearby superconducting loop is used to tune $E_\text{J}/E_\text{C}$. \B{(Right)} The ``gatemon'' approach~\cite{de2015realization,larsen2015semiconductor}, where a gate voltage is used to adjust $E_\text{J}$ instead.}
		\end{figure}
	
		We discussed earlier in section~\ref{sec:two_qubit_gates} that two superconducting qubits can be capacitively coupled to each other when their impedances ($Z=\sqrt{8E_\text{C}/E_\text{J}}$) are made to match for a limited amount of time\footnote{In more general terms, an impedance mismatch is when two things take a different amount of time to compress and then relax, which leads to a less efficient collision.}.
		To achieve this, either the capacitance or the Josephson coupling of one or both of the qubits needs to be changed.
		While the capacitance is literally hardwired into the circuit, we can quite easily control the effective Josephson coupling of two parallel junctions by changing the magnetic field that is imposed on the area between them, see Fig.~\ref{fig:tunable_transmon} (left).
		In such a SQUID geometry, the superconducting condensate will enforce quantization of the penetrating flux towards the nearest integer multiple of the flux quantum by generating clockwise or anti-clockwise currents.
		So while the maximum critical current $I_\text{c,0}$ is fixed by the physical dimensions of the two junctions, the effective critical current $I_\text{c}=I_\text{c,0}|\cos\phi_\text{ext}|$ and therefore the effective coupling can be tuned~\cite{dicarlo2009demonstration}:
		\begin{equation}\Ha_\text{transmon}=4E_\text{C}\,n^2-\dfrac{I_\text{c,0}\Phi_0}{2\pi}\cos(\phi\Bgreen{\,+\,\phi_\text{ext}}).\end{equation}
		This design has its drawbacks however, as flux lines of nearby qubits can interact and need to be corrected for each other, and currents linear in the number of qubits will need to be supplied to the chip (as we calculated before, kilo-Amps in the case of megaqubit devices).
		This is why some groups~\cite{de2015realization,larsen2015semiconductor} have started exploring the option of replacing the SQUID and flux line by a single, non-dissipative gate-tuned semiconducting junction, shown in Fig.~\ref{fig:tunable_transmon} (right).
		This gatemon design is what is pursued in this thesis as well.
		
		While the above-mentioned implementations have been very successful in demonstrating proof of principle with relatively long coherence times, both relied on InAs nanowires for their semiconducting junctions.
		This choice is well-motivated by the absence of any Schottky barriers and the resulting high transparency at the S/Sm interfaces~\cite{doh2005tunable,ebel2005supercurrent}, especially when the Al/InAs interface can be grown epitaxially~\cite{krogstrup2015epitaxy}, but fabrication techniques involving nanowires have so far been difficult to scale.
		This barrier to scalability can be overcome by lithographically patterning the semiconducting junctions in planar geometries instead~\cite{vigneau2019germanium,aggarwal2020enhancement}.
		
\begin{figure}
			\centering			
			\begin{tikzpicture}[x=1.4cm,y=1.4cm]
			\draw[fill={gray!30!white}] (0,2) to[out=0,in=130] (1,1.6) -- (1,0) -- (0,0) -- cycle;
			\draw[fill={gray!30!white}] (1,0.4) to[out=-50,in=180] (2,0) -- (3,0) -- (3,0) -- (1,0) -- cycle;
			\draw[fill={gray!30!white}] (3,0.4) to[out=-130,in=0] (2,0) -- (1,0) -- (1,0) -- (3,0) -- cycle;
			\draw[fill={gray!30!white}] (4,2) to[out=180,in=50] (3,1.6) -- (3,0) -- (4,0) -- cycle;
			
			\draw[thick] (1,0) -- (1,2);
			\draw[thick] (3,0) -- (3,2);
			\draw (0.5,0.75) node[anchor=south]{S};
			\draw (2,0.75) node[anchor=south]{Sm};
			\draw (3.5,0.75) node[anchor=south]{S};
			
			\draw[style=thick] (1.35,2.5) -- (1.35,2) -- (2.65,2) -- (2.65,2.5);
			\draw (2,2) node[anchor=south] {$V_\text{g}>0$};
			\draw (2,2.05) node[anchor=north] {$+$};

			\draw[fill={gray!30!white}] (5,2) to[out=0,in=110] (6,1.2) -- (6,0) -- (5,0) -- cycle;
			\draw[fill={gray!30!white}] (6,0.8) to[out=-50,in=175] (8,0.1) -- (8,0) -- (6,0) -- cycle;
			\draw[fill={gray!30!white}] (8,0.8) to[out=-130,in=5] (6,0.1) -- (6,0) -- (8,0) -- cycle;
			\draw (6,0.8) to[out=-50,in=175] (8,0.1);
			\draw[fill={gray!30!white}] (9,2) to[out=180,in=80] (8,1.2) -- (8,0) -- (9,0) -- cycle;
			
			\draw[thick] (6,0) -- (6,2);
			\draw[thick] (8,0) -- (8,2);
			\draw (5.5,0.75) node[anchor=south]{S};
			\draw (7,0.75) node[anchor=south]{Sm};
			\draw (8.5,0.75) node[anchor=south]{S};
			
			\draw[style=thick] (6.35,2.5) -- (6.35,2) -- (7.65,2) -- (7.65,2.5);
			\draw (7,2) node[anchor=south] {$V_\text{g}<0$};
			\draw (7,2.05) node[anchor=north] {$-$};
			
			\draw[->] (0,-0.2) -- (0,2.3) node[anchor=south] {$\Delta(x)$};
			\draw[->] (-0.2,0) -- (9.2,0) node[anchor=west] {$x$};
			\node[white,anchor=east] at (-0.2,0) {$x$};
			\end{tikzpicture}
			\caption{\label{fig:cartoon_order_parameter}A cartoon illustration of the effect of applying a gate voltage to a semiconducting barrier between two superconductors. Shown is the superconducting order parameter, or effective gap, throughout the junction: its induction in the semiconducting (Sm) channel is called the proximity effect, while the suppression within the superconducting leads is the inverse proximity effect. There are two separate processes through which the proximity effect can be enhanced~\cite{clark1980feasibility}: the barrier transparency can be increased, or the conductive and diffusive properties of the channel itself can be changed.}
		\end{figure}
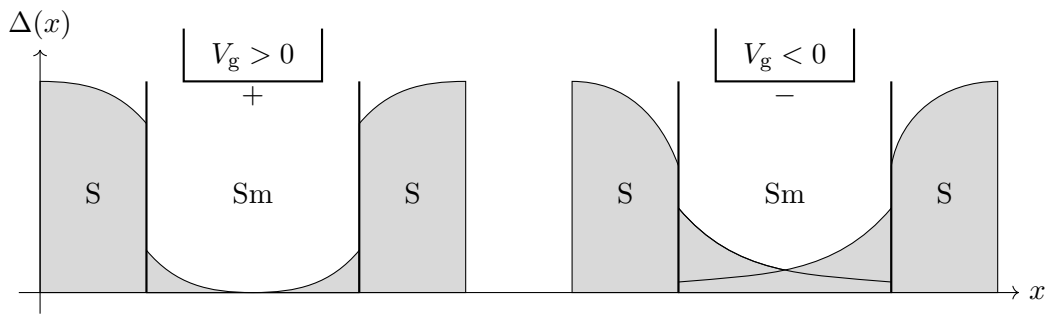

		As shown in Fig.~\ref{fig:cartoon_order_parameter}, the electrostatic field generated by the gate voltage (here shown for a p-doped channel) will attract charge carriers (holes in this case).
		In short junctions, where the critical current is proportional to the normal-state conductance~\cite{clark1980feasibility}, this increase in carrier density will strengthen the Josephson coupling by $E_\text{J}\propto G_N\propto n^{2/3}$.
		In longer junctions, where the proximity effect decays exponentially over a length scale roughly proportional to the carrier density~\cite{volkov1996effect}, this dependence can be even stronger.
		By attracting carriers, the electric field also lifts up or pushes down the Fermi level inside the semiconductor, changing the mismatch in work function with the metallic superconducting leads, and thus modifying the height of the Schottky barrier.
		Increasing the carrier density will also reduce the width of the barrier by more effectively shielding the charges on the metallic side.
		This provides an additional means of modifying the critical current by reducing the tunnel barrier in non-Ohmic Schottky-barrier devices.
		Taken together, these various effects mean that though phenomenological and qualitative understanding exists, the exact relationship between the applied voltage and the critical current is device-specific, and can at best be determined only experimentally.
		We therefore give the gatemon Hamiltonian only in its general form,
		\begin{equation}\Ha_\text{gatemon}=4E_\text{C}\,n^2-\dfrac{\Phi_0}{2\pi}\,\Bgreen{I_\text{c}(V_\text{g})}\,\cos(\phi),\end{equation}
		and will explore in the coming sections these diverse phenomena of superconducting transport that together determine $I_\text{c}(V_\text{g})$.
		
\section{\label{sec:superconducting_transport}Superconducting transport across semiconducting junctions}
	
\begin{figure}
		\centering
		\includegraphics[width=0.6\textwidth]{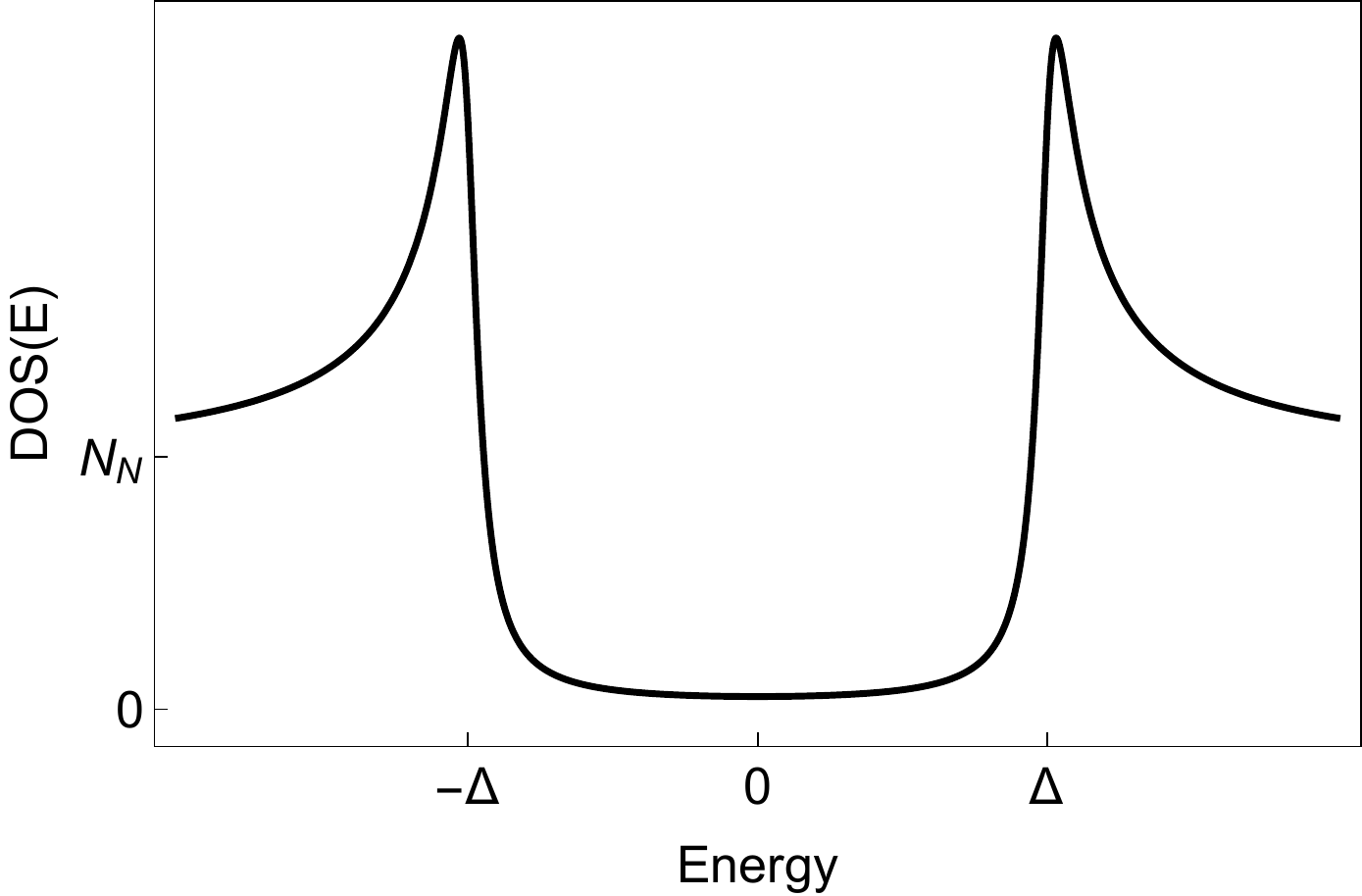}
		\caption{\label{fig:dynes_plot}The quasiparticle density of states in a superconductor as a function of energy relative to the Fermi level, calculated from eq.~\eqref{eq:dynes_dos}. The Dynes parameter $\Gamma$ is set at $5\%$ of the superconducting gap.}
	\end{figure}

	Within the superconductors on either side of the Josephson junction, there is a phonon-mediated attractive force between electrons that provides entangled states~\cite{hofstetter2009cooper} with lower energy~\cite{bardeen1957theory,bardeen1957microscopic}.
	It was shown that due to this attractive interaction, the elementary quasiparticle excitations (``Bogoliubons''\footnote{Evidence for the existence of these combined particle-hole excitations~\cite{bardeen1961chapter} comes from Cooper pairs decaying not just into two electrons, but into both electrons \emph{and} holes~\cite{beasley2010michael}.}) acquire energies of
	\begin{equation}E_{\B{k}}=\sqrt{\xi_{\B{k}}^2+|\Delta|^2},\end{equation}
	where $\xi_{\B{k}}$ is the kinetic plane-wave energy (usually measured from the Fermi level), and $\Delta$ the energy associated with the coupling, that also determines the energy band within which these quasiparticles are expected to form Cooper pairs~\cite{cooper1956bound}.
	Since the number of states per unit momentum is the same whether the material is superconducting (S) or not (N, for ``normal''), we can write
	\begin{equation}\dfrac{dN_\text{S}}{dk}=\dfrac{dN_\text{N}}{dk},\end{equation}
	and because in the normal state $\Delta=0$, we can also assume that $E_{\B{k}}=\xi_{\B{k}}$.
	Then, after multiplying both sides by $dk/dE$ and integrating with respect to energy, we find
	\begin{equation}\dfrac{dN_\text{S}}{dE}\dfrac{dE}{dk}=\dfrac{dN_\text{N}}{d\xi}\dfrac{d\xi}{dk}\quad\Rightarrow\quad N_\text{S}(E)=N_\text{N}(E)\,\dfrac{d\xi}{dE}=N_\text{N}\,\dfrac{d\sqrt{E^2-\Delta^2}}{dE}.\end{equation}
	We can thus derive the density of quasiparticle states per unit energy, where we'll introduce the Dynes parameter $\Gamma$~\cite{dynes1984tunneling}, such that $E\rightarrow E-i\Gamma\Delta$, to account for inelastic scattering events\footnote{In case you're curious about the appearance of the absolute value: this is to make sure that we have a positive number of states at energies below the Fermi level. As far as I can tell, the origin of the missing sign is in the substitution of $\sqrt{h_\text{k}^2}\rightarrow h_\text{k}$ in eq.~(2.41) of Ref.~\citenum{bardeen1957theory}.}:
	\begin{equation}\label{eq:dynes_dos}N_\text{S}=N_\text{N}\left|\text{Re}\left(\dfrac{E-i\Gamma}{\sqrt{(E-i\Gamma)^2-\Delta^2}}\right)\right|.\end{equation}
	This density of states, shown in Fig.~\ref{fig:dynes_plot}, has a energy gap of $2\Delta=3.53k_\text{B}T_\text{c}$ at the Fermi level~\cite{glover1956transmission,glover1957conductivity,bardeen1957theory}, which means that no states for single quasiparticles are available.
	
	When a metal, where the density of states is to good approximation flat around the Fermi energy, contacts this superconductor, it is therefore impossible for single electrons to enter individually.
	Instead, transport across the interface happens through a process called \emph{Andreev reflection}~\cite{andreev1964thermal}, where an incoming electron is reflected back into the metal as a hole that coherently traces back the electron's path, while a Cooper pair that emerges on the other side ensures that charge is conserved.
	This process effectively leads to a ``leaking'' of the superconductivity into the normal metal, referred to as the proximity effect, where it induces a smaller \emph{minigap}~\cite{belzig1996local,belzig1999quasiclassical}.
	
	If the metal is contacted on the other side by a second superconductor, there will now be two distinct ways that charge can be transported across the junction.
	Either the coherence of the electron and hole persist throughout, in which case the probability of entering the superconductor on the other side depends on the phase difference between the two condensates~\cite{andreev1966electron,belzig1999quasiclassical}, or the electron-hole pair decoheres, and the current is partially carried by regular individual quasiparticles.
	While the former creates the desired Josephson coupling between the two superconducting leads~\cite{cottet2002implementation}, the latter could add an undesirable energy relaxation mechanism.
	Additionally, at higher temperatures it becomes possible for quasiparticles to enter the superconductor directly without Andreev reflection, causing a different type of decoherence called quasiparticle poisoning~\cite{catelani2011relaxation,sun2012measurements}.
	
	In the doped silicon channels of the Schottky-barrier MOSFETs that we study, there are relatively many impurities, and we expect the transport through the junction to be diffusive rather than ballistic ($\ell_\text{mfp}\approx\SI{10}{\nano\meter}\ll L_\text{channel}\approx\SI{100}{\nano\meter}$).
	In this case the spatial decay of the proximity effect depends on how the diffusion time of a charge across the junction compares to the decoherence timescales of the many disruptive physical processes that cause inelastic scattering.
	Before we discuss this in more detail, we will first briefly review the principles of Andreev reflection and diffusion.
	
	\subsection{Andreev reflection and the proximity effect}
		
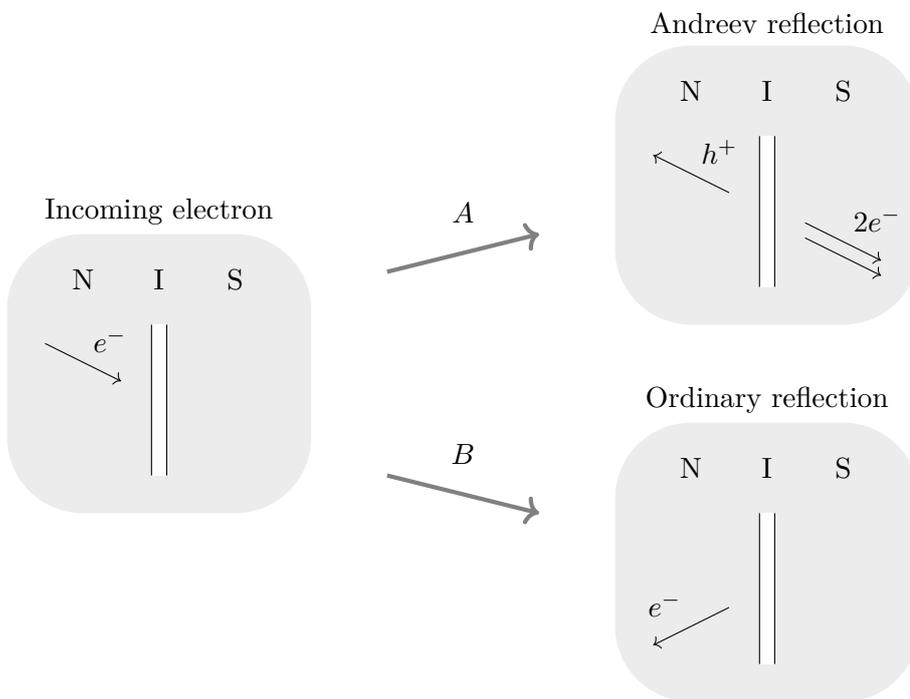
\begin{figure}
			\centering
			\begin{tikzpicture}
				\fill[rounded corners=1cm,gray!15!white] (-6,-1.5) rectangle (-2,2.2);
				\node at (-4,2.5) {Incoming electron};
				\node at (-5,1.6) {N};
				\node at (-4,1.6) {I};
				\node at (-3,1.6) {S};
				\fill[white] (-4.1,1) rectangle (-3.9,-1);
				\draw (-4.1,1) -- (-4.1,-1);
				\draw (-3.9,1) -- (-3.9,-1);
				\draw[->] (-5.5,0.75) -- (-4.5,0.25) node[midway,anchor=south west] {$e^{-}$};
				
				\draw[ultra thick,gray,->] (-1,1.7) -- (1,2.2) node[black,midway,anchor=south,yshift=0.25cm] {$A$};
				\draw[ultra thick,gray,->] (-1,-1) -- (1,-1.5) node[black,midway,anchor=south,yshift=0.25cm] {$B$};
				
				\fill[rounded corners=1cm,gray!15!white] (2,1) rectangle (6,4.7);
				\node at (4,5) {Andreev reflection};
				\node at (3,4.1) {N};
				\node at (4,4.1) {I};
				\node at (5,4.1) {S};
				\fill[white] (4.1,1.5) rectangle (3.9,3.5);
				\draw (3.9,3.5) -- (3.9,1.5);
				\draw (4.1,3.5) -- (4.1,1.5);
				\draw[->] (3.5,2.75) -- (2.5,3.25) node[midway,anchor=south west] {$h^+$};
				\draw[->] (4.5,2.15) -- (5.5,1.65);
				\draw[->] (4.5,2.35) -- (5.5,1.85) node[midway,anchor=south west] {$2e^-$};

				\fill[rounded corners=1cm,gray!15!white] (2,-0.3) rectangle (6,-4);
				\node at (4,0) {Ordinary reflection};
				\node at (3,-0.9) {N};
				\node at (4,-0.9) {I};
				\node at (5,-0.9) {S};
				\fill[white] (4.1,-1.5) rectangle (3.9,-3.5);		
				\draw (3.9,-1.5) -- (3.9,-3.5);
				\draw (4.1,-1.5) -- (4.1,-3.5);
				\draw[->] (3.5,-2.75) -- (2.5,-3.25) node[midway,anchor=south east] {$e^-$};
			\end{tikzpicture}
			\caption{\label{fig:andreev_reflection}For incident particles with energies below the gap, $E<\Delta$, two processes can occur: Andreev reflection ($A$) and ordinary reflection ($B$). Their respective probabilities are a function of $E/\Delta$ and the barrier strength $Z$~\cite{blonder1982transition}. Note that in the former case, the hole is \B{not} reflected specularly, but instead traces back the path of the incident electron~\cite{andreev1964thermal}.}
		\end{figure}

		After a condensate has formed inside a superconductor, no single-particle states near the Fermi level will be available anymore.
		This seemingly poses a problem for an incoming electron or hole from an adjacent normal material, which at low temperatures necessarily moves near the Fermi energy.
		Luckily, a second-order process is possible, whereby the incident electron can form a Cooper pair as it enters the superconductor by picking up an additional electron near the interface, at the cost of back-scattering a hole with opposite spin into the normal lead.
		This is illustrated in Fig.~\ref{fig:andreev_reflection}, which assumes that the voltage bias across the insulating barrier (I) is less than the pair potential inside the superconductor (S), such that the incoming electron has an energy $E<\Delta$.

		As an historical note, it is interesting that the notion of a proximity effect in fact predates this description.	
		The leaking of the pair potential across the interface can already be derived with a phenomenological approach\footnote{By ``phenomenological'' we mean assuming that whatever causes superconductivity, it is described by a wave function, and leave it at that.
		This simple approach~\cite{landau1950k,landau2009theory} recovers many of the relevant features~\cite{tinkham2004introduction}.}, which also gives rise to different coherence lengths on the normal and superconducting sides, over which the proximity effect and its inverse decay~\cite{de1964boundary}, 
		The microscopic description was found later when Andreev solved the Gor'kov equations~\cite{gor1958energy}\footnote{The Schr\"{o}dinger equation with interactions only in a narrow band around the Fermi energy, i.e. the ``quasiclassical'' approximation that is in general valid for superconductors.} for this system~\cite{andreev1964thermal}.
		He assumed solutions of the form
		\begin{equation}\begin{array}{r@{\;}c@{\;}l@{\quad}l}
			f	& =	& e^{i(\B{k}\cdot\B{r}-Et/\hbar)}\,\eta(\B{r})&\text{(particle)},\\\\
			\varphi	& =	& e^{i(\B{k}\cdot\B{r}-Et/\hbar)}\,\chi(\B{r})&\text{(hole)},
		\end{array}\end{equation}
		where $\eta$ and $\chi$ are the envelopes of the wave packets~\cite{griffiths2005introduction}, and arrived at kinetic equations of the form
		\begin{equation}\begin{array}{l@{\;}l@{\;}c@{\;}l}
			\left(\dfrac{i\hbar v_\text{F}}{m}\,\B{n}\cdot\nabla+E\right)\eta&-i\Delta(\B{r})\chi	& =	& 0,\\\\
			\left(\dfrac{i\hbar v_\text{F}}{m}\,\B{n}\cdot\nabla-E\right)\chi&-i\Delta(\B{r})\eta	& =	& 0,
		\end{array}\end{equation}
		with $\B{n}$ the direction of the incident particle.
		What is remarkable, is that when these are solved on the normal side, where $\Delta(\B{r})=0$, the solution is
		\begin{equation}\Pm{\eta\\\chi}=A\Pm{1\\0}e^{i\B{k_\text{\B{N}}}\cdot\B{r}}+B\Pm{0\\1}e^{-i\B{k_\text{\B{N}}}\cdot\B{r}},\end{equation}
		which are a particle and a hole moving with precisely opposite momenta,
		\begin{equation}\B{k_\text{\B{N}}}=\B{n}E/\hbar v_\text{F}.\end{equation}
		In the superconductor we use instead $\Delta(\B{r})=\Delta_0$, and obtain
		\begin{equation}\Pm{\eta\\\chi}=\dfrac{C}{\sqrt{2}}\Pm{\sqrt{1+\hbar v_\text{F}\B{n}\cdot\B{k_\text{\B{S}}}/E}\\-i\sqrt{1-\hbar v_\text{F}\B{n}\cdot\B{k_\text{\B{S}}}/E}}e^{i\B{k_\text{\B{S}}}\cdot\B{r}},\end{equation}
		with a momentum ($+$ for an incident particle, $-$ for an incident hole)
		\begin{equation}\B{k_\text{\B{S}}}\cdot\B{n}=\dfrac{\pm 1}{\hbar v_\text{F}}\,\sqrt{E^2-\Delta_0^2}\,.\end{equation}
		This is a Cooper pair with a momentum that in the plane of the interface is identical to that of the incident particle, but that in the orthogonal direction depends on the energy of that particle relative to the gap $\Delta$.
		Instead of seeing charge transmission across the interface forbidden by the lack of available states, Andreev found that it could in fact be enhanced to above unity!
		
		Although it was not yet possible within this treatment to find the incidence, reflection and transmission coefficients $A$, $B$ and $C$ for arbitrary energy $E$, it did establish the now-famous mechanism that is truly quantum mechanical in nature.		
		The predicted non-specular reflection (``back-scattering'' or ``retroreflection'') of a hole, and the resulting doubling of the current~\cite{beenakker1992quantum}, was confirmed by an experiment with a quantum point contact on the back of a silver crystal (N) coated with lead (S)~\cite{benistant1985angular}.
		Later simulations of this experiment, and variants thereof, further corroborated the tracing out by the hole of the incident electron's path, and found that this back-focusing is independent of the angle of incidence, any particularities of the interface, and even of the number of impurities in the normal material~\cite{de1994andreev,de1994andreevsup}.
		A more quantitative description of the scattering rates was given in 1982 by Blonder, Tinkham and Klapwijk~\cite{blonder1982transition}, which we will treat in more detail in chapter~\ref{sec:jofets}.
		
		The situation is somewhat more complicated in semiconductors, for which there is no single overarching description~\cite{likharev1979superconducting,kroger1980josephson}.
		While highly (``degenerately'') doped semiconductors may behave as metals~\cite{huang1974josephson}\footnote{``Degenerate'' refers to the overlap of orbitals of nearby donors or acceptors at high doping levels ($\gtrapprox\SI{1E18}{\per\centi\meter\cubed}$), and the resulting hybridization of their energy levels. Conductance \emph{increases} with temperature in non-degenerate semiconductors, while it \emph{decreases} in degenerate ones (which is why they are often referred to as ``metallic'').}, albeit with a much shorter coherence length~\cite{likharev1979superconducting}, materials with lower doping levels will be insulating after the dopants have frozen out~\cite{clark1980feasibility,simoen1989freeze}.
		An additional mechanism may provide a means of transport at intermediate doping.
		At sufficiently high impurity densities and low temperatures, paths across periodically spaced dopants will generate what are called ``resonance-percolation trajectories''~\cite{lifshitz1979tunnel}, which can become the dominant contribution to coherent transport~\cite{aslamazov1982resonant}.
		In general, the supercurrent through a semiconducting weak link will have to overcome both imperfect transmission at the interfaces, and decoherence in the channel itself~\cite{aslamazov1981temperature}.
		Intuitively, for short junctions or high doping, the critical current will be limited by the interfaces and depend mostly on the properties of the superconductors, while for longer junctions or lower doping it will depend mainly on the properties of the weak link~\cite{aslamazov1981temperature,kleinsasser1990crossover,volkov1996effect}.
		
		The clear advantage of semiconductors is that, unlike both insulators and metals, their properties can be altered with an electrostatic field; by accumulating carriers~\cite{aggarwal2020enhancement}, the diffusion can be enhanced~\cite{volkov1996effect} and the effective Schottky barrier width at the interfaces can be reduced~\cite{kleinsasser1990crossover}.
		Since Andreev reflection is a second-order process, this reduction of the interface barrier strength has an even greater impact on superconducting transport than it does on normal transmission.

	\subsection{\label{sec:diffusion_equation}Diffusion}
		
		Diffusion is important not just in the context of the proximity effect in dirty semiconductors, but underlies all physical processes where scattering happens on length scales much smaller than the system size.
		This is the case for solid state reactions such as silicidation and oxidation (to be discussed in chapter~\ref{sec:silicides}), where atoms move erratically over distances of a few ångstr\"om, as well as for electronic transport in most devices, where carriers typically travel tens of nanometers between collisions (chapter~\ref{sec:jofets}).

		\begin{figure}
			\centering
			\begin{subfigure}[c]{0.6\textwidth}
				\includegraphics[width=0.95\textwidth]{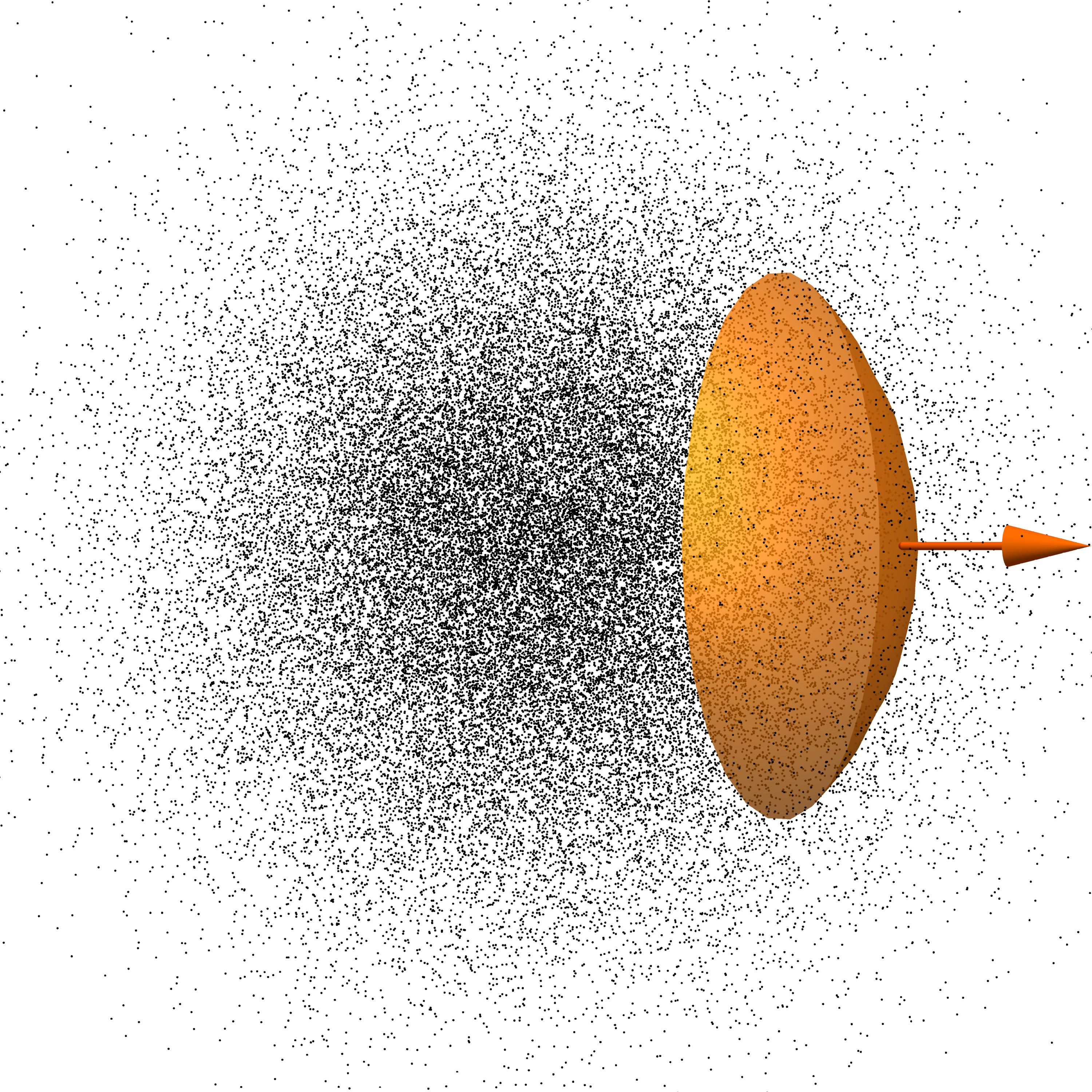}
				\hfill
			\end{subfigure}\begin{subfigure}[c]{0.4\textwidth}
				\hfill
				\includegraphics[width=0.95\textwidth]{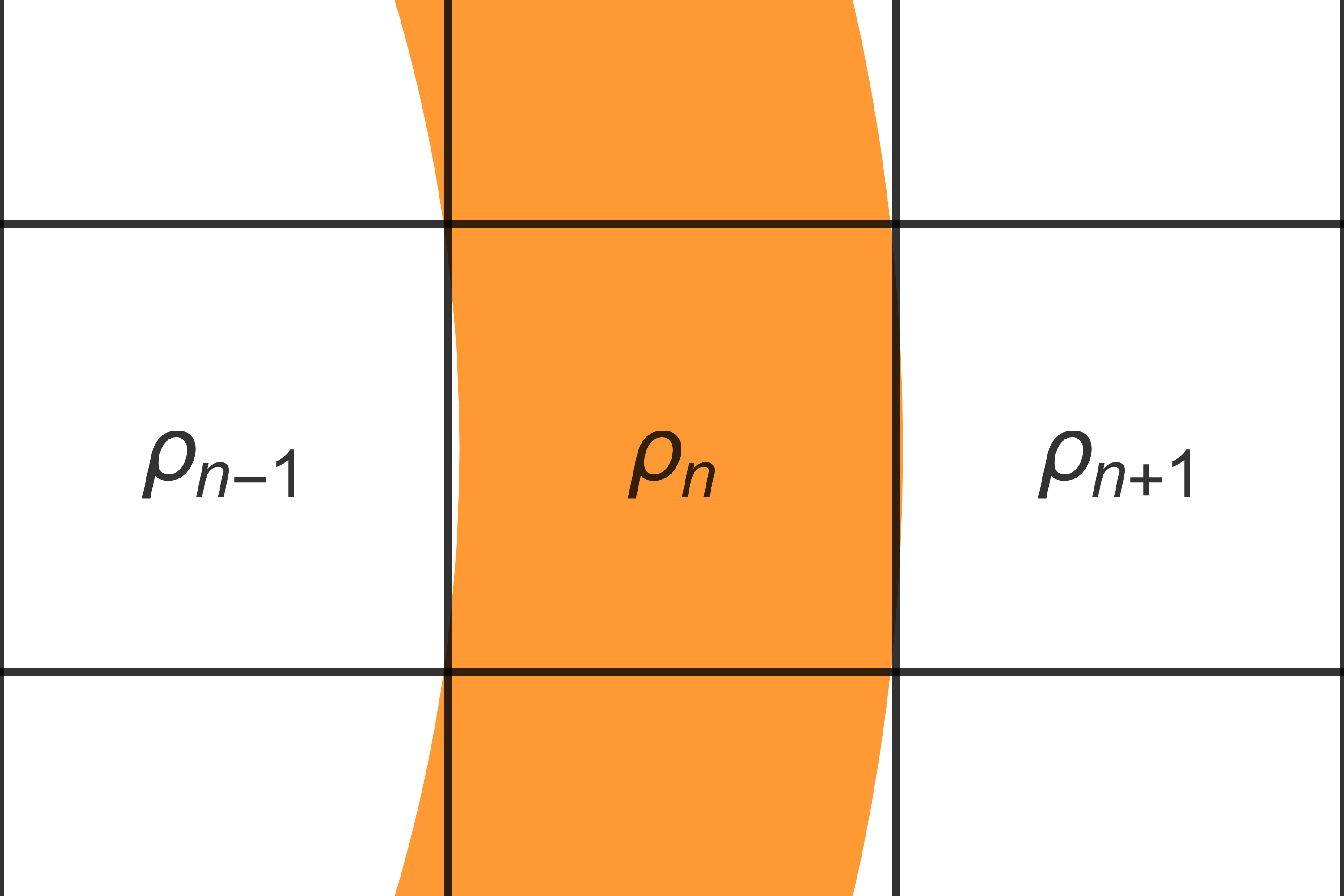}
			\end{subfigure}
			\caption{\label{fig:diffusion}\B{(Left)} When a collection of particles are allowed to diffuse from the origin, the density profile will be spherically symmetric. The concentration will be constant along some surface at fixed distance from the origin, and we can define a diffusion direction orthogonal to it. \B{(Right)} We can now figure out the time evolution of the system by analyzing a one-dimensional array of positions along the diffusion direction.}
		\end{figure}
		
		To illustrate this process, consider a large collection of particles that at time $t=0$ are placed at the origin, and are allowed to perform random walks for some time, such that they slowly spread out over space.
		A typical distribution is drawn on the left in Fig.~\ref{fig:diffusion}, where the section of a sphere indicates a plane of constant density.
		Since the density is constant along the indicated contour, we can assume that there is no net flow of particles to neighboring regions above and below, nor to those in and out of the page.
		The change in the density $\rho_n$ per time step $\Delta t$ can now be calculated from the densities in cells $n-1$ and $n+1$.
		If we choose our $\Delta t$ such that there is enough time for the density at point~$n$ to become the average of what the surrounding densities were at the previous time step, then the change in density is
		\begin{equation}\Delta_t\rho_n= \dfrac{1}{2}\left[(\rho_{n+1}-\rho_{n})-(\rho_{n}-\rho_{n-1})\right].\end{equation}
		These steps in time and space are of course derivatives,
		\begin{equation}
			\dfrac{\Delta_t\rho}{\Delta t} = \underbrace{\dfrac{1}{2}\,\dfrac{(\Delta x)^2}{\Delta t}}_{D}\,\dfrac{\Delta_x^2\,\rho}{(\Delta x)^2},
		\end{equation}
		where the prefactor $D$ on the right-hand side is called the \emph{diffusion constant}.
		We can rewrite this with continuous variables by letting $\Delta t,\;\Delta x\rightarrow 0$, such that\footnote{Note that the only difference in form between the diffusion equation and Schr\"odinger's equation for a free particle is a factor $i$ in front of the time derivative, accounting for the latter's phase evolution:
		\begin{equation*}i\hbar\,\dfrac{\partial\,\Psi}{\partial\,t}=-\dfrac{\hbar^2}{2m}\,\nabla^2\,\Psi.\end{equation*}}
		\begin{equation}\partial_t\rho_n=D\,\nabla^2\rho.\end{equation}
		For electrons we have
		\begin{equation}D=\dfrac{v_\text{F}\ell_e}{3},\end{equation}
		where $v_\text{F}$ is the Fermi velocity and $\ell_e$ the mean free path.
		
		The important thing to notice in this equation is that it contains a \emph{first} order derivative to time, but a \emph{second} order derivative to space: the change in the local density of particles is proportional to the \emph{difference in difference} in density between pairs of cells.
		We can thus immediately guess that the time it takes to diffuse a particle a certain distance $L$ is proportional to the \emph{square} of that distance, $t\propto L^2$, and vice versa; that the spread of a collection of particles grows only as the square root of time.
		A more hand-waving and intuitive argument can be made based on the statistics of random walks alone.
		After a particle has taken $N$ steps of size $\ell$, it will have covered a path of total length $N\ell$, and on average it will end up right where it started.
		But since the uncertainty in the average of $N$ uncorrelated values is proportional to $1/\sqrt{N}$, our estimate of the \emph{sum} of all the steps in the particle's path will be accurate only to the order of $N\ell/\sqrt{N}=\sqrt{N}\,\ell$.
		Roughly speaking, we thus expect the particle to travel a net distance $\sqrt{N}\,\ell$ during $N$ steps in time\footnote{Random walks and their square laws are surprisingly ubiquitous in science and math, appearing even in the remote field of Economics. For example, two Bayesians $A$ and $B$ who have disagreeing prior estimates $X_A$ and $X_B$ of some $X\in[0,1]$ because they each have different information, will each update their estimates upon hearing the other's (and thus learning about their information) in a random-walk fashion, such that they need $1/\epsilon^2$ updates to agree within $|X_A-X_B|\leq\epsilon$~\cite{aaronson2005complexity}.
		Even more interestingly, \emph{quantum} random walks \emph{do not} follow this square law, and so can explore graphs in linear time, providing a quadratic speedup compared to classical algorithms for stochastic processes~\cite{grover2015quantum,childs2020power}.}: $L\propto\sqrt{t}$.
		Inversely, if the step size $\ell$ were to decrease, the time needed to travel a distance $L$ would go up as $t\propto(L/\ell)^2$.
	
		We will also encounter this relationship in the formation of silicides during heating: by checking if the growth rate of a new phase is linear in, or proportional to the square root of the annealing time, we can determine if the reaction is limited by the nucleation of the new compound at the interface or by the diffusion of the atoms through the forming silicide.
	
	\subsection{Decoherence}
		
		While an electron tracing back its own path would reverse any phase shift it had acquired, the back-scattered hole will compound it, so that after a path of length $L$ covering an area\footnote{Here, $A$ would be the net area enclosed by the path after the start and end points are connected by a straight line. To see why, imagine instead of the hole tracing back the same path, the electron moving back along the trajectory mirrored in that line. Clearly that path would enclose $2A$, and half the phase shift would be acquired along the trajectory we are considering.} $A$ orthogonal to the field $\B{B}$, the electron and hole phases would be off by~\cite{van1992excess}\footnote{The prefactor $2\pi$, rather than $4\pi$ as used in the cited work, is due to our choice for the flux quantum of $\Phi_0=h/2e$ instead of $\Phi_0=h/e$.}
		\begin{equation}\Delta\phi=\dfrac{2EL}{\hbar v_\text{F}}+2\pi\,\dfrac{\B{B}\cdot\B{\hat{n}}A}{\Phi_0},\quad\text{with}\quad\Phi_0\equiv\dfrac{h}{2e}.\end{equation}
		Here we assume for simplicity that both are at energies $E\ll\Delta$, and ignore the small phase shift acquired during the reflection itself~\cite{de1994andreev,de1994andreevsup}.
		If only a single trajectory is involved, the coupling could be enhanced again as $\Delta\phi$ approaches $2\pi$.
		More realistically, many different paths contribute, each with different lengths $L$ for the same depth into the semiconductor, and this relative dephasing will cause the effective coupling to decay exponentially inside the semiconductor, where the order parameter will evolve as
		\begin{equation}|\psi|^2=\psi_{h^+}^\dagger\psi_{e^-}\propto e^{-\braket{\Delta\phi}}.\end{equation}
		Apart from pure dephasing, decoherence can be caused by a broadening of the individual energy distributions of either particle due to a coupling to other degrees of freedom.
		
\begin{figure}
			\centering
			\includegraphics[width=\textwidth]{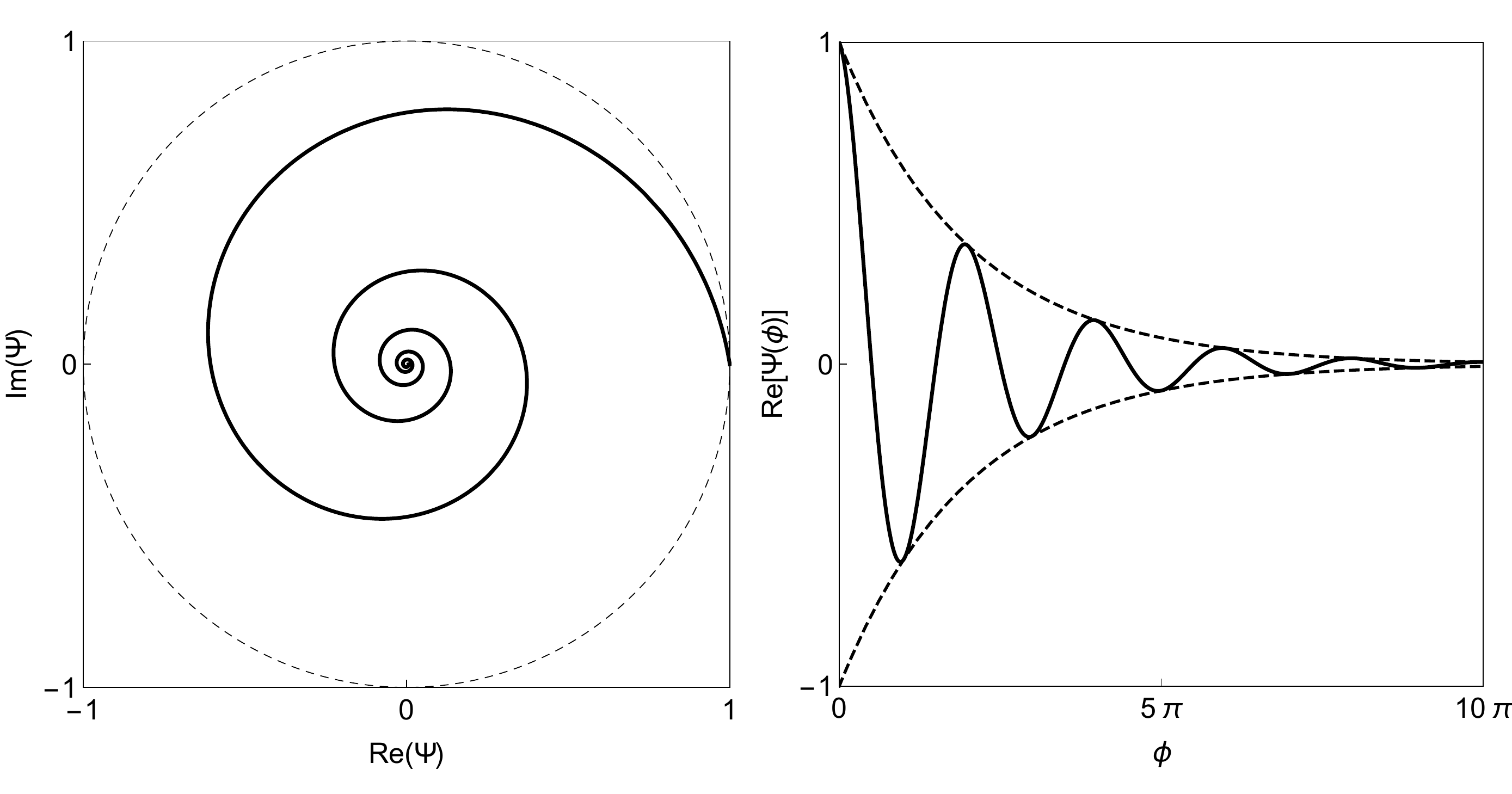}
			\caption{\label{fig:decoherence}Decoherence of a wave function $\Psi=e^{i(\phi+i\Gamma)}$. \B{(Left)} the rotation of the function in the complex plane, which can be also taken as the equatorial plane of the Bloch sphere for a dephasing qubit. \B{(Right)} The decay of the real part of the wave function with growing $\phi$.}
		\end{figure}

		For the purpose of what we discuss here, we can reasonably approximate the electrons and holes that travel through the semiconducting channel as plane waves~\cite{griffiths2005introduction},
		\begin{equation}\label{eq:plane_wave}\Psi(\B{r},t)=e^{i(\B{k}\cdot\B{r}-E(\B{k})t)},\end{equation}
		which are eigenstates of the momentum operator.
		Each of these momentum states has this same form $e^{i\phi}$, which Euler taught us to decompose into a sine and a cosine,
		\begin{equation}e^{i\phi} = \Sum_{n=0}^\infty\dfrac{(i\phi)^n}{n!}=\Sum_{n=0}^\infty\, \underbrace{(-1)^n\,\dfrac{\phi^{2n}}{(2n)!}}_{\cos(\phi)} + \,\underbrace{i\,(-1)^n\,\dfrac{\phi^{2n+1}}{(2n+1)!}}_{i\sin(\phi)},\end{equation}
		which oscillate out of phase and orthogonal to each other in the complex plane.
		If the energy term in eq.~\eqref{eq:plane_wave} is completely real, it will only determine the rotation in the complex plane, but if there are imaginary terms such that $\B{k}\cdot\B{r}+Et=A+iB$, we find that
		\begin{equation}e^{i(A+iB)}=e^{iA}e^{-B}\quad\text{and}\quad|\Psi|^2=(e^{iA}e^{-B})(e^{-iA}e^{-B})=e^{-2B},\end{equation}
		and the wave function will be suppressed exponentially.
		This is shown graphically in Fig.~\ref{fig:decoherence}.

		\emph{Elastic} scattering events, such as the majority of those causing diffusion in a silicon channel (collisions off heavy objects such as dopants or grain boundaries), add only real terms to the energy, making it possible for a particle to change momentum direction hundreds or thousands of times before coherence is lost~\cite{van1992excess}.
		The decoherence comes from \emph{inelastic} scattering events, where carriers lose some of their kinetic energy as they interact with lighter objects (such as other carriers~\cite{dynes1984tunneling}).
		In such inelastic interactions, two quantum systems couple their energy degrees of freedom, speeding up interactions with the rest of the environment, which randomize their wave functions' phases.
		Decoherence due to phase randomization is hard to appreciate in wave function notation, so consider the simple density matrix of a superposition
		\begin{equation}\begin{array}{r@{\;}c@{\;}l}
			\ket{\Psi}	& =	& \alpha\ket{0}+\beta\ket{1}=a\ket{0}+be^{i\phi}\ket{1},\\\\
			\rho		& =	& \ket{\Psi}\bra{\Psi}=\Pm{a^2&ab\,\Bred{e^{-i\phi}}\\ab\,\Bred{e^{i\phi}}&b^2}.
		\end{array}\end{equation}
		Once the phase $\phi$ randomizes, the off-diagonal elements average to zero, and all that remains is a classical probability distribution on the diagonal: no wave function can reconstruct a purely diagonal density matrix with multiple nonzero entries.
		Moreover, since it is the off-diagonal components of the density matrix that describe the time evolution, the particle has ceased to oscillate between its basis states.
		
		In short, the electron-hole correlations will decohere due to an accumulating phase difference even if their constituent quasiparticles remain coherent.
		Inelastic scattering events that couple the degrees of freedom of the quasiparticle to the environment through energy exchange add an additional source of decoherence~\cite{zeh1970interpretation,schlosshauer2019quantum} that prevents them from interfering constructively.
		Together, dephasing and inelastic scattering limit coherent transport across the junction.
		
	\subsection{\label{sec:energy_length_time}Energy, length and time scales}
		
\begin{figure}
			\centering
			\begin{subfigure}[b]{0.45\textwidth}
				\centering
				\includegraphics[width=0.9\textwidth]{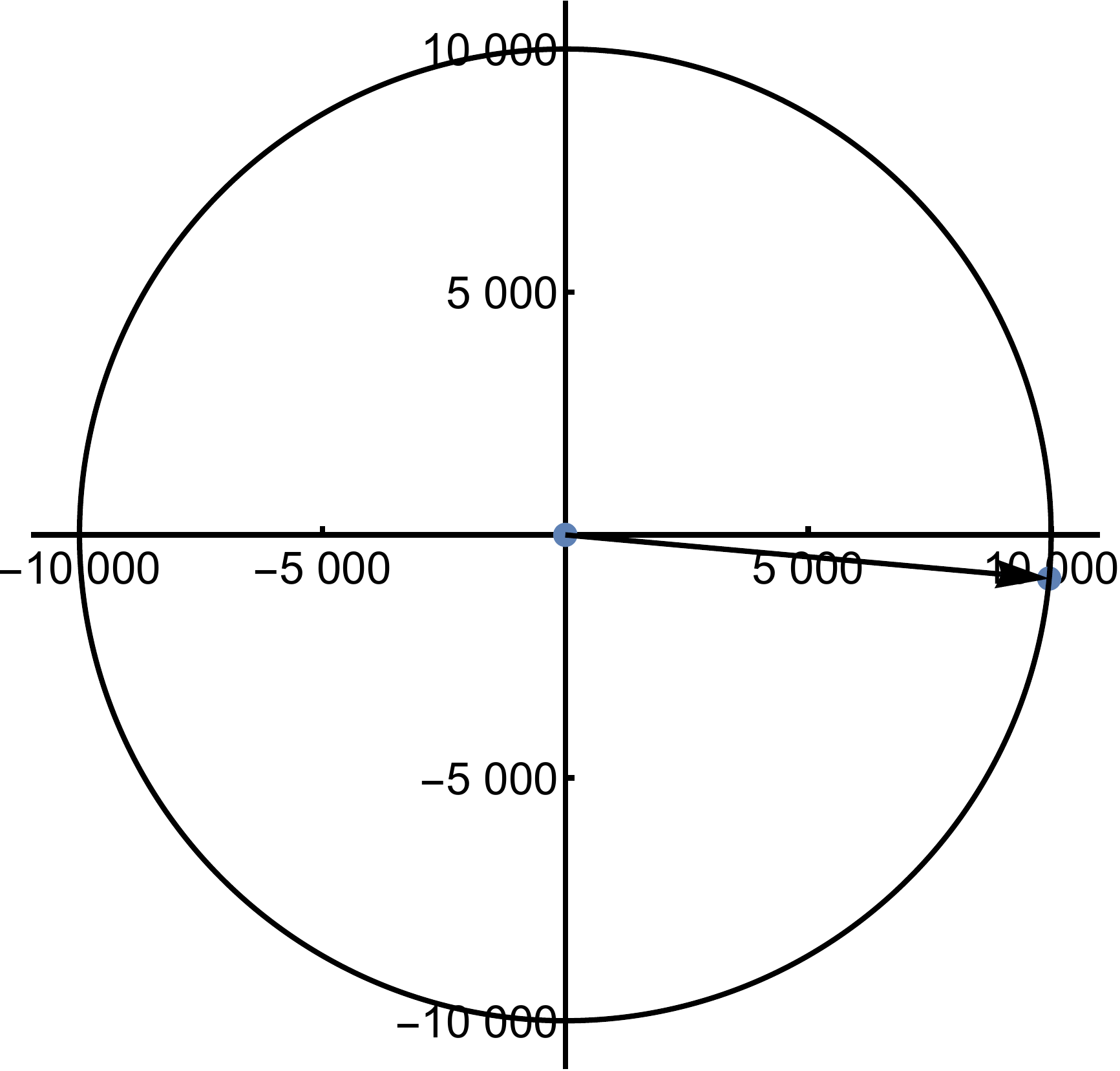}
			\end{subfigure}\begin{subfigure}[b]{0.45\textwidth}
				\centering
				\includegraphics[width=0.9\textwidth]{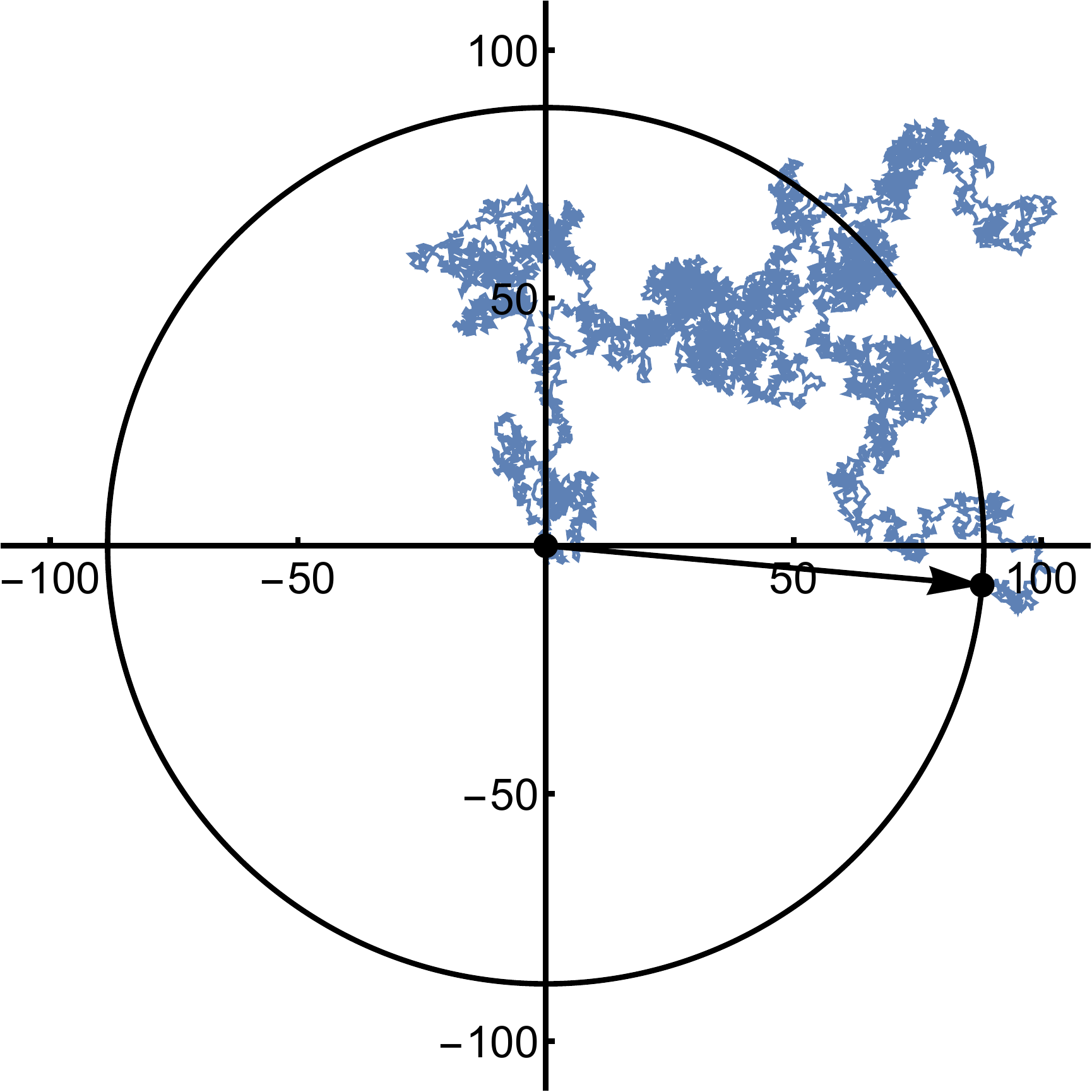}
			\end{subfigure}
			\caption{\label{fig:ballistic_vs_diffusion}The motion of a particle at a speed of one unit distance per unit time, over $10\,000$ units of time. \B{(Left)} Ballistic transport is free of scattering, so the distance traveled is proportional to the time $t$. \B{(Right)} In the diffusive limit, the direction of motion is randomized many times, and the average distance from the origin is proportional to $\sqrt{t}$ (there is a statistical prefactor $<1$). The same power law holds for diffusion in three dimensions.}
		\end{figure}

		If no scattering occurs at all before a particle traverses the channel, then the total distance that it covers is simply proportional to the time:
		\begin{equation}L_\text{ballistic}=\vF t.\end{equation}
		At the other extreme, if the scattering occurs over distances much smaller than the channel length and movement becomes Brownian, then the particle is said to diffuse, and the length it is expected to have traveled is proportional to the \emph{square root} of the time:
		\begin{equation}L_\text{diffusive}=\sqrt{D\,t},\end{equation}
		where the diffusion constant is defined as $D=\vF\Le/3$. The difference between these two limits is illustrated in Fig.~\ref{fig:ballistic_vs_diffusion}.
		Consequently, the shorter the mean free path $\Le$, the longer it takes to cover some distance $L$, one of the concerns when increasing the doping.
		
\begin{figure}
			\centering
			\includegraphics[width=0.6\textwidth]{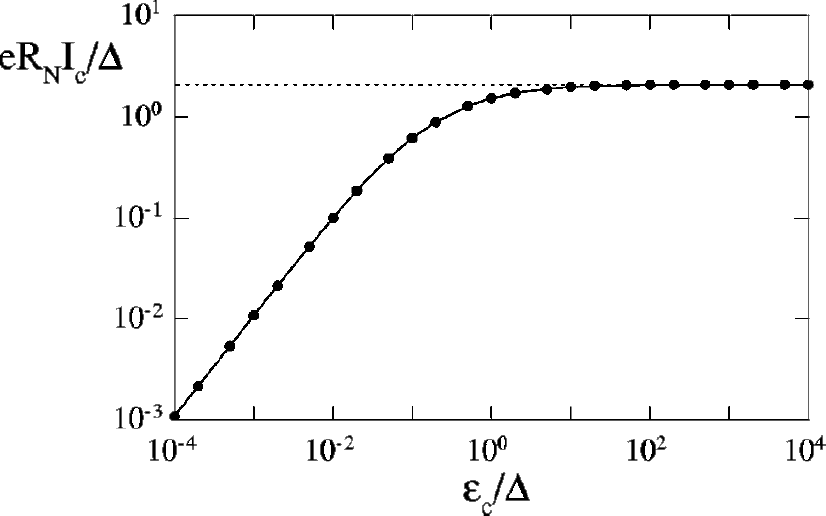}
			\caption{\label{fig:dubos2001josephson_fig1}Fig.~1 from Ref.~\citenum{dubos2001josephson}: on the $x$-axis it the normalized Thouless energy $\varepsilon_c/\Delta=E_\text{Th}/\Delta$, on the $y$-axis the normalized $I_\text{c}R_\text{N}$ product. This product, the voltage that would have to be applied in the normal state to get $I_\text{N}=I_\text{S,c}$, and a common experimental measure of the quality of a junction, is limited by the Thouless energy if the weak link is long ($E_\text{Th}<\Delta$), but by the gap for short junctions ($E_\text{Th}>\Delta$).}
		\end{figure}

		As the wavefunction spreads out through the channel, it interacts with different potentials at different positions, which each change the effective $E(t)$ of the part of the wavefunction that passes there.
		In the case of coherent transport, the spread in phase between the paths that this causes should be less than $\pi$,\begin{equation}\Delta\phi=\dfrac{t\,\delta E}{\hbar}\lessapprox \pi.\end{equation}
		This allows us to directly relate the time that the particle spends in the channel to the energy required to affect its coherence, the Thouless energy $\ETh$.
		In the ballistic case, since the time spent in the channel is simply proportional to its length,
		\begin{equation}E_\text{Th,ballistic}=\dfrac{\hbar\,\vF}{L_\text{channel}},\end{equation}
		while in the diffusive limit the transfer time goes as the \emph{square} of the length~\cite{edwards1972numerical}:
		\begin{equation}E_\text{Th,diffusive}=\dfrac{\hbar\,D}{L_\text{channel}^2}.\end{equation}
		This provides a characteristic energy scale for the junction: any kind of interaction on the order $E>\ETh$ will cause decoherence before the particle can reach the other end\footnote{Here $E$ represents the energy \emph{difference} between the different paths along the junction. If all paths are affected by identical energy shifts, no broadening of $\phi$ will occur and so neither will any decoherence.}. Conversely, we can associate any scattering energy with a coherence length~\cite{de1964boundary},\footnote{A factor $2\pi$ is introduced to scale $\xi_\text{E}$ such that $|\psi(L)|^2\propto e^{-\xi/\xi_\text{E}}$~\cite{dubos2001josephson}.}
		\begin{equation}\xi_\text{E,ballistic}=\dfrac{\hbar\,\vF}{2\pi\,E},\qquad \xi_\text{E,diffusive}=\sqrt{\dfrac{\hbar\,D}{2\pi\,E}}.\end{equation}
		Finally, we can think of these processes in terms of coherence times $\tau$, and compare them to the time it takes to cross the channel:
		\begin{equation}\tau_E=\dfrac{\hbar}{E},\qquad t_{ballistic}=\dfrac{L}{\vF},\qquad t_\text{diffusive}=\dfrac{L^2}{D}.\end{equation}
		In the end, of course all these ways of looking at it are equivalent; interactions will suppress superconducting transport exponentially if they cause decoherence over length scales shorter than the channel, or if they do so in less time than it takes to cross it, and the transport will be limited by the Thouless energy.
		If the channel is traversed faster than this, transport will instead be limited by the properties of the interface and the magnitude of the superconducting gap.
		This was captured most clearly in 2001 with an experimental and theoretical study of SNS junctions of varying length~\cite{dubos2001josephson}, the first figure of which is reproduced in Fig.~\ref{fig:dubos2001josephson_fig1}.
		
\section{\label{sec:transistors}Transistors}
	
	For a qubit platform to be ``scalable'' means in broad terms that the resources needed for both the operation and the fabrication of circuits does not grow too quickly with the total number of physical qubits.
	In section~\ref{sec:quantum_harmonic_oscillator} we discussed how operation at scale has led to the choice of gate-tunable transmons, while fabrication could be scaled by moving towards an all-CMOS qubit.
	CMOS technology has a strong track record: without much change in the total amount of time spent in fabricating a single die, the transistor count has increased exponentially from 4 in the late 50's to more than $10^{10}$ at the time of writing~\cite{owid2013technological}.
	If gatemons could be fabricated using this same technology, not only could large numbers be patterned simultaneously and reproducibly, it would also allow for the on-chip integration of the classical logic circuits that are necessary for real-time error correction.
	
\begin{figure}
			\centering
			\begin{tikzpicture}[x=\columnwidth/30,y=\columnwidth/30]
				\fill[V3Si] (-8,1) -- (8,1) -- (8,3) -- (-8,3) -- cycle;
				\fill[V3Si] (-3,3.5) -- (3,3.5) -- (3,5.5) -- (-3,5.5) -- cycle;
				\path\Sip (-3,2.5) -- (3,2.5) -- (3,3.5) -- (-3,3.5) -- cycle;
				\fill[SiN] (-4,2) -- (-3,2) -- (-3,5.5) to[out=220,in=90] (-4,2);
				\fill[SiN] (4,2) -- (3,2) -- (3,5.5) to[out=-40,in=90] cycle;
				\fill[SiO2] (-3,2) -- (3,2) -- (3,2.5) -- (-3,2.5) -- cycle;
				\fill[Si] (-8,0) -- (8,0) -- (8,1) -- (3,1) to[out=180,in=270] (2,2) -- (-2,2) to[out=270,in=0] (-3,1) -- (-8,1) -- cycle; 	
				\fill[SiO2] (-8,0) -- (8,0) -- (8,-2) -- (-8,-2) -- cycle; 
				\fill[Si] (-8,-2) -- (8,-2) -- (8,-3) -- (-8,-3) -- cycle;
				
				\fill[gray!15] 	(9,6) -- (16,6) -- (16,-3) -- (9,-3) -- cycle;
				\fill[V3Si] 			(10,5) -- (12,5) -- (12,4) -- (10,4) -- cycle;
				\node[anchor=west] at 	(12.5,4.5) {\ce{PtSi}};
				\fill[SiN] 				(10,3) -- (12,3) -- (12,2) -- (10,2) -- cycle;
				\node[anchor=west] at 	(12.5,2.5) {\ce{SiN}};
				\fill[SiO2] 			(10,1) -- (12,1) -- (12,0) -- (10,0) -- cycle;
				\node[anchor=west] at 	(12.5,0.5) {\ce{SiO2}};
				\fill[Si] 				(10,-1) -- (12,-1) -- (12,-2) -- (10,-2) -- cycle;
				\node[anchor=west] at 	(12.5,-1.5) {\ce{Si}};
	
				\node[anchor=center] at (-6,2) {\large\B{S}};
				\node[anchor=center,white] at (-6.05,2.05) {\large\B{S}};
				\node[anchor=center] at (6,2) {\large\B{S}};
				\node[anchor=center,white] at (5.95,2.05) {\large\B{S}};
				\node[anchor=center] at (0,1) {\large\B{Sm}};
				\node[anchor=center,white] at (-0.05,1.05) {\large\B{Sm}};
			\end{tikzpicture}
			\vspace*{0.5\baselineskip}
			
			\caption{\label{fig:jofet}A Josephson field effect transistor (JoFET) has superconducting source and drain, here made of \ce{PtSi}. It functions like a transistor, except that the current that is modulated is a resistanceless supercurrent.}
		\end{figure}
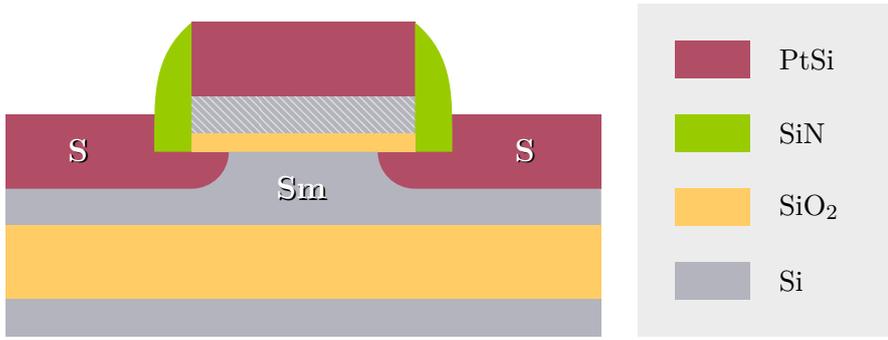

	The central idea in this approach is to use a modified transistor as gate-tunable Josephson junction, an example of which is shown in Fig.~\ref{fig:jofet}.
	Below we will discuss some of their principles.
	
	\subsection{\label{sec:field_effect}The field effect}

		To understand the effect that an electrostatic field has on the current through a junction, consider first the simplified case in which there are no barriers to overcome, and an infinite reservoir of carriers is available to be attracted underneath the gate oxide.
		The total current that will flow from source to drain is then proportional to the charge accumulated in the channel, times the rate at which that charge moves crosses the device~\cite{hook2010solid},
		\begin{equation}I_\text{d}=\underbrace{C(V_\text{g}-V_\text{th})}_{\text{Q/A}}\overbrace{W\braket{v}}^{A/t}.\end{equation}
		Applying a gate voltage $V_\text{g}$ then attracts charges proportional to the oxide's capacitance, which will linearly increase the current in the case that the drift and diffusion rates remain constant.
		
		In reality this behavior is only observed above some threshold gate voltage $V_\text{th}$ where all the carriers are of the same sign.
		Below this voltage, small numbers $n_\text{e}$ and $n_\text{h}$ of both electrons and holes are present, the product of which is constant at a fixed temperature, and equal to the square of the intrinsic carrier density $n_\text{i}$ of the semiconductor,
		\begin{equation}\label{eq:electron_hole_concentration_constant}n_\text{e}n_\text{h}=n_\text{i}^2.\end{equation}
		This number of intrinsic carriers in turn depends on the temperature of the material~\cite{green1990intrinsic},
		\begin{equation}n_\text{i}^2=C_\text{n}\,T^3\,e^{-E_\text{g}/k_\text{B}T},\end{equation}
		where $C_\text{n}$ is some constant such that the intrinsic concentration $n_\text{i}$ of silicon is $10^10\si{\per\centi\meter\cubed}$ at room temperature ($T=\SI{300}{\kelvin}$)~\cite{sproul1991improved}, and $E_\text{g}$ the band gap of silicon.
		When a material is doped, the Fermi level is moved towards the conduction or valence band for (n and p doping, respectively), changing the electron and hole concentrations.
		For example, in the case of a p-doped (p for positive) channel with boron doping~\cite{chrzanowska1989bilow},
		\begin{equation}n_\text{h}=n_\text{B},\qquad n_\text{e}=\dfrac{n_\text{i}^2}{n_\text{B}},\end{equation}
		where a wide range of $n_\text{B}$ from \SI{1E13}{\per\centi\meter\cubed} to \SI{1E18}{\per\centi\meter\cubed} has been used in industry~\cite{colinge2005physics}.
		This relationship $n_\text{h}=n_\text{B}$ does not hold at all temperatures however, since at low temperatures dopants can freeze out, which is an especially serious concern for holes in silicon~\cite{lengeler1974semiconductor}.

		\begin{figure}
			\centering
			\includegraphics[width=0.6\textwidth]{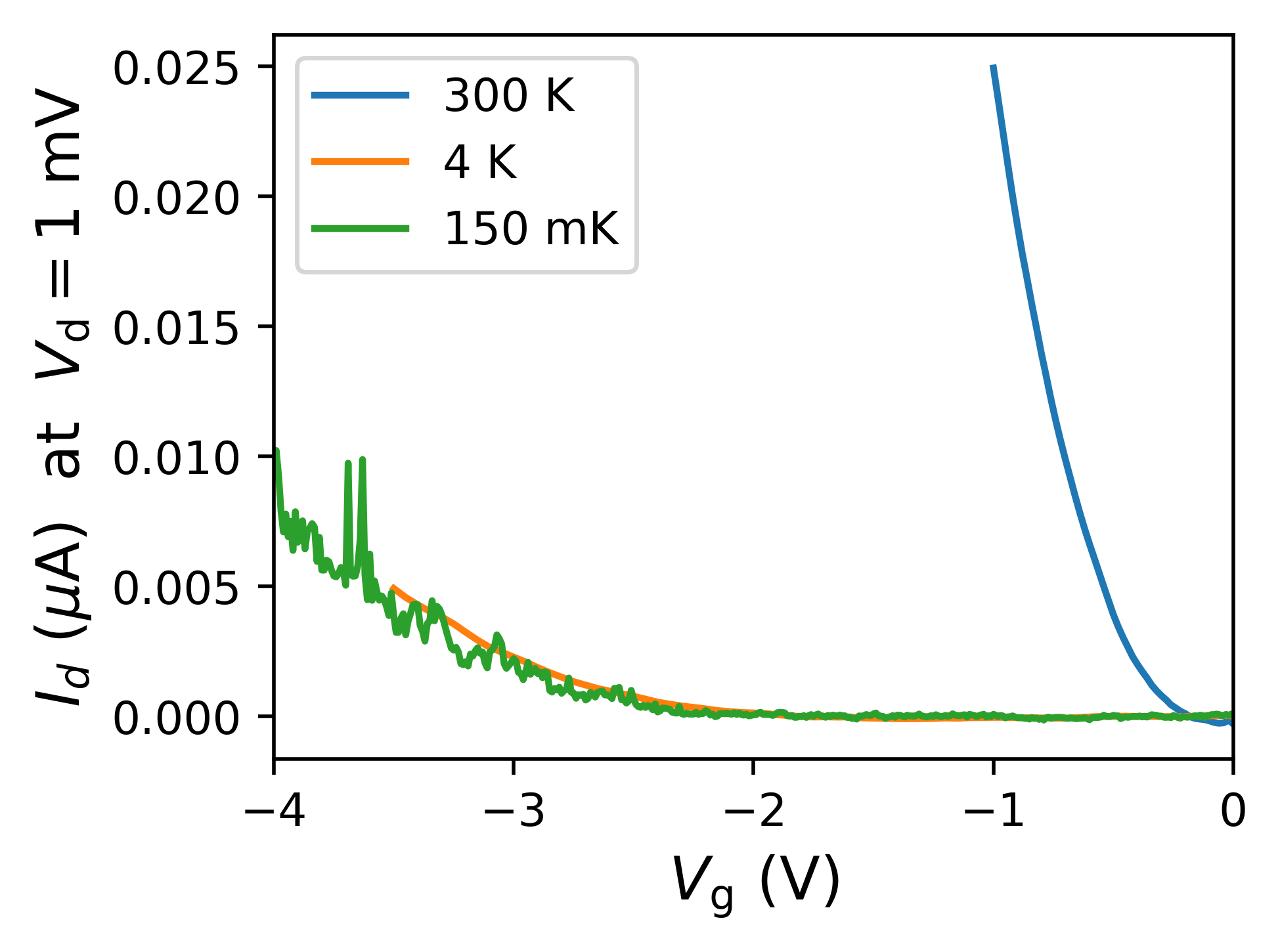}
			\caption{\label{fig:d84d4gdiff}The field effect measured at different temperatures in a \SI{50}{\nano\metre} long, \SI{2.5}{\micro\metre} wide SBMOSFET with PtSi contacts. We observe a shift in the threshold voltage $V_\text{th}$ due to the freezing out of intrinsic carriers, as well as the charging of individual dopants at low temperatures.}
		\end{figure}
		
		Below the threshold voltage, when the channel is essentially depleted, charge can still cross the channel through thermionic emission proportional to $\exp(eV_\text{g}/k_\text{B}T)$.
		This gives an exponentially decaying tail to the drain current as a function of gate voltage,
		\begin{equation}I_\text{d}(V_\text{g})\propto \exp\left(\dfrac{eV_\text{g}}{k_\text{B}T(1+\alpha)}\right).\end{equation}
		Here $\alpha$ is called the ``level-arm parameter'', and quantifies the reduction in the field effect of the gate due to a parasitic capacitance between the channel and the substrate,
		\begin{equation}\alpha=\dfrac{C_\text{channel-substrate}}{C_\text{channel-gate}}.\end{equation}
		In general it is desirable to have only a narrow range of gate voltages where a sub-threshold current $I_\text{off}$ is observed (and thus a small $\alpha$).
		The degree to which this is achieved in a device is often expressed in the ``sub-threshold swing'', which is the increase in gate voltage required to multiply the drain current ten-fold.
		We calculate this as the inverse of the slope of the logarithm of the drain current versus the gate voltage,
		\begin{equation}S=\left(\dfrac{d\log_{10}(I_\text{d})}{d|V_\text{g}|}\right)^{-1}=\ln(10)\dfrac{k_\text{B}T}{e}(1+\alpha).\end{equation}
		In the ideal case where $\alpha=0$, this gives a sub-threshold swing of \SI{59.6}{\milli\volt\per\decade} at \SI{300}{\kelvin}.
		At low temperatures the freezing out of dopant carriers and the suppression of thermionic emission mean that both the total current is reduced, and the threshold voltage is shifted.
		This is shown in Fig.~\ref{fig:d84d4gdiff} for a device that we will discuss further in chapter~\ref{sec:jofets}.
			
	\subsection{Schottky barriers}
		
		\begin{figure}
			\centering
			\begin{tikzpicture}[x=\columnwidth/12,y=\columnwidth/12]
				\fill[gray!30!white] (-4,0) -- (4,0) -- (4,1.1) -- (-4,1.1) -- cycle;
				\fill[gray!30!white, path fading=north] (-4,1.1) -- (4,1.1) -- (4,3) -- (-4,3) -- cycle;
				\fill[white] (4,2) -- (2,2) to[out=180,in=60] (0,1) -- (0,3) to [out=60,in=180] (2,4) -- (4,4) -- cycle;
				\draw[ultra thick] (0,1) to [out=60,in=180] (2,2) -- (4,2);
				\draw[ultra thick] (0,3) to [out=60,in=180] (2,4) -- (4,4);
				\draw[dashed,very thick] (-4,2.1) -- (4,2.1) node[anchor=west] {$E_\text{F}$};
				\node[white,anchor=east] at (-4,3) {$E_\text{F}$};
				\draw[->,ultra thick] (-4.2,0) -- (4.2,0) node[right]{$x$};
				\draw[->,ultra thick] (-4,-0.2) -- (-4,4.8) node[above]{$E(x)$};
				
				\draw[ultra thick] (-4.1,4.5) node[anchor=east] {$E_\text{ionize}$} -- (-4,4.5);
				
				\draw[ultra thick, dashed] (-1.5,1) -- (0,1);
				\draw[<->,ultra thick] (-1,1) -- (-1,2.1);
				\node[anchor=east] at (-1,1.55) {$\Phi_\text{B}$};
				\draw[red,fill=red] (-0.3,1.9) circle (0.75ex);
				\draw[->,ultra thick,red] (-0.1,1.9) -- (1.6,1.9);
				
				\draw[ultra thick, dashed] (-1.5,4.5) -- (0.5,4.5);
				\draw[<->,ultra thick] (-1,3) -- (-1,4.5);
				\node[anchor=east] at (-1,4) {$\Phi_\text{metal}$};
				\draw[<->,ultra thick] (0,3) -- (0,4.5);
				\node[anchor=west] at (0,4) {$\Phi_\text{semi}$};
			\end{tikzpicture}
			\vspace*{0.5\baselineskip}
			
			\caption{\label{fig:schottky}The formation of a Schottky barrier for holes with a height $\Phi_\text{B}$ at the interface between a metal and a p-doped semiconductor. Shown in red is a hole that tunnels through the barrier.}
		\end{figure}
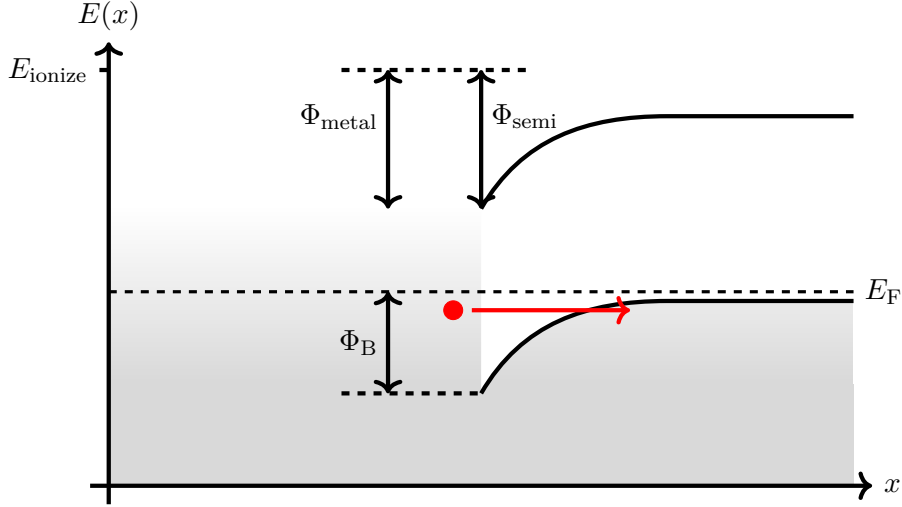
		
		A phenomenon that becomes especially relevant at low temperatures is the formation of a Schottky barrier at the interface between a doped semiconductor (such as a p-doped channel) and a metal (such as PtSi contacts).
		The appearance of this tunnel barrier can be understood as follows.
		Since the Fermi level is the energy up to which the states are occupied (give or take $k_\text{B}T$), it needs to be constant across a device with finite resistivity.
		When two materials with different Fermi levels are brought together, a particle occupying a higher-energy state on the one side could gain energy by moving to the other, which ensures that the maximum occupied energy on either side of the interface equals out (the Fermi levels are ``pinned'' to each other).
		At the same time, since the two materials are now conductively coupled, a charge from either side of the interface can be moved through the other, which means that the total energy of removing a charge in either material (the work function $\Phi$) also needs to match close to the interface.
		We saw above that doping the semiconductor will shift the Fermi level within the band gap closer to either the valence or conduction band, while the work function of the material remains unchanged.
		For both the Fermi levels and the work functions to match at the interface, the valence and conduction bands thus have to be bent~\cite{schottky1942vereinfachte}, as shown in Fig.~\ref{fig:schottky}.
		Different heuristics for predicting the degree and direction of band bending have been proposed, taking into account charge buildup at the interface and focusing on the work functions~\cite{mott1939theory} or including also surface states that effectively pin the Fermi level~\cite{bardeen1947surface}, neither of which fully predict Schottky barrier heights at metal/semiconductor interfaces~\cite{tersoff1984schottky}.
		
		At higher temperatures, the Schottky barrier can be overcome by thermionic emission, but as the device is cooled down, $k_\text{B}T$ becomes smaller than the Schottky barrier height (SBH) $\Phi_\text{B}$, and it will act as a tunneling barrier for the electrons or holes.
		Since this barrier needs to be crossed twice in the case of Andreev reflection, while it is traversed only once by an individual quasiparticle, Schottky barriers suppress supercurrent more than they do normal transport, and are therefore usually undesirable in JoFETs where coupling to quasiparticles should be minimized.
		
	\subsection{\label{sec:jofet_considerations}JoFET considerations}
		
		Since we can always use the gate voltage to completely deplete the channel and switch off the Josephson coupling, it is in general our goal to fabricate devices with large critical currents $I_\text{c}$.
		Note that since we also need a very large capacitance $C$ in a transmon qubit, we are not in principle limited by the size of the device, and could achieve this by making junctions arbitrarily wide.
		Assuming a dielectric constant of $\kappa=3.9$ for \ce{SiO2}~\cite{kingon2000alternative}, a gate oxide between 5 and \SI{20}{\nano\meter} thick and a target charging energy of 100 to \SI{300}{\mega\hertz} (65 to \SI{200}{\nano\electronvolt}), we find that channels can be up to 60 -- \SI{700}{\micro\meter\squared} large.
		There are other reasons why you may want to avoid too large devices though, such as flux noise sensitivity, gate leakage, and losses to parasitic two-level systems in the oxide~\cite{martinis2005decoherence,koch2007charge}.
		It is therefore preferable to derive the charging energy from a separate shunting capacitance made of a material with a lower loss tangent~\cite{koch2007charge} (proportional to the density of two-level systems~\cite{martinis2005decoherence}) than \ce{SiO2}.
		In practice then, we are looking for transistor devices with as large a critical current as possible, while limiting its size.
		This implies a preference for high-mobility semiconducting materials~\cite{casparis2018superconducting} and transparent interfaces between the superconductor and channel~\cite{vigneau2019germanium}.
		
		It is less straightforward to determine what level of doping should be aimed for in a specific system.
		On the one hand, the supercurrent carrying capacity of a junction is closely related to its normal-state conductance (especially for short junctions, where $I_\text{c}\propto G_\text{N}\Delta$)~\cite{de1964boundary,de1966superconductivity,kleinsasser1991critical}, which in turn is the product of mobility and carrier concentration, suggesting that increasing the doping level of the channel could improve its behavior.
		Even in longer junctions one can expect diffusion to be aided by higher carrier densities~\cite{volkov1996effect}, despite a reduction in mean free path~\cite{aslamazov1981temperature}.
		Moreover, dopants could provide a resonance-percolation trajectories~\cite{lifshitz1979tunnel} when other modes of transport have already been suppressed~\cite{aslamazov1982resonant}.
		Higher doping concentrations would in turn also reduce the width of the Schottky barrier~\cite{mott1938note,harrison2013effective}, exponentially increasing its tunneling probability at low temperatures.
		For devices with silicon channels, it may therefore be necessary to have higher doping, especially near the interface with the metal contacts.
		On the other hand, the scattering sites that this introduces may in fact bring down the total transmission through the channel by reducing the mobility~\cite{sammak2019shallow}.
		
		A different set of concerns is raised by the choice of contacting metal.
		Just as the coupling between two transmons depends on the matching of their impedances, so does the transmission probability of a quasiparticle through an interface depend on the matching of its inertia on either side, given by the electron effective mass~\cite{harrison2013effective}.
		Differences in lattice spacing between the two materials will also induce a shift in momentum as the incoming and outgoing particles occupy different Bloch states, both creating an effective potential and narrowing the recombination window by causing inelastic scattering across the interface.
		Other factors that can limit the transparency are the Schottky barrier height that we discussed above, as well as the general quality of the interface in terms of grain boundaries, impurities and oxide layers.
		The application of the transistor for superconducting transport adds the requirement that the contacting metal have a high superconducting critical temperature.
		In the next chapter we will see how all of these aspects are addressed by fabricating the contacts with superconducting silicides.
	
\printbibliography

\end{refsection}

\begin{refsection}
	\graphicspath{{img/ch3/}}
	\setcounter{chapter}{2}
\chapter{\label{sec:silicides}Silicides}

\section{Introduction}
	
	We saw in the previous chapter that silicon-based transistors could serve as the weak links in superconducting qubits.
	Certainly silicon is not the \emph{ideal} material for Josephson junctions, with its significant Schottky barriers at any useful doping level, low mobility and early onset of dopant freeze-out.
	It is, however, \emph{compatible}.
	A device based on silicon can, within some constraints on materials and fabrication techniques, readily be integrated alongside existing cryo-CMOS circuits~\cite{patra2017cryo,le2020low,le2020FDSOI}, and leverage fabrication techniques that have been honed over decades of innovation.
	The conventional MOSFET design relies on silicides, either by depositing pure metal that then reacts with the underlying silicon, or co-depositing both metal and silicon in the right stoichiometry~\cite{theuerer1964getter,michikami1982v3si,murarka1983high}.
	While the former has the benefit of self-aligning the formed silicide with the exposed patches of silicon (known as the SALICIDE process) and diffusing any impurities introduced at the metal-silicon interface, the latter allows to select silicide phases that are otherwise unlikely to form, and often requires lower formation temperatures~\cite{murarka1995silicide}.
	In either case, the resulting compound's metal content ensures high conductance, while the incorporation of silicon generally provides a low contact resistance to the channel.
	By selecting silicides that also happen to be superconducting, these methods can be applied directly to the fabrication of JoFETs.
	The two materials that we will focus on in this chapter are \ce{V3Si}, for its superconducting properties, and \ce{PtSi}, which boasts established integration processes.
	
	We will begin with \ce{V3Si}, which is interesting first of all for its promise in JoFET applications; it has the highest superconducting critical temperature of any silicide that we are aware of~\cite{hardy1953superconducting,hardy1954superconductivity,blumberg1960correlations}, and has excellent lattice matching to (111) Si~\cite{zur1985transition} --- off by only 0.4\%, much better than the best that can be achieved with PtSi (3.0\%)~\cite{ben1978crystallography}.
	Second, it is also interesting for the highly unusual dependence of its thermal, conductive and superconducting properties on the strain that it is under~\cite{testardi1970unusual,testardi1971unusual95,testardi1971unusual}.
	Third, in systems of vanadium and silicon, this is the most elusive of the congruently forming phases, as it is neither the first phase to form~\cite{tu1973formation,krautle1974kinetics,chu1974identification}, nor does it have the greatest enthalpy of formation (and is therefore not thermodynamically favorable)~\cite{pretorius1993thin}.
	The first of these justifies our efforts, while the challenge of the other two makes it all worth our while.
	
	\begin{figure}
		\begin{subfigure}[t]{0.475\textwidth}
			\centering
			\begin{tikzpicture}[x=0.75cm,y=0.75cm]
				\path[fill=redbg,rounded corners=10pt] (-5,-2) -- (5,-2) -- (5,2) -- (-5,2) -- cycle;
				
				\fill[Si] (-3,-1.5) -- (-1.2,-1.5) -- (-1.2,-0.5) -- (-3,-0.5) -- cycle;
				\node[anchor=east] at (-3,-1) {Si};
				\fill[SiO2] (-3,-0.5) -- (-1.2,-0.5) -- (-1.2,0.2) -- (-3,0.2) -- cycle;
				\node[anchor=east] at (-3,-0.15) {\ce{SiO2}};
				\fill[Va] (-3,0.2) -- (-1.2,0.2) -- (-1.2,1.5) -- (-3,1.5) -- cycle;
				\node[anchor=east] at (-3,0.85) {\ce{V}};
				
				\draw[ultra thick,green!60!black,->,>=stealth] (-0.7,0) -- (0.7,0);
				
				\fill[Si] (3,-1.5) -- (1.2,-1.5) -- (1.2,-0.5) -- (3,-0.5) -- cycle;
				\fill[V3Si] (3,-0.5) -- (1.2,-0.5) -- (1.2,0.49) -- (3,0.49) -- cycle;
				\node[anchor=west] at (3,0) {\ce{V3Si}};
				\fill[VO] (3,0.49) -- (1.2,0.49) -- (1.2,1.21) -- (3,1.21) -- cycle;
				\node[anchor=west] at (3,0.85) {\ce{VO_x}};
			\end{tikzpicture}
			\caption{\label{fig:four_methods_v3si1}Sputtering of V on \ce{SiO2}~\cite{krautle1974reactions}.}
		\end{subfigure}
		\hfill
		\begin{subfigure}[t]{0.475\textwidth}
			\centering
			\begin{tikzpicture}[x=0.75cm,y=0.75cm]
				\path[fill=white,rounded corners=10pt] (-5,-2) -- (5,-2) -- (5,2) -- (-5,2) -- cycle;
	
				\fill[Si] (-3,-1.5) -- (-1.2,-1.5) -- (-1.2,0.17) -- (-3,0.17) -- cycle;
				\fill\SiOdoped (-3,-0.5) -- (-1.2,-0.5) -- (-1.2,0.17) -- (-3,0.17) -- cycle;
				\node[anchor=east] at (-3,-0.17) {Si:O};
				\fill[Va] (-3,0.17) -- (-1.2,0.17) -- (-1.2,1.5) -- (-3,1.5) -- cycle;
				\node[anchor=east] at (-3,0.85) {\ce{V}};
				
				\draw[ultra thick,green!60!black,->,>=stealth] (-0.7,0) -- (0.7,0);
				
				\fill[Si] (3,-1.5) -- (1.2,-1.5) -- (1.2,-0.5) -- (3,-0.5) -- cycle;
				\fill[V3Si] (3,-0.5) -- (1.2,-0.5) -- (1.2,1.17) -- (3,1.17) -- cycle;
				\node[anchor=west] at (3,0.33) {\ce{V3Si}};
				\fill[VO] (3,1.15) -- (1.2,1.15) -- (1.2,1.25) -- (3,1.25) -- cycle;
				\node[anchor=west] at (3,1.25) {\ce{VO_x}};
			\end{tikzpicture}
			\caption{\label{fig:four_methods_v3si2}Sputtering of V on O-rich aSi and cSi~\cite{schutz1979formation}.}
		\end{subfigure}
		
		\vspace*{\baselineskip}

		\begin{subfigure}[b]{0.475\textwidth}
			\centering
			\begin{tikzpicture}[x=0.75cm,y=0.75cm]
				\path[fill=greenbg,rounded corners=10pt] (-5,-2) -- (5,-2) -- (5,2) -- (-5,2) -- cycle;
				
				\fill[Si] (-3,-1.5) -- (-1.2,-1.5) -- (-1.2,0.17) -- (-3,0.17) -- cycle;
				\fill[V3Si] (-3,-0.5) -- (-1.2,-0.5) -- (-1.2,1.5) -- (-3,1.5) -- cycle;
				\node[anchor=east] at (-3,0.5) {\ce{aV3Si}};
				
				\draw[ultra thick,green!60!black,->,>=stealth] (-0.7,0) -- (0.7,0);
				
				\fill[Si] (3,-1.5) -- (1.2,-1.5) -- (1.2,-0.5) -- (3,-0.5) -- cycle;
				\fill[V3Si] (3,-0.5) -- (1.2,-0.5) -- (1.2,1.4) -- (3,1.4) -- cycle;
				\node[anchor=west] at (3,0.45) {\ce{cV3Si}};
			\end{tikzpicture}
			\caption{\label{fig:four_methods_v3si4}Sputtering from a \ce{V3Si} target~\cite{michikami1982v3si}.}
		\end{subfigure}
		\hfill		
		\begin{subfigure}[b]{0.475\textwidth}
			\centering
			\begin{tikzpicture}[x=0.75cm,y=0.75cm]
				\path[fill=white,rounded corners=10pt] (-5,-2) -- (5,-2) -- (5,2) -- (-5,2) -- cycle;
				
				\fill[Si] (-3,-1.5) -- (-1.2,-1.5) -- (-1.2,0.17) -- (-3,0.17) -- cycle;
				\path\SiVseq (-3,-0.5) -- (-1.2,-0.5) -- (-1.2,1.5) -- (-3,1.5) -- cycle;
				\node[anchor=east] at (-3,0.5) {Si, V};
				
				\draw[ultra thick,green!60!black,->,>=stealth] (-0.7,0) -- (0.7,0);
				
				\fill[Si] (3,-1.5) -- (1.2,-1.5) -- (1.2,-0.5) -- (3,-0.5) -- cycle;
				\fill[V3Si] (3,-0.5) -- (1.2,-0.5) -- (1.2,1.21) -- (3,1.21) -- cycle;
				\node[anchor=west] at (3,0.33) {\ce{V3Si}};
			\end{tikzpicture}
			\caption{\label{fig:four_methods_v3si3}Sequential sputtering of V and Si~\cite{de1985preparation}.}
		\end{subfigure}

		\caption{\label{fig:four_methods_v3si}A selection of methods described in the literature to form \ce{V3Si} thin films, one of which has been reproduced without (indicated in red) and another with success (green).}
	\end{figure}
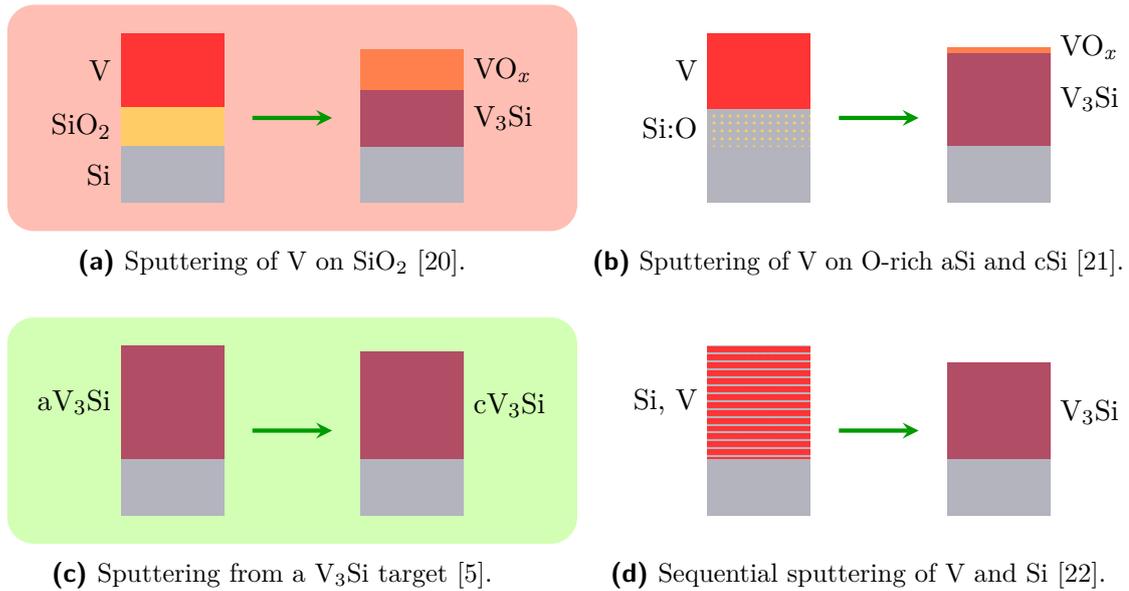
	
	Two approaches were tried in forming \ce{V3Si} thin films for our studies.
	The first took place at Uppsala Universitet, where the readily installed vanadium target allowed us to perform depositions on a variety of substrates.
	Reports in the literature had suggested that the formation of the right phase could be triggered by methods that can broadly be grouped into three mechanisms.
	Number one would see silicon, the dominant diffusing species in \ce{VSi2} during its formation, captured with oxygen either in the form of a native oxide~\cite{schutz1979formation,krautle1974kinetics,chu1978identification,oya1982superconducting} (see Fig.~\ref{fig:four_methods_v3si1}), or with impurities at a concentration of a few \%at in deposited silicon~\cite{schutz1979formation} (see Fig.~\ref{fig:four_methods_v3si2}), effectively decoupling the silicide from the Si reservoir and thus allowing the otherwise much more slowly diffusing vanadium to gain a foothold.
	This result of inverting the relative mobilities of the two species this way may be interpreted in the context of the \ce{Cu3Au} effect: \emph{``the first phase to nucleate is the phase rich in the high-mobility constituent''}, or vice versa, \emph{``the mobility is higher for the majority constituent''}~\cite{d1986kinetics}.
	In the second method, the Si substrate is amorphized prior to deposition~\cite{schutz1979formation,psaras1984sequential}, likely aiding \ce{V3Si} formation by lowering the energy barrier to nucleation (see  also Fig.~\ref{fig:four_methods_v3si2}).
	The third would overcome the limitation posed by vanadium's slow diffusion by immediately distributing the two species in the right stoichiometry~\cite{michikami1982v3si,ferdeghini2009superconducting} or at least approaching this with thin alternating layers of Si and V~\cite{de1985preparation} (see Figs.~\ref{fig:four_methods_v3si3} and~\ref{fig:four_methods_v3si4}).
	Since the target available in Uppsala was pure metal, and the second (and last) target slot in the deposition tool was to be used for titanium to allow for in-situ capping with TiN, the first campaign was limited to deposition of pure V on various substrates.	

\FloatBarrier
\section{V3Si formation by pure metal deposition}
		
		Perhaps an equal amount of time was spent preparing and analyzing the samples where pure vanadium was deposited (the ``Uppsala samples''), as was used for those with \ce{V3Si} deposition from the compound target (the ``Grenoble samples'').
		However, because it was ultimately concluded that we may only have successfully formed \ce{V3Si} on a single sample from this first batch (one that was annealed within hours of deposition), little of scientific value came out, and so this section will be short.
		Yet, precisely because this turned out to be a dead end, and because so much time was lost, it is worth tracing our steps, and we will conclude with some valuable lessons about methodology.
		
		A total of 14 silicon wafers were prepared in Grenoble, 8 of which with a diameter of \SI{300}{\milli\metre} and 6 of \SI{200}{\milli\metre}, each with different surface preparations, giving us access to a range of different substrate parameters.
		This set was intentionally designed to be broader than that which could be fully analyzed, such that a series of quick tests in Uppsala could narrow down the wafers of interest.
		Eventually a total of five wafers were selected for deposition, shown in Table~\ref{tab:wafers_uppsala}, allowing us to study the effects of oxide (both native and thermal), oxygen implantation, substrate pre-amorphization and a combination of the latter two.
		A wafer with silicon nitride was included to test the viability of this material as a spacer in CMOS transistors with \ce{V3Si} contacts (any sign of reaction with the silicide would rule it out).
		Since vanadium is known to oxidize extremely rapidly, it was necessary to cap the metal in-situ, for which we chose a \SI{15}{\nano\metre} layer of TiN.
		This material was chosen not just for its availability in Uppsala, but also because consultation of binary phase diagrams for V-N and V-Ti~\cite{enomoto1996n} indicated stability up to extremely high temperatures of \SI{3290}{\celsius} for TiN, while \ce{V_xN_{1-x}} and \ce{V_xTi_{1-x}} were only stable up to 2340 and \SI{850}{\celsius}, respectively.
		This relative stability of TiN was taken to be an indicator of thermodynamic favorability, and thus unlikelihood of reaction with the underlying V, which assured us of its suitability.
		This conclusion was further supported by its routine use as a capping material for the silicidation of other metals, notably Pt, which will be discussed in section~\ref{sec:ptsi}.
		As we will see later, our confidence in this choice was disastrously misplaced.

		\begin{table}
			\centering
			\caption{\label{tab:wafers_uppsala}A short summary of the wafer types analyzed for pure vanadium deposition in Uppsala. Treatments in parentheses are only performed on some samples.}
			\rowcolors{2}{gray!15}{white}
			\begin{tabular}{l l l l l}
				\rowcolor{gray!30}\hline
				Wafer 			& Substrate 								& Treatment							& Deposition\\\hline\hline
				A				& bulk Si									& (HF)								& V (+ TiN)\\
				B			& Si + \SI{20}{\nano\metre} thermal \ce{SiO2}		& (\ce{HfO2} ALD)				& V (+ TiN)\\
				C				& amorphized SOI with O doping (0.1--5\%)	& HF									& V + TiN\\
				D			& Si + \SI{40}{\nano\metre} \ce{Si3N4}		& None							& V + TiN\\
				E				& SOI with O doping (0.1--10\%)			& HF									& V + TiN\\\hline
			\end{tabular}
		\end{table}
		
		\FloatBarrier
		\subsection{Early encouragement}
			
			Based on multiple reports in the literature, it was expected that vanadium deposited onto a \SI{20}{\nano\metre} layer of thermal \ce{SiO2} would upon thermal processing lead to the formation of a \ce{V3Si} layer.
			Whether the vanadium was deposited onto heated substrates~\cite{de1985preparation}, or at room temperature and annealed later~\cite{oya1982superconducting}, a maximum superconducting critical temperature was obtained at deposition/annealing temperatures of around \SI{840}{\celsius}.
			During this reaction, the oxygen is transported to the top of the vanadium layer much faster than the \ce{V3Si} forms~\cite{krautle1974kinetics}, forming there a vanadium oxide layer of about the same thickness as the consumed \ce{SiO2},
			\begin{equation}\ce{SiO2} + \left(3+\dfrac{2}{x}\right) V \rightarrow \ce{V3Si} + \dfrac{2}{x}\, \ce{VO_x}.\end{equation}
			If $x=1$, which it will approach when the reaction is limited by the availability of \ce{SiO2}~\cite{krautle1974kinetics}, then \SI{20}{\nano\metre} of \ce{SiO2} will react with \SI{36.8}{\nano\metre} of V to form \SI{28.0}{\nano\metre} of \ce{V3Si} and \SI{20.5}{\nano\metre} of \ce{VO}.
			After the \ce{SiO2} is fully consumed, the reaction may continue by consuming the underlying Si substrate~\cite{oya1982superconducting}, though it will slow down~\cite{hugunin1995superconductor}.
			
			To reproduce these results, layers of \SI{80}{\nano\metre} of vanadium, capped with \SI{15}{\nano\metre} of TiN, were deposited onto oxidized wafers, and immediately (within an hour) annealed during two minutes at \SI{840}{\celsius}, with a high ramp rate of \SI{20}{\celsius/\second}.
			Sheet resistance measurements were taken before and after deposition (see Table~\ref{tab:samples_uppsala}), showing an increase from \SI{8.05\pm0.16}{\ohm} to \SI{11.39\pm0.25}{\ohm}.
			Sheet resistance values taken from calibration samples where 5, 10, 20, 40 and \SI{80}{\nano\metre} of V was deposited, each followed by \SI{15}{\nano\metre} of TiN, indicated that neither the oxidized substrate, nor the as-deposited TiN layer was highly resistive ($R_\square\gg\SI{100}{\ohm}$), so that these values can be taken to be accurate for the layers of vanadium and its reaction products.
			In the most optimistic case, where the entire vanadium layer was consumed during reaction with the substrate (consuming first the oxide, then around \SI{21}{\nano\metre} of silicon), around \SI{83}{\nano\metre} of \ce{V3Si} would have formed, in which case the measured sheet resistance would translate to a resistivity of \SI{94\pm2}{\micro\ohm\centi\metre}.
			Though this is only a rough estimate, it is within the expected range for crystalline \ce{V3Si}, which for high-quality samples can be as low as \SI{77}{\micro\ohm\centi\metre}~\cite{taub1974electrical,caton1978normal,caton1982analysis,zhang2021thin}.
			
			\begin{table}
				\centering
				\caption{\label{tab:samples_uppsala}Sheet resistance values and detected XRD peaks for samples that were annealed at \SI{840}{\celsius} for 2 minutes under a \ce{N2} atmosphere within hours or days after deposition. All samples were capped with \SI{15}{\nano\metre} TiN.}
				\resizebox{\columnwidth}{!}{%
				\rowcolors{2}{gray!15}{white}
				\begin{tabular}{p{2.5cm} l l l p{6.5cm}}
					\rowcolor{gray!30}\hline
					Substrate 														& V (nm)	& as-dep $R_\square$	& annealed $R_\square$	& XRD peaks detected\\\hline\hline
					Si (HF)															& 80				& \SI{6.06\pm0.28}{\ohm}	& \SI{3.95\pm0.11}{\ohm}	& 39.6 (VTi or \ce{VSi2}), 42.2 (V or \ce{VSi2}), 42.8 (VTi or \ce{V3Si}), 49.2 (\ce{VSi2})\\
					
					Si (no HF)															& 80				& \SI{8.81\pm0.29}{\ohm}	& \SI{4.87\pm0.27}{\ohm}& 39.6 (VTi or \ce{VSi2}), 42.2 (V or \ce{VSi2}), 42.8 (VTi or \ce{V3Si}, much smaller), 49.2 (\ce{VSi2}, smaller)\\
					
					\SI{20}{\nano\metre} \ce{SiO2}											& 80				& \SI{8.05\pm0.16}{\ohm}	& \SI{11.39\pm0.25}{\ohm}	& 38.0 (\ce{VN}, \ce{V3Si}), 44.0 (VN), 81.0 (\ce{VN}, \ce{V3Si})\\
					
					\SI{20}{\nano\metre} \ce{SiO2} + \SI{10}{\nano\metre} \ce{HfO2}	& 100				&	\SI{11.86\pm0.11}{\ohm}	& \SI{15.3\pm4.1}{\ohm}	& No XRD performed\\
					
					\SI{40}{\nano\metre} \ce{Si3N4}											& 100				&	\SI{6.06\pm0.17}{\ohm}	& \SI{3.95\pm0.11}{\ohm}	& 37.9 (\ce{VN}, \ce{V3Si}), 44.0 (\ce{VN})\\\hline
				\end{tabular}%
				}
			\end{table}

			XRD analysis in a Bragg-Brentano configuration (a first scan in the 35--50° range, a second between 72 and 98°) indicated two peaks at 38.0 and 81.0, that could correspond to the (200) and (400) peaks of \ce{V3Si}, expected at 38.0 and 81.4.
			The absence of other XRD peaks of \ce{V3Si} in the $2\theta=35$--50° range, notably the (210) and (211) peaks at 42.7° and 47.1°, could be interpreted as strong epitaxial alignment of \ce{V3Si} with the substrate after full consumption of the \ce{SiO2} layer.
			The large error in the position of the second peak was at the time ignored, while there are in fact many vanadium nitride compounds with different stoichiometries, two (VN and \ce{VN_{0.35}}) with peaks right at 39.8 and 81.4 (PDF reference codes 00-025-1252 and 00-006-0624), and another with a peak at 81.0 (\ce{N_{0.9}V2}, reference code 00-030-1420).
			As mentioned earlier, TiN is well established as a capping material, with an extremely stable chemistry, so while binary compounds of V and N were indeed cross-checked during the XRD analysis in Uppsala, our confirmation bias (based on the \ce{V3Si} formation literature) led us to discard the hypothesis that vanadium nitrides had formed.
 			
 			A comparison was made with samples where \SI{80}{\nano\metre} of vanadium was deposited onto either HF-cleaned silicon (samples A42A,B,C,D, see Appendix.~\ref{sec:samples_v}), or silicon with remaining native oxide (A32A,B,C,D).
 			After annealing at \SI{840}{\celsius}, both showed three distinct peaks that could be attributed to \ce{VSi2}, two of which (42.8 and 49.2°) were smaller on the sample with native oxide than they were on the one that was cleaned with HF, suggesting that HF limits \ce{VSi2} growth.
 			On all samples, a broad peak centered around 42.50 associated with pure vanadium was identified prior to annealing, which disappeared afterwards, confirming that the V was consumed.
 			Furthermore, drops in sheet resistance after annealing (see Table~\ref{tab:samples_uppsala}) were also consistent with the formation of \ce{VSi2}.
 			Around \SI{232}{\nano\metre} would form by complete consumption of the vanadium, giving a resistivity of \SI{93\pm3}{\micro\ohm\centi\metre}, close to the \SI{80}{\micro\ohm\centi\metre} reported elsewhere for films cured for \SI{30}{\minute} at \SI{800}{\celsius}~\cite{nava1986electrical}.
 			A smaller drop in sheet resistance on the uncleaned sample is consistent with the smaller XRD peaks observed and could indicate that the \ce{VSi2} formation is slowed down by oxygen~\cite{schutz1979formation}.
 			In short, the evidence points towards \ce{VSi2} forming on a silicon substrate, and the inhibition of this reaction by the presence of a native oxide.
			
			\begin{figure}
				\centering
				\includegraphics[width=0.8\textwidth]{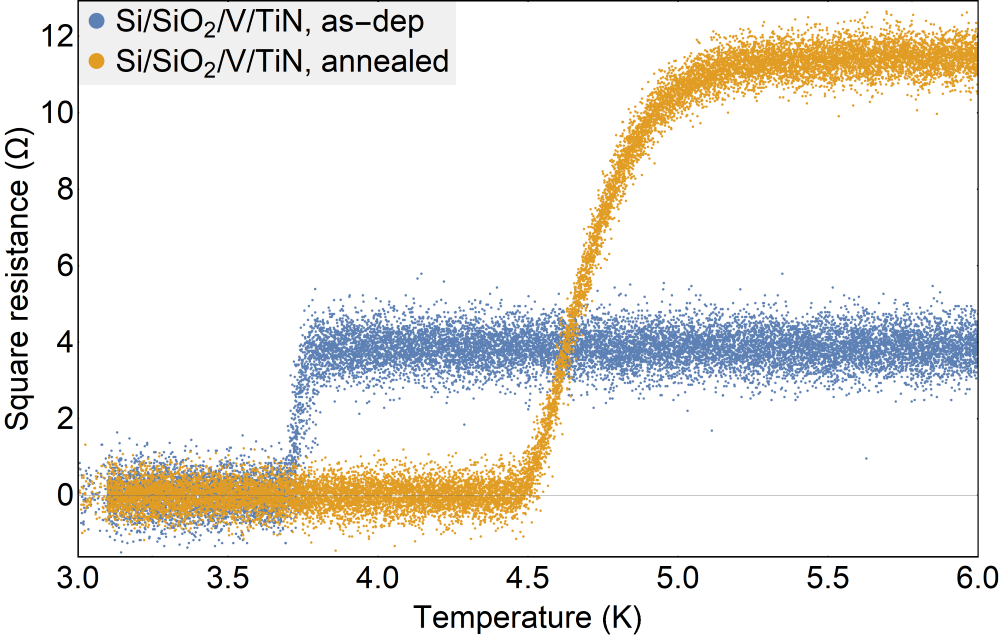}
				\caption{\label{fig:b54ab}The resistance measured during slow heating from 3 to \SI{6}{\kelvin} on an as-dep sample (blue) and a sample that was annealed in Uppsala at \SI{840}{\celsius} (orange). The critical temperature is found to increase after annealing.}
			\end{figure}
			
			More apparent evidence of indeed having grown \ce{V3Si} on samples where V was deposited onto \ce{SiO2} came in the form of low-temperature measurements performed later in Grenoble.
			Monocrystalline vanadium has a superconducting critical temperature of around \SI{5.4}{\kelvin}~\cite{radebaugh1966low,sekula1972magnetic,khlyustikov2021surface}, while that of high-quality TiN optimized for its superconductivity can be up to \SI{4.8}{\kelvin}~\cite{vissers2013characterization,yu2005fabrication}, though polycrystalline thin films of either are expected to have critical temperatures well below these values.
			Two samples with \SI{80}{\nano\metre} of \ce{V} deposited onto \SI{20}{\nano\metre} of oxide were measured at low temperature.
			One of these was as-deposited, while the other had been annealed at \SI{840}{\celsius} for 2 minutes (with a ramp rate \SI{20}{\celsius/\second}) under an \ce{N2} atmosphere.
			The critical temperatures were determined by following the resistance of $4\times\SI{10}{\milli\metre\squared}$ pieces during both cooling and heating at cryogenic temperatures (see Fig.~\ref{fig:b54ab}).
			The first of these samples (blue curve) showed a critical temperature of \SI{3.70}{\kelvin} with a transition width\footnote{The temperature difference between the points at which 10 and 90\% of the normal-state resistance is lost.} of \SI{0.09}{\kelvin}, which increased to \SI{4.69}{\kelvin} with a width of \SI{0.44}{\kelvin} for the second.
			
			There are many ways that this increase in $T_\text{c}$ can be explained.
			Critical temperatures of around \SI{5}{\kelvin} have been reported even for unannealed e-beam evaporated vanadium films of around \SI{80}{\nano\metre} thick~\cite{alekseevskii1976superconducting}, so it is possible that the vanadium simply improved in quality through grain growth.
			The presence of the broad V peak at 42.5° on as-deposited samples indicates that the film is already crystalline before annealing, though the grain size is still on the order of the X-ray wavelength (a few times $\lambda=\SI{1.54}{\angstrom}$).
			The same could be true for the TiN capping layer, and of course it is possible that a vanadium nitride film has formed.
			However, reports in the literature suggest that the reaction of V with \ce{SiO2} to form \ce{V3Si} should occur at 650~\cite{oya1982superconducting} or \SI{700}{\celsius}~\cite{tu1973formation}, and should be fast enough to consume at least half of the \SI{80}{\nano\metre} film at \SI{840}{\celsius}~\cite{krautle1974kinetics}.
			Moreover, it is a priori not likely that a thermodynamically stable material like TiN would react at a temperature in this range at all.
			In short, with sheet resistance, XRD and critical temperature results all in principle consistent with the formation of \ce{V3Si}, there was good reason to be hopeful that we had indeed been successful in obtaining this phase.
		
		\FloatBarrier	
		\subsection{Signs of trouble}
			
			\begin{figure}
				\centering
				\includegraphics[width=0.8\textwidth]{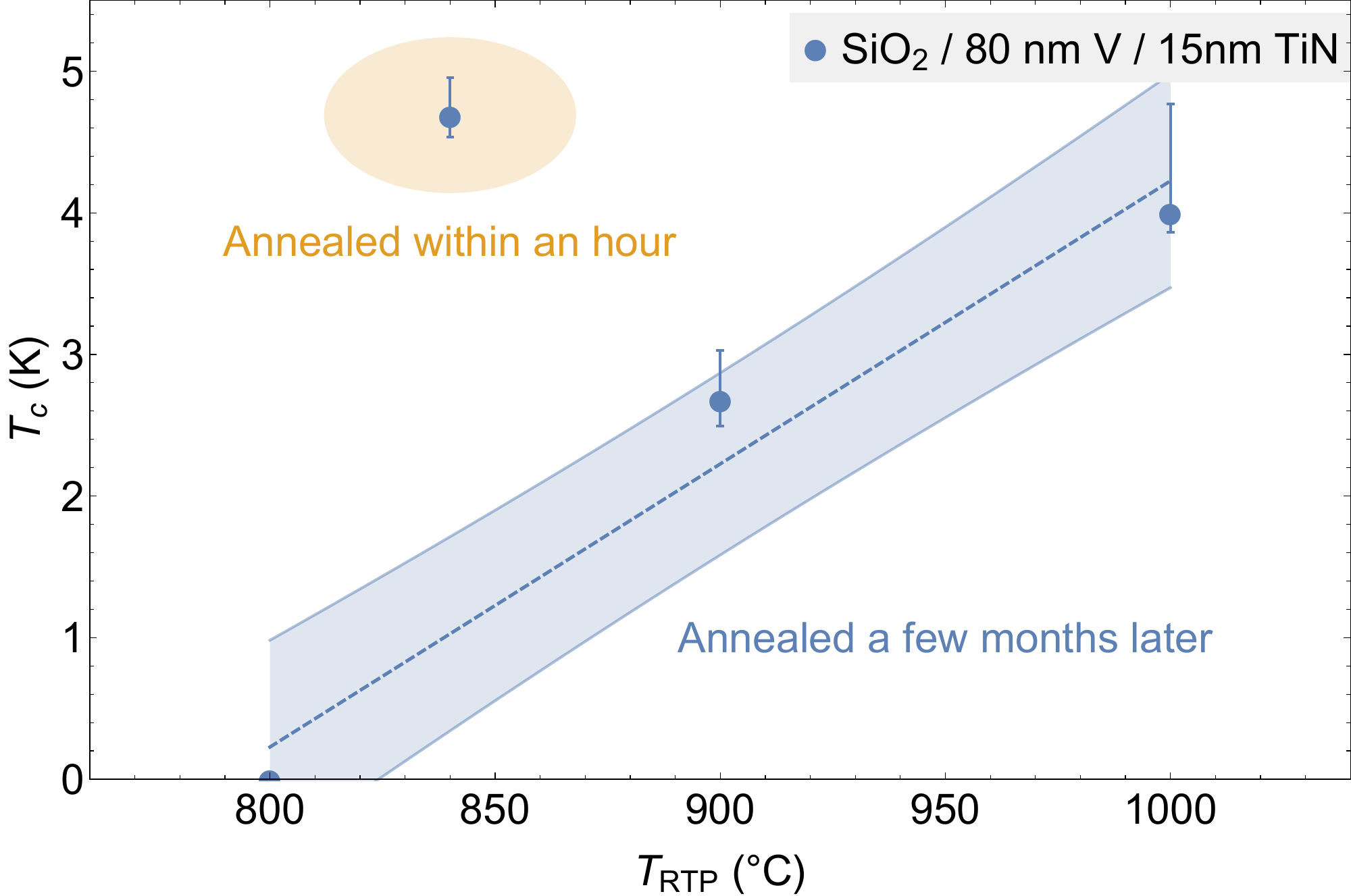}
				\caption{\label{fig:VdepSiO280nmTiNRTATcPlot}While samples that were stored for a few months (blue) also showed an increase in $T_\text{c}$ after annealing, these critical temperatures were in stark contrast to the result obtained earlier.}
			\end{figure}			
			
			Stimulated by the early results on the samples that were processed in Uppsala, we then moved on to evaluate a wider range of annealing temperatures in Grenoble.
			Five samples with \ce{SiO2} substrates were selected: two with \SI{40}{\nano\metre} of deposited vanadium, and three with \SI{80}{\nano\metre}.
			The thinner films were annealed for 2 minutes under a nitrogen atmosphere at 700 or \SI{800}{\celsius}, while the thicker ones were annealed at 800, 900 or \SI{1000}{\celsius}, after which the critical temperature of each was determined.
			Neither of the samples with \SI{40}{\nano\metre} of vanadium exhibited superconductivity, while critical temperatures of \SI{2.68}{\kelvin} (width of \SI{0.54}{\kelvin}) and \SI{4.00}{\kelvin} (0.91) were obtained after annealing at 900 and \SI{1000}{\celsius} (see Fig.~\ref{fig:VdepSiO280nmTiNRTATcPlot}).
			These results are clearly different from those obtained on the sample annealed in Uppsala, and could be explained either by different annealing conditions, or aging effects.
			
			\begin{figure}
				\centering
				\includegraphics[width=0.8\textwidth]{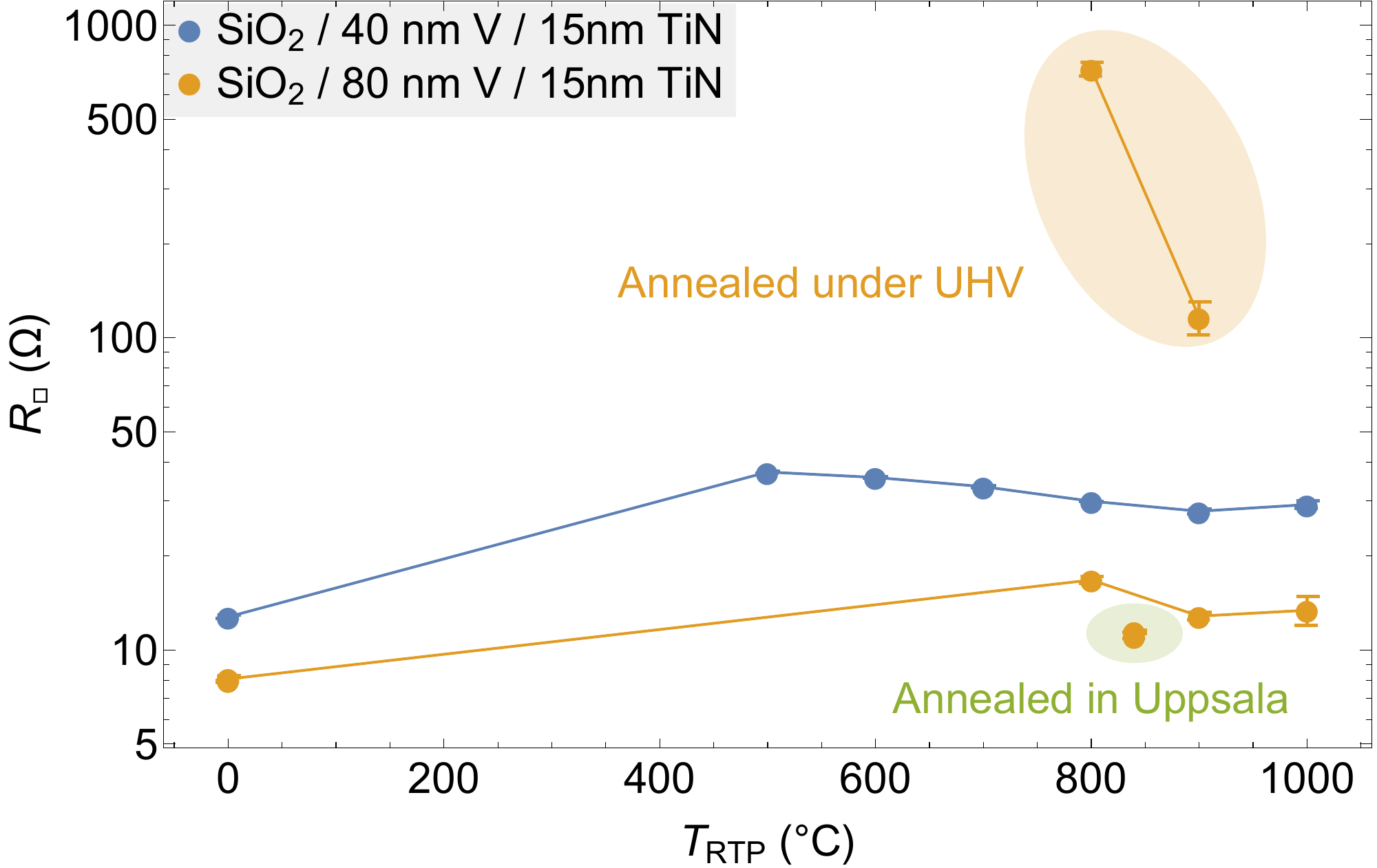}
				\caption{\label{fig:VdepSiO240nm80nmTiNRTARsPlot}Sheet resistance of samples where 40 or \SI{80}{\nano\metre} of V was deposited on \SI{20}{\nano\metre} of \ce{SiO2}. Three groups are indicated: samples annealed under \ce{N2} atmosphere within an hour in Uppsala (green background), samples annealed under \ce{N2} atmosphere a few months later in Grenoble (no colored background), and samples that were annealed under ultra high vacuum (orange background).}
			\end{figure}
			
			To gain more insight into the reactions that occur in this system, four more samples with \SI{40}{\nano\metre} of vanadium were annealed, at temperatures both lower (500 and \SI{600}{\celsius}), and higher (900 and \SI{1000}{\celsius}) than the previous two.
			An increase in sheet resistance relative to that of the as-deposited samples was already observed at \SI{500}{\celsius} (see Fig.~\ref{fig:VdepSiO240nm80nmTiNRTARsPlot}), far lower than \SI{650}{\celsius}, the lowest temperature at which a reaction between V and \ce{SiO2} has been recorded~\cite{oya1982superconducting,krautle1974reactions,krautle1974kinetics}.
			No such increase in sheet resistance would be expected if the deposited vanadium would remain unreacted, and only increase its grain size.
			
			\begin{figure}
				\centering
				\includegraphics[width=0.8\textwidth]{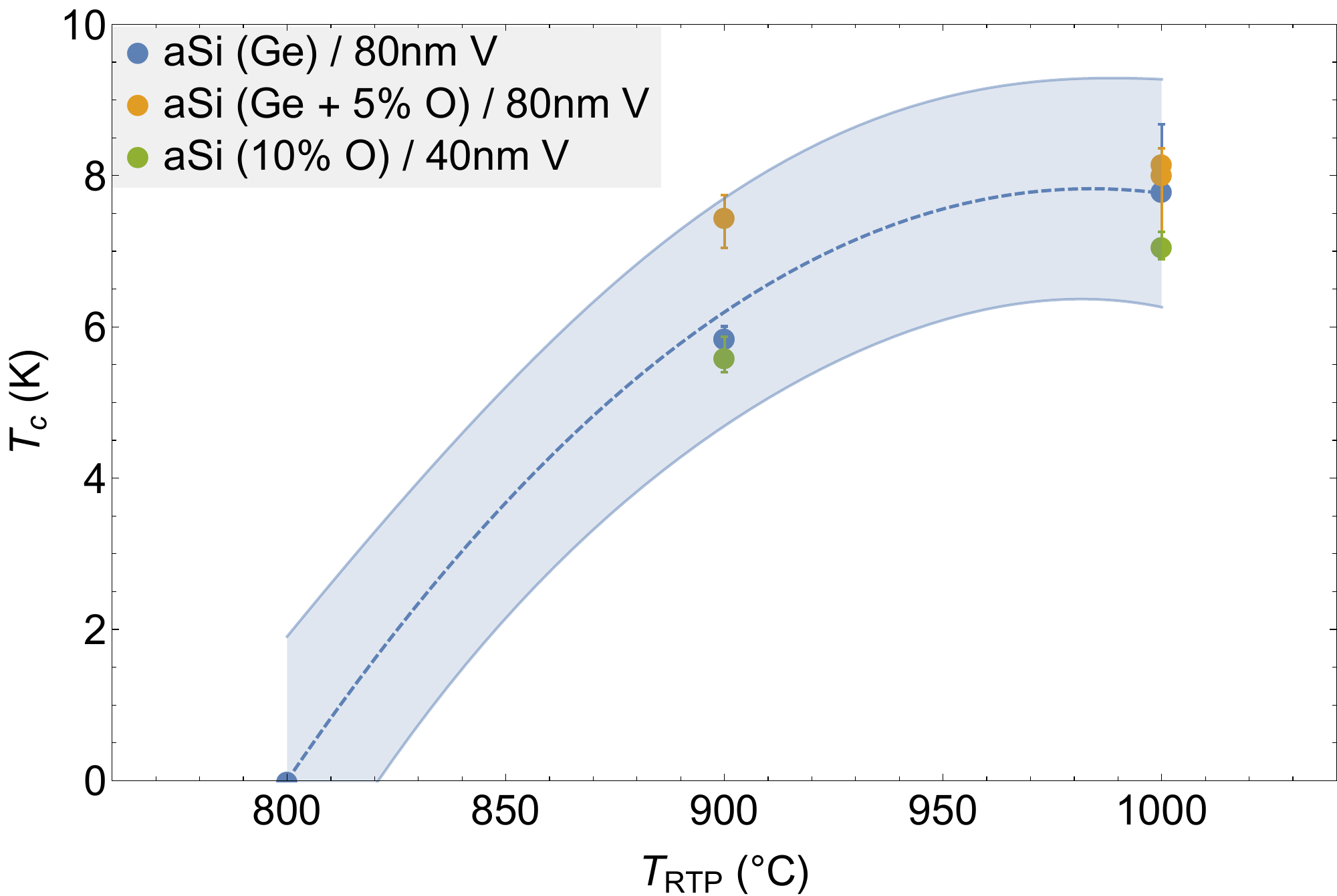}
				\caption{\label{fig:VdepaSiRTPTcPlot}Critical temperatures measured on samples where the substrate was amorphized by germanium and/or oxygen implantation.}
			\end{figure}			
			
			In a separate set of experiments, reactions of vanadium on amorphized silicon were studied.
			It has been reported that both amorphizing the silicon, and introducing small amounts of oxygen at less than 10\%at could trigger the growth of \ce{V3Si} instead of \ce{VSi2}~\cite{schutz1979formation}.
			To reproduce this, a set of thinly oxidized \SI{300}{\milli\metre} Si wafers were prepared in Grenoble, where oxygen was implanted to a final atomic concentration of 0, 0.1, 0.5, 1, 5 or 10\% just below the native oxide.
			A second round of implantations followed with heavier Ge atoms, which reached a maximum concentration of 1.4\% at a depth of \SI{15}{\nano\metre}, amorphizing it down to \SI{18}{\nano\metre} below the wafer surface.
			Layer of vanadium and TiN were later deposited in Uppsala, after which they were thermally processed in Grenoble.
			As on the \ce{SiO2} substrates, no superconductivity was observed after annealing at \SI{800}{\celsius}, while the critical temperature increased with processing temperature up to a maximum of \SI{8.17}{\kelvin} (see Fig.~\ref{fig:VdepaSiRTPTcPlot}).
			This $T_\text{c}$ is well above that reported for either vanadium or TiN, although it unfortunately does not surpass that of \ce{VN_x}, which can reach \SI{9.25}{\kelvin} when $x$ approaches unity~\cite{zhao1984superconducting}.

			\begin{figure}
				\centering
				\includegraphics[width=0.8\textwidth]{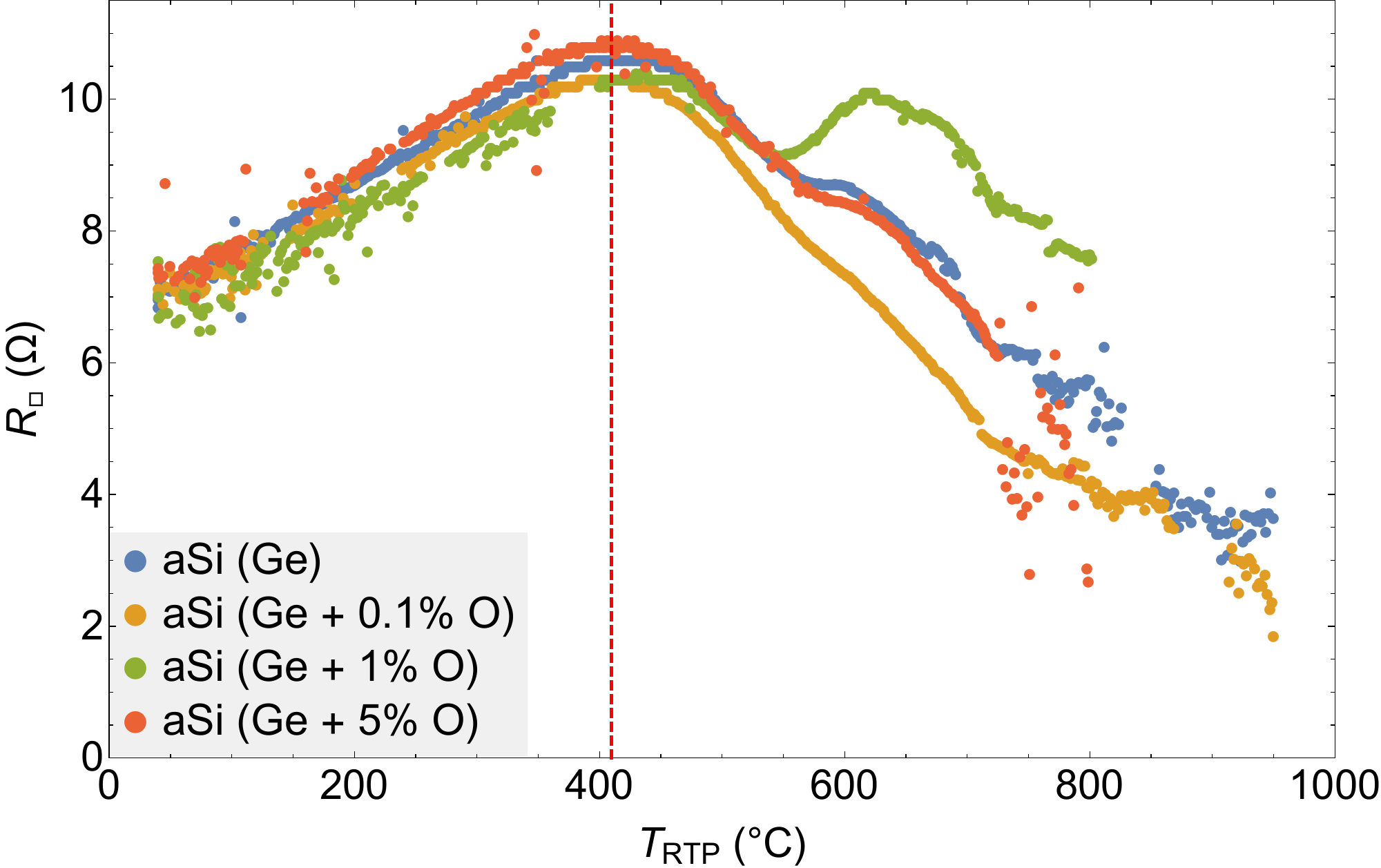}
				\caption{\label{fig:In_situ_sheet_resistance_aSi}In-situ sheet resistance $R_\square$ measured by four-point probe during the heating of $2\times\SI{2}{\centi\metre\squared}$ samples. No clear relationship is found between the oxygen content in the substrate, and the variation in $R_\square$.}
			\end{figure}	
			
			The sheet resistances of samples with amorphized silicon with various oxygen concentrations were measured in-situ during gradual heating with a ramp rate of \SI{1}{\celsius/\second} by master student Reda Alwaradi (see Fig.~\ref{fig:In_situ_sheet_resistance_aSi}).
			It was found that while there were variations in sheet resistance development between these samples, no clear correlation with the oxygen content could be deduced.
			More importantly, the maximum sheet resistance\footnote{The initial rise is due to an increase in dissolved atoms (in this case, N), as both the solubility limit and the diffusion rate grow with time.} is reached already at \SI{410}{\celsius}, far below\footnote{Note that these measurements are done at a high ramp rate, which means that when the reaction starts, the diffusion time through the thin film is longer than the time scale over which the diffusivity itself grows.
			Since~\cite{krautle1974kinetics}
			\begin{equation}D(T)=D_0\,e^{-\Delta E/k_\text{B} T},\end{equation}
			an increase by a factor $e$ in $\Delta T$ would occur over a change in temperature $\Delta T$ such that \mbox{$D(T+\Delta T)/D(T)=e$} (usually on the order of 10--$10^2\si{\kelvin}$),
			\begin{equation}\Delta T=\dfrac{\Delta E k_\text{B}T}{\Delta E-k_\text{B}T}-k_\text{B}T\approx \dfrac{(k_\text{B}T)^2}{\Delta E},\end{equation}
			which for a ramp rate $dT/dt$ gives a characteristic time $\tau$,
			\begin{equation}\tau\approx \dfrac{(k_\text{B}T)^2}{\Delta E}\left(k_\text{B}\,\dfrac{dT}{dt}\right)^{-1}.\end{equation}
			We will thus have a time lag in detecting the start of the reaction when the diffusion time is longer than $\tau$,
			\begin{equation}t_\text{diffusion}=\dfrac{d_\text{film}^2}{D(T)}>\dfrac{(k_\text{B}T)^2}{\Delta E}\left(k_\text{B}\,\dfrac{dT}{dt}\right)^{-1}.\end{equation}
			Typically, this leads to a few tens of degrees shift between the point at which an in-situ measurement detects a reaction~\cite{zhou1999kinetics}, and the temperature at which one would see it in samples that were annealed at constant temperatures for a few minutes, making the above observation even more at odds with \ce{V3Si} formation.} the temperature at which \ce{V3Si} is expected to form by reacting with an aSi substrate~\cite{schutz1979formation}.
			Later XRD analysis showed that on amorphized samples with anywhere between 0 and 10\%at. of O, each annealed at \SI{1000}{\celsius}, both VN and \ce{TiVN2} had consistently formed.
		
		\FloatBarrier
		\subsection{Lessons learned the hard way}
			
			\begin{figure}
				\centering
				\begin{tikzpicture}
					\node[anchor=center,inner sep=0] at (0,0) {
						\includegraphics[width=0.72\textwidth]{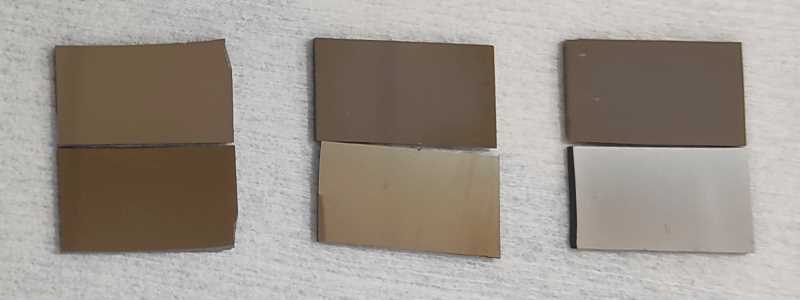}
						};
					\node[anchor=center] at (-0.23*\textwidth,0.17*\textwidth) {\SI{800}{\celsius}};
					\node[anchor=center] at (0*\textwidth,0.17*\textwidth) {\SI{900}{\celsius}};
					\node[anchor=center] at (0.23*\textwidth,0.17*\textwidth) {\SI{1000}{\celsius}};
					
					\node[anchor=west] at (-0.5*\textwidth,0.05*\textwidth) {\ce{N_2}};
					\node[anchor=west] at (-0.5*\textwidth,-0.05*\textwidth) {Vacuum};
					\node[anchor=east,white] at (0.5*\textwidth,0) {phantom};
					
					\node[anchor=center] at (-0.23*\textwidth,0.05*\textwidth) {A33B1};
					\node[anchor=center,white] at (-0.23*\textwidth-0.2,0.05*\textwidth+0.1) {A33B1};
					\node[anchor=center] at (-0.23*\textwidth,-0.05*\textwidth) {A33B2};
					\node[anchor=center,white] at (-0.23*\textwidth-0.2,-0.05*\textwidth+0.1) {A33B2};
					
					\node[anchor=center] at (0*\textwidth,0.05*\textwidth) {A33C1};
					\node[anchor=center,white] at (0*\textwidth-0.2,0.05*\textwidth+0.1) {A33C1};
					\node[anchor=center] at (0*\textwidth,-0.05*\textwidth) {A33C2};
					\node[anchor=center,white] at (0*\textwidth-0.2,-0.05*\textwidth+0.1) {A33C2};
					
					\node[anchor=center] at (0.23*\textwidth,0.05*\textwidth) {A33D1};
					\node[anchor=center,white] at (0.23*\textwidth-0.2,0.05*\textwidth+0.1) {A33D1};
					\node[anchor=center] at (0.23*\textwidth,-0.05*\textwidth) {A33D2};
					\node[anchor=center,white] at (0.23*\textwidth-0.2,-0.05*\textwidth+0.1) {A33D2};
				\end{tikzpicture}
				\caption{\label{fig:a33bcd}Samples with silicon substrates, \SI{40}{\nano\metre} V, capped with \SI{15}{\nano\metre} TiN, annealed under \ce{N2} (top) or vacuum (bottom). Even from just looking at the color of the samples, it is clear that the annealing atmosphere matters. Note that XRD analysis did not pick up on such change, and is thus at most an indicative, rather than a conclusive method of characterization (see Table~\ref{tab:xrd_stephane}).}
			\end{figure}
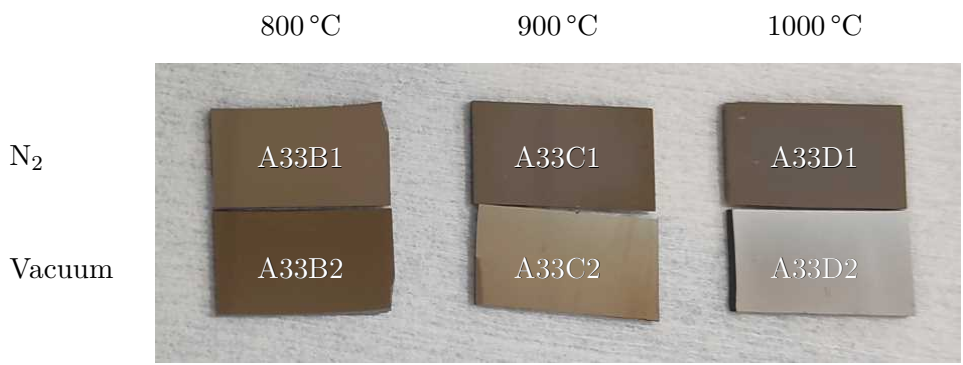
			
			\begin{table}
				\centering
				\caption{\label{tab:xrd_stephane}The results of XRD analysis on various samples. No \ce{V3Si} was formed anywhere, but a rich variety of other compounds was.}
				\resizebox{\columnwidth}{!}{%
				\rowcolors{2}{gray!15}{white}
				\begin{tabular}{l l l l l l l}
					\rowcolor{gray!30}\hline
					Sample		& Substrate	& V thickness	& Cap	& $T_\text{RTP}$	& RTP atmosphere	& XRD\\\hline\hline
					A33B1		& Si			& 40			& TiN	& 800				& \ce{N2}		& \ce{VSi2}, \ce{TiN}\\
					A33B2		& Si			& 40			& TiN	& 800				& UHV			& \ce{VSi2}, \ce{TiN}\\
					A33C1		& Si			& 40			& TiN	& 900				& \ce{N2}		& \ce{VSi2}, \ce{TiN}\\
					A33C2		& Si			& 40			& TiN	& 900				& UHV			& \ce{VSi2}, \ce{TiN}\\
					B56I1A		& \ce{SiO2}		& 100			& ---	& 800				& \ce{N2}		& \ce{V2O5}\\
					B74A		& \ce{SiO2}		& 80			& TiN	& 800				& UHV			& not \ce{V3Si}\\
					B74D		& \ce{SiO2}		& 80			& TiN	& 900				& UHV			& not \ce{V3Si}\\
					C26C		& aSi (Ge)		& 80			& TiN	& 800				& \ce{N2}		& \ce{VN}, \ce{TiVN2}\\
					C65C		& aSi (Ge + 5\% O)& 80			& TiN	& 800				& UHV			& \ce{VN}, \ce{TiVN2}\\
					E55C		& aSi (10\% O)	& 40			& TiN	& 900				& \ce{N2}		& \ce{VN}, \ce{TiVN2}\\\hline
				\end{tabular}%
				}
			\end{table}
			
			After it was found that the superconductivity on samples with amorphized silicon was due to the presence of VN, other samples were annealed under ultra-high vacuum instead of under \ce{N2} flux (see Table~\ref{tab:xrd_stephane}).
			No \ce{V3Si} was found on any of these samples either, suggesting that the capping layer may already have let through gases during storage, which would be consistent with the difference between the sample that was annealed in Uppsala immediately after deposition, and those that were annealed later in Grenoble.
			It is clear however, from the large change in sheet resistance upon switching from \ce{N2} to UHV (see Fig.~\ref{fig:VdepSiO240nm80nmTiNRTARsPlot}), that at least \emph{some} \ce{N2} diffusion through the TiN layer occurs during annealing under \ce{N2} flux.
			Though this may be due to the particular composition of the TiN used in Uppsala (calibrated to around 1:1 Ti:N), and may not apply to TiN with different stoichiometries, the sheer ubiquity of this material as a capping layer in the semiconductor industry makes this fact worth exploring further.
			XRD analysis on other samples, where no capping layer was deposited at all, showed that \ce{V3O5} had formed.
			It is thus imperative to find a different material that can be used to cap deposited layers of vanadium, before \ce{V3Si} can be formed by metal deposition.

			\begin{figure}
				\centering
				\includegraphics[width=0.8\textwidth]{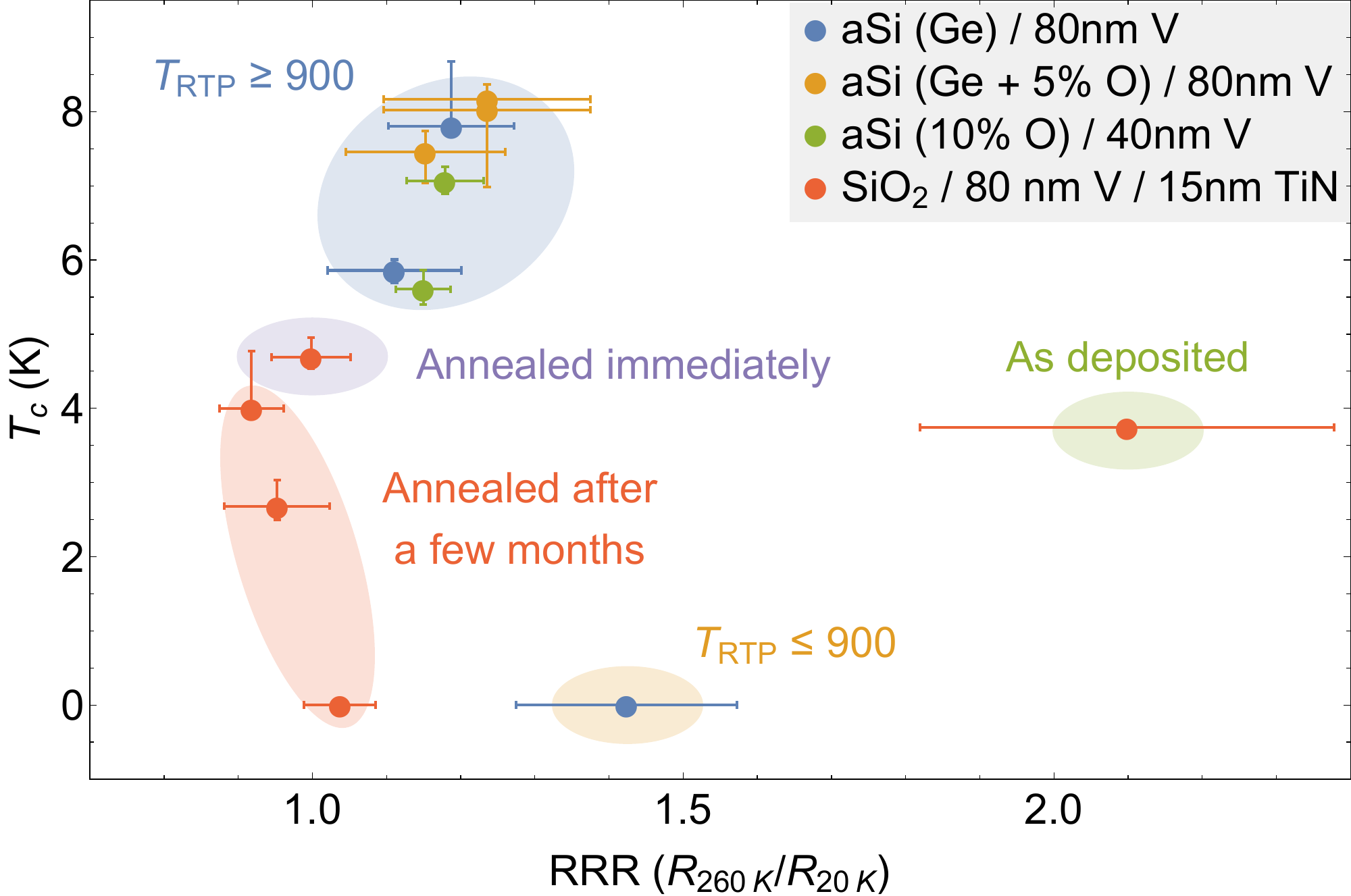}
				\caption{\label{fig:VdepCombinedRRRTcaSiSiO2Plot}The resistance ratio (RRR) vs the critical temperature ($T_\text{c}$) of a selection of samples that were measured at low temperature. Based on analysis of XRD, sheet resistance and low-temperature data, five groups of similar material properties are identified.}
			\end{figure}
			
			Though the compounds detected by XRD (see table~\ref{tab:xrd_stephane}) are all thought to be products of a reaction between vanadium and either the TiN capping layer or \ce{N2} in the storage and annealing atmospheres, the substrate is found to be of some influence.
			As can be seen in Fig.~\ref{fig:VdepCombinedRRRTcaSiSiO2Plot}, there are large differences between the samples where the vanadium was deposited on amorphous silicon, and those with \SI{20}{\nano\metre} of \ce{SiO2}.
			
			A few clear lessons can be drawn from our experience with this project, in which disproportionately much time was spent on what later turned out to be a dead end:
			\begin{enumerate}
				\item \B{Identify first, \emph{then} characterize.} 
				Early results were consistent with the formation of \ce{V3Si}, though none actually confirmed it.
				Characterization of various physical properties on a wide range of samples continued for more than a year before identification methods (XRD) became available.
				This time could have been better spent.
				\item \B{Verify.} 
				It was assumed that TiN would not react with the vanadium or the atmosphere, and we expected that repeating the methods described in the literature would reliably produce \ce{V3Si}.
				Fabrication is a complex process, with many unknowns.
				\item \B{Diversify.} Ultimately, we were able to recover lost ground thanks to the arrival of a 3:1 V:Si deposition target.
			\end{enumerate}

\clearpage
\FloatBarrier
\section{\ce{V3Si} formation by deposition from a compound target}
	
	Though silicidation, the reaction of a deposited metal with the underlying silicon, is a common technique for silicide formation in CMOS devices (notably through the SALICIDE technology), we saw in the previous section that it did not work out for \ce{V3Si}.
	The main difficulty is that this phase is unlikely to form in the traditional scenario of metal deposition onto the monocrystalline exposed source/drain silicon, which instead leads to \ce{VSi2}.
	Besides lowering the activation energy of \ce{V3Si}, or slowing down the dominant diffusing species in \ce{VSi2} formation, a third option is to directly prepare the deposited film with the right atomic ratio (see Fig.~\ref{fig:four_methods_v3si4}).
	Thanks to a target provided by JX Nippon, this last option was put into practice to great success.
	
	\FloatBarrier
	\subsection{\label{sec:strain_sound}Strain, sound velocity and a drop in critical temperature}
		
		One important fact that has become central in our work, is that strain has a large effect on the physical properties of \ce{V3Si}, just as it does on other A-15 compounds.
		This topic was researched extensively in the 60s and 70s~\cite{smith1970pressure,smith1971variation,fawcett1971low,sham1971thermodynamics,shelton1975pressure,smith1975superconductivity}, mainly at Bell Labs~\cite{batterman1964crystal,testardi1965lattice,batterman1966low,testardi1967lattice402,testardi1967lattice,testardi1970unusual,testardi1971unusual95,testardi1971unusual,testardi1972structural,testardi1975structural,testardi1977anomalous}, and ultimately relates to a sudden weakening of the restoring force for shear deformation at low temperatures~\cite{testardi1965lattice}.
		Since the potential energy stored in a spring is proportional to this restoring force (the ``spring constant''), a weakening is associated with a change in the lowest-energy configuration of the atoms, and the \emph{a priori} cubic \ce{V3Si} becomes tetragonal, with two of the axes (henceforth $a$ and $b$) contracting, while the other ($c$) extends~\cite{batterman1964crystal}.
		A nice corollary is that the spring constant also determines the resonant frequency (see section~\ref{sec:quantum_harmonic_oscillator}), and thus the rate at which displaced atoms bounce back, allowing for the study of the mechanical properties of the material by sound velocity measurements.
		
		Without doing any justice to the profundity of Testardi's analysis, which involves arguments about the thermodynamic, electrical and superconducting properties of A-15 compounds like \ce{V3Si} (references above for the avid reader), I will now boil down the relation between strain and a change in superconducting critical temperature to four hand-waving lines of equations.
		First, let us introduce the \emph{stiffness parameters} $c_{ij}$, which are components of the elastic tensor that relate the stresses $\sigma_{ij}$ to the strain $\epsilon_{ij}$, which due to symmetries can be reduced to a matrix~\cite{alberts2002ultrasonic}\footnote{The $i,j=1,2,3$ terms represent the coupling of normal stress components, while the $i,j=4,5,6$ are the shear components. The zeros imply that normal stresses lead to normal strains, and shear stresses lead to shear strains~\cite{eberhardt2017stress}.},
		\begin{equation}\sigma_{ij}=c_{ij}\epsilon_j,\qquad c_{ij}=
		\Pm{c_{11}&c_{12}&c_{13}&&&\\
		c_{21}&c_{22}&c_{23}&&\emptyset&\\
		c_{31}&c_{32}&c_{33}&&&\\
		&&&c_{44}&&\\
		&\emptyset&&&c_{55}&\\
		&&&&&c_{66}}.\end{equation}
		Like an ordinary one-dimensional spring constant, they are the factor of proportionality between the force and the displacement\footnote{To be more precise, in this three-dimensional case between the \emph{force per unit area} or \emph{pressure} and the displacement.}, the latter of which in the case of crystals corresponds to the strain $\epsilon=(d-d_0)/d_0$.
		Zooming in on a single unit of volume (to get the right units for the argument),
		\begin{equation}\vec{F}_i=-\Sum_{j}c_{ij}\epsilon_j,\quad V=-\Int\vec{F}\cdot d\vec{r}=\dfrac{1}{2}\Sum_{ij}c_{ij}^2\epsilon_{j}\quad\Rightarrow\quad c_{ij}=\left.\left(\dfrac{\partial^2V}{\partial\epsilon_i\partial\epsilon_j}\right)\right|_T,\end{equation}
		relating the stiffness parameters to the energy stored in the mechanical strain.
		In superconductors, the energy associated with condensing into the superconducting state is proportional to $E_\text{cond}\propto(T_\text{c}-T)^2$~\cite{tinkham2004introduction}, which means that
		\begin{equation}\partial_TE_\text{cond}\propto T_\text{c}\quad\Rightarrow\quad\dfrac{\partial^2T_c}{\partial\epsilon_i\partial\epsilon_j}\propto\dfrac{\partial}{\partial T}\left(\dfrac{\partial E_\text{cond}}{\partial \epsilon_i\partial\epsilon_j}\right)=\dfrac{\partial c_{ij}}{\partial T}=\rho\dfrac{\partial v_{\text{sound},ij}}{\partial T}.\end{equation}
		So if a reduction in $T_\text{c}$ with strain $\vec{\epsilon}$ is indeed linked to a change in crystal stiffness, then sound velocity measurements should match the prefactors in the series expansion of $T_\text{c}(\epsilon)$,
		\begin{equation}\label{eq:tc_strain}T_\text{c}(\epsilon)-T_\text{c}(0)=\Sum_i\Gamma_i\epsilon_i+\dfrac{1}{2}\Sum_i\Sum_j\Delta_{ij}\epsilon_i\epsilon_j+\mathcal{O}(\epsilon^3),\end{equation}
		with e.g. for the parallel sound velocity $\rho\partial_Tv_{\text{sound},\parallel}^2=\Delta_{11}$.
		Measurements confirmed this hypothesis~\cite{testardi1965lattice,testardi1967lattice,testardi1967lattice402,testardi1971unusual95}, providing values for these prefactors of $|\Gamma|<\SI{50}{\kelvin}$, $\Delta_{11}=-\SI{2.4E5}{\kelvin}$, $\Delta_{12}=-\SI{5E4}{\kelvin}$ and $\Delta_{44}=\SI{-1E4}{\kelvin}$, while all other $\Delta_{ij}=0$~\cite{testardi1970unusual}.
	
	\FloatBarrier
	\subsection{Substrate-induced thermal stress}
		
\begin{figure}
			\centering
			\includegraphics[width=0.8\textwidth]{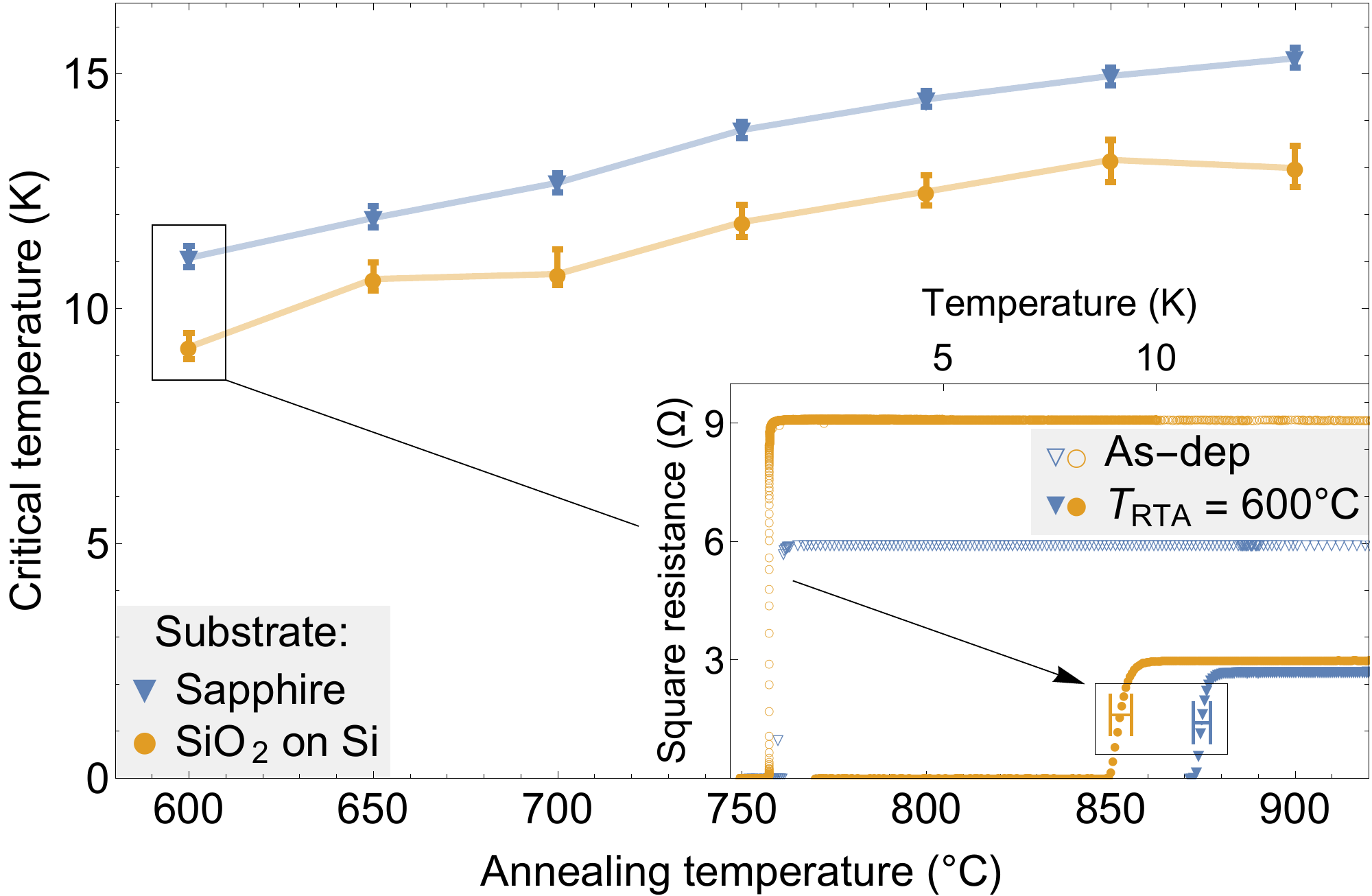}
			\caption{\label{fig:jap_fig1}The critical temperature ($T_\text{c}$) of samples with a \SI{200}{\nano\metre} thin film of \ce{V3Si} is plotted versus the temperature at which each sample was annealed, with error bars indicating the temperature at which t10 and 90\% of the normal-state resistance was lost. \B{Inset:} $R(T)$ measurements on the first two samples annealed at \SI{600}{\celsius}, as well as on an unannealed sample for each substrate. The horizontal error bars in the inset correspond to the vertical bars in the main graph.}
		\end{figure}

		One way that strain can be introduced into the silicide, is by attaching it to a material with a different thermal expansion coefficient.
		As the substrate-film couple is heated during crystallization annealing, several sequential stress developments occur that depend strongly on the expansion of the film relative to the substrate (more on this later), which are compounded by later cooling to cryogenic temperatures.
		This strain development was studied by depositing \SI{200}{\nano\metre} \ce{V3Si} films from a compound target onto substrates made of sapphire and silicon with \SI{20}{\nano\metre} of \ce{SiO2}, the details of which are published elsewhere~\cite{vethaak2021influence}.
		As can be seen in Fig.~\ref{fig:jap_fig1}, the differences in accumulated strain between these two substrates leads to a relative reduction in $T_\text{c}$ on silicon.
		The absence of any relationship between the residual resistance ratio and the critical temperature provides evidence that this relatively low $T_\text{c}$ is not due to either a chemical or morphological dependence on the substrate\footnote{Early on in this project, it was already known from the literature that no detrimental chemical reaction should occur between \ce{V3Si} and \ce{SiO2}, and that improvements in crystallinity and thus $T_\text{c}$ due to lattice matching~\cite{shiino2010improvement} are unlikely to appear on hexagonal sapphire (\ce{V3Si} is cubic, and the numbers don't add up even for weak matching once every few cells).
		Nonetheless, arguments by yours truly that strain should be responsible for the differences in $T_\text{c}$ between the silicon and sapphire substrates took a while to be taken seriously, and models that later turned out to align with those of Testardi were initially disregarded in favor of ``interface effect'' interpretations.
		Given the overwhelming evidence that strain \emph{is} the most relevant variable in this system~\cite{smith1970pressure,smith1971variation,fawcett1971low,sham1971thermodynamics,shelton1975pressure,smith1975superconductivity,batterman1964crystal,testardi1965lattice,batterman1966low,testardi1967lattice402,testardi1967lattice,testardi1970unusual,testardi1971unusual95,testardi1971unusual,testardi1972structural,testardi1975structural,testardi1977anomalous,vethaak2021influence}, this now-irrelevant concern will not be further discussed here.}.
		
		All of these interactions with the substrate are simplified a bit by symmetry: both the silicon and sapphire were monocrystalline and oriented such --- (100) and (0001) respectively --- that their expansion was isotropic in the plane.
		Furthermore, though \ce{V3Si} also expands when heated~\cite{testardi1972structural}, at any fixed temperature its volume remains constant under stress~\cite{batterman1966low,testardi1970unusual}, which means that an in-plane compression is associated with a predictable out-of-plane expansion.
		Given some in-plane strains $\epsilon_1=\epsilon_2$, we can thus directly compute
		\begin{equation}\epsilon_3=\dfrac{-2\epsilon_1-\epsilon_1^2}{(1+\epsilon_1)^2}\;\stackrel{\epsilon_i\ll1}{\approx}\;-2\epsilon_1.\end{equation}
		This then gives a strain vector of
		\begin{equation}\vec{\epsilon}=\left(\epsilon_1,-\dfrac{\epsilon_1}{2},-\dfrac{\epsilon_1}{2},0,0,0\right),\quad\text{where}\quad\epsilon_1=\dfrac{2}{3}\left(\dfrac{c}{a}-1\right),\end{equation}
		with $a=b$ and $c$ the in-plane and out-of-plane lattice parameters, respectively, as mentioned in section~\ref{sec:strain_sound}.
		We can now simplify the general expression that we found in eq.~\eqref{eq:tc_strain}, to
		\begin{equation}\label{eq:testardi_simplified}T_\text{c}(\vec{\epsilon})-T_\text{c}(0)= \dfrac{3}{4}\,\epsilon_1^2\left(\Delta_{11}-\Delta_{12}\right),\end{equation}
		such that measurement of the strain along a single direction is enough to determine the expected reduction in $T_\text{c}$.
		
		\begin{figure}
			\centering
			\includegraphics[width=0.8\textwidth]{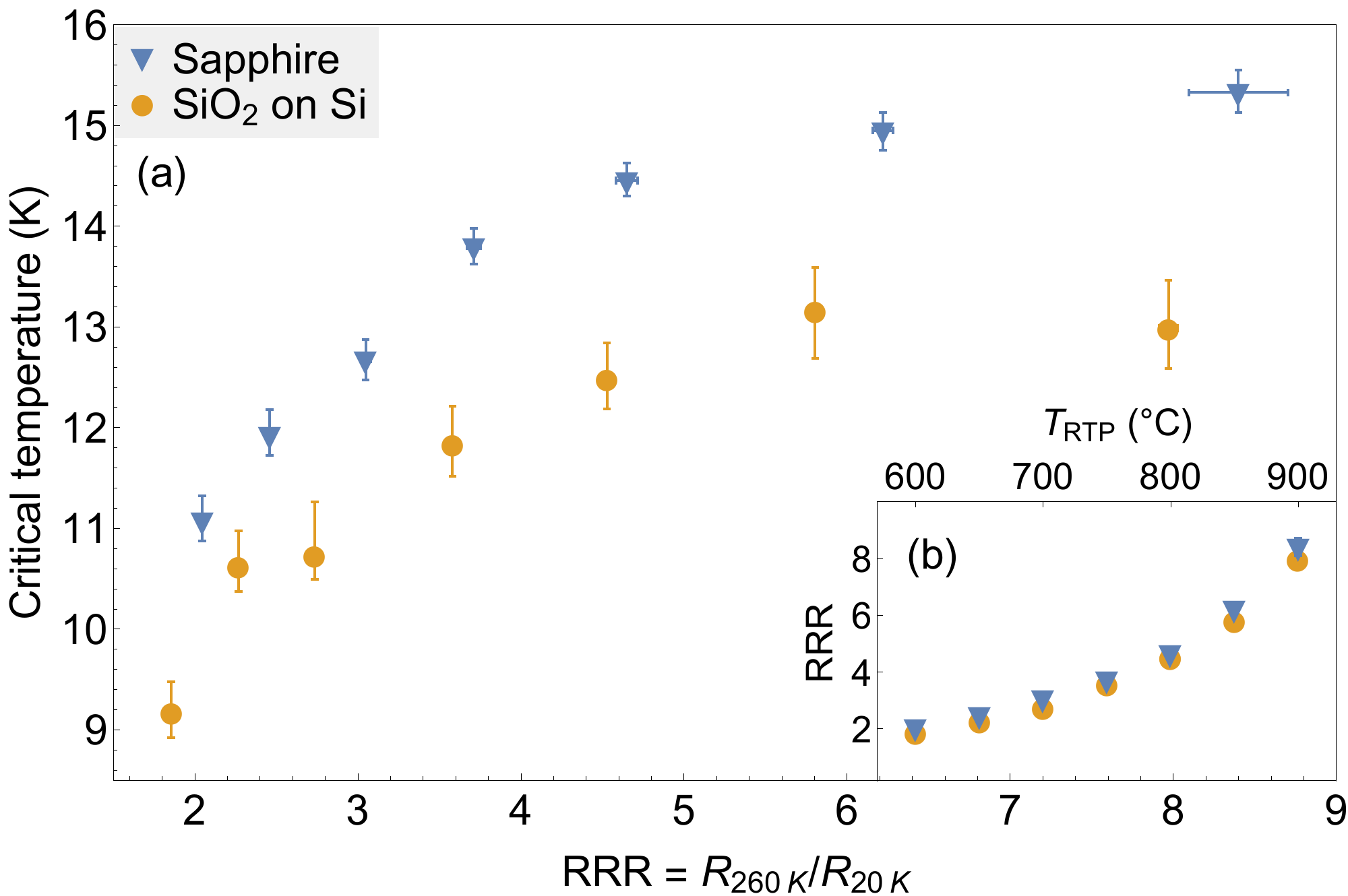}
			\caption{\label{fig:jap_fig2}\B{(a)} The residual resistance ratio $\text{RRR}=R_\text{\SI{260}{\kelvin}}/R_\text{\SI{20}{\kelvin}}$ is a measure of material quality, where any resistance that is \emph{residual} after the thermal electron-phonon scatterings are suppressed, is assumed to be due to impurities and defects. For any given RRR, there is a large difference in $T_\text{c}$ between the two substrates, indicating that it is affected by something other than the quality of the \ce{V3Si}. \B{(b)} A near-perfect match between the RRRs measured on sapphire and silicon shows that the substrate has little influence on the quality of the \ce{V3Si} film.}
		\end{figure}	

		\begin{table}%
			\centering%
			\caption{\label{tab:induced_contractions}The estimated contractions of sapphire~\cite{wachtman1962linear}, silicon~\cite{lyon1977linear} and \ce{V3Si}~\cite{testardi1972structural,smith1975superconductivity} from 300 to \SI{16}{\kelvin} (valid both in-plane and out-of-plane), and the induced out-of-plane expansion of \ce{V3Si}.}%
			\rowcolors{2}{gray!15}{white}
			\begin{tabular}{l@{\;\;\;}l@{\;\;\;}l@{\;\;\;}l}
				\rowcolor{gray!30}\hline
				Material	& Contraction 				& Relative to \ce{V3Si}	& Induced $\Delta\epsilon_1$\\\hline\hline
				Sapphire
							& $-\SI{6.2\pm0.8E-4}{}$	& $+\SI{5.8\pm1.3E-4}{}$& $-\SI{1.2\pm0.3E-3}{}$\\
				Silicon
							& $-\SI{2.33E-4}{}$			& $+\SI{9.7\pm1E-4}{}$	& $-\SI{1.9\pm0.2E-3}{}$\\
				\ce{V3Si}
							& $-\SI{1.2\pm0.1E-3}{}$	& ---					& ---\\\hline
			\end{tabular}
		\end{table}
		
		It was found (see Fig.~\ref{fig:jap_fig1}) that the superconducting transition occurs at temperatures \SI{1.9\pm0.3}{\kelvin} lower on a silicon substrate than it does on sapphire.
		In the most naive interpretation (useful to get a sense of the orders of magnitude), where only one of the two substrates imposes a strain, this would correspond to an out-of-plane deformation of \SI{3.7\pm0.3E-3}{}, or about 0.4\%.
		The reality is more complex however, and we will need to take into account both the strain built up during the thermal processing, and that induced by the relative contraction of the silicide during subsequent cooling to cryogenic temperatures.
		The last cooling step from \SI{300}{\kelvin} to \SI{16}{\kelvin} is in fact not that important, as there is little difference in this temperature range between the thermal strains induced by the sapphire and silicon substrates (see Table~\ref{tab:induced_contractions}).
		Instead, more attention should be paid to the intricate pattern of stress developments during thermal processing, which was studied in detail with two separate experiments.
		
\begin{figure}
			\centering
			\includegraphics[width=0.8\textwidth]{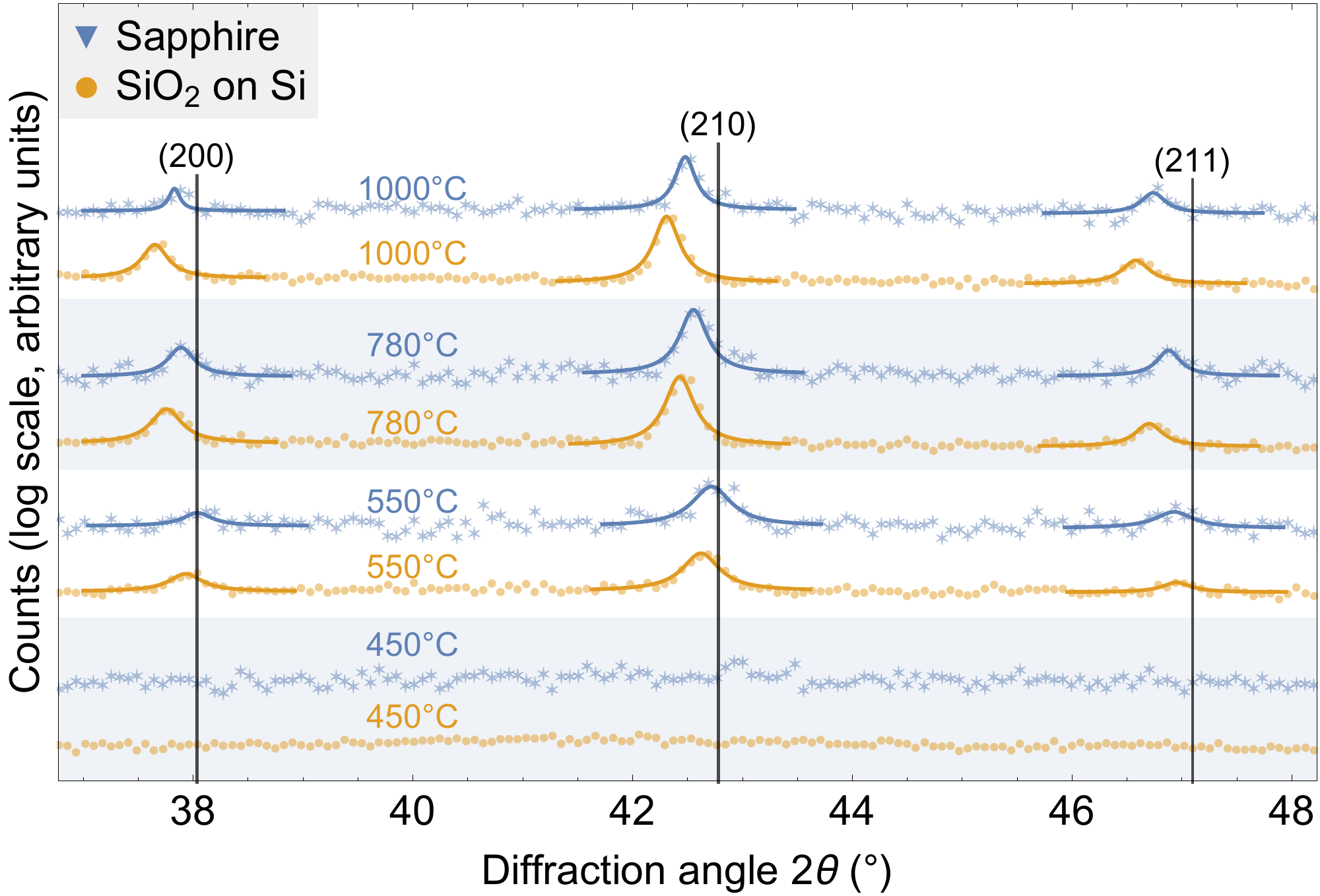}
			\caption{\label{fig:jap_fig3}A selection of XRD $\theta/2\theta$ curves, where each peak corresponds to a reflecting plane of the cubic \ce{V3Si} crystal (hkl indices for expected peak positions at \SI{300}{\kelvin} indicated above). No peaks are observed at all below \SI{500}{\celsius}, indicating that the silicide is amorphous, while they become higher and sharper as the temperature is increased. A leftward shift towards smaller angles is due to the thermal expansion of the crystal.}
		\end{figure}

\begin{figure}
			\centering
			\includegraphics[width=0.8\textwidth]{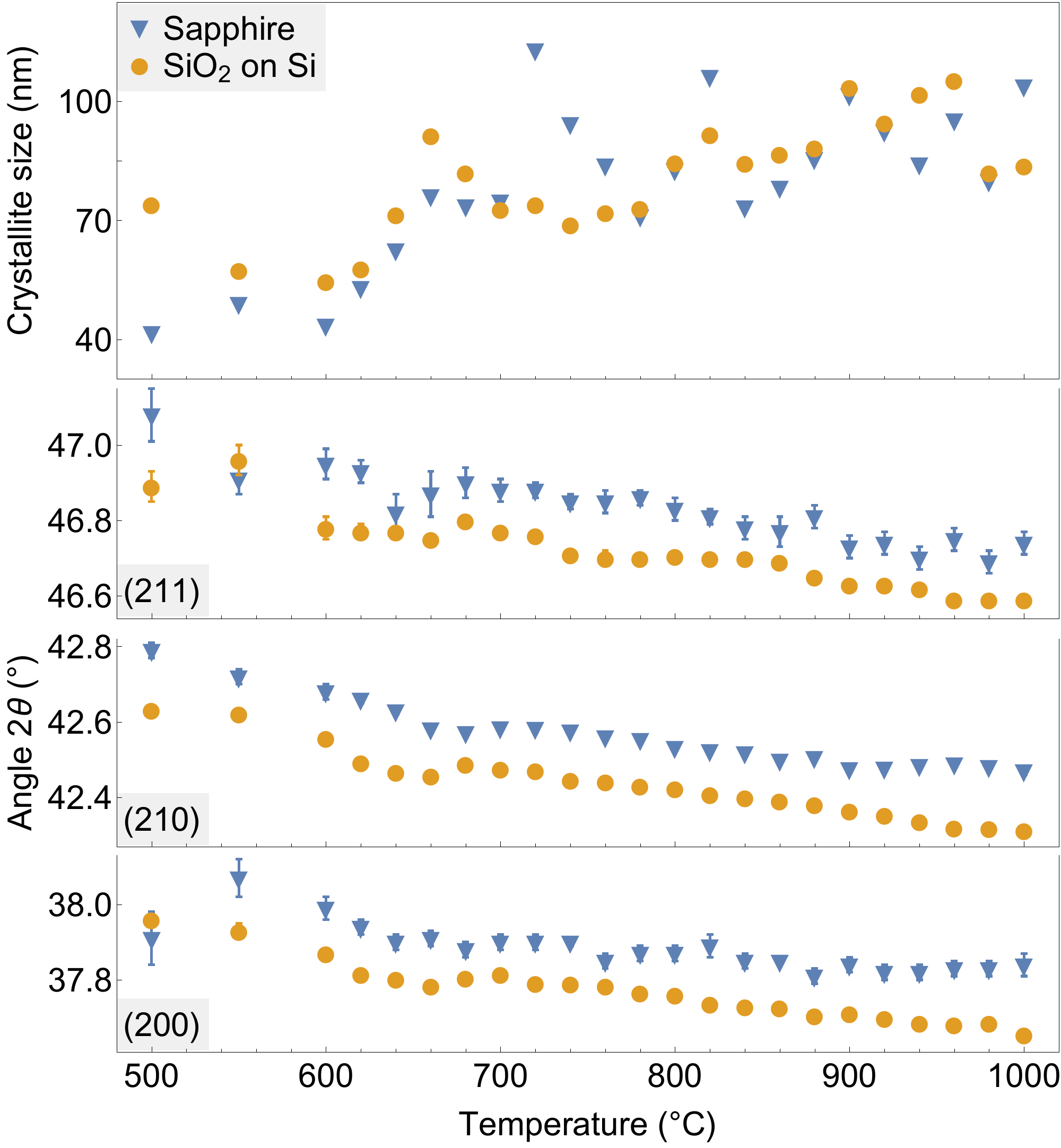}
			\caption{\label{fig:jap_fig4} \B{(top)} A rough estimate of the crystallite size can be obtained from the peak width using the Scherrer equation~\cite{scherrer1918nachr,cullity2013elements}. The measurement on the sapphire sample was performed with a slightly smaller slit size, leading to less instrumental line broadening and thus a larger apparent grain size at high temperatures. \B{(bottom panels)} Plotted is the shift in the peak position with temperature (see Fig.~\ref{fig:jap_fig3}), from which the out-of-plane lattice parameter can then be extracted (see Fig.~\ref{fig:jap_fig5}).}
		\end{figure}

\begin{figure}
			\centering
			\includegraphics[width=0.8\textwidth]{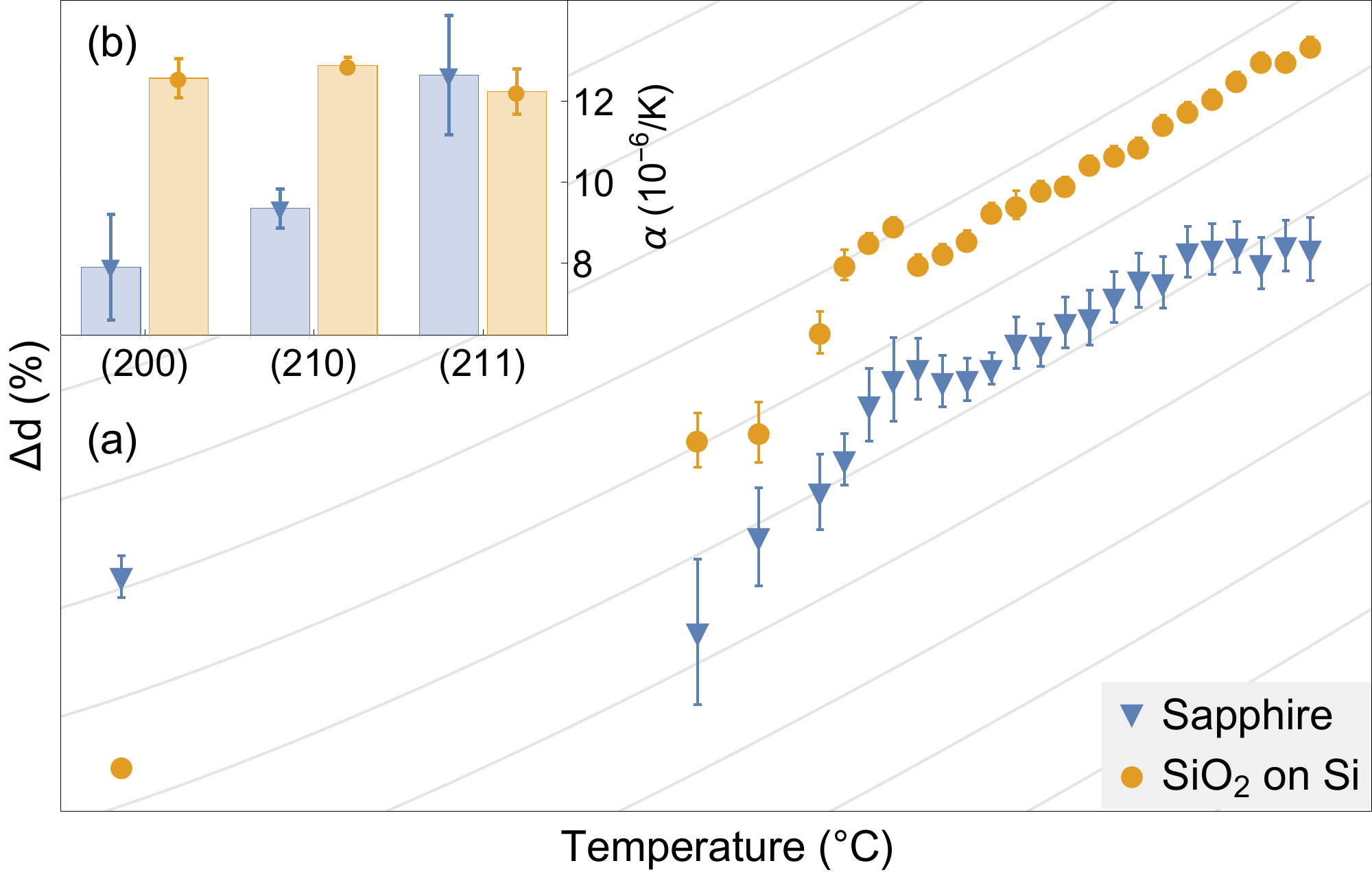}
			\caption{\label{fig:jap_fig5}\B{(a)} Using data from Fig.~\ref{fig:jap_fig4}, the change in out-of-plane lattice parameter during heating is plotted versus the temperature. Another data point was taken after the samples had cooled down to \SI{30}{\celsius}. Gray inclined lines in the background show the thermal expansion that would be observed on free-standing crystalline bulk \ce{V3Si}~\cite{testardi1972structural}, with different offsets.
			A clear difference can be observed between the two substrates: the silicide starts out more strained on sapphire right after crystallization at \SI{500}{\celsius}, while it ends up more strained on silicon at the end of the heating cycle.
			\B{(b)} Effective out-of-plane thermal expansion coefficients are extracted for each XRD peak from linear regressions between 700 and \SI{1000}{\celsius} on the peak positions in Fig.~\ref{fig:jap_fig4}. While the expansion rate is independent of crystallite orientation on silicon, tilted planes expand faster on sapphire.}
		\end{figure}

		The first of these was the slow, step-wise heating of two \emph{a priori} unannealed samples (one for each substrate) under secondary vacuum up to \SI{1000}{\celsius} while in-situ XRD scans were performed (see Figs.~\ref{fig:jap_fig3},~\ref{fig:jap_fig4} and~\ref{fig:jap_fig5}).
		On each $\theta/2\theta$ scan, Lorentzian distributions were fitted using Mathematica around the expected $2\theta$ values of the (200), (210) and (211) peaks (drawn superposed in Fig.~\ref{fig:jap_fig3}), giving fit parameters of both the position and width of each peak (shown in Fig.~\ref{fig:jap_fig4}).
		The slow shifts in the position of the three peaks were then weighted by peak intensity (number of counts), and used to estimate the crystallite expansion $\Delta d$ out of plane (Fig.~\ref{fig:jap_fig5}).	
		
		No significant difference in grain growth was observed between the two substrates (estimated from crystallite size, see Fig.~\ref{fig:jap_fig4}), consistent with the earlier conclusion from RRR measurements that there is no dependence of the material quality on the substrate.
		However, as shown in Fig.~\ref{fig:jap_fig5}, there is a large difference in the out-of-plane strain that was developed.
		Most importantly, the strain difference inverted after the samples were cooled down, leading to an out-of-plane strain of only $\epsilon_1=+\SI{0.2\pm0.4E-3}{}$ on sapphire, and $\epsilon_1=-\SI{3.3\pm0.1E-3}{}$ on silicon.
		Note that this implies that a stress relaxation occurred on sapphire at some point during the cooling from \SI{1000}{\celsius} to room temperature, since the higher thermal expansion coefficient of \ce{V3Si} ($\alpha_\text{\ce{V3Si}}\approx\SI{7.5E-6}{\per\kelvin}$~\cite{testardi1972structural}, while $\alpha_{s\text{sapphire}\perp(0001)}\approx\SI{5.0E-6}{\per\kelvin}$~\cite{dobrovinskaya2009properties}) would otherwise have caused the out-of-plane lattice parameter to shrink \emph{faster} than free-standing \ce{V3Si}, giving its trajectory a larger slope than the indicated gray lines in Fig.~\ref{fig:jap_fig5}.

		\begin{figure}
			\centering
			\begin{subfigure}[b]{0.475\textwidth}
				\centering
				\begin{tikzpicture}[x=1cm,y=1cm,thick]
					\draw[white] (0,-3.4) -- (0,0);
					\draw[blue!80!black,ultra thick,opacity=0.4] (-2.9,0.7) -- (-0.5,0.7);
					\draw[blue!80!black,ultra thick,opacity=0.4] (-2.9,-0.7) -- (-0.5,-0.7);
					\draw (-1,-0.7) rectangle ++(-1.4,1.4);
					\draw[ultra thick,green!60!black,->,>=stealth] (-0.5,0) -- (0.5,0);
					\draw[blue!80!black,ultra thick,opacity=0.4] (2.9,0.6) -- (0.5,0.6);
					\draw[blue!80!black,ultra thick,opacity=0.4] (2.9,-0.6) -- (0.5,-0.6);
					\draw[thick] (0.8,-0.6) rectangle ++(1.8,1.2);
				\end{tikzpicture}
				\caption{\label{fig:100cell}A unit cell aligned with the substrate can stretch in plane and shrink out of plane without tilting any right angles.}
			\end{subfigure}
			\hfill
			\begin{subfigure}[b]{0.475\textwidth}
				\centering
				\begin{tikzpicture}[x=1cm,y=1cm,thick]
					\path[fill=redbg!50!white,rounded corners=5pt] (-3.7,-1.2) -- (3.7,-1.2) -- (3.7,1.2) -- (-3.7,1.2) -- cycle;
					\draw[blue!80!black,ultra thick,opacity=0.4] (-3.5,1) -- (-0.5,1);
					\draw[blue!80!black,ultra thick,opacity=0.4] (-3.5,-1) -- (-0.5,-1);
					\draw (-3,0) -- (-2,-1) -- (-1,0) -- (-2,1) -- cycle;
					\draw[ultra thick,red!80!black,->,>=stealth] (-0.5,0) -- (0.5,0);
					\draw[red!80!black] (-0.2,-0.2) -- (0.2,0.2);
					\draw[red!80!black] (-0.2,0.2) -- (0.2,-0.2);
					\draw[blue!80!black,ultra thick,opacity=0.4] (3.5,0.8) -- (0.5,0.8);
					\draw[blue!80!black,ultra thick,opacity=0.4] (3.5,-0.8) -- (0.5,-0.8);
					\draw[red!80!black, very thick] (0.92,-0.14) to[out=45,in=-45] (0.92,0.14);
					\draw[red!80!black, very thick] (3.08,-0.14) to[out=135,in=-135] (3.08,0.14);
					\draw (3.3,0) -- (2,-0.8) -- (0.7,0) -- (2,0.8) -- cycle;
				\end{tikzpicture}
				
				\vspace*{1.25\baselineskip}
							
				\begin{tikzpicture}[x=1cm,y=1cm,thick]
					\path[fill=greenbg!50!white,rounded corners=5pt] (-3.7,-1.4) -- (3.7,-1.4) -- (3.7,1.4) -- (-3.7,1.4) -- cycle;
					\draw[blue!60!black,ultra thick,opacity=0.4] (-3.5,1) -- (-0.5,1);
					\draw[blue!60!black,ultra thick,opacity=0.4] (-3.5,-1) -- (-0.5,-1);
					\draw (-3,0) -- (-2,-1) -- (-1,0) -- (-2,1) -- cycle;
					\draw[ultra thick,green!60!black,->,>=stealth] (-0.5,0) -- (0.5,0);
					\draw[blue!60!black,ultra thick,opacity=0.4] (3.5,1.2) -- (0.5,1.2);
					\draw[blue!60!black,ultra thick,opacity=0.4] (3.5,-1.2) -- (0.5,-1.2);
					\draw[green!60!black, very thick] (1.02,-0.22) -- (1.24,0) -- (1.02,0.22);
					\draw[green!60!black, very thick] (2.98,-0.22) -- (2.76,0) -- (2.98,0.22);
					\draw (3.2,0) -- (2,-1.2) -- (0.8,0) -- (2,1.2) -- cycle;
				\end{tikzpicture}
				\caption{\label{fig:110cell}A unit cell misaligned with the substrate cannot be simultaneously stretched in plane and compressed out of plane without bending any right angles.}
			\end{subfigure}
			\vspace*{0.2\baselineskip}
			
			\caption{\label{fig:tilted_unit_cell}The more the unit cell is tilted, the less able it is to conserve the unit cell volume. From this it follows that (200) planes should expand less out of plane than (210) and (211), which explains the behavior seen in Fig.~\ref{fig:jap_fig5}b.}
		\end{figure}
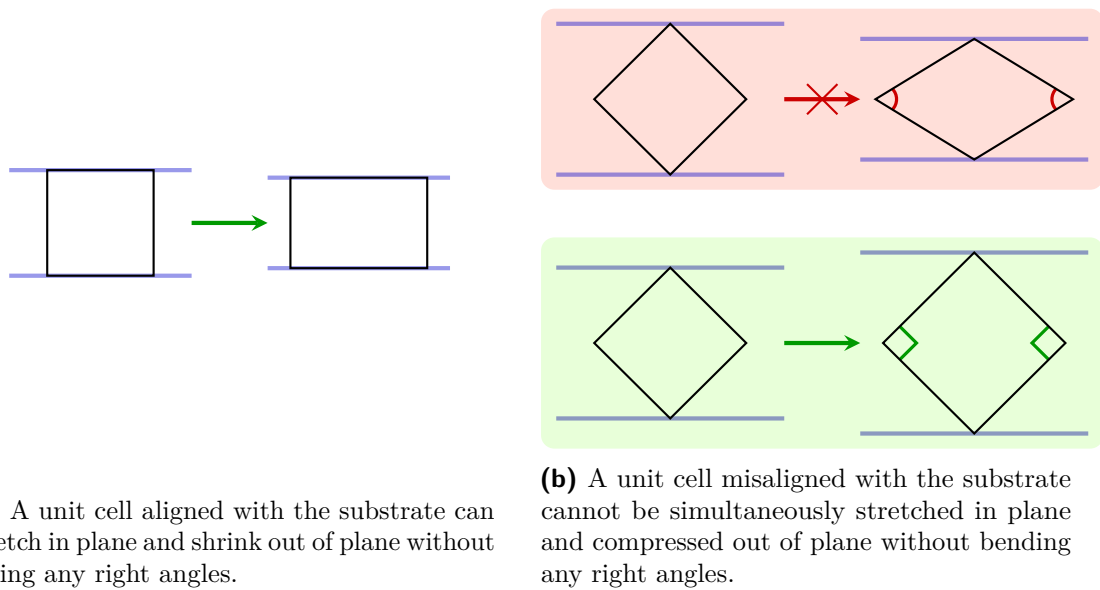
		
		When extrapolating these strains down to cryogenic temperatures using the values in Table~\ref{tab:induced_contractions}, the following estimates are obtained:
		\begin{equation}\label{eq:strain_sapphire_si}\begin{array}{r@{\;}c@{\;}l}
			\epsilon_1^\text{sapphire}	& =	& -\SI{0.9\pm0.5E-3}{},\\\\
			\epsilon_1^\text{silicon}	& =	& -\SI{5.2\pm0.2E-3}{}.
		\end{array}\end{equation}
		If the strain would be distributed perfectly homogeneously across crystallite orientations (which it isn't, see Fig.~\ref{fig:110cell}), and the films behaved entirely thermoelastically on either substrate (which they don't~\cite{testardi1970unusual,batterman1966low,batterman1964crystal}), then eq.~\eqref{eq:testardi_simplified} tells us that such strains would cause reductions in $T_\text{c}$ of between 0 and \SI{0.3}{\kelvin} on sapphire, and between 3.5 and \SI{4.2}{\kelvin} on silicon.
		We would then expect a relative reduction in $T_\text{c}$ on silicon of anywhere between 3.2 and \SI{4.2}{\kelvin}.
		Instead, as we saw in Fig.~\ref{fig:jap_fig1}, the critical temperature is reduced on silicon only by about \SI{1.9}{\kelvin}, a factor two less.
		Vice versa, the measured reduction in critical temperature should be associated with a difference in strain of \emph{at most}\footnote{There is a quadratic relation between $T_\text{c}$ and $\epsilon$, so the required $|\epsilon_1^\text{sapphire}-\epsilon_1^\text{silicon}|$ is smaller when $\min(|\epsilon_1^\text{sapphire}|,|\epsilon_1^\text{silicon}|)$ is larger.} \SI{3.7\pm0.3E-3}{}.
		
		This is where \ce{V3Si} becomes truly fascinating: it turns out that the weakening of the restoring force for shear deformation that we discussed earlier also causes the crystal to undergo a Martensitic transformation\footnote{``Martensitic'' refers to a change in local crystal structure where the individual atoms move around by less than the lattice parameter (it is diffusionless). To test the reversibility of such a transition (if any occurred in our films), a sample with \SI{200}{\nano\metre} of \ce{V3Si} on silicon with thermal oxide was annealed at \SI{900}{\celsius} under \ce{N2} flow for two minutes, and then cycled six times in a cryostat between 11 and \SI{30}{\kelvin}, during which no change in $T_\text{c}$ was observed.}.
		It has been observed on free-standing pieces of \ce{V3Si} that the \emph{a priori} cubic crystal undergoes a tetragonal deformation at temperatures of around 20--\SI{21}{\kelvin}, just above the superconducting transition~\cite{testardi1970unusual,batterman1966low,batterman1964crystal}.
		This ever-so-slight change is associated with a ``strain''\footnote{Sticking to our earlier definition for the benefit of consistency, $\epsilon=(1-d_0)/d_0$ along any direction, with $d_0$ the spacing expected for \emph{cubic} \ce{V3Si}. It may be more technically correct to use the post-transition tetragonal structure as zero-strain reference point, but for our purposes that would be as correct as it would be useless.} of up to $\epsilon_1=-2\epsilon_{2,3}=+\SI{1.7E-3}{}$ (expansion along one axis, contraction along the other two), with higher strain values on samples with higher RRR~\cite{batterman1966low}.
		
		This Martensitic transformation could also occur in polycrystalline thin films, in which case its effect on the critical temperature of a grain would depend on the orientation of the crystal relative to the substrate.
		As shown in eq.~\eqref{eq:strain_sapphire_si}, the \ce{V3Si} is expected to be compressed out of plane due to thermal strain.
		Grains that have their crystal structure aligned with the substrate would thus (before the Martensitic transition) be subject to tensile strain along two axes, and compressive strain along the other.
		A tetragonal deformation with expansion along the out-of-plane axis could compensate this thermal strain, lessening the reduction in $T_\text{c}$.
		If this deformation would instead provide an expansion along one of the in-plane directions, the critical temperature would be further reduced.
		Since the Martensitic transition is thought to occur precisely because the crystal is more stable under a certain amount of strain, it is likely that the transformation would predominantly stretch crystals out of plane where they were strongly compressed before.
		On the other hand, it is possible that the transformation would instead align itself to \emph{increase} the total strain in grains that are strained \emph{less} than the stable equilibrium.
		This provides a mechanism for the strain on the two substrates to converge towards a common value, thus explaining the reduction in $T_\text{c}$ on silicon of only \SI{1.9}{\kelvin} relative to the sapphire substrate.
		Moreover, since compensation of in-plane tensile stress by tetragonal deformation is only possible in grains that are aligned with the substrate, 
		such a mechanism should cause the shift in $T_\text{c}$ to be smaller in tilted grains, and thus result in an overall broadening of the superconducting transition.
		As can be seen in Fig.~\ref{fig:jap_fig1}, broader transitions are indeed observed on silicon, on which the strain is expected to be larger than on sapphire.
		
		Plans to measure the strain in-situ by cooling \ce{V3Si} samples in a cryostat to liquid helium temperatures (\SI{4}{\kelvin}) were abandoned once it transpired that the cryo-enabled XRD setup had broken down while moving the laboratory to a new building.
	
	\FloatBarrier
	\subsection{Strain development during thermal processing}	
		
\begin{figure}
			\centering
			\includegraphics[width=0.8\textwidth]{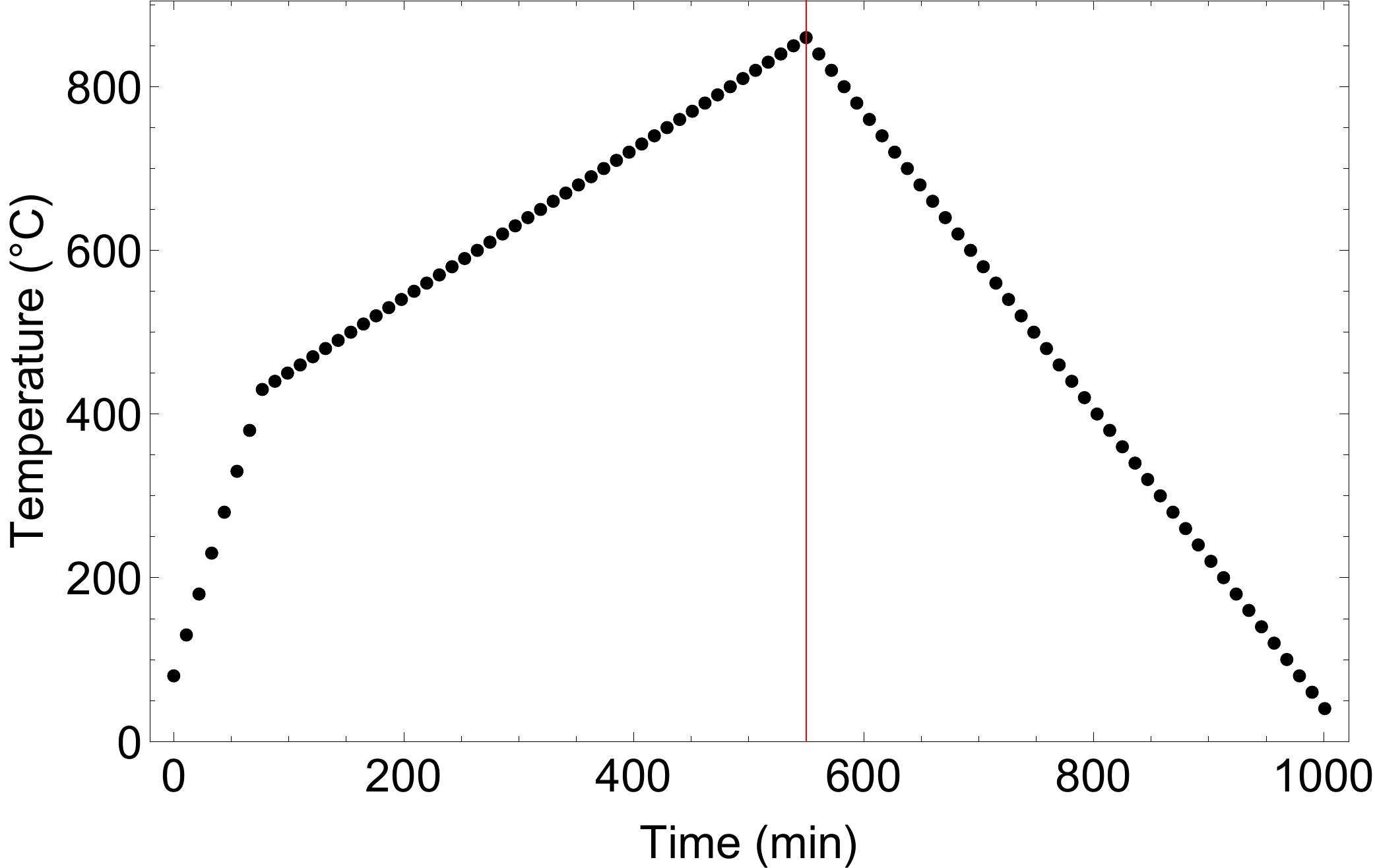}
			\caption{\label{fig:mam_xrd_t}Temperature profile versus annealing time during the second set of in-situ XRD experiments. The red line in this and the following graphs in this section indicates the maximum temperatures reached, after which cooling starts.}
		\end{figure}

		To gain a better understanding of the stress development during thermal processing, a second set of in-situ XRD measurements was performed, using again samples where \SI{200}{\nano\metre} of \ce{V3Si} had been deposited onto oxidized Si wafers with \SI{20}{\nano\metre} of thermal \ce{SiO2}, and sapphire.
		This time, a different setup was used where peak positions could be tracked more accurately both in plane and out of plane, though no good vacuum could be obtained (the previous experiment was performed under secondary vacuum).
		To avoid oxidation, a purified nitrogen atmosphere was therefore used, at a pressure slightly above ambient, i.e. \SI{1.25}{\bar}.
		Fearing nitridation, as well as reactions with residual traces of other gases, it was decided to stop the temperature ramp at \SI{860}{\celsius}\footnote{A trial run up to \SI{1000}{\celsius} with a silicon substrate sample caused a visible degradation of the surface. A similar reaction may have occurred during an in-situ sheet resistance measurement (see Fig.~\ref{fig:In_situ_sheet_resistance_aSi}), where the furnace was opened and air was let in before the cooling was completed, which caused the surface of the sample to turn into a fine dust within seconds.}, after which the samples were slowly cooled down (see Fig.~\ref{fig:mam_xrd_t}).

		\begin{figure}
			\centering
			\includegraphics[width=0.8\textwidth]{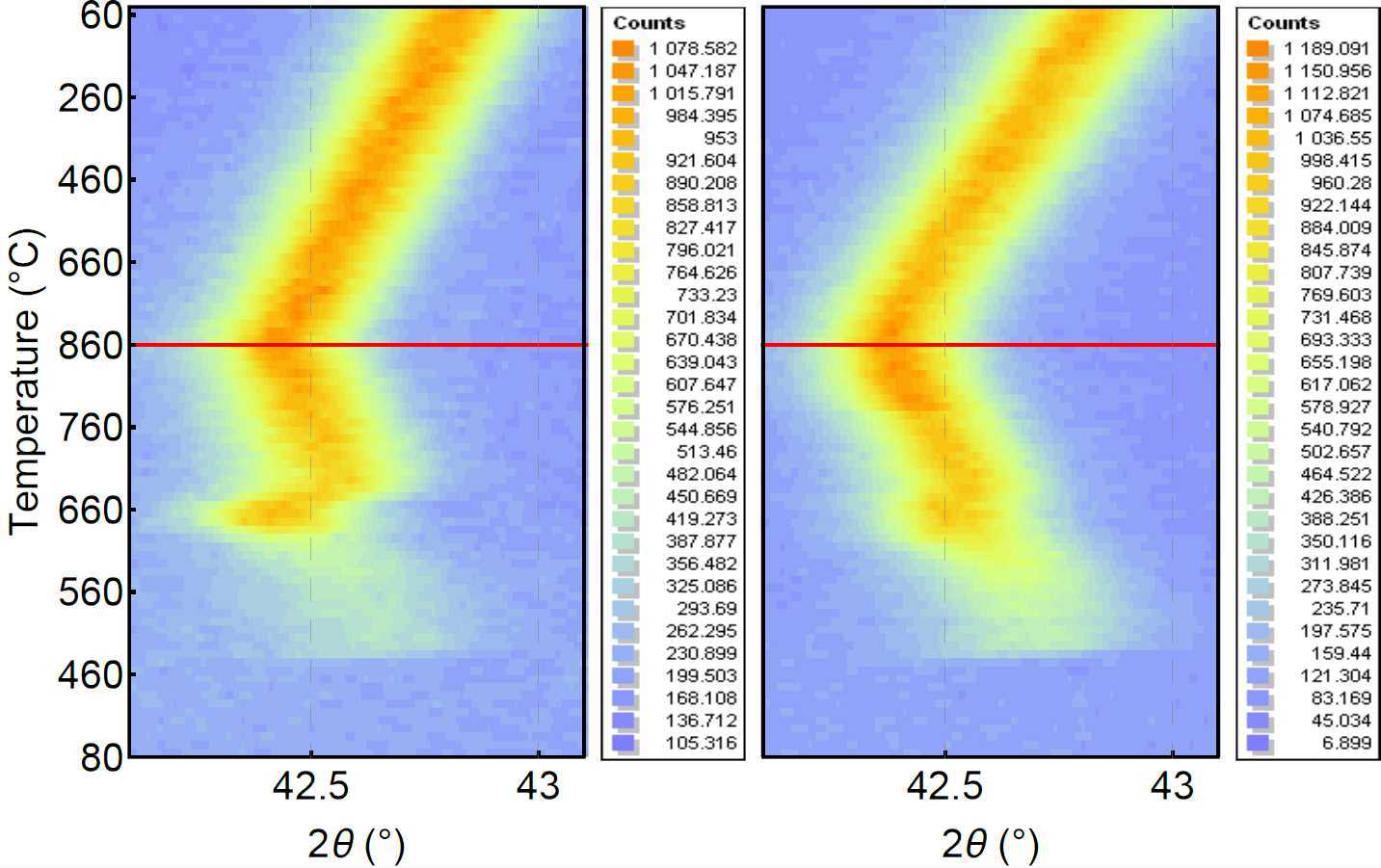}
			\caption{\label{fig:mam_xrd_oop}Contour maps of the \emph{out-of-plane} (210) peak observed on samples with \SI{200}{\nano\metre} of \ce{V3Si} deposited onto substrates of \B{(left)} sapphire and \B{(right)} silicon.}
		\end{figure}
	
		\begin{figure}
			\centering
			\includegraphics[width=0.8\textwidth]{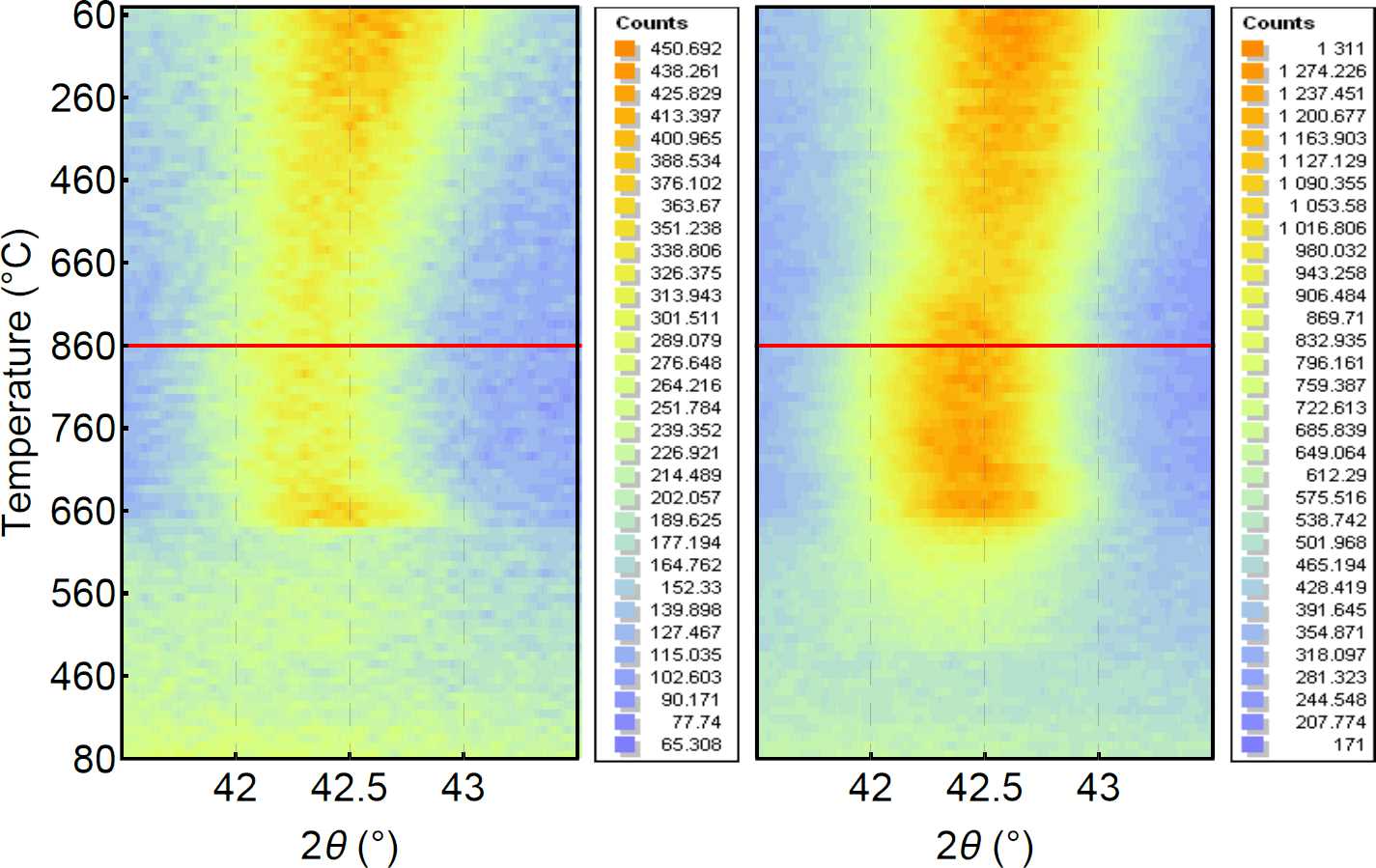}
			\caption{\label{fig:mam_xrd_ip}Contour maps of the \emph{in-plane} (210) peak observed on the same samples as in Fig.~\ref{fig:mam_xrd_oop}.}
		\end{figure}
		
\begin{figure}
			\centering
			\includegraphics[height=0.9\textheight]{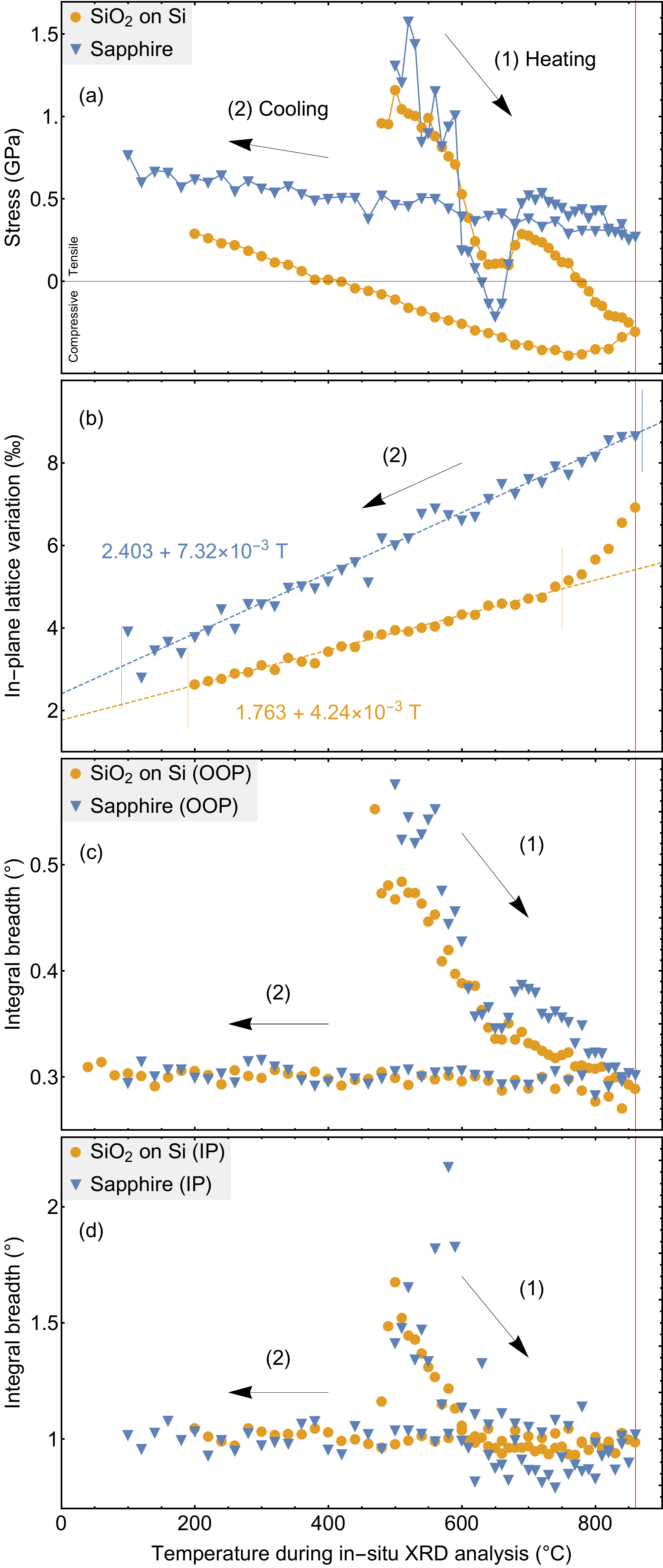}
			\caption{\label{fig:mam_xrd_strain}\B{(a)} Stress evolution versus annealing temperature for both the sapphire and the silicon substrates. \B{(b)} The strain of the \ce{V3Si} film versus the annealing temperature, shown only for the part of the temperature ramp where the samples are cooling. \B{(c,d)} OOP and IP integral breadth of the \ce{V3Si} (210) peak versus temperature.}
		\end{figure}

		The XRD measurements were optimized for the detection of only the position of the (210) peak, which is the largest in the \ce{V3Si} spectrum (see Fig.~\ref{fig:jap_fig3}).
		Distinct behaviors were observed during the out-of-plane (see Fig.~\ref{fig:mam_xrd_oop}) and in-plane (see Fig.~\ref{fig:mam_xrd_ip}) measurements, showing a more non-linear development of the peak position out of plane.
		Combining the data from the in-plane (IP) and out-of-plane (OOP) measurements, it is then possible to extract the stress using the $\sin^2\psi$ methodology~\cite{noyan2013residual} (assuming stress isotropy in the plane),
		\begin{equation}\label{eq:noyan2013}\epsilon_\psi=\dfrac{(d_\psi-d_0)}{d_0}=\dfrac{1}{2}S_2(hkl)\sigma\sin^2\psi+2S_1(hkl)\sigma.\end{equation}
		Here the stress-free (210)-plane spacing $d_0$ is first calculated from a weighted mean of the detected OOP and IP spacings $d_\perp$ and $d_\parallel$,
		\begin{equation}d_0=\dfrac{d_\perp-Ad_\parallel}{1-A},\quad\text{where}\quad A=\dfrac{4S_1(hkl)}{S_2(hkl)+4S_1(hkl)},\end{equation}
		with $A\approx-1$ a negative number that gives the relative weights of $d_\perp$ and $d_\parallel$, and which depends on the (hkl) index-dependent elastic constants $S_1$ and $S_2$ of \ce{V3Si}~\cite{murray2013equivalence,gaillac2016elate}.
		It was calculated that for the (210) plane, these constants are
		\begin{equation}\begin{array}{r@{\;}c@{\;}l}
			S_1(210)	& =	& -\SI{1.66E-6}{\per\mega\pascal},\quad\text{and}\\\\
			S_2(210)	& =	& +\SI{1.338E-7}{\per\mega\pascal},
		\end{array}\end{equation}
		respectively, which gives $A=-0.985$ and thus $d_0\approx(d_\perp+d_\parallel)/2)$.
		This then allows us to calculate the in-plane stress $\sigma$ from eq.~\eqref{eq:noyan2013},
		\begin{equation}\sigma=\dfrac{2}{S_2(hkl)}\left(\dfrac{d_\parallel-d_\perp}{d_0}\right),\end{equation}
		which is plotted in Fig.~\ref{fig:mam_xrd_strain}a.
		
		\begin{figure}
			\centering
			\includegraphics[width=0.8\textwidth]{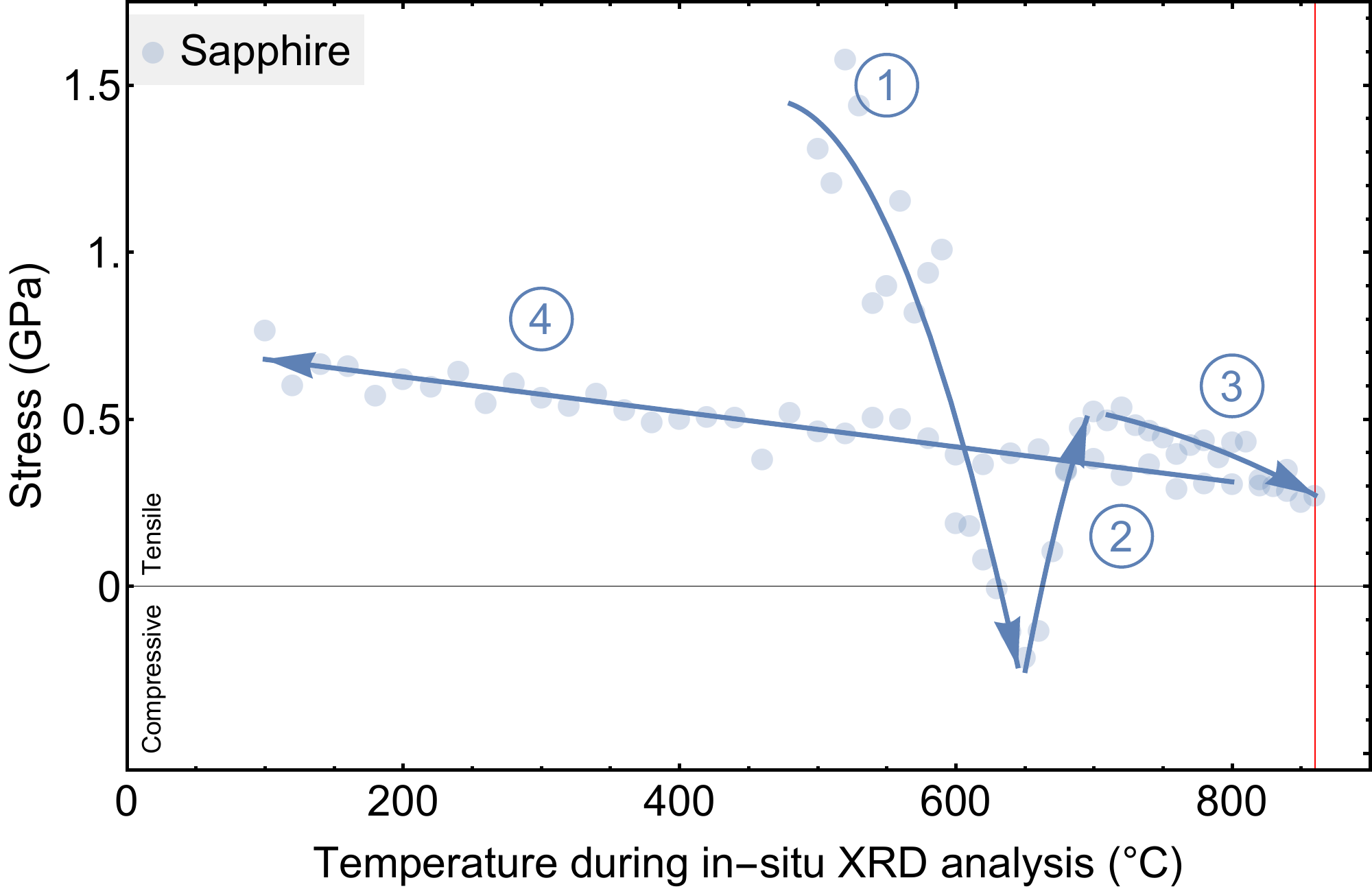}
			\caption{\label{fig:sapphire_stress_analysis}Four stages can be identified in the IP stress development on the sapphire substrate.}
		\end{figure}
		
		To illustrate the various processes that occur during the heating, a simplified plot of the stress development on the sapphire substrate is shown in Fig.~\ref{fig:sapphire_stress_analysis}.
		Initially, as the \emph{a priori} amorphous \ce{V3Si} crystallizes around \SI{500}{\celsius}, it becomes more compact while the substrate maintains its volume, leading to an in-plane tensile stress.
		During step \circled{1}, this stress relaxes due to a combination of thermally activated plastic deformation and a mismatch in thermal expansion coefficients, which around room temperature are \SI{7.5E-6}{\per\kelvin} for \ce{V3Si}~\cite{testardi1972structural} and \SI{5.0E-6}{\per\kelvin} in-plane for (0001) sapphire~\cite{dobrovinskaya2009properties}.
		Once the grains become too large to allow for relaxation by plastic deformation around \SI{650}{\celsius}, a second increase in tensile stress occurs during step \circled{2} due to grain growth.
		This volume reduction due to grain growth slows down at around \SI{700}{\celsius} as the film nears its maximum packing density, giving way to \circled{3} a second compression due to thermal strain.
		Neither plastic deformation nor grain growth occur after the maximum temperature of \SI{1000}{\celsius} has been reached (see Fig.~\ref{fig:mam_xrd_strain}c,d), and \circled{4} during cooling the in-plane stress then develops proportional to the thermoelastic strain imposed by the substrate (see Fig.\ref{fig:mam_xrd_strain}b).
		
		Note that not all of these steps (\circled{1},\circled{2},\circled{3}) will have occurred on samples annealed by rapid thermal processing (RTP) at lower temperatures, such as those reported on in Figs.~\ref{fig:jap_fig1} and~\ref{fig:jap_fig2}.
		Specifically, the fact that the dip in in-plane stress around \SI{650}{\celsius} (see both Figs.~\ref{fig:jap_fig5} and~\ref{fig:mam_xrd_strain}) is deeper on sapphire than it is on silicon, could explain why there is less of a difference in $T_\text{c}$ between the sapphire and Si samples annealed at this temperature.
	
	\FloatBarrier
	\subsection{Stability of \ce{V3Si} thin films during thermal processing}
		
		Boundaries between grains, like the silicide-vacuum and silicide-substrate interfaces at the top and bottom of the film, have a positive surface energy associated with them.
		When enough energy is provided in the form of heat to activate the diffusion of atoms, the system will generally move towards a state in which the total surface energy is minimized, leading to a competition in area reduction between the different interfaces.
		
\begin{figure}
			\centering
			\begin{subfigure}[t]{0.5\textwidth}
				\centering
				\includegraphics[width=\textwidth]{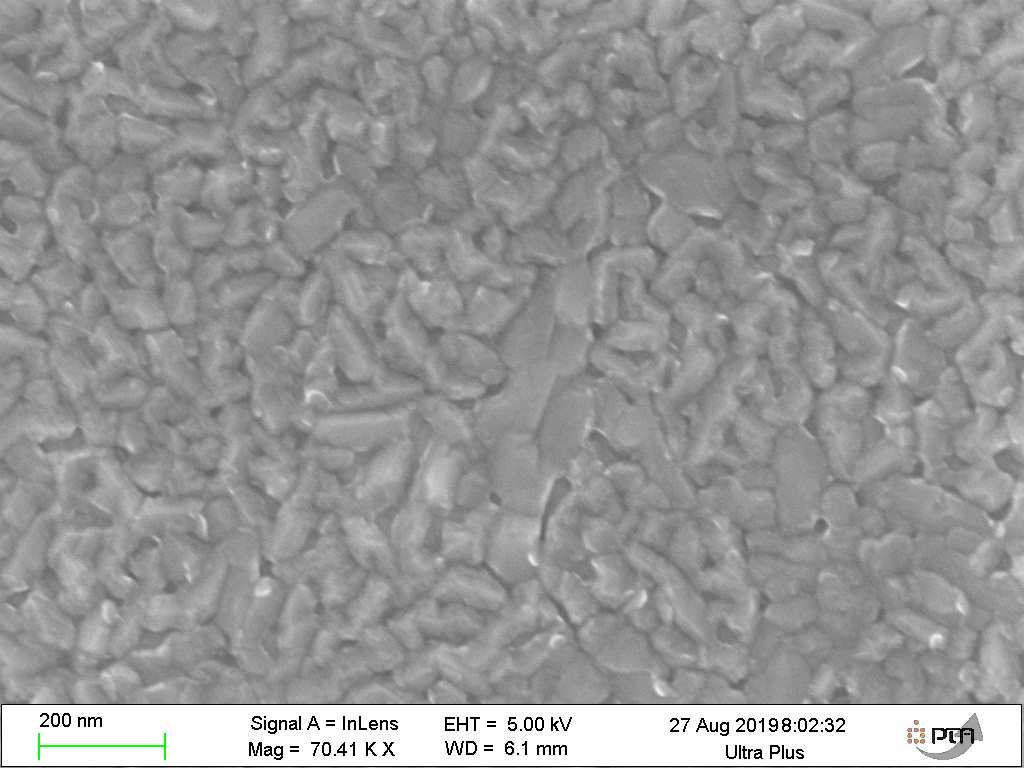}
				\caption{\SI{600}{\celsius}, $R_\square=\SI{150\pm5}{\ohm}$.}
			\end{subfigure}\begin{subfigure}[t]{0.5\textwidth}
				\centering
				\includegraphics[width=\textwidth]{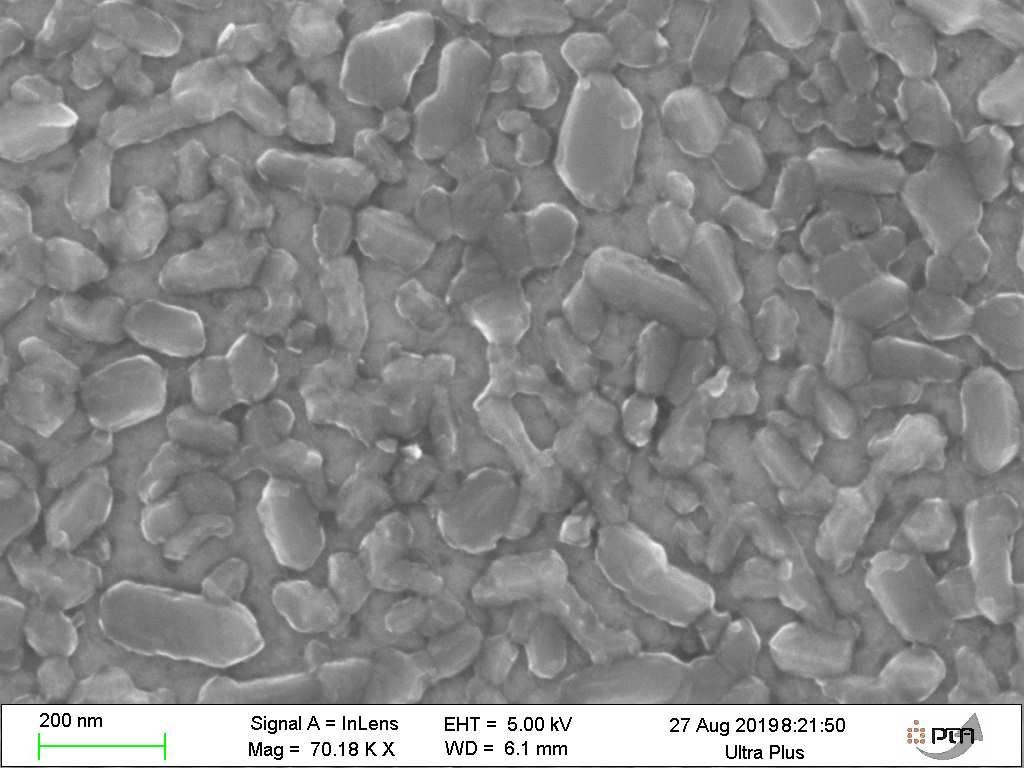}
				\caption{\SI{700}{\celsius}, $R_\square=\SI{135\pm3}{\ohm}$.}
			\end{subfigure}
			\vspace*{0.5\baselineskip}
			
			\begin{subfigure}[t]{0.5\textwidth}
				\centering
				\includegraphics[width=\textwidth]{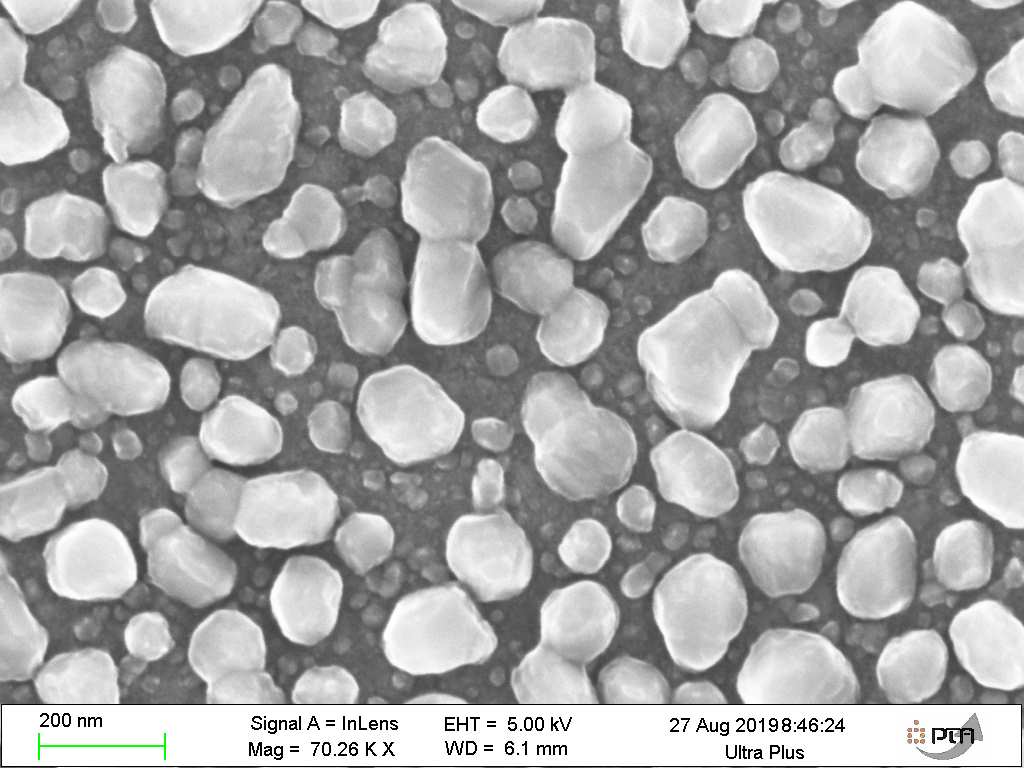}
				\caption{\SI{800}{\celsius}, $R_\square=\SI{1.04\pm0.18E3}{\ohm}$.}
			\end{subfigure}\begin{subfigure}[t]{0.5\textwidth}
				\centering
				\includegraphics[width=\textwidth]{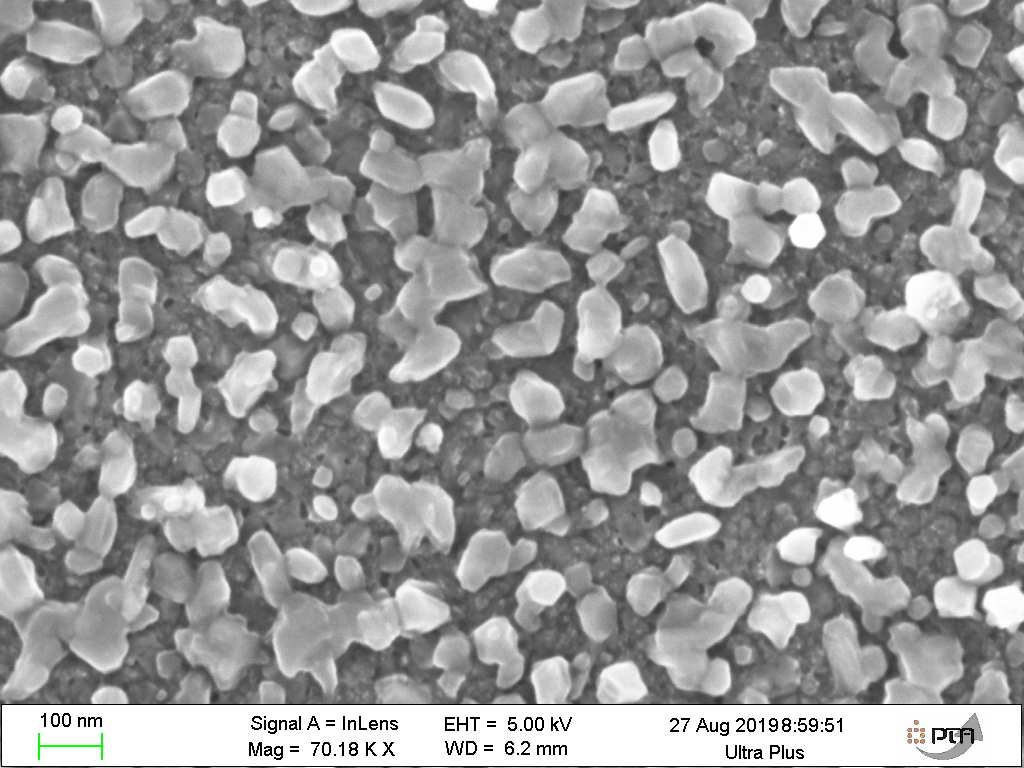}
				\caption{\SI{900}{\celsius}, $R_\square=\SI{2.8\pm1.6E3}{\ohm}$.}
			\end{subfigure}
			\caption{\label{fig:sem_test_wafer_d}Scanning electron microscope images of thermally processed samples with a \SI{17}{\nano\metre} layer of \ce{V3Si} deposited onto \SI{300}{\nano\metre} of \ce{SiO2}. The samples were annealed at the four indicated temperatures under vacuum during two minutes, after which the sheet resistance was measured. The deposition conditions were different from those used for the rest of the samples discussed in this chapter: the layers were RF sputtered (as opposed to DC) at a pressure of \SI{100}{\micro\bar}. The sputtering power (\SI{200}{\watt}) and argon flow (\SI{50}{\sccm}) were the same as those used elsewhere.}
		\end{figure}

		Shown in Fig.~\ref{fig:sem_test_wafer_d} are planar view scanning electron microscope (SEM) images of samples with a \SI{17}{\nano\metre} \ce{V3Si} film on \ce{SiO2} annealed under vacuum at different temperatures.
		As can be seen from the gradual island formation with increasing temperature, the surface energy density of the \ce{SiO2}/vacuum interface is smaller than the sum of the energy densities of the \ce{SiO2}/\ce{V3Si} and \ce{V3Si}/vacuum interfaces~\cite{srolovitz1995thermodynamics,leroy2016control}, causing it to de-wet.
		We will see more of this in the context of PtSi island formation in section~\ref{sec:ptsi}.
		
\begin{figure}
			\centering
			\begin{subfigure}[t]{0.5\textwidth}
				\centering
				\includegraphics[width=\textwidth]{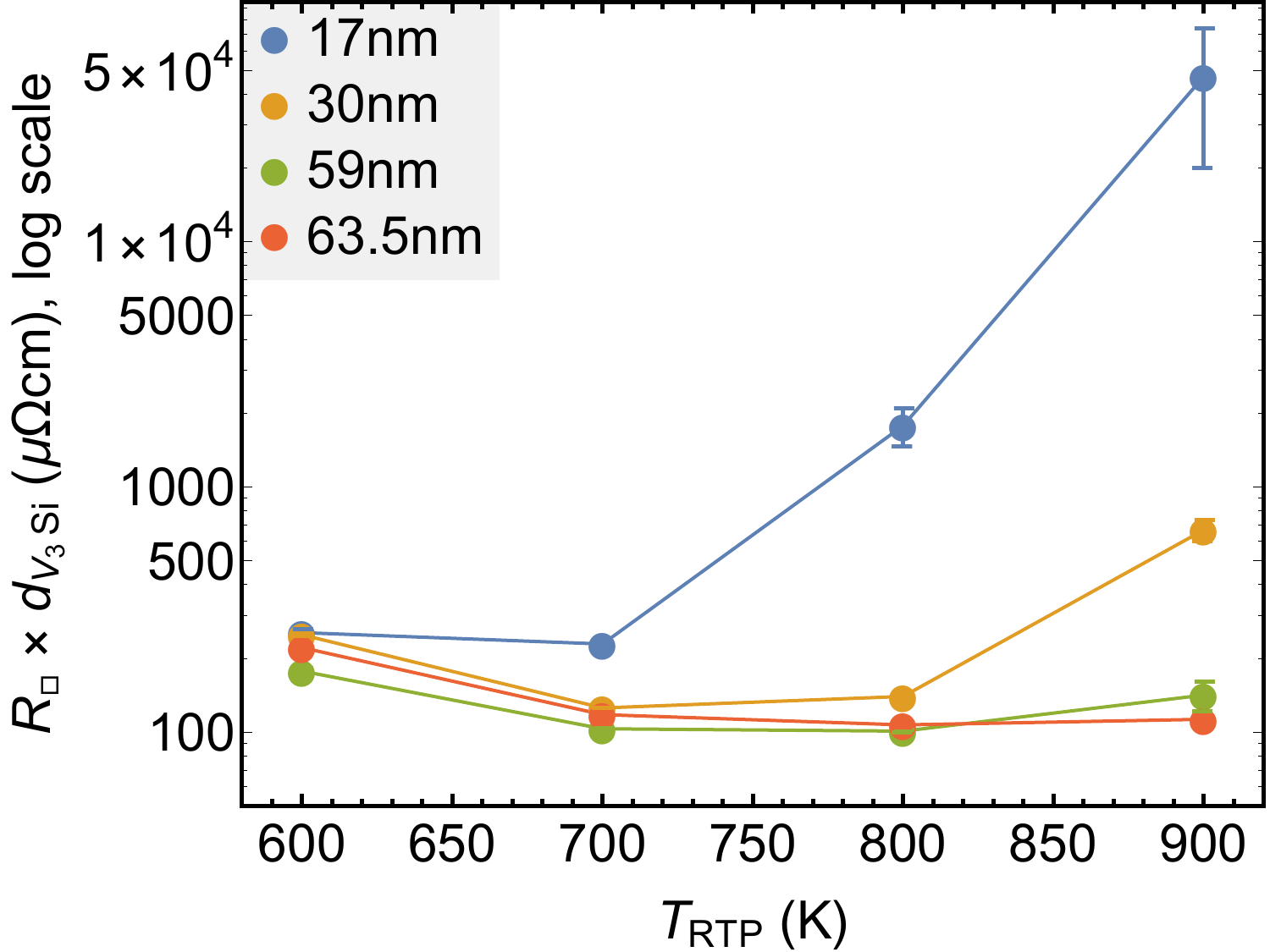}
				\caption{Normalized sheet resistance vs $T_\text{RTP}$ for selected thicknesses.}
			\end{subfigure}\begin{subfigure}[t]{0.5\textwidth}
				\centering
				\includegraphics[width=\textwidth]{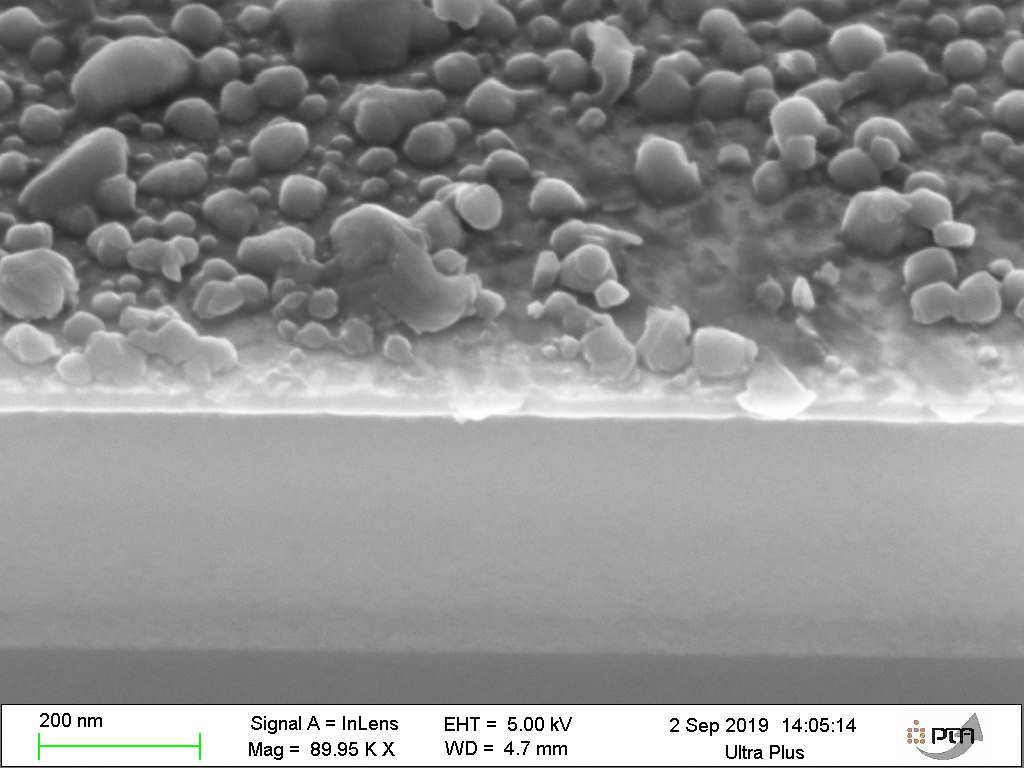}
				\caption{\SI{30}{\nano\metre} \ce{V3Si} on \SI{300}{\nano\metre} \ce{SiO2}, $T_\text{RTP}=\SI{800}{\celsius}$.}
			\end{subfigure}
			\vspace*{0.5\baselineskip}
			
			\begin{subfigure}[t]{0.5\textwidth}
				\centering
				\includegraphics[width=\textwidth]{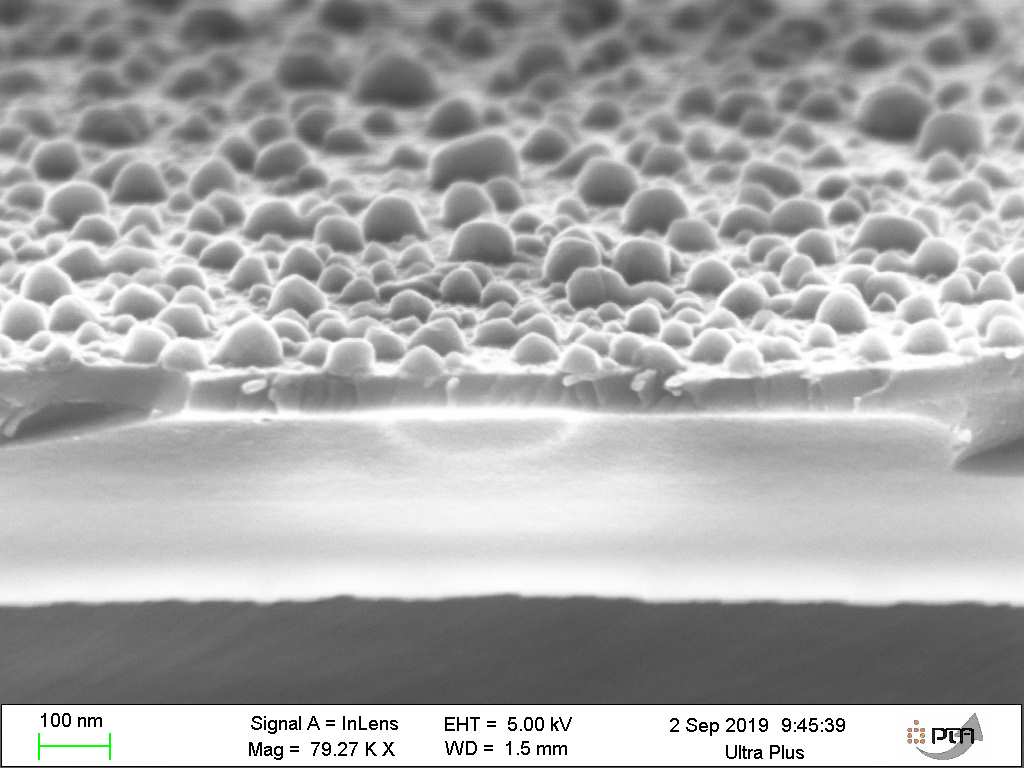}
				\caption{\SI{63.5}{\nano\metre} \ce{V3Si} on \SI{300}{\nano\metre} \ce{SiO2}, $T_\text{RTP}=\SI{800}{\celsius}$.}
			\end{subfigure}\begin{subfigure}[t]{0.5\textwidth}
				\centering
				\includegraphics[width=\textwidth]{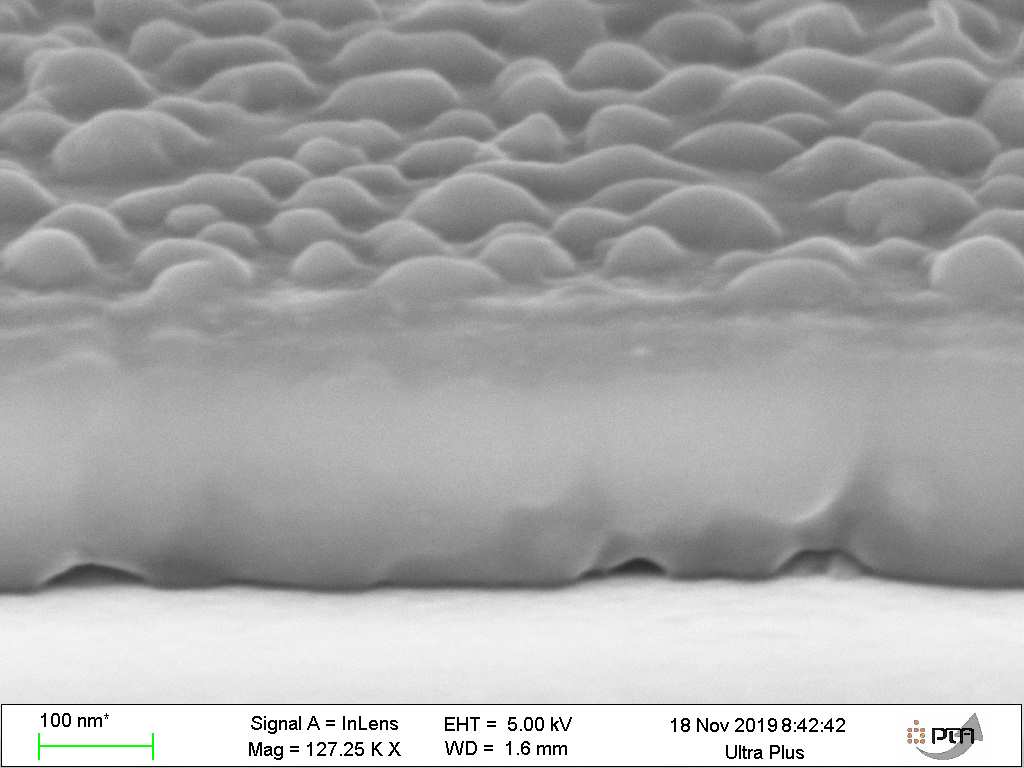}
				\caption{\SI{200}{\nano\metre} \ce{V3Si} on \SI{300}{\nano\metre} \ce{SiO2}, $T_\text{RTP}=\SI{800}{\celsius}$.}
			\end{subfigure}
			\caption{\label{fig:instability_thin_films} \B{(a,b)} A combination of grain growth and de-wetting causes films to become discontinuous after annealing, leading to a divergence in sheet resistance (normalized by deposited \ce{V3Si} thickness). Thicker films remain continuous up to higher temperatures. \B{(c,d)} On thicker films of 63.5 and \SI{200}{\nano\metre}, only a thin surface layer with protrusions appears after annealing at \SI{800}{\celsius}.}
		\end{figure}

		The temperature up to which a film of \ce{V3Si} remains stable depends on the thickness, since higher diffusion rates are required to overcome the energy barrier of increasing the film/vacuum surface area before the substrate can be exposed~\cite{leroy2016control}.
		Fig.~\ref{fig:instability_thin_films} shows a collection of SEM cross-sections of samples with various thicknesses of \ce{V3Si} on \ce{SiO2} substrates that were annealed under vacuum, with in the top-left the normalized sheet resistances versus annealing temperature.
		A divergence in the sheet resistance indicates the formation of gaps or even islands in the \ce{V3Si} film, which occurs at higher temperatures on thicker films.
		Since the quality of the film improves with annealing temperature (see Figs.~\ref{fig:jap_fig1} and~\ref{fig:jap_fig2}), the film thickness thus determines the thermal budget and limits the maximum attainable critical temperature.
		It is important to note that in all these samples the \ce{V3Si} was deposited onto an oxide, rather than HF-cleaned silicon, which likely gives a different surface energy density than a \ce{V3Si}/Si interface would.
		
\begin{figure}
			\centering
			\begin{tikzpicture}[x=0.816cm,y=0.816cm]
				\node[anchor=north west,inner sep=0] at (0,0) {
					\includegraphics[width=0.48\textwidth]{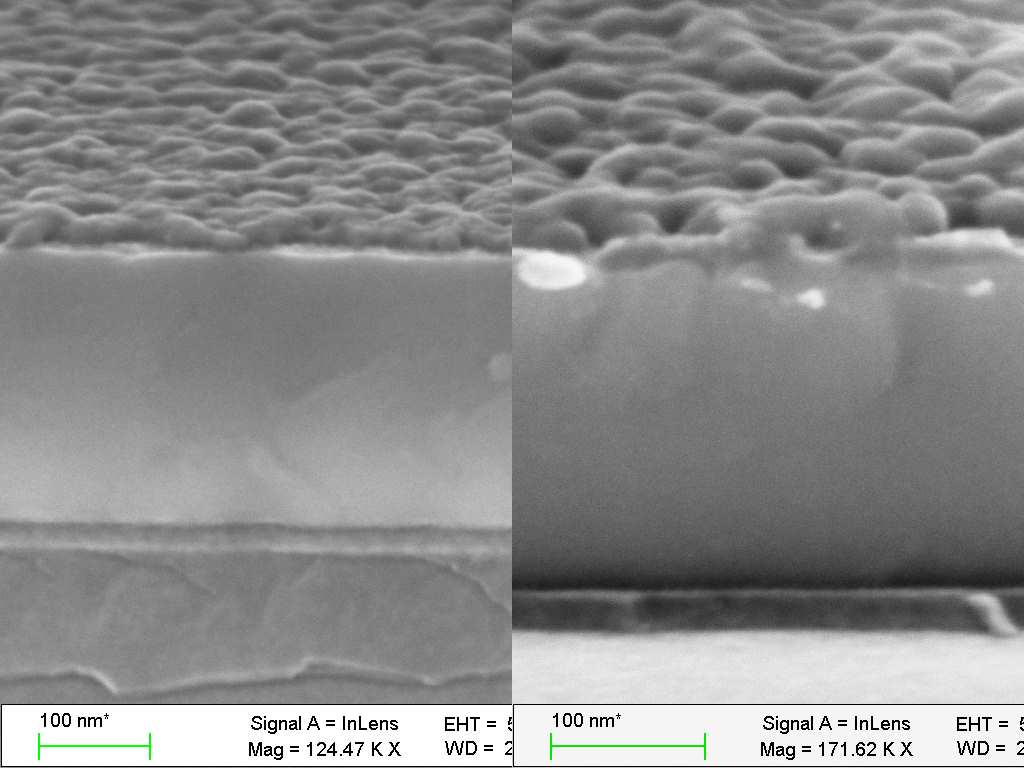}
					};
				\node[color=black,anchor=north west,yshift=-0.1cm] at (0.12,-0.01) {\Large \B{600°C}};
				\node[color=white,anchor=north west,yshift=-0.1cm] at (0.1,0) {\Large \B{600°C}};
				\node[color=black,anchor=north west,yshift=-0.1cm] at ($(0.25\textwidth,0)+(0.12,-0.01)$) {\Large \B{650°C}};
				\node[color=white,anchor=north west,yshift=-0.1cm] at ($(0.25\textwidth,0)+(0.1,0)$) {\Large \B{650°C}};
				
				\draw[thick, red] (0.3,-4.4) -- (1.3,-4.43);
				\draw[thick, red] (0.3,-4.62) -- (1.3,-4.65);
				\node[black] at (2.0,-3.3) {\large \B{\ce{V3Si}}};
				\node[white] at (2.02,-3.31) {\large \B{\ce{V3Si}}};
				\node[black] at (2.0,-4.54) {\large \B{\ce{SiO2}}};
				\node[white] at (2.02,-4.55) {\large \B{\ce{SiO2}}};
				\node[black] at (2.0,-5.4) {\large \B{\ce{Si}}};
				\node[white] at (2.02,-5.41) {\large \B{\ce{Si}}};
				
				\draw[thick, red] ($(0.25\textwidth,0)+(0.3,-5.0)$) -- ($(0.25\textwidth,0)+(1.3,-5.0)$);
				\draw[thick, red] ($(0.25\textwidth,0)+(0.3,-5.33)$) -- ($(0.25\textwidth,0)+(1.3,-5.33)$);
				\node[black] at ($(0.25\textwidth,0)+(2.0,-3.66)$) {\large \B{\ce{V3Si}}};
				\node[white] at ($(0.25\textwidth,0)+(2.02,-3.67)$) {\large \B{\ce{V3Si}}};
				\node[black] at ($(0.25\textwidth,0)+(2.0,-5.16)$) {\large \B{\ce{SiO2}}};
				\node[white] at ($(0.25\textwidth,0)+(2.02,-5.17)$) {\large \B{\ce{SiO2}}};
				\node[white] at ($(0.25\textwidth,0)+(2.0,-5.66)$) {\large \B{\ce{Si}}};
				\node[black] at ($(0.25\textwidth,0)+(2.02,-5.67)$) {\large \B{\ce{Si}}};
			\end{tikzpicture}\hfill\begin{tikzpicture}[x=0.816cm,y=0.816cm]
				\node[anchor=north west,inner sep=0] at (0,0) {
					\includegraphics[width=0.48\textwidth]{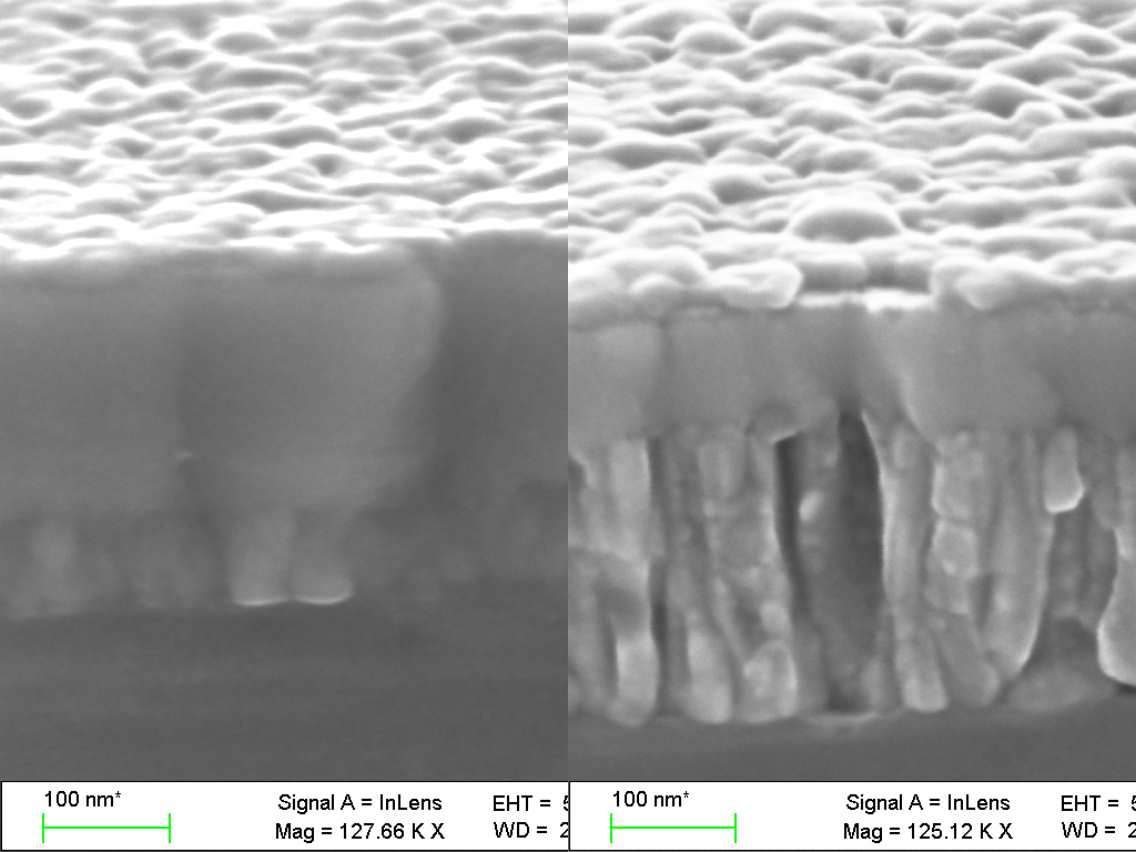}
				};
				\node[color=white,anchor=north west,yshift=-0.1cm] at (0.12,-0.01) {\Large \B{600°C}};
				\node[color=black,anchor=north west,yshift=-0.1cm] at (0.1,0) {\Large \B{600°C}};
				\node[color=white,anchor=north west,yshift=-0.1cm] at ($(0.25\textwidth,0)+(0.12,-0.01)$) {\Large \B{650°C}};
				\node[color=black,anchor=north west,yshift=-0.1cm] at ($(0.25\textwidth,0)+(0.1,0)$) {\Large \B{650°C}};
				
				\draw[thick, red] (0.3,-3.9) -- (1.3,-3.9);
				\draw[thick, red] (0.3,-4.66) -- (1.3,-4.66);
				\node[black] at (2.0,-3.1) {\large \B{\ce{V3Si}}};
				\node[white] at (2.02,-3.11) {\large \B{\ce{V3Si}}};
				\node[black] at (2.0,-4.34) {\large \B{\ce{VSi2}}};
				\node[white] at (2.02,-4.35) {\large \B{\ce{VSi2}}};
				\node[black] at (2.0,-5.3) {\large \B{\ce{Si}}};
				\node[white] at (2.02,-5.31) {\large \B{\ce{Si}}};
				
				\draw[thick, red] ($(0.25\textwidth,0)+(0.3,-3.3)$) -- ($(0.25\textwidth,0)+(1.3,-3.3)$);
				\draw[thick, red] ($(0.25\textwidth,0)+(0.3,-5.5)$) -- ($(0.25\textwidth,0)+(1.3,-5.5)$);
				\node[black] at ($(0.25\textwidth,0)+(2.0,-2.86)$) {\large \B{\ce{V3Si}}};
				\node[white] at ($(0.25\textwidth,0)+(2.02,-2.87)$) {\large \B{\ce{V3Si}}};
				\node[black] at ($(0.25\textwidth,0)+(2.0,-4.46)$) {\large \B{\ce{VSi2}}};
				\node[white] at ($(0.25\textwidth,0)+(2.02,-4.47)$) {\large \B{\ce{VSi2}}};
				\node[black] at ($(0.25\textwidth,0)+(2.0,-5.75)$) {\large \B{\ce{Si}}};
				\node[white] at ($(0.25\textwidth,0)+(2.02,-5.76)$) {\large \B{\ce{Si}}};
			\end{tikzpicture}
			\caption{\label{fig:sem_sio2_si}SEM images of cleaved samples with \SI{200}{\nano\metre} of \ce{V3Si} deposited onto either oxidized or HF-cleaned silicon, and annealed at 600 or \SI{650}{\celsius}. \B{(Left)} A silicon substrate with \SI{20}{\nano\metre} of thermal oxide. No formation of \ce{VSi2} observed. \B{(Right)} When an equal thickness of \ce{V3Si} is deposited on oxide-free silicon and annealed at \SI{600}{\celsius} and \SI{650}{\celsius}, a \ce{VSi2} layer with distinct columnar morphology appears. It was confirmed by XRD analysis that \ce{V3Si} is present after deposition, and that \ce{VSi2} appears after annealing.}
		\end{figure}
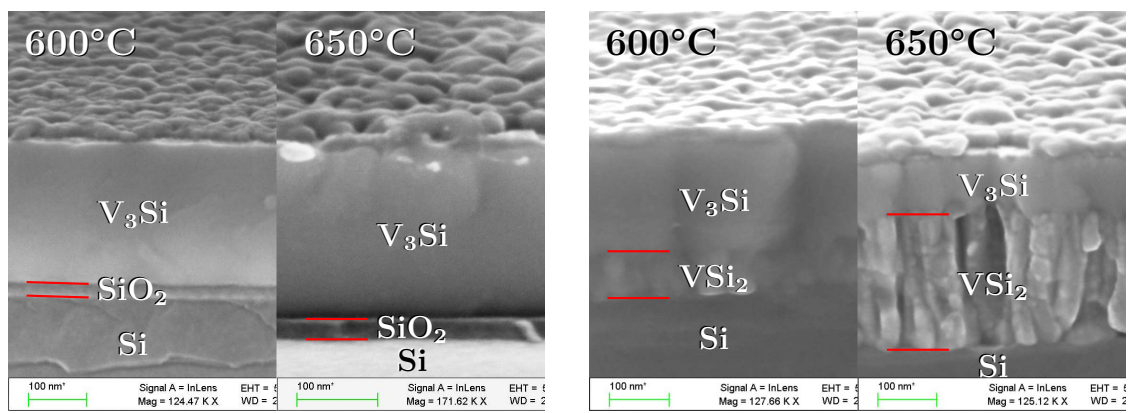

		A second type of thermodynamic instability occurs when \ce{V3Si} is deposited onto a silicon substrate.
		Whereas \ce{V3Si} does not react with \ce{SiO2} due to the strong binding between silicon and oxygen, \ce{VSi2} is likely to form at a \ce{V3Si}/Si interface (see Fig.~\ref{fig:sem_sio2_si}).
		Though this new phase will likely nucleate at higher temperatures even at a perfectly sharp interface between crystalline \ce{V3Si} and Si, its formation is greatly aided by the conditions under which the \ce{V3Si} is sputtered onto the film.
		
		As the V and Si atoms arrive on the Si surface, their kinetic and condensation energies together with the heating by the argon plasma cause an intermixing layer to form that typically depends in thickness on the total time of deposition.
		In this mixed zone there will be a smooth distribution in V and Si atomic concentrations, from 25\% Si within the deposited layer, to 100\% Si in the substrate.
		If \ce{V3Si} were the energetically most favorable state with the highest energy gain per bound V atom, then there would be a lower chemical potential for these atoms closer to the \ce{V3Si} layer, providing a mechanism for asymmetric diffusion.
		Alas, although \ce{V3Si} (\ce{V_{0.75}Si_{0.25}}) has a larger effective heat of formation (EHF) from pure V and Si per mole of atoms involved \emph{in total}~\cite{pretorius1993thin}, the energy gain \emph{per vanadium atom} is greater for \ce{VSi2} (\ce{V_{0.33}Si_{0.67}}), of which three times as many molecules can be formed for a fixed amount of vanadium.
		Therefore, unless the intermixing layer is so thin as to make the appearance of a new phase energetically unfavorable due to the disproportionately large surface energy, it is to be expected that \ce{VSi2} will form in such an intermixing layer.
				
\begin{figure}		
			\begin{subfigure}[b]{0.475\textwidth}
				\centering
				\begin{tikzpicture}[x=0.7cm,y=0.7cm]
					\path[fill=greenbg,rounded corners=10pt] (-5,-2) -- (5,-2) -- (5,2) -- (-5,2) -- cycle;
					
					\fill[Si] (-3,-1.5) -- (-1.2,-1.5) -- (-1.2,0.5) -- (-3,0.5) -- cycle;
					\fill[V3Si] (-3,0.5) -- (-1.2,0.5) -- (-1.2,1.5) -- (-3,1.5) -- cycle;
					\node[anchor=east] at (-3,0.8) {\ce{aV3Si}};
					
					\draw[ultra thick,green!60!black,->,>=stealth] (-0.7,0) -- (0.7,0);
					
					\fill[Si] (3,-1.5) -- (1.2,-1.5) -- (1.2,0.5) -- (3,0.5) -- cycle;
					\fill[V3Si] (3,0.5) -- (1.2,0.5) -- (1.2,1.45) -- (3,1.45) -- cycle;
					\node[anchor=west] at (3,0.775) {\ce{cV3Si}};
				\end{tikzpicture}
			\end{subfigure}
			\hfill
			\begin{subfigure}[b]{0.475\textwidth}
				\centering
				\begin{tikzpicture}[x=0.7cm,y=0.7cm]
					\path[fill=redbg,rounded corners=10pt] (-5,-2) -- (5,-2) -- (5,2) -- (-5,2) -- cycle;
					
					\fill[Si] (-3,-1.5) -- (-1.2,-1.5) -- (-1.2,0.1) -- (-3,0.1) -- cycle;
					\fill[V3Si] (-3,0.1) -- (-1.2,0.1) -- (-1.2,1.5) -- (-3,1.5) -- cycle;
					\node[anchor=east] at (-3,0.8) {\ce{aV3Si}};
					
					\draw[ultra thick,green!60!black,->,>=stealth] (-0.7,0) -- (0.7,0);
					
					\fill[Si] (3,-1.5) -- (1.2,-1.5) -- (1.2,-1.23) -- (3,-1.23) -- cycle;
					\fill[VSi2] (3,-1.23) -- (1.2,-1.23) -- (1.2,0.37) -- (3,0.37) -- cycle;
					\node[anchor=west] at (3,-0.43) {\ce{VSi2}};
					\fill[V3Si] (3,0.37) -- (1.2,0.37) -- (1.2,1.07) -- (3,1.07) -- cycle;
					\node[anchor=west] at (3,0.72) {\ce{V3Si}};
				\end{tikzpicture}
			\end{subfigure}

			\vspace*{1.2\baselineskip}
			
			\begin{subfigure}[t]{\textwidth}
				\centering
				\begin{tikzpicture}[x=0.4cm,y=0.4cm]
					\fill[V3Si] (-8,1) -- (8,1) -- (8,3) -- (-8,3) -- cycle;
					\fill[V3Si] (-3,3.5) -- (3,3.5) -- (3,5.5) -- (-3,5.5) -- cycle;
					\path\Sip (-3,2.5) -- (3,2.5) -- (3,3.5) -- (-3,3.5) -- cycle;
					\fill[SiN] (-4,2) -- (-3,2) -- (-3,5.5) to[out=220,in=90] (-4,2);
					\fill[SiN] (4,2) -- (3,2) -- (3,5.5) to[out=-40,in=90] cycle;
					\fill[SiO2] (-3,2) -- (3,2) -- (3,2.5) -- (-3,2.5) -- cycle;
					\fill[Si] (-8,0) -- (8,0) -- (8,1) -- (3,1) to[out=180,in=270] (2,2) -- (-2,2) to[out=270,in=0] (-3,1) -- (-8,1) -- cycle; 	
					\fill[SiO2] (-8,0) -- (8,0) -- (8,-2) -- (-8,-2) -- cycle; 
					\fill[Si] (-8,-2) -- (8,-2) -- (8,-3) -- (-8,-3) -- cycle;
					
					\fill[VSi2] (-8,0.5) -- (-2.5,0.5) to[out=0,in=270] (-1,2) -- (1,2) to[out=270,in=180] (2.5,0.5) -- (8,0.5) -- (8,1.5) -- (3.5,1.5) to[out=180,in=270] (2.5,2) -- (-2.5,2) to[out=270,in=0] (-3.5,1.5) -- (-8,1.5) -- cycle; 	
					\fill[VSi2] (-3,3) -- (3,3) -- (3,4) -- (-3,4) -- cycle;
					
					\fill[gray!15!white] 	(9,6) -- (23,6) -- (23,-3) -- (9,-3) -- cycle;
					\fill[V3Si] 			(10,5) -- (12,5) -- (12,4) -- (10,4) -- cycle;
					\node[anchor=west] at 	(12.5,4.5) {\ce{V3Si}};
					\fill[VSi2] 			(17,5) -- (19,5) -- (19,4) -- (17,4) -- cycle;
					\node[anchor=west] at 	(19.5,4.5) {\ce{VSi2}};
					\fill[SiN] 				(10,3) -- (12,3) -- (12,2) -- (10,2) -- cycle;
					\node[anchor=west] at 	(12.5,2.5) {\ce{SiN}};
					\fill[SiO2] 			(10,1) -- (12,1) -- (12,0) -- (10,0) -- cycle;
					\node[anchor=west] at 	(12.5,0.5) {\ce{SiO2}};
					\fill[Si] 				(10,-1) -- (12,-1) -- (12,-2) -- (10,-2) -- cycle;
					\node[anchor=west] at 	(12.5,-1.5) {\ce{Si}};
				\end{tikzpicture}
			\end{subfigure}
			\vspace*{0.4\baselineskip}
			
			\caption{\label{fig:v3si_jofet}The annealing trade-off: while the \ce{V3Si} is crystallized and improves in superconducting properties, a \ce{VSi2} layer is formed below until no \ce{V3Si} is left and superconductivity vanishes.}
		\end{figure}
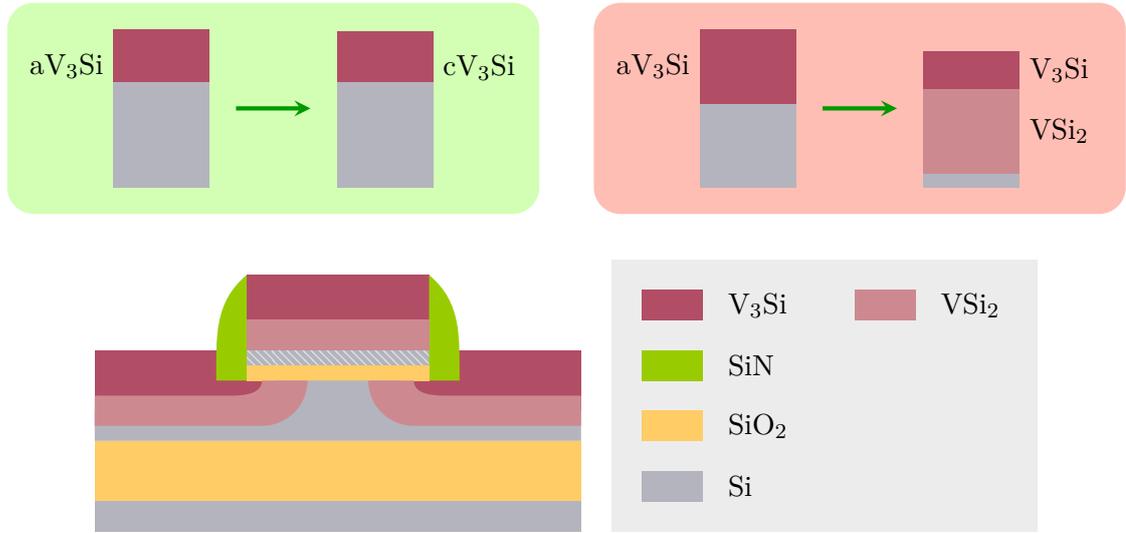

		Once \ce{VSi2} has formed, it will begin to compete with the formation of \ce{V3Si}.
		Given that a reservoir of Si is present in the substrate, while no more V exist than what is deposited in the amorphous layer of \ce{V_{0.75}Si_{0.25}}, the relevant energy is the heat of formation per atom of vanadium, which is minimized in the following reaction:
		\begin{equation}\label{eq:v3si_vsi2_reaction}\underbrace{\ce{V3Si}}_{\SI{-45.2}{\kilo\joule/\mole}\text{ at}} + \underbrace{\phantom{\!\!\!\!_0}5\ce{Si}\phantom{\!\!\!\!_0}}_{\phantom{\si{\per\mole}}0\phantom{\si{\per\mole}}} \rightarrow \underbrace{3\ce{VSi2}}_{\SI{-40.2}{\kilo\joule/\mole}\text{ at}},\end{equation}
		where an energy of $\SI{181}{\kilo\joule/\mole}$ is gained per mole of \ce{V3Si} that is transformed.
		Since we are aiming to fabricate a JoFET with \ce{V3Si} source and drain contacts (see Fig.~\ref{fig:v3si_jofet}), it is imperative that if this reaction cannot be prevented, it is at least controlled.

		To find a process window within which \ce{V3Si} can be crystallized to improve its electrical and superconducting properties, while also limiting the formation of \ce{VSi2}, a set of samples was prepared where different thicknesses of \ce{V3Si} (20, 50, 100 and \SI{200}{\nano\metre}) were deposited onto HF-cleaned substrates.
		Though removal of the native oxide is advisable in general to obtain good ohmic contacts, it may also aid \ce{VSi2} formation (which is slowed down by oxygen~\cite{schutz1979formation}).
		This is only a minor concern however: superconductivity disappears between 650 and \SI{700}{\celsius} on samples with \SI{200}{\nano\metre} \ce{V3Si} on Si, whether it was cleaned with HF or not.
		
		\begin{figure}
			\centering
			\includegraphics[height=0.9\textheight]{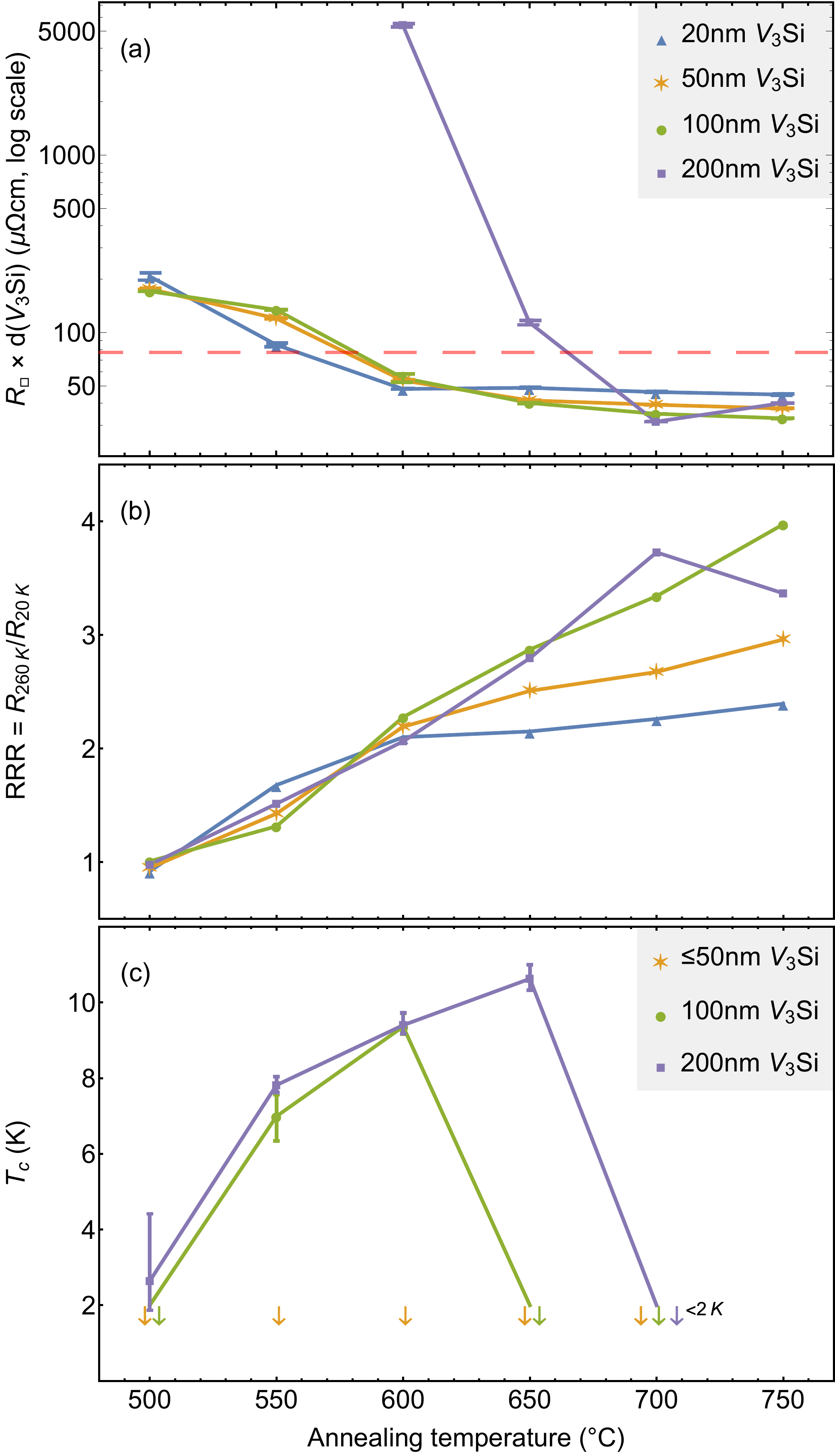}
			\caption{\label{fig:mam_vsi_competition}\B{(a)} The resistivity vs annealing temperature for each film thickness, the red dashed line indicates the lowest resistivity expected for V3Si (see main text). \B{(b)} The residual resistance ratio (RRR). \B{(c)} The critical temperature. Arrows down ($\downarrow$) indicate that the critical temperature (if any) is below \SI{2}{K}.}
		\end{figure}
		
		For each thickness of \ce{V3Si}, six samples were annealed at temperatures between 500 and \SI{750}{\celsius}, after which their sheet resistance, residual resistance ratio (RRR) and critical temperature were determined (see Fig.~\ref{fig:mam_vsi_competition}).
		Samples with thinner layers ($\leq\SI{100}{\nano\metre}$) annealed at \SI{500}{\celsius} showed a resistivity of \SI{180\pm10}{\micro\ohm\centi\metre}, similar to that of as-deposited amorphous \ce{V3Si}, while the sheet resistance of the \SI{200}{\nano\metre} layer remained out of range for our instruments up to annealing temperatures of \SI{600}{\celsius}.
		Indicated in Fig.~\ref{fig:mam_vsi_competition}a is a red dashed line at \SI{77.1}{\micro\ohm\centi\metre}, which corresponds to the resistivity measured on a sample with \SI{200}{\nano\metre} of \ce{V3Si} on a sapphire substrate, annealed at \SI{900}{\celsius} under vacuum.
		Since this is the lowest resistivity that we have measured for \ce{V3Si} on any sample where we are sure that no chemical reaction had occurred with the substrate (XRD confirmed that \ce{V3Si} was the only V-rich compound), any value below this dashed line is taken to indicate the presence of \ce{VSi2}.
		
		As shown in Fig.~\ref{fig:mam_vsi_competition}b, at high annealing temperatures the RRR is higher for thicker layers, which can be explained by thickness-limited grain growth and is consistent with the lower quality of thinner films reported elsewhere~\cite{michikami1982v3si}.
		An interesting inversion of the ordering of the RRR values for the three first thicknesses occurs at \SI{550}{\celsius}, where the \SI{20}{\nano\metre} film shows the highest RRR.
		This could be explained by the early homogenization of the forming \ce{VSi2} film, which would imply a Si diffusivity of around \SI{1.8}{\nano\metre\squared\per\second} through the \ce{VSi2} layer~\cite{chu1974identification}, a value that has been reported elsewhere only for annealing temperatures of \SI{650}{\celsius}~\cite{tu1973formation}.
		
		The crystallization of \ce{V3Si} is observed on thicker films of at least \SI{100}{\nano\metre} by critical temperature measurements (Fig.~\ref{fig:mam_vsi_competition}c).
		It is important to note that the critical temperatures obtained for \SI{200}{\nano\metre} films on HF-cleaned Si are within experimental error identical to those obtained on \ce{SiO2} (see data points for 600 and \SI{650}{\celsius} in Fig.~\ref{fig:jap_fig1}), which means that the critical temperature does \emph{not} depend on the thickness of the \ce{V3Si} film itself, but rather on the initial thickness of the as-deposited layer.
	
	\FloatBarrier
	\subsection{Conclusion}
		
		\ce{V3Si} is certainly a promising material for the CMOS integration of JoFETs, with a relatively low Schottky barrier to n-doped Si~\cite{hugunin1995superconductor}, a critical temperature that can reach over \SI{15}{\kelvin} in thin films, and near-perfect lattice matching.
		Two challenges to its integration have however been identified; a negative impact of strain that is built up during thermal processing and compounded during cooling to cryogenic temperatures, and the appearance of \ce{VSi2} at the \ce{V3Si}/Si interface at higher temperatures.
		Luckily, both of these can mediated with the appropriate choice of annealing time and temperature.
		
		The strain built up in the \ce{V3Si} thin film appears to depend mainly on the maximum temperature reached during annealing, at which point the atoms are most mobile.
		Since any solid-state reaction or diffusion is enhanced exponentially with temperature, relatively little plastic deformation will occur during and after cooling, such that the system is deformed only elastically and essentially retains a ``memory'' of this highest-temperature moment.
		Similarly, any reaction with the underlying silicon will be limited by atomic diffusion rates, and can therefore be controlled with thermal processing.
		Since the total distance that needs to be traveled by either species (V or Si) inside the initially amorphous \ce{V3Si} film to arrange itself in a crystalline manner is much smaller than that which needs to be crossed to consume the film by \ce{VSi2} formation, it should be possible to find some time and temperature range within which to optimize the trade-off between \ce{V3Si} crystallinity and \ce{VSi2} formation.
		It is not yet clear whether it is preferable to aim for some minimal critical temperature, and then minimize the thickness of the \ce{VSi2} layer within that constraint, or if more robust superconductivity is required.
		As-deposited \ce{V3Si} already superconducts on a silicon substrate at \SI{0.9}{\kelvin} (and at \SI{1.2}{\kelvin} on sapphire), similar to what is observed for traditional aluminum thin films.
		Nor is it clear that the thickness of the \ce{VSi2} layer should be minimized at all.
		For longer or more diffusive (highly doped) silicon channels, the precise details of the contact materials will not be relevant anymore, and the critical current will mainly be limited by the length of the semiconducting part of the channel.
		Schottky barriers will appear at the \ce{VSi2}/Si interface, rather than at the initial \ce{V3Si}/Si boundary, and will move underneath the gate electrode, where their effective width can be controlled (see chapter~\ref{sec:jofets}).
		Even when taking into account that every nanometer of consumed Si will give way to $\SI{2.3}{\nano\metre}$ of \ce{VSi2}, the proximity effect will likely be suppressed less in such a metallic region.
		
		In the end, whichever combination of \ce{VSi2} formation and \ce{V3Si} crystallization turns out to be optimal in JoFET devices, the sensitivity of this reaction to heat enforces a stringent thermal budget on subsequent processing steps.
		Such constraints can be avoided entirely by selecting a superconducting silicide that actually \emph{is} the thermodynamically favored phase within its system, and therefore stable.
		We find this property in the more conventional \ce{PtSi}, a material that has already been used in MOSFET fabrication for decades, and which we will discuss in the next section.

\FloatBarrier
\clearpage
\section{\label{sec:ptsi}PtSi}
	
	While \ce{V3Si} is exciting for its novelty in the context of CMOS technology, PtSi is important precisely because it has already been successfully integrated in multiple generations of VLSI devices.
	What is unknown however, is how this material can be best used in superconducting applications such as JoFETs.
	To study this aspect, we need to move from unpatterned ``blanket'' thin films to actual devices.
	By patterning transmission line measurement (TLM) structures, we can measure such things as the contact and channel resistance, and ultimately determine the superconducting properties of the PtSi/Si interface at low temperatures.
	
	The main concern in JoFET fabrication is that the proximity effect will be suppressed, part at the superconductor/semiconductor (S/Sm) interface and part within the semiconducting channel, to the point that Josephson coupling can no longer occur between the superconducting leads.
	While the precise details of electron-hole pair decoherence in silicon depend on such experimental variables as doping level, surface oxide scattering, and the interaction with charging levels of dopants and the channel itself, we can to first order assume that it decays exponentially with the length $L$ of the junction~\cite{volkov1996effect},
	\begin{equation}I_\text{c}\propto e^{-L/\xi}.\end{equation}
	Expecting $\xi$ to be on the order of tens of nanometers, we therefore aim for devices with the shortest channel possible, so that we will be limited only by the interface transparency.
	One way to push this length down, is by having the contact silicide encroach underneath the gate spacers (see Fig.~\ref{fig:v3si_jofet}).
	Since dopants in the channel can aggregate at the PtSi/Si interface~\cite{takai1985effects}, this gives some control over the Schottky barrier.
	By bringing the S/Sm interface within reach of the electrostatic gate field, it further provides control of the barrier by the gate.
	
	Encroachment, in turn, brings along a new risk.
	In the self-aligned silicide (SALICIDE) process, the metal is deposited once the contact openings to the monocrystalline silicon have been defined, after which rapid thermal processing activates the silicidation reaction.
	In the case of PtSi formation, which is diffusion controlled~\cite{takai1985effects}, planar silicide layers grow with a thickness proportional to the square root of time.
	Encroachment starts once the silicide extends below the spacers\footnote{This can be delayed by epitaxial growth of Si in the contact openings.}, and then suddenly speeds up once the silicide reaches the buried oxide below the contacts.
	This sudden increase in horizontal silicide growth underneath the spacer and gate can be understood as follows.
	The reaction is limited by the rate at which Si atoms (the dominant diffusing species in PtSi formation~\cite{bindell1976analytical}) arrive at the Pt or \ce{Pt2Si} reservoir, where they bind to Pt and form PtSi.
	This arrival rate then balances the rate at which Si atoms dissolve into the forming PtSi from the other side of the film, which means that in a planar configuration the Si/PtSi interface recedes downward at a pace proportional to the upward extension of the PtSi/\ce{Pt2Si} or PtSi/Pt interface.
	Once the Si on the bottom of the contacts is fully consumed however, the surface area of the Si/PtSi interface is given only by the vertical sides of the Si channel on which the silicide is encroaching.
	Balancing the arrival rate of Si atoms at the large planar Pt reservoir now requires a much higher absorption rate of Si at the smaller, vertical Si/PtSi interface on the side of the channel.
	At this point, the distance of horizontal encroachment becomes the effective diffusion length that limits the reaction as Si atoms diffuse through this narrow estuary, and the $\ell\propto\sqrt{t}$ behavior begins anew.
	
	To further expound on the difficulty of controlling encroachment, consider the respective numbers of Pt and Si atoms involved.
	While the thickness of planar PtSi films can be easily controlled by precisely tuning the amount of deposited Pt, any metal that remains after full consumption of the Si in the contacts will be incomparably large compared to the volume of silicon in the channel, providing a practically inexhaustible metal reservoir.

	\begin{table}
		\centering
		\caption{\label{tab:ptsi_splits_overview}An overview of the lots prepared for the study of PtSi formation.}
		\resizebox{\columnwidth}{!}{%
		\rowcolors{2}{gray!15}{white}
		\begin{tabular}{l l l l}
			\rowcolor{gray!30}\hline
			Name	& Reference	& Wafer types				& Split type \\\hline\hline
			Lot 1	& D16S0667A	& blanket, patterned (TLM)	& Pt thickness\\
			Lot 2	& D16S0667B	& blanket, patterned (TLM)	& RTP temperature, time\\
			Lot 3	& D19S0918	& blanket					& RTP temperature, \# steps completed\\
			Lot 4	& D19S1723	& blanket, patterned (TLM)	& Pt thickness\\\hline
		\end{tabular}}
	\end{table}
	
	This brings us to the question that we will try to answer in this section: how can we optimize the formation of PtSi in JoFETs, such that we simultaneously obtain a high superconducting critical temperature, a moderate amount of encroachment, and highly transparent interfaces?
	Four lots\footnote{In this context, a ``lot'' is a batch of 25 wafers. From a common Germanic root, the same word exists in French; in algorithm~\ref{alg:tp} you could find yourself assessing a \emph{lot de 6 rouleaux}.} were prepared to answer this question, listed in Table.~\ref{tab:ptsi_splits_overview}.
	In each of these, a \emph{split} was made, in which the parameters of only a few of the process steps vary across the wafers.
	Since the 200 and \SI{300}{\milli\metre} industrial-cleanroom fabrication technologies employed at the CEA LETI are highly reproducible, we can then with high confidence assume that any variations observed in the properties of these wafers are due to the differences designed in the split\footnote{To get a sense of the reproducibility of CMOS technology: for our TLM mask set it takes 111 process steps to make a ``base wafer'' ready for metallization, and a further 72 steps to finish the back-end. Still, broken devices are usually due only to the steps intentionally varied in the split. Even wafers with actual transistors, which require hundreds more steps, typically have 80--90\% yields.}.
	
	\FloatBarrier
	\subsection{\label{sec:split1}Split 1: platinum deposition}
	
		The first lot, detailed in Table~\ref{tab:lot1}, considers a variation in deposited Pt thickness.
		Though perhaps at first sight a mundane parameter, its choice in fact has wide-ranging consequences for the properties of the final device.
		First, this parameter gives direct control over the superconducting critical temperature, which depends strongly on both the quality and thickness of the film.
		Second, the duration of the silicidation and thickness of the silicide is likely to affect the morphology of the interface~\cite{ru2002surface}, which in turn influences the S/Sm interface transparency.
		Third, the amount of available platinum determines the depth to which the silicon in the contacts will be consumed, and thus how far the superconducting contacts will be from the channel.

		\begin{table}
			\centering
			\caption{\label{tab:lot1}Lot 1 (D16S0667A): A set of both patterned and blanket wafers are prepared, with deposited Pt thicknesses varying from 5 to \SI{25}{\nano\metre}. A double is included for the patterned wafer with median thickness of \SI{15}{\nano\metre}, to perform in-line XRD.}
			\rowcolors{2}{white}{gray!15}
			\begin{tabular}{l c c c c c c | c c c c c}
				\rowcolor{gray!30}\hline
				\B{Lot 1}								& \multicolumn{11}{c}{Wafer \#}\\
				\rowcolor{gray!30}
														& 1		& 2		& 3		& 4		& 5		& 6		& 7		& 8		& 9		& 10	& 11	 \\\hline\hline
				Blanket bulk Si wafer					& \oDOT	& \oDOT	& \oDOT	& \oDOT	& \oDOT	& \oDOT	& \DOT	& \DOT	& \DOT	& \DOT	& \DOT	\\
				Patterned SOI wafer						& \DOT	& \DOT	& \DOT	& \DOT	& \DOT	& \DOT	& \oDOT	& \oDOT	& \oDOT	& \oDOT	& \oDOT	\\
				Deposit $<>$ \si{\nano\metre} Pt + TiN	& 5		& 10	& 15	& 15	& 20	& 25	& 5		& 10	& 15	& 20	& 25	\\
				RTA	\SI{500}{\celsius} 120s				& \DOT	& \DOT	& \DOT	& \DOT	& \DOT	& \DOT	& \DOT	& \DOT	& \DOT	& \DOT	& \DOT	\\
				Etch TiN								& \DOT	& \DOT	& \DOT	& \DOT	& \DOT	& \DOT	& \DOT	& \DOT	& \DOT	& \DOT	& \DOT	\\
				Etch Pt									& \DOT	& \DOT	& \DOT	& \DOT	& \DOT	& \DOT	& \DOT	& \DOT	& \DOT	& \DOT	& \DOT	\\
				XRD										& \oDOT	& \oDOT	& \oDOT	& \DOT	& \oDOT	& \oDOT	& \DOT	& \DOT	& \DOT	& \DOT	& \DOT	\\
				Complete back-end						& \DOT	& \DOT	& \DOT	& \oDOT	& \DOT	& \DOT	& \oDOT	& \oDOT	& \oDOT	& \oDOT	& \oDOT	\\
			\end{tabular}
		\end{table}
		
\begin{figure}
			\centering
			\includegraphics[height=0.8\textheight]{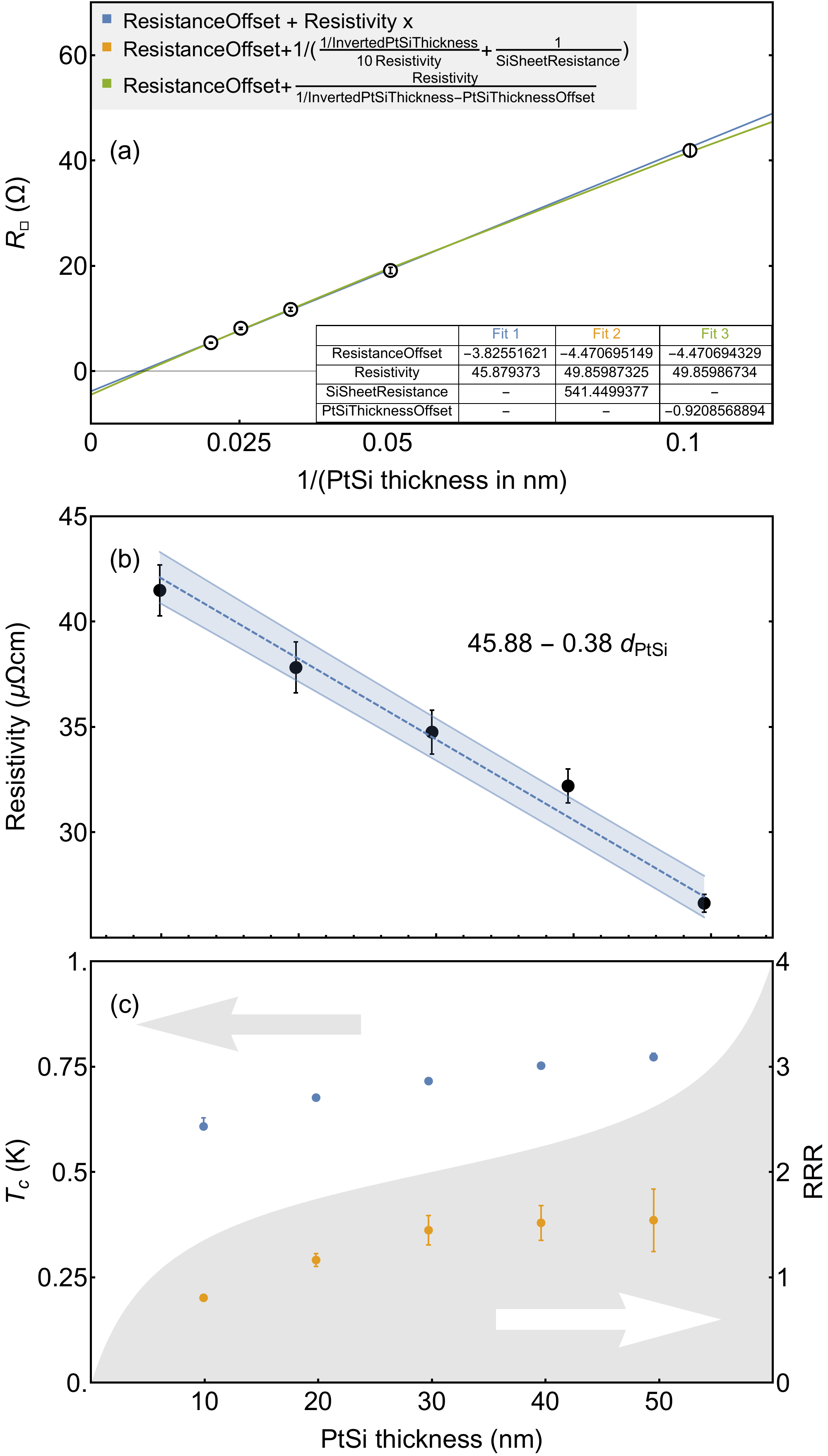}
			\caption{\label{fig:D16S0667A}\B{(a)} It is customary to plot the sheet resistance versus the inverse of the film thickness (\SI{1.98}{\nano\metre} of PtSi is formed per nm of deposited Pt), such that the resistivity of the material is given by the slope. If this method is used, which assumes that the resistivity of the film is independent of the film thickness, we have to conclude that the measurement apparatus has an offset. \B{(b)} If instead, the resistivity is calculated for each individual film, and then plotted versus the thickness, we find a more likely explanation: the film quality improves with thickness. The blue dashed line is a linear regression with $2\sigma$ error margin. \B{(c)} Both the residual resistance ratio (RRR) and the critical temperature $T_\text{c}$ improve with film thickness, in line with the improvement in room-temperature resistivity.}
		\end{figure}

		As discussed before in the context of \ce{V3Si}, resistivity is a good proxy for the overall material quality.
		By assuming that all the Pt had reacted in a 1:1 ratio with the underlying silicon to form PtSi, this resistivity can be calculated directly from the sheet resistance of the bulk wafers (P07 -- P11 in Table~\ref{tab:lot1}).
		As shown in Fig.~\ref{fig:D16S0667A}, consistently higher quality films are obtained by depositing thicker layers of platinum.
		Such a decrease in $R_\square$ with film thickness has been observed before, with resistivities reducing from 59 to \SI{32}{\micro\ohm\centi\metre} by going from 14 to \SI{250}{\nano\metre} of PtSi~\cite{das1994thickness}, and has been attributed to increases in grain size.
		This quality is reflected also in the residual resistance ratio, indicating fewer non-thermal scattering events in thicker films.
		Such scattering impacts superconductivity in the thinner films~\cite{ferdeghini2009superconducting}, which is further affected by the finite thickness itself~\cite{romijn1982critical}.

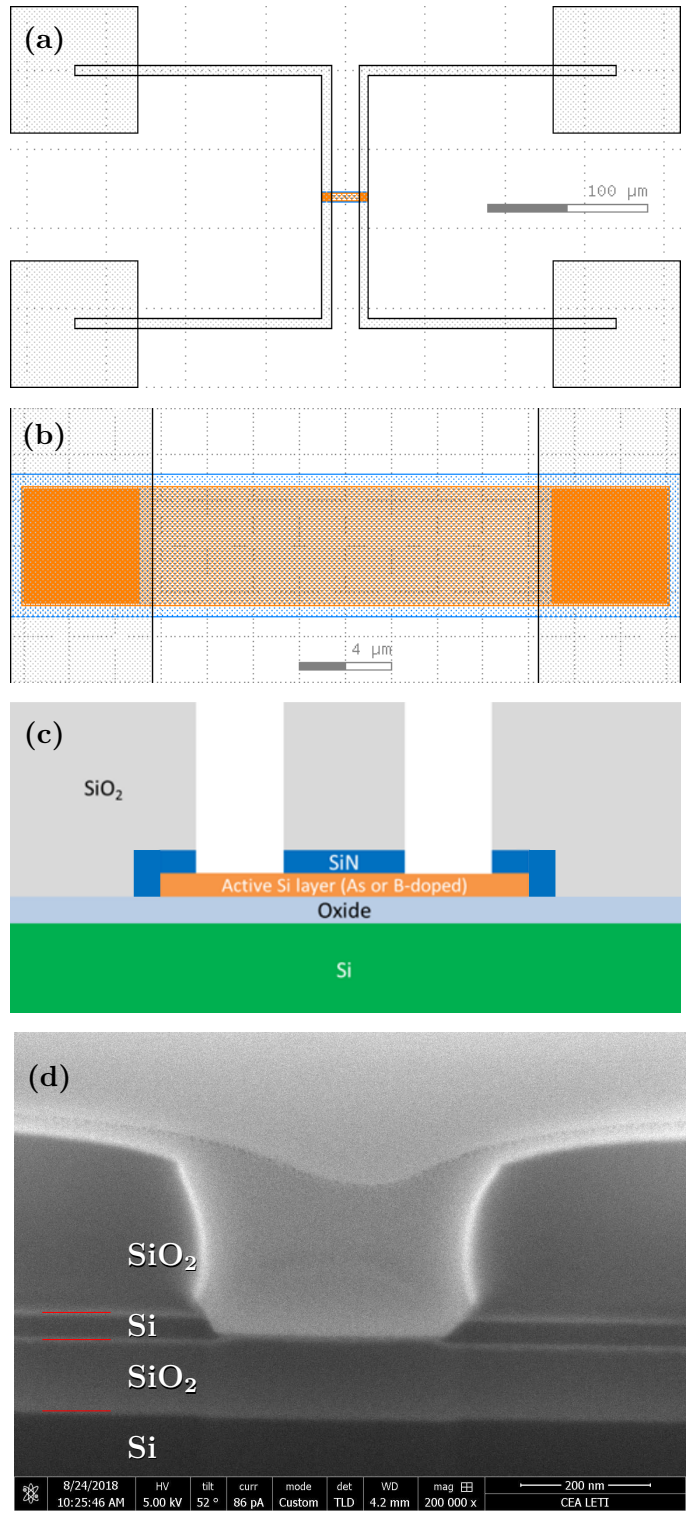
\begin{figure}
			\centering
			\begin{subfigure}{0.6\textwidth}
				\centering
				\begin{tikzpicture}[x=0.9cm,y=0.9cm]
					\node[anchor=center,inner sep=0] at (0,0) {%
						\includegraphics[width=\textwidth]{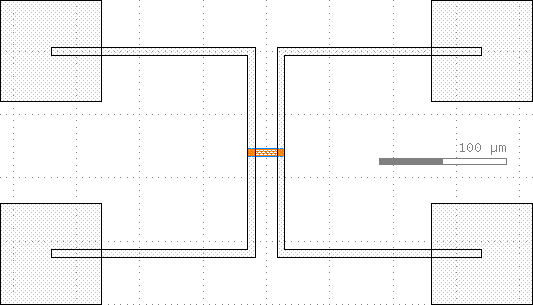}};
						\node at (-4.4,2.3) {\B{(a)}};
				\end{tikzpicture}
			\end{subfigure}
			\vspace*{0.5\baselineskip}
			
			\begin{subfigure}{0.6\textwidth}
				\centering
				\begin{tikzpicture}[x=0.9cm,y=0.9cm]
					\node[anchor=center,inner sep=0] at (0,0) {%
						\includegraphics[width=\textwidth]{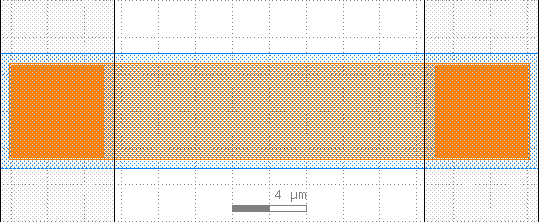}};
						\node at (-4.4,1.6) {\B{(b)}};
				\end{tikzpicture}
			\end{subfigure}
			\vspace*{0.5\baselineskip}
			
			\begin{subfigure}{0.6\textwidth}
				\centering
				\begin{tikzpicture}[x=0.9cm,y=0.9cm]
					\node[anchor=center,inner sep=0] at (0,0) {%
						\includegraphics[width=\textwidth]{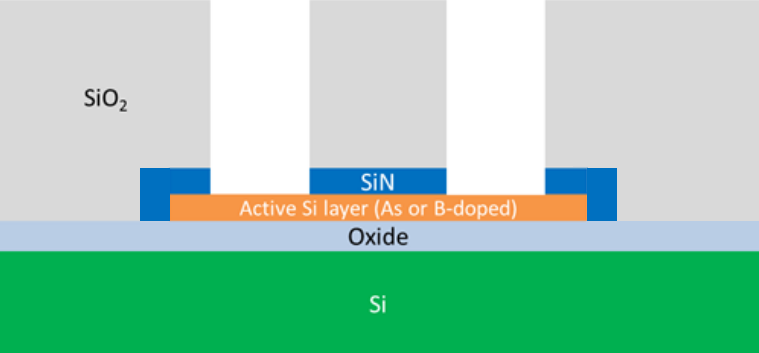}};
						\node at (-4.4,1.8) {\B{(c)}};
				\end{tikzpicture}
			\end{subfigure}
			\vspace*{0.5\baselineskip}
			
			\begin{subfigure}{0.6\textwidth}
				\centering
				\begin{tikzpicture}[x=0.9cm,y=0.9cm]
					\node[anchor=center,inner sep=0] at (0,0) {%
						\includegraphics[width=\textwidth]{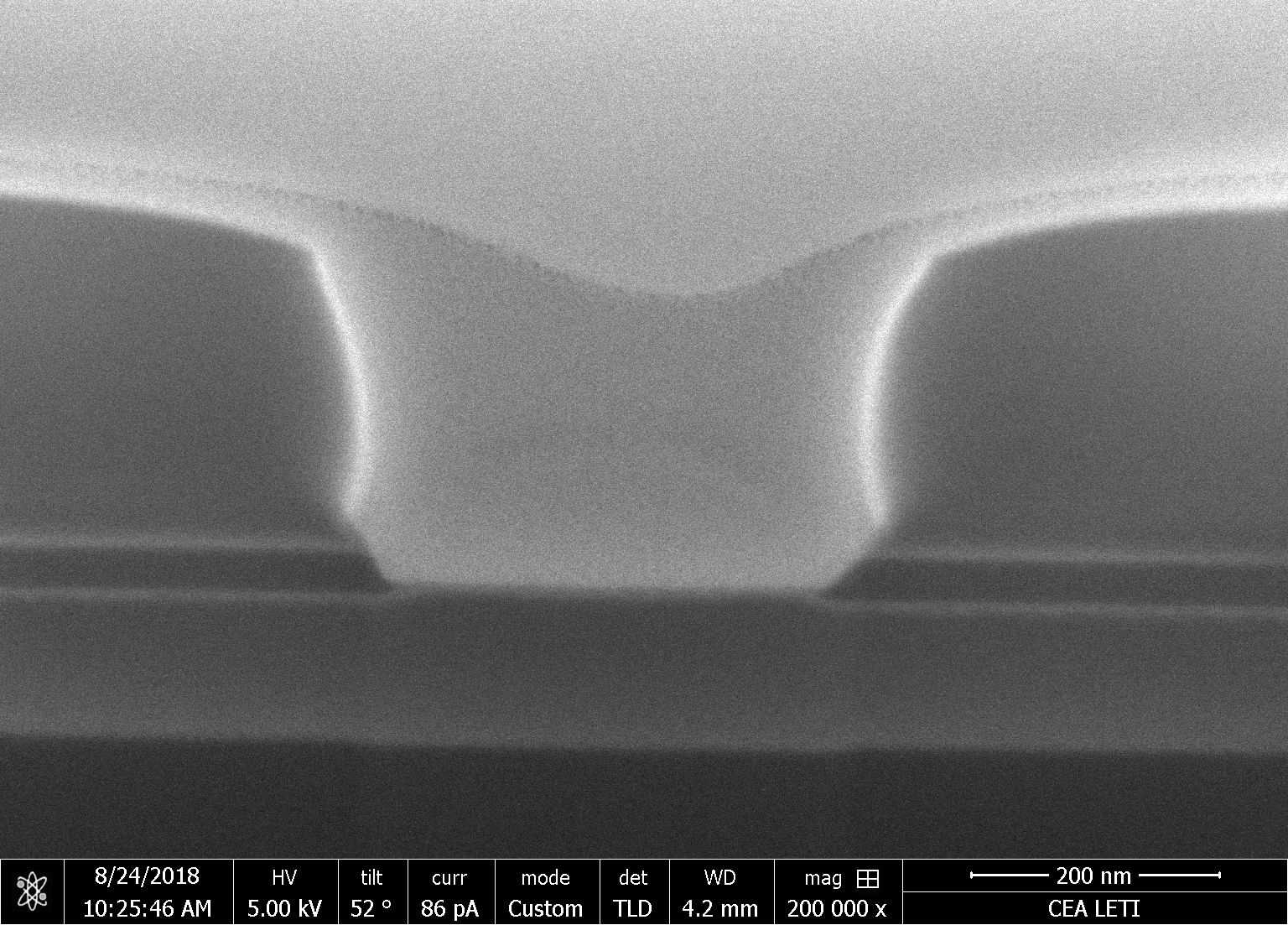}};
						\node at (-4.4,2.85) {\B{(d)}};
						\draw[red] (-4.5,-0.6) -- (-3.5,-0.6);
						\draw[red] (-4.5,-1) -- (-3.5,-1);
						\draw[red] (-4.5,-2.05) -- (-3.5,-2.05);
						
						\node[anchor=west] at (-3.4,-0.8) {\large \B{Si}};
						\node[anchor=west,white] at (-3.41,-0.78) {\large \B{Si}};
						
						\node[anchor=west] at (-3.4,0.2) {\large \B{\ce{SiO2}}};
						\node[anchor=west,white] at (-3.41,0.22) {\large \B{\ce{SiO2}}};
						
						\node[anchor=west] at (-3.4,-1.6) {\large \B{\ce{SiO2}}};
						\node[anchor=west,white] at (-3.41,-1.58) {\large \B{\ce{SiO2}}};
						
						\node[anchor=west] at (-3.4,-2.6) {\large \B{Si}};
						\node[anchor=west,white] at (-3.41,-2.58) {\large \B{Si}};
				\end{tikzpicture}
			\end{subfigure}
			\caption{\label{fig:DS16S0667AP20FIBSEM}\B{(a)} The design of a TLM structure: in gray are shown the bonding pads and lines leading to the two PtSi contacts on either side of a silicon channel. \B{(b)} Zoom on the channel itself. \B{(c)} A cross-section of the structure at the stage that the wafer was analyzed: before platinum deposition. \B{(d)} A FIB-SEM image taken of wafer 20 from base lot D16S0667, the lot that the patterned wafers in lots~1 and~2 were taken from. Shown is a contact opening for a TLM structure, where it is clear that there is no remaining Si in the contact. This explains why no working devices were found.}
		\end{figure}

		Low-temperature measurements on TLM structures are required to further investigate the superconducting behavior of PtSi/Si junctions.
		Unfortunately, it was found that none of the wafers in lot~1 (nor lot~2, for that matter) had any working devices.
		This was ultimately traced back to faulty base wafers, where the Si in the contacts had been over-etched during the last process step prior to metallization (see Fig.~\ref{fig:DS16S0667AP20FIBSEM}).
		A new lot with a split identical to lot~1 (patterned wafers only) was launched to correct for this mistake, which was analyzed by master student Axel Leblanc and is discussed briefly in section~\ref{sec:tasp}\footnote{\label{foot:industrial_slow}Industrial 200 and \SI{300}{\milli\metre} cleanroom technology is certainly more reliable, and in many aspects more advanced than fabrication processes in academic cleanrooms. It is also much slower. After the conclusive FIB-SEM analysis shown in Fig.~\ref{fig:DS16S0667AP20FIBSEM}, it took from 2018-09-04 until 2020-01-27, a total of 1 year, 4 months and 23 days, before a replacement for lot~1 was finished (this replacement, lot~4, was in fact given priority in the cleanroom). This delay partially explains the relative emphasis on material studies in this manuscript.}.
		
	\FloatBarrier
	\subsection{\label{sec:lot2}Split 2: rapid thermal processing (RTP)}
		
		Like the thickness of deposited Pt, the thermal processing that follows can be used to control a wide range of parameters.
		In the context of JoFETs, the most important of these is perhaps again the encroachment.
		Annealing can be done in a single high-temperature step, immediately forming the final silicide phase, as was done in the split discussed in section~\ref{sec:split1}.
		Since the grain size and thus the quality of the final silicide improves with temperature~\cite{das1994thickness}, it is necessary to go up to perhaps 450 or \SI{500}{\celsius}.
		This is about 150 to \SI{200}{\celsius} higher than the temperature at which \ce{PtSi} nucleates at the silicon-metal interface, leading to full metal consumption within seconds.
		Needless to say, this makes it impossible to control the encroachment.
		One solution is to divide the annealing process into two: a first step at a lower temperature, where diffusion is slow enough to control the lateral formation, followed by the removal of unreacted metal and \ce{Pt2Si} by selective etch and a second heating at \SI{500}{\celsius} to reach the desired material quality.

		\bgroup
		\setlength\tabcolsep{0.30em}
		\begin{table}
			\centering
			\caption{\label{tab:lot2}The split in lot~2 (D16S0667B), with a variation in the temperature and time of the first annealing step.}
			\rowcolors{2}{white}{gray!15}
			\begin{tabular}{l c c c c c c | c c c c c c c}
				\rowcolor{gray!30}\hline
				\B{Lot 2}								& \multicolumn{13}{c}{Wafer \#}\\
				\rowcolor{gray!30}
														& 1		& 2		& 3		& 4		& 5		& 6		& 7		& 8		& 9		& 10	& 11	& 12	& 13	\\\hline\hline
				Blanket bulk Si wafer					& \DOT	& \DOT	& \DOT	& \DOT	& \DOT	& \DOT	& \oDOT	& \oDOT	& \oDOT	& \oDOT	& \oDOT	& \oDOT	& \oDOT	\\
				Patterned SOI wafer						& \oDOT	& \oDOT	& \oDOT	& \oDOT	& \oDOT	& \oDOT	& \DOT	& \DOT	& \DOT	& \DOT	& \DOT	& \DOT	& \DOT	\\
				\makecell{Deposit\\
				 \SI{25}{\nano\metre} Pt + \SI{10}{\nano\metre} TiN}	
				 										& \DOT	& \DOT	& \DOT	& \DOT	& \DOT	& \DOT	& \DOT	& \DOT	& \DOT	& \DOT	& \DOT	& \DOT	& \DOT	\\
				RTP 1: Temperature (\si{\celsius})		& 300	& 300	& 300	& 350	& 350	& 350 	& 300	& 300	& 300	& 300	& 350	& 350	& 350	\\
				\phantom{RTP 1: }time (\si{\second})	& 10	& 30	& 60	& 10	& 30	& 60	& 10	& 30	& 60	& 60	& 10	& 30	& 60	\\
				Etch TiN								& \DOT	& \DOT	& \DOT	& \DOT	& \DOT	& \DOT	& \DOT	& \DOT	& \DOT	& \DOT	& \DOT	& \DOT	& \DOT	\\
				Etch Pt									& \DOT	& \DOT	& \DOT	& \DOT	& \DOT	& \DOT	& \DOT	& \DOT	& \DOT	& \DOT	& \DOT	& \DOT	& \DOT	\\
				RTP 2: \SI{500}{\celsius} 120s			& \DOT	& \DOT	& \DOT	& \DOT	& \DOT	& \DOT 	& \DOT	& \DOT	& \DOT	& \DOT	& \DOT	& \DOT	& \DOT	\\
				XRD										& \DOT	& \DOT	& \DOT	& \DOT	& \DOT	& \DOT	& \oDOT	& \oDOT	& \oDOT	& \DOT	& \oDOT	& \oDOT	& \oDOT	\\
				Complete back-end						& \oDOT	& \oDOT	& \oDOT	& \oDOT	& \oDOT	& \oDOT	& \DOT	& \DOT	& \DOT	& \oDOT	& \DOT	& \DOT	& \DOT	\\\hline
			\end{tabular}
		\end{table}
		\egroup
		
		As shown in Table~\ref{tab:lot2}, six conditions were tested for the first annealing step, with temperatures of 300 or \SI{350}{\celsius}, and durations between 10 and \SI{60}{\second}, after which all samples were annealed a second time at \SI{500}{\celsius} for a longer period of 2~minutes.
		A relatively large amount of Pt (\SI{25}{\nano\metre}) was deposited, about 30\% more than what is required to consume the \SI{25}{\nano\metre} of Si on the bottom of the contact openings\footnote{Of course this estimate for the available volume of Pt is not accurate for trenches in patterned wafers, ignoring both the reduced Pt thickness on the bottom due to shadow effects, and the available volume of Pt on the side walls. In any case, 30\% overhead is enough to assume that the reaction will not be limited by the amount of Pt available.}, to ensure that the reactions would be limited by the temperature, rather than the available reagents.
		Previous studies on thicker (\SI{120}{\nano\metre} Pt) films with annealing times of multiple hours had shown that \ce{Pt2Si} forms at \SI{260}{\celsius}, while PtSi appears around \SI{312}{\celsius}~\cite{pan1984situ}.
		Multiple phases are unlikely to appear simultaneously in metal-silicon systems, especially in thin films~\cite{gas1993formation}, and \ce{Pt2Si} usually occurs before the nucleation of \ce{PtSi}~\cite{naem1988platinum}, though this first phase may be skipped altogether in thinner films or higher temperatures.
		Short heating at \SI{300}{\celsius} during the first RTP is thus expected to lead to relatively slow and controllable \ce{Pt2Si} formation by Pt diffusion into the Si~\cite{tu1975selective}, followed by Si diffusion into the \ce{Pt2Si} and the nucleation of \ce{PtSi} during RTP~2~\cite{mcleod1992marker}.
		At a higher temperature of \SI{350}{\celsius} during RTP~1, the \ce{Pt2Si} phase is expected to be either quickly transformed into \ce{PtSi}, or skipped entirely, leading to the immediate appearance of the final phase.
		Since \SI{350}{\celsius} is still far lower than the temperature used during single-step annealing (see Table~\ref{tab:lot1}), it is hoped that some degree of control over the Pt consumption remains.

		\begin{figure}
			\centering
			\includegraphics[width=0.8\textwidth]{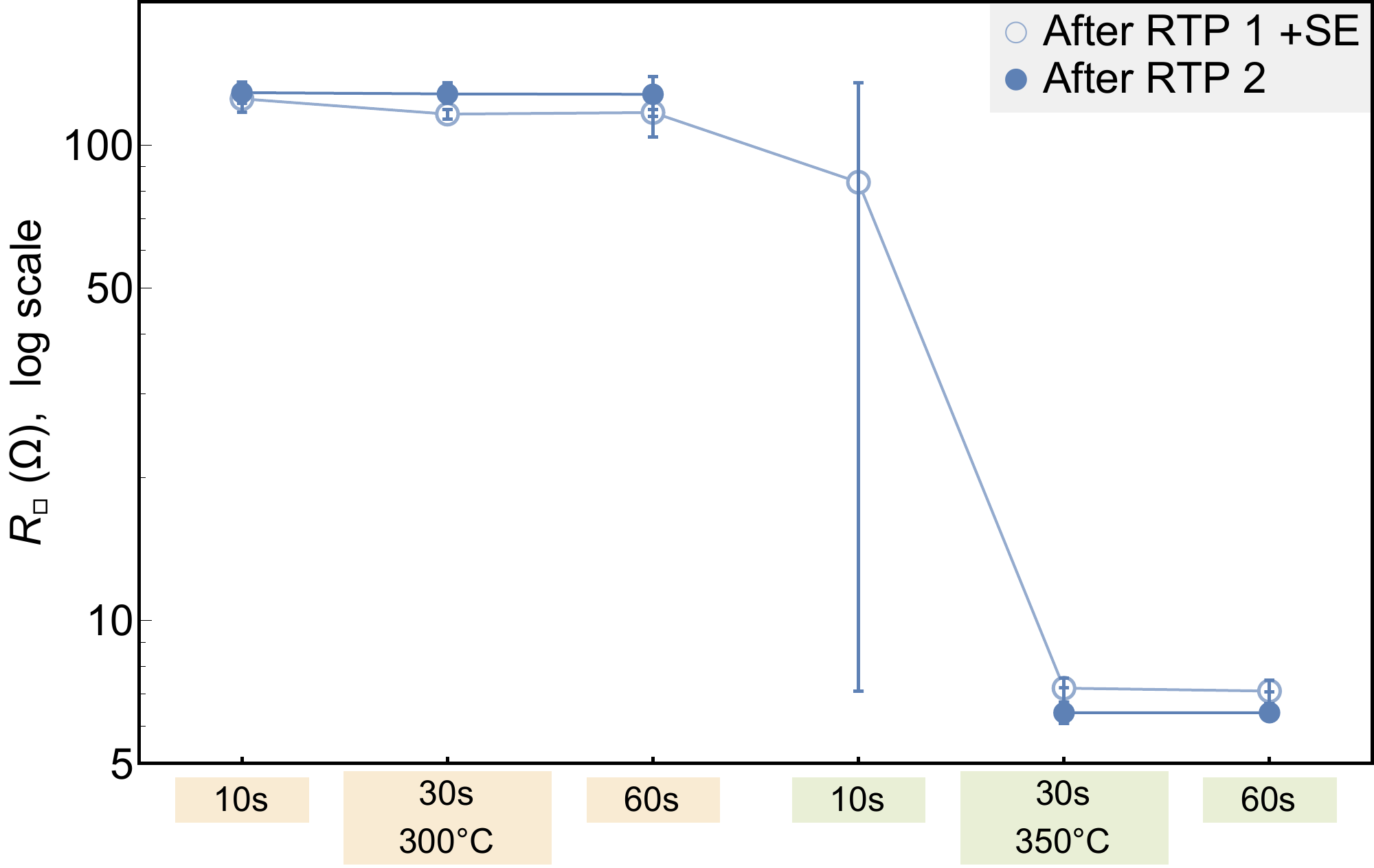}
			\caption{\label{fig:D16S0667RTPRsPlot}Sheet resistance measurements after RTP~1 and the selective etch (open symbols), and after RTP~2 (closed symbols). A clear distinction is visible between the wafers annealed at \SI{300}{\celsius}, and those annealed at \SI{350}{\celsius}. Wafer P04, annealed at \SI{350}{\celsius} for \SI{10}{\second}, had a variation in color, with a darker spot near the center, where the $R_\square$ was lower --- logs of the RTP show a $\sim\SI{10}{\celsius}$ higher overshoot on the thermometer below the center of the wafer than on those near the edge.}
		\end{figure}
		
		The subsequent selective etch (SE) of the TiN capping layer and unreacted Pt is designed to leave behind both \ce{PtSi} and \ce{Pt2Si}.
		Any \ce{Pt2Si} that formed during the first RTP (and remained after the SE), will be fully transformed into \ce{PtSi} during later thermal processing at \SI{500}{\celsius}, consuming a roughly equal volume of Si.
		In case PtSi already appeared (in which case no \ce{Pt2Si} is expected to remain~\cite{gas1993formation}), no further Si consumption should follow during RTP~2.
		
		As with lot~1, no working devices were found on the patterned wafers (see Fig.~\ref{fig:DS16S0667AP20FIBSEM}).
		Shown in Fig.~\ref{fig:D16S0667RTPRsPlot} are sheet resistance measurements on the blanket wafers averaged over 49 equally spaced locations on the wafer, taken after both RTP~1 (and SE) and RTP~2.
		The large difference in sheet resistance after RTP~1 between the two sets of three wafers annealed at different temperatures suggests that different reactions occurred, likely the formation of \ce{Pt2Si} at \SI{300}{\celsius}, and \ce{PtSi} at \SI{350}{\celsius}.
		The resistivity obtained after RTP~1 at \SI{350}{\celsius} for \SI{30}{\second} on P06 is \SI{35\pm2}{\micro\ohm\centi\metre}, already close to the \SI{26.6\pm0.4}{\micro\ohm\centi\metre} obtained after single-step annealing at \SI{500}{\celsius} (see Table~\ref{tab:lot1}, wafer~11 and Fig.~\ref{fig:D16S0667A}).
		The high sheet resistance after RTP~1 on the wafers annealed at \SI{300}{\celsius} cannot be explained only by the formation of \ce{Pt2Si}, which has a similar resistivity to \ce{PtSi} of around \SI{32}{\micro\ohm\centi\metre}~\cite{faber2011kinetics}.
		Instead, we now know that the selective etch with aqua regia (\ce{HNO3}:HCl:\ce{H2O} with volume ratios 2:2:4, at \SI{60}{\celsius}) that was employed for Pt removal, is likely to have etched \ce{Pt2Si} as well, and may even have transformed some of the \ce{Pt2Si} at the interface with Si into PtSi~\cite{jin1999microstructural}.

		\begin{figure}
			\centering
			\begin{subfigure}[b]{0.46\textwidth}
				\centering
				\begin{tikzpicture}
					\node[anchor=north west,inner sep=0] at (0,0) {
					\includegraphics[width=\textwidth]{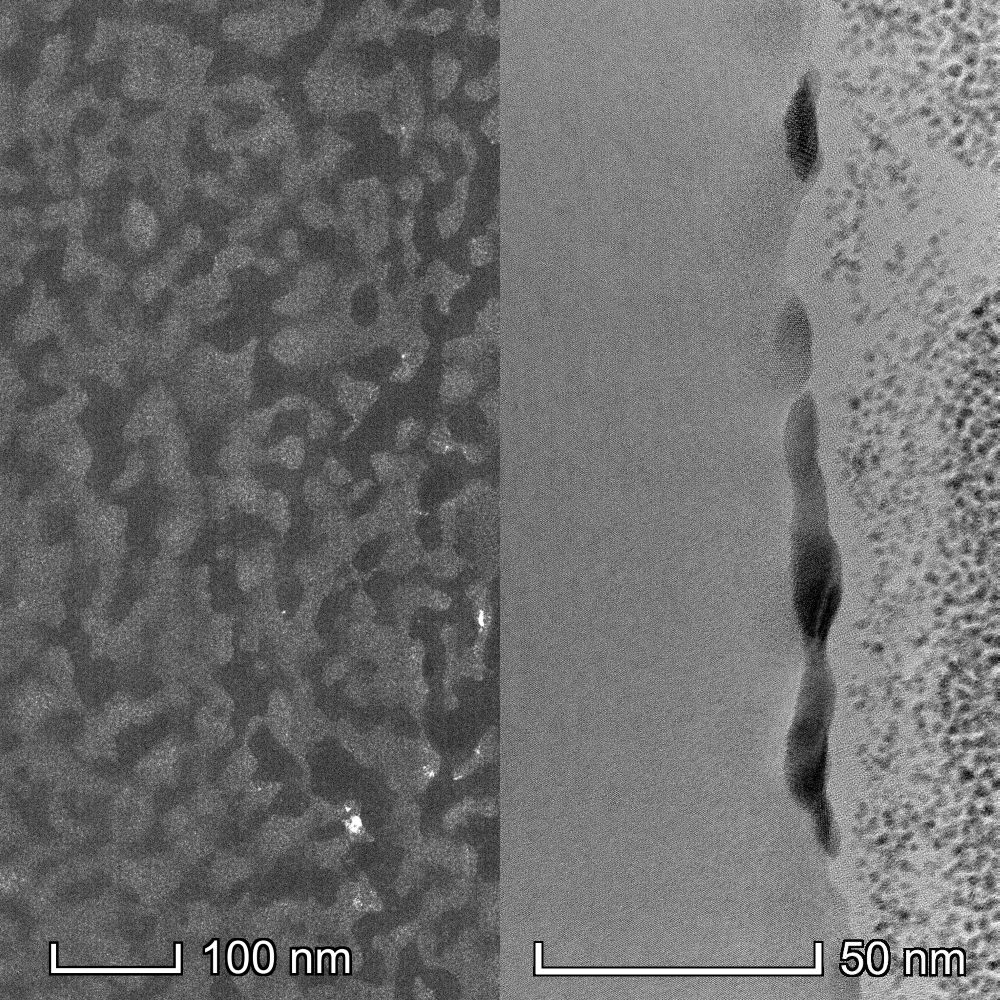}
						};
					
					\node at (1.3,-1.3) {\Large \B{(a)}};
					\node[white] at (1.28,-1.28) {\Large \B{(a)}};
					\node at ($(1.3,-1.3)+(0.5\textwidth,0)$) {\Large \B{(b)}};
					\node[white] at ($(1.28,-1.28)+(0.5\textwidth,0)$) {\Large \B{(b)}};
				\end{tikzpicture}
			\end{subfigure}\hfill\begin{subfigure}[b]{0.51\textwidth}
				\centering
				\includegraphics[width=\textwidth]{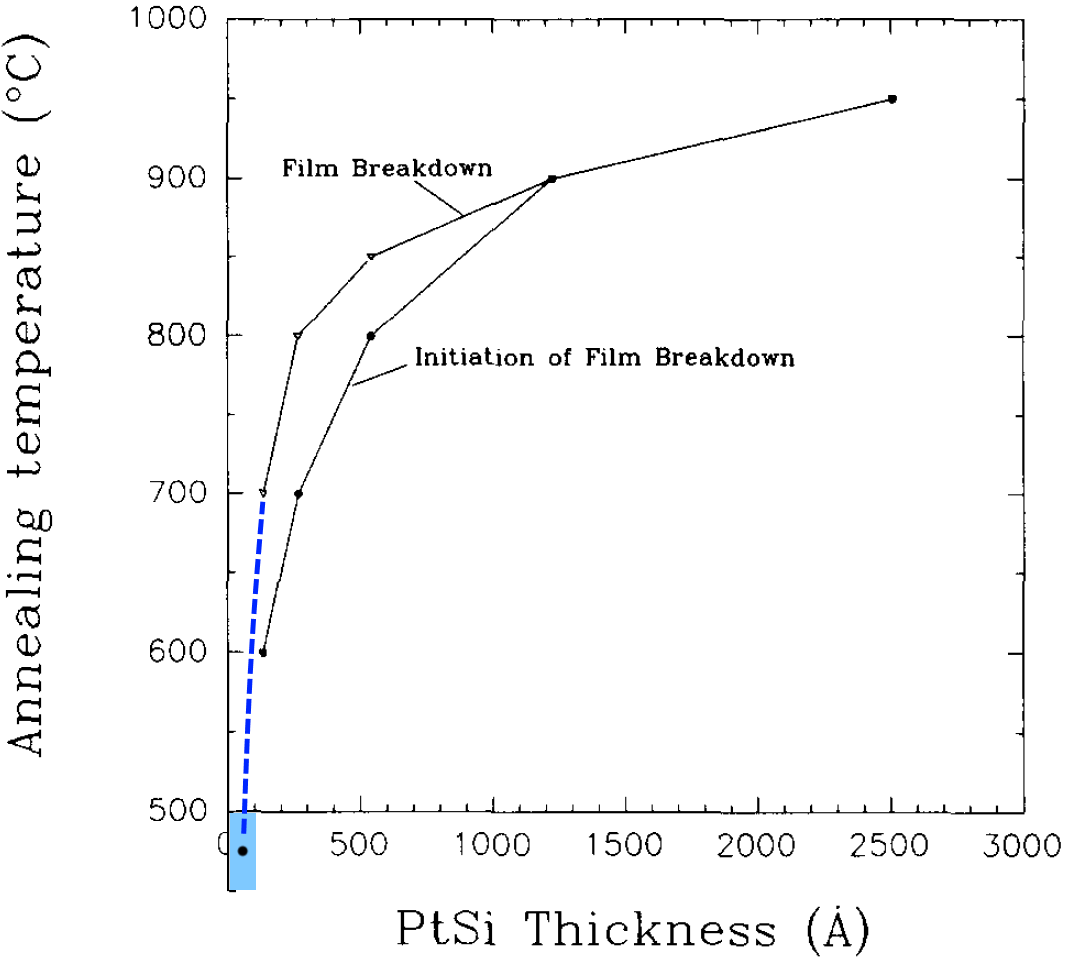}
			\end{subfigure}
			\caption{\label{fig:island_formation_das1994thickness}\B{(Left, a)} Plane view SEM and \B{(b)} Cross-section TEM of lot~2 (D16S0667B) P01. The lighter meanders on the left are PtSi (insulating Si appears darker under SEM), which is visible as darker regions on the right (heavier Pt is darker under TEM), as confirmed by EDS analysis. XRD analysis found a grain size of around 3--\SI{4}{\nano\metre}, on the same order as the size of the islands. \B{(Right)} Reproduced from Ref.~\citenum{das1994thickness}, blue dashed line added. Shown is the temperature at which PtSi films of different thicknesses break down, which is attributed to Pt diffusion into Si, disintegration of PtSi layer and simultaneous formation of a \ce{Pt3Si} phase.}
		\end{figure}
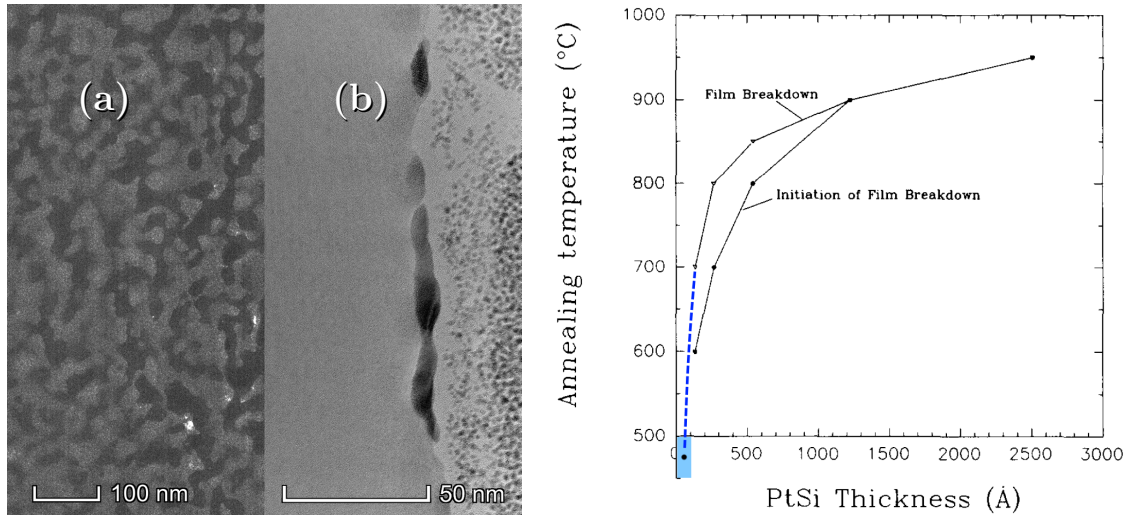
		
		Fig.~\ref{fig:island_formation_das1994thickness} (left) shows SEM and TEM analysis of a sample annealed at \SI{300}{\celsius} during \SI{10}{\second}, followed by SE, and then at \SI{500}{\celsius} during \SI{2}{\minute}.
		The film has likely broken up during the second RTP at \SI{500}{\celsius}, which would be consistent with earlier studies on the stability of PtSi thin films~\cite{das1994thickness}.
		As shown in the right panel of Fig.~\ref{fig:island_formation_das1994thickness}, though not reported before, the breakdown of a few-nanometer film into islands is likely to occur already at a temperature below \SI{500}{\celsius}.
		Evidently, only a very small amount of Pt remains in the system, suggesting that most or all of the \ce{Pt2Si} that had formed after RTP~1 had indeed been etched by aqua regia as reported elsewhere~\cite{jin1999microstructural}.

		\begin{figure}
			\centering
			\includegraphics[height=0.7\textheight]{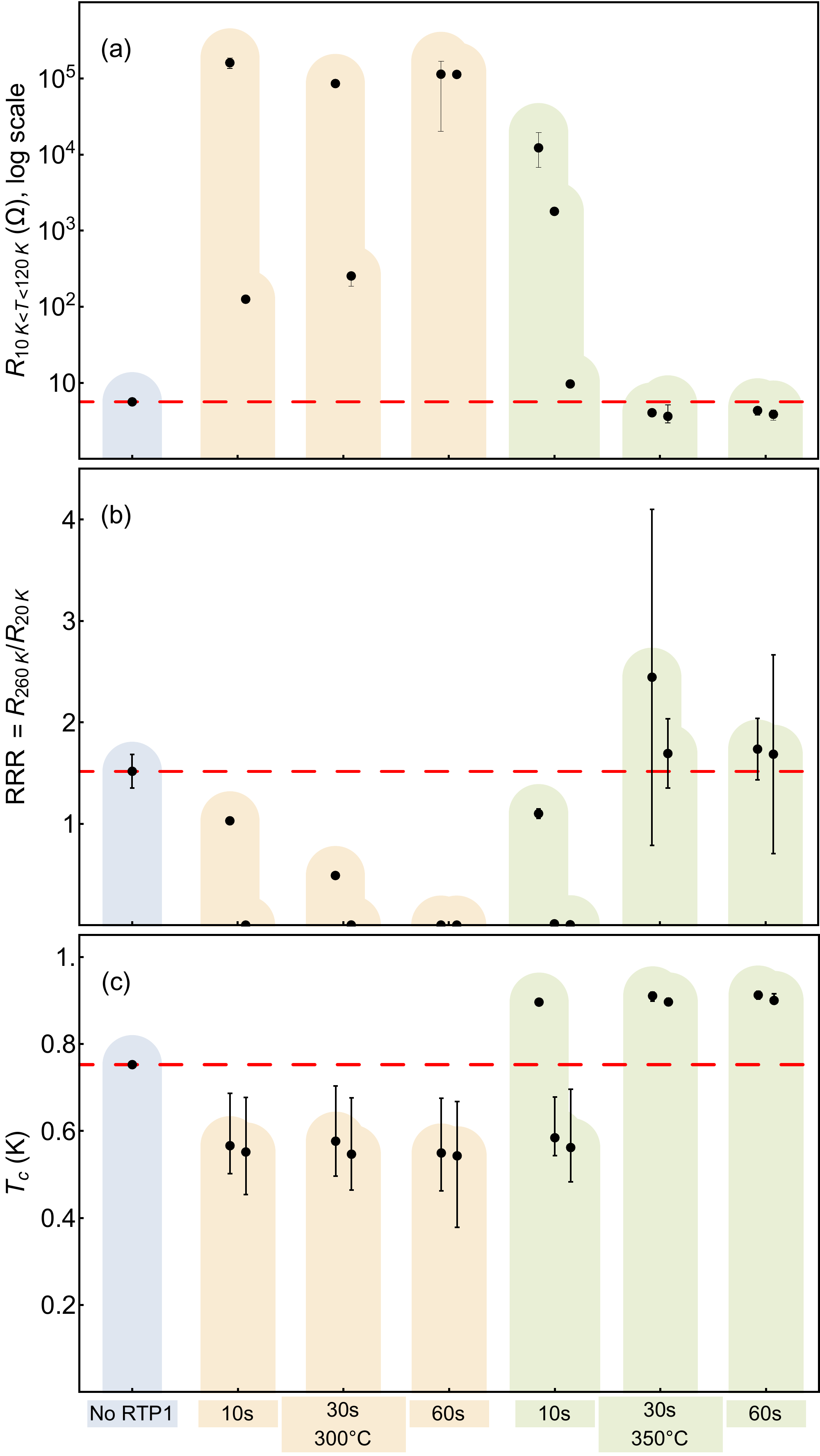}
			\caption{\label{fig:D16S0667BCombinedPlot}Low-T behavior of lot~2 (averaged between 10 and \SI{120}{\kelvin}, see Fig.~\ref{fig:LogRTPlotP15P03P06}), in blue on the left lot~1 P11 (identical Pt thickness, annealed once at \SI{500}{\celsius}, see Table~\ref{tab:lot1}). Multiple measurements performed on each wafer: first data point within weeks after RTP~2, second a few months later (degradation associated with Si depletion of PtSi by \ce{SiO2} formation). Three data points on P04 (\SI{350}{\celsius}, \SI{10}{\second}) each during the first round of measurements, on samples taken at increasing distances from the dark spot. \B{(a)} Sheet resistance between \SI{10}{\kelvin} and \SI{120}{\kelvin}, in which all samples had relatively flat $R(T)$. \B{(b)} The residual resistance ratio. \B{(c)} The critical temperature, measured with a bias current on the order of \SI{1}{\micro\ampere}.}
		\end{figure}
		
		\begin{figure}
			\centering
			\includegraphics[width=0.8\textwidth]{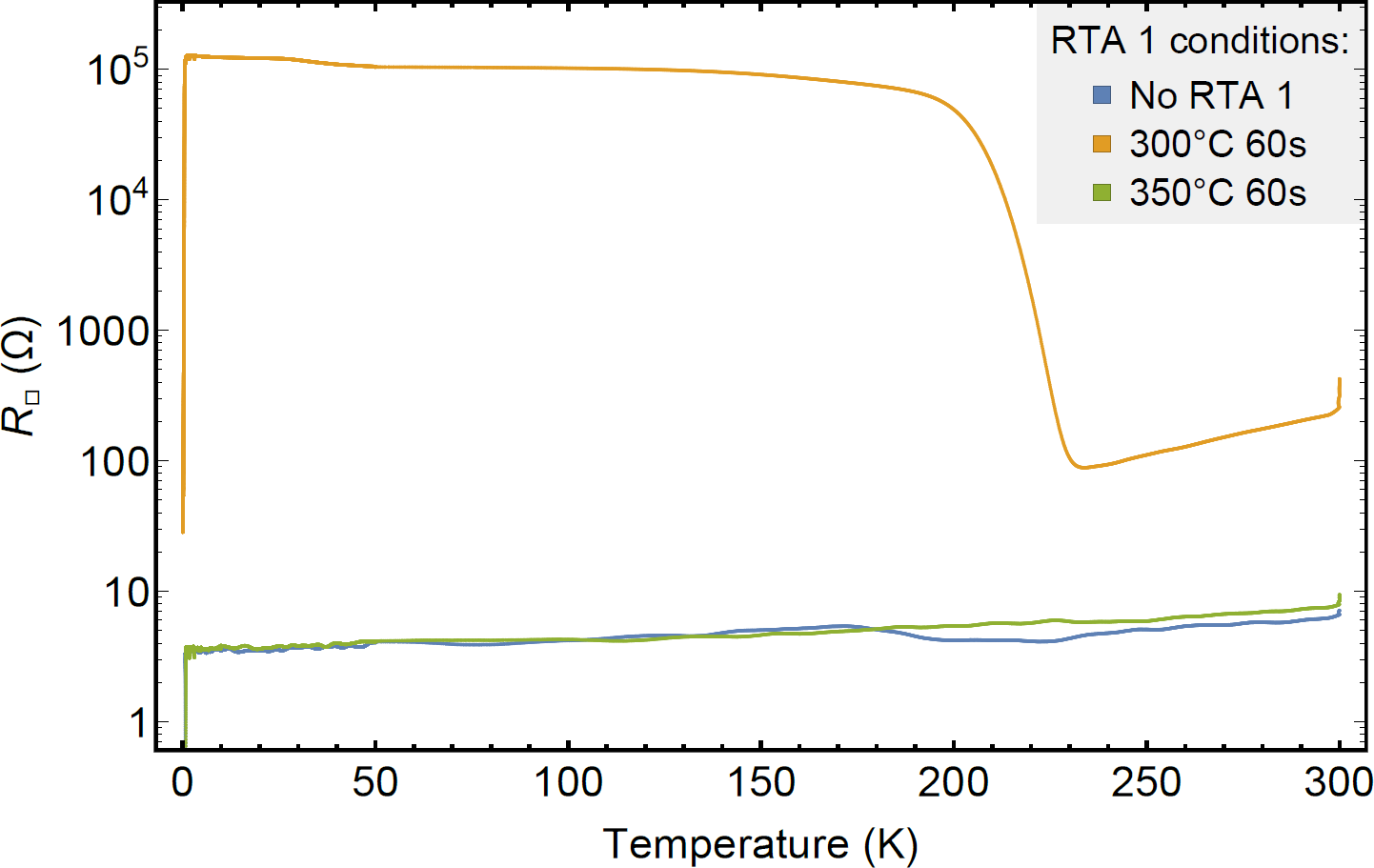}
			\caption{\label{fig:LogRTPlotP15P03P06}Logarithmic plot of the sheet resistance of three wafers, with either no low-temperature annealing step at all, RTP~1 at \SI{300}{\celsius} or RTP~1 at \SI{350}{\celsius}. The wafers that underwent RTP~1 subsequently had unreacted Pt and \ce{Pt2Si} removed by selective etch, after which all three underwent a final annealing step at \SI{500}{\celsius}. The film on the \textcolor{blue!70!gray}{\B{first}} of these, only annealed once at \SI{500}{\celsius}, has a metallic behavior: the resistance decreases as the wafer is cooled. This is consistent with the thick, \SI{50}{\nano\metre} \ce{PtSi} film that we expect. The \textcolor{orange!90!black}{\B{second}} wafer underwent RTP~1 at a temperature too low to transform the entire Pt layer to \ce{PtSi}, such that only a small amount of metal remained after selective etch. The ensuing high-temperature anneal then likely caused de-wetting of the film into disconnected islands, which explains the insulating behavior at temperatures below the freeze-out of intrinsic carriers in the silicon substrate. A slightly higher temperature during RTP~1, as experienced by the \textcolor{green!60!black}{\B{third}} sample, allows for full Pt consumption by PtSi formation during RTP~1, such that the resulting film remains stable during RTP~2, and metallic behavior is again recovered.}
		\end{figure}

		The results of low-temperature measurements on the six blanket wafers in lot~2 are shown in Fig.~\ref{fig:D16S0667BCombinedPlot}.
		Samples that we annealed only at \SI{300}{\celsius} during RTP~1 had high sheet resistances at room temperature, and show even higher sheet resistance below temperatures at which carriers introduced by dopants in Si freeze out ($\lesssim\SI{150}{\kelvin}$).
		These low-temperature resistance values are five orders of magnitude above that obtained on the samples annealed during RTP~1 at \SI{350}{\celsius}, a difference that is expressed also in the residual resistance ratio.
		This is a behavior that is not expected of continuous metallic films, and indicates that at least part of the current path passes through insulating regions, likely the substrate itself.
		
\begin{figure}
			\centering
			\begin{subfigure}[t]{0.55\textwidth}
				\centering
				\includegraphics[width=\textwidth]{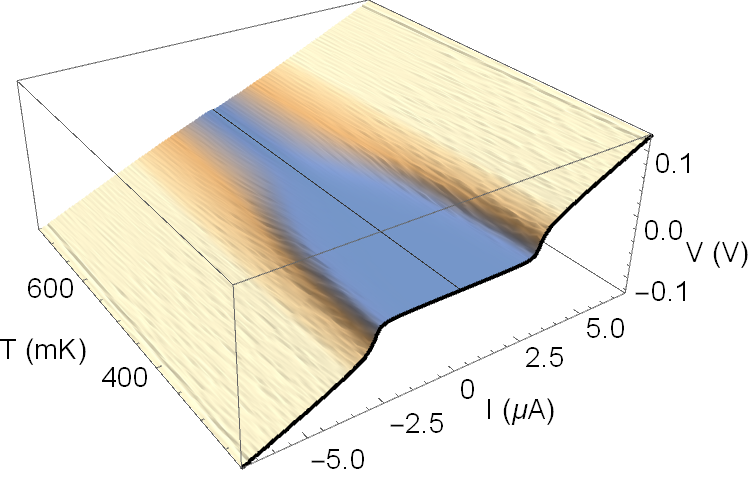}
			\end{subfigure}\hfill\begin{subfigure}[t]{0.4\textwidth}
				\centering
				\includegraphics[width=\textwidth]{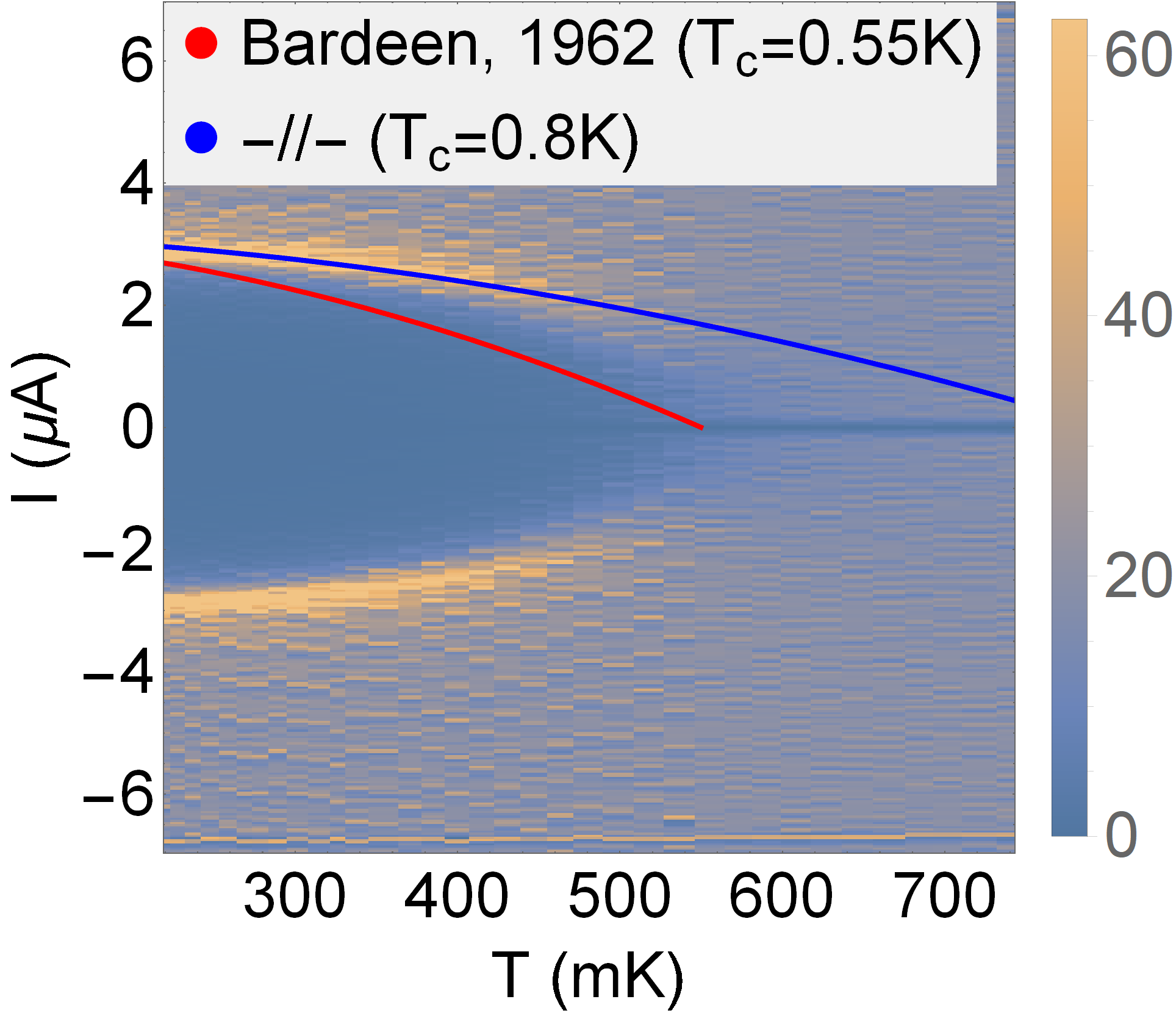}
			\end{subfigure}
			\caption{\label{fig:switchingcurrent}Critical (switching) current measured on lot~2 P01. \B{(Left)} Measured voltage as the bias current is swept from 0 to $\pm\SI{7}{\micro\ampere}$. The inflection points indicate the switching current. \B{(Right)} Differential resistance (\si{\kilo\ohm}) extracted from the plot on the left, with the predicted temperature dependence of the switching current for thin films superposed in red, using $T_\text{c}=\SI{800}{\milli\kelvin}$ and $I_\text{c}(T=0)=\SI{3.2}{\micro\ampere}$.}
		\end{figure}

		Further, Fig.~\ref{fig:LogRTPlotP15P03P06}c shows that the samples with only meanders/islands of PtSi (RTP~1 at \SI{300}{\celsius} for any duration, or \SI{350}{\celsius} for \SI{10}{\second}), have both lower $T_\text{c}$ and broader superconducting transitions.
		This can be explained by the reduced thickness of the films (see Fig.~\ref{fig:D16S0667A}), as well as their inhomogeneity.
		Measurements of the critical current between aluminum bonding wires with roughly $50\times\SI{50}{\micro\metre\squared}$ contacts to the PtSi film on P01 from lot~2 are shown in Fig.~\ref{fig:switchingcurrent}.
		The temperature dependence of the critical current does not follow the usual $(1-T/T_\text{c})^{2/3}$ behavior for thin films~\cite{romijn1982critical,bardeen1962critical,tinkham2004introduction}, and instead drops off more sharply close to the effective $T_\text{c,eff}\approx\SI{0.6}{\kelvin}$.
		It is likely that the thicker regions in these films have higher critical temperatures of up to perhaps \SI{0.8}{\kelvin}, while the effective $T_\text{c}$ is given by the weakest links in any current path.
		
	\FloatBarrier
	\subsection{Split 3: A look under the hood}
	
		\bgroup
		\setlength\tabcolsep{0.38em}
		\begin{table}%
			\centering
			\caption{\label{tab:lot3}The split for lot 3. Four annealing temperatures between 250 and \SI{350}{\celsius} are evaluated for the first RTP. For each temperature, three wafers are stopped at different stages.}
			\rowcolors{2}{white}{gray!15}
			\resizebox{\columnwidth}{!}{%
			\begin{tabular}{l c c c c c c c c c c c c c c c}
			\rowcolor{gray!30}\hline
			\B{Lot 3}		& \multicolumn{15}{c}{Wafer \#}\\
			\rowcolor{gray!30}
							& 01	& 02	& 03	& 04	& 05	& 06	& 07	& 08	& 09	& 10	& 11	& 12	& 13	& 14	& 15	\\\hline\hline
\makecell{Deposit\\\SI{10}{\nano\metre} Pt + \SI{10}{\nano\metre} TiN}	& \DOT	& \DOT	& \DOT	& \DOT	& \DOT	& \DOT	& \DOT	& \DOT	& \DOT	& \DOT	& \DOT	& \DOT	& \DOT	& \DOT	& \DOT	\\
Anneal \SI{250}{\celsius} 60s	& \oDOT	& \oDOT	& \oDOT	& \DOT	& \DOT	& \DOT	& \oDOT	& \oDOT	& \oDOT	& \oDOT	& \oDOT	& \oDOT	& \oDOT	& \oDOT	& \oDOT	\\
Anneal \SI{300}{\celsius} 60s	& \oDOT	& \oDOT	& \oDOT	& \oDOT	& \oDOT	& \oDOT	& \DOT	& \DOT	& \DOT	& \oDOT	& \oDOT	& \oDOT	& \oDOT	& \oDOT	& \oDOT	\\
Anneal \SI{325}{\celsius} 60s	& \oDOT	& \oDOT	& \oDOT	& \oDOT	& \oDOT	& \oDOT	& \oDOT	& \oDOT	& \oDOT	& \DOT	& \DOT	& \DOT	& \oDOT	& \oDOT	& \oDOT	\\
Anneal \SI{350}{\celsius} 60s	& \oDOT	& \oDOT	& \oDOT	& \oDOT	& \oDOT	& \oDOT	& \oDOT	& \oDOT	& \oDOT	& \oDOT	& \oDOT	& \oDOT	& \DOT	& \DOT	& \DOT	\\
Etch TiN						& \DOT	& \DOT	& \DOT	& \DOT	& \DOT	& \DOT	& \DOT	& \DOT	& \DOT	& \DOT	& \DOT	& \DOT	& \DOT	& \DOT	& \DOT	\\
Etch Pt (and \ce{Pt2Si})		& \oDOT	& \DOT	& \DOT	& \oDOT	& \DOT	& \DOT	& \oDOT	& \DOT	& \DOT	& \oDOT	& \DOT	& \DOT	& \oDOT	& \DOT	& \DOT	\\
Anneal \SI{500}{\celsius} 120s	& \oDOT	& \oDOT	& \DOT	& \oDOT	& \oDOT & \DOT	& \oDOT	& \oDOT	& \DOT	& \oDOT	& \oDOT	& \DOT	& \oDOT	& \oDOT	& \DOT	\\\hline	
			\end{tabular}}
		\end{table}
		\egroup
		
		A third split was prepared (see table~\ref{tab:lot3}) to better understand what happened during the first annealing in split~2 discussed in the previous section.
		In addition to the temperatures used for RTP~1 in this earlier split, two more were added: 250 and \SI{325}{\celsius}.
		To simplify the comparison, no further split in annealing times was used, and instead all wafers were either not annealed at all or for the same duration of \SI{60}{\second}.
		One of the questions raised during the previous campaign was whether the island formation had occurred during RTP~2.
		To answer this, three wafers were prepared for each RTP~1 condition: one where only the protective capping layer of TiN was removed by selective etch, another where also the unreacted Pt and \ce{Pt2Si} were removed, and a last wafer that further underwent the final annealing step at \SI{500}{\celsius} during \SI{2}{\minute}.
		Unfortunately, a mistake was made in the planning of the Pt deposition, and \SI{10}{\nano\metre} was deposited instead of 25, making direct comparison to lot~2 difficult.
		We will therefore mainly limit the discussion to comparisons between wafers of lot~3.
		
		\begin{figure}
			\centering
			\includegraphics[width=0.8\textwidth]{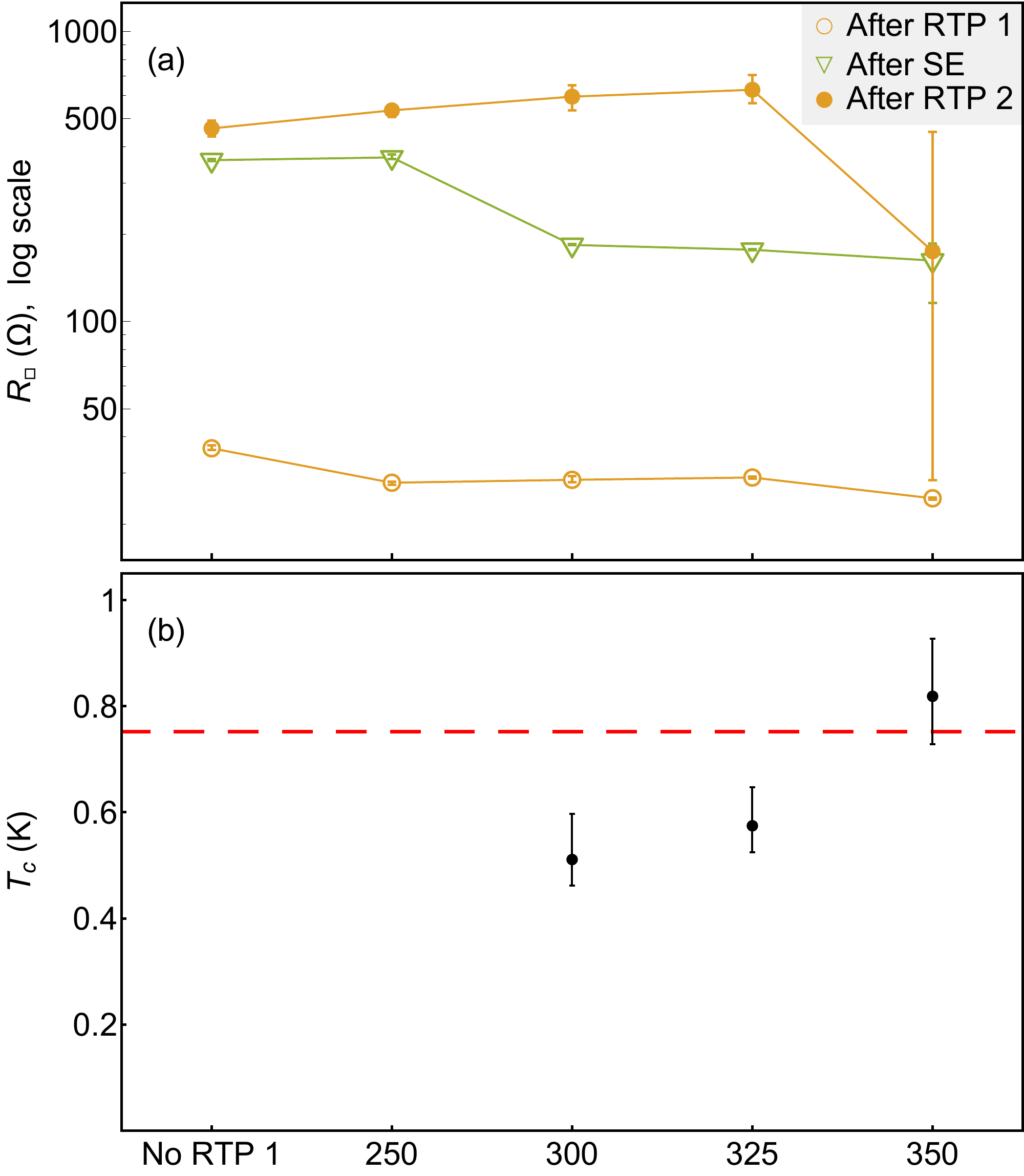}
			\caption{\label{fig:D19S0918RsPlot}\B{(a)} Logarithmic plot of the sheet resistances of each wafer in Table~\ref{tab:lot3}, grouped by temperature during RTP~1. A large increase in sheet resistance is observed on all wafers after SE with aqua regia, with a subsequent increase after RTP~2. \B{(b)} Critical temperature measured on wafers P09, P12 and P15, showing a positive dependence on the temperature during RTP~1. The dashed red line indicates the $T_\text{c}$ of reference sample P11 from lot~1, where \SI{25}{\nano\metre} of Pt was deposited, and which was annealed once at \SI{500}{\celsius}.}
		\end{figure}
		
		In Fig.~\ref{fig:D19S0918RsPlot}a are shown the sheet resistances for each of the wafers listed in Table~\ref{tab:lot3}.
		The wafers are grouped on the $x$-axis by the temperature used during RTP~1, such that an consistent rise in sheet resistance during the fabrication process becomes apparent.
		Most striking is that for all RTP~1 temperatures, a large increase in sheet resistance by roughly an order of magnitude can be observed immediately after the selective etch with aqua regia (green curve), indicating the removal of most of the material present after RTP~2.
		This is consistent with the hypothesis that aqua regia also etches \ce{Pt2Si}~\cite{jin1999microstructural}.
		While \ce{PtSi} was already found after annealing at \SI{350}{\celsius} for \SI{60}{\second} in lot~2, it is unlikely that the entire volume of Pt had been consumed to form this phase on the present lot.
		When \SI{25}{\nano\metre} of Pt was deposited on the wafers of lot~2, a large difference was seen between the sheet resistances of wafers annealed at \SI{300}{\celsius} and those heated to \SI{350}{\celsius} (see Fig.~\ref{fig:D16S0667RTPRsPlot}).
		Further, PtSi films of \SI{20}{\nano\metre} should be stable up to \SI{600}{\celsius}, and not form any islands until \SI{700}{\celsius} (see Fig.~\ref{fig:island_formation_das1994thickness})~\cite{das1994thickness}.
		However, when the wafer heated to \SI{350}{\celsius} during RTP~1 was subsequently annealed at \SI{500}{\celsius}, a large variation in sheet resistance across the wafer was observed (solid orange curve in Fig.~\ref{fig:D19S0918RsPlot}), suggesting the onset of the film breakdown.
		All other wafers, annealed at \SI{325}{\celsius} or less during RTP~1, saw a further large increase in sheet resistance after RTP~2, which is expected if de-wetting has occurred.
		
\begin{figure}
			\centering
			\begin{subfigure}[b]{0.33\textwidth}
				\centering
				\includegraphics[width=\textwidth]{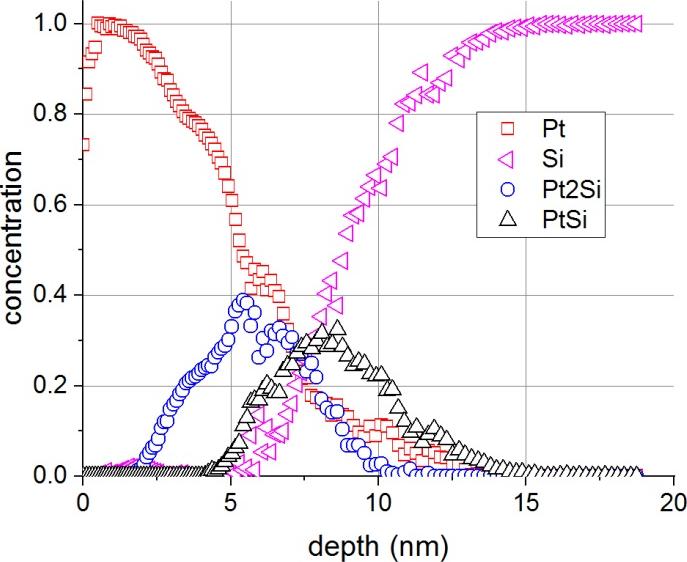}
				\caption{\label{fig:plasmon_depth_p01}P01: No RTP~1.}
			\end{subfigure}\hfill\begin{subfigure}[b]{0.33\textwidth}
				\centering
				\includegraphics[width=\textwidth]{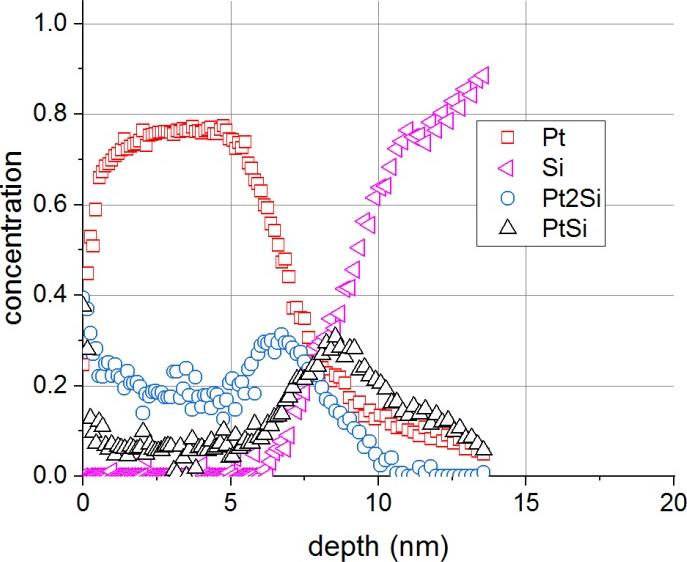}
				\caption{P04: RTP~1 at \SI{250}{\celsius}.}
			\end{subfigure}\begin{subfigure}[b]{0.33\textwidth}
				\centering
				\includegraphics[width=\textwidth]{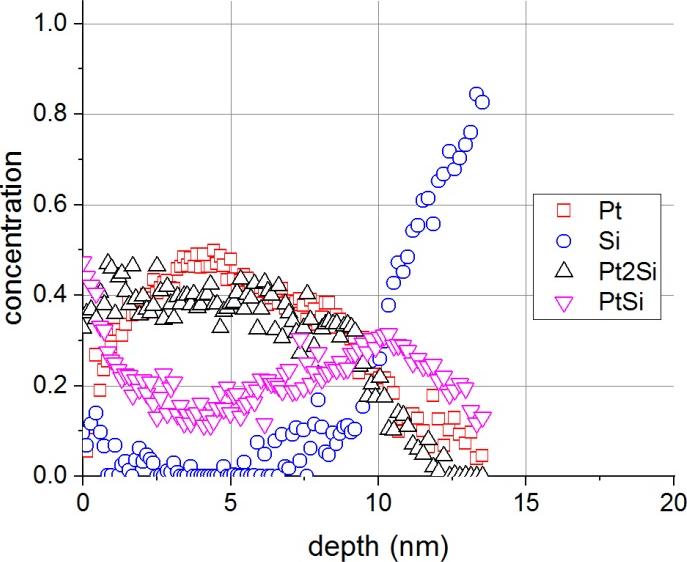}
				\caption{P13: RTP~1 at \SI{350}{\celsius}.}
			\end{subfigure}
			\caption{\label{fig:plasmon_depth}Combining plasmon spectroscopy and XPS measurements, it was possible to estimate the variation with depth in the relative densities of chemical bonds associated with Pt, \ce{Pt2Si}, PtSi and Si. It is clear that PtSi and \ce{Pt2Si} co-exist after RTP~1 up to \SI{350}{\celsius}.}
		\end{figure}

\begin{figure}
			\centering
			\includegraphics[width=0.8\textwidth]{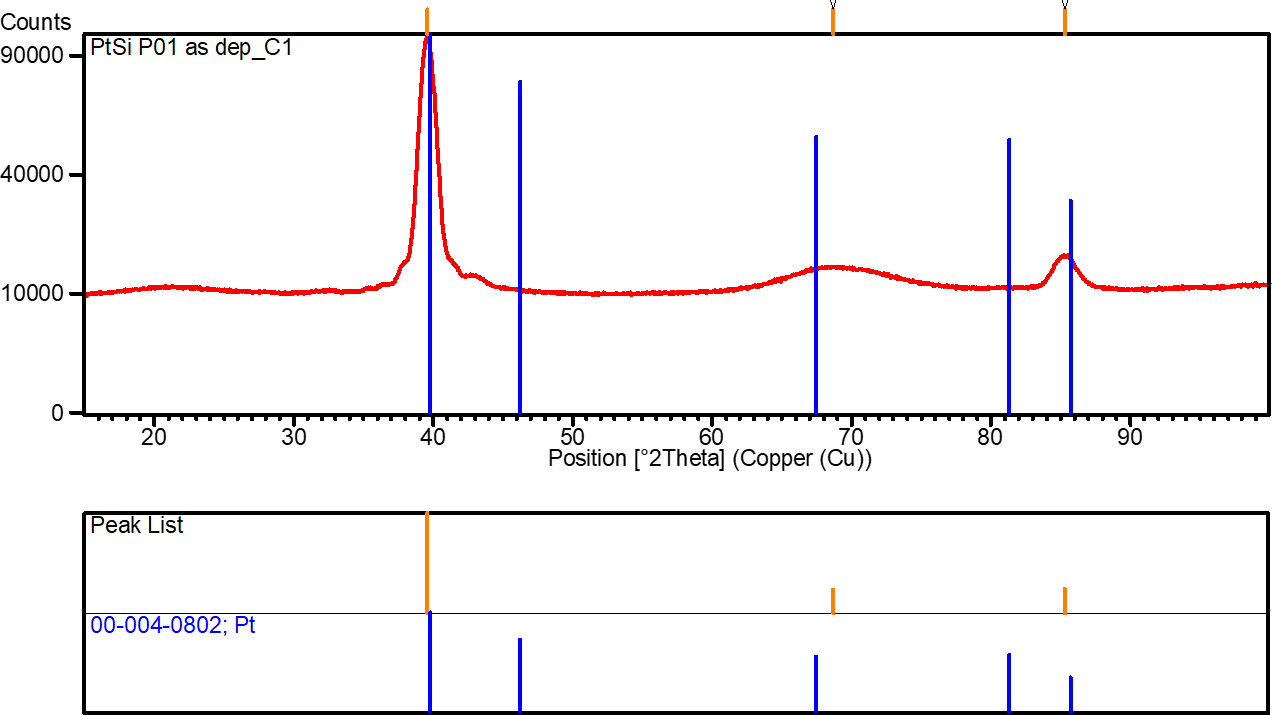}
			\vspace*{0.1\baselineskip}
			
			\includegraphics[width=0.8\textwidth]{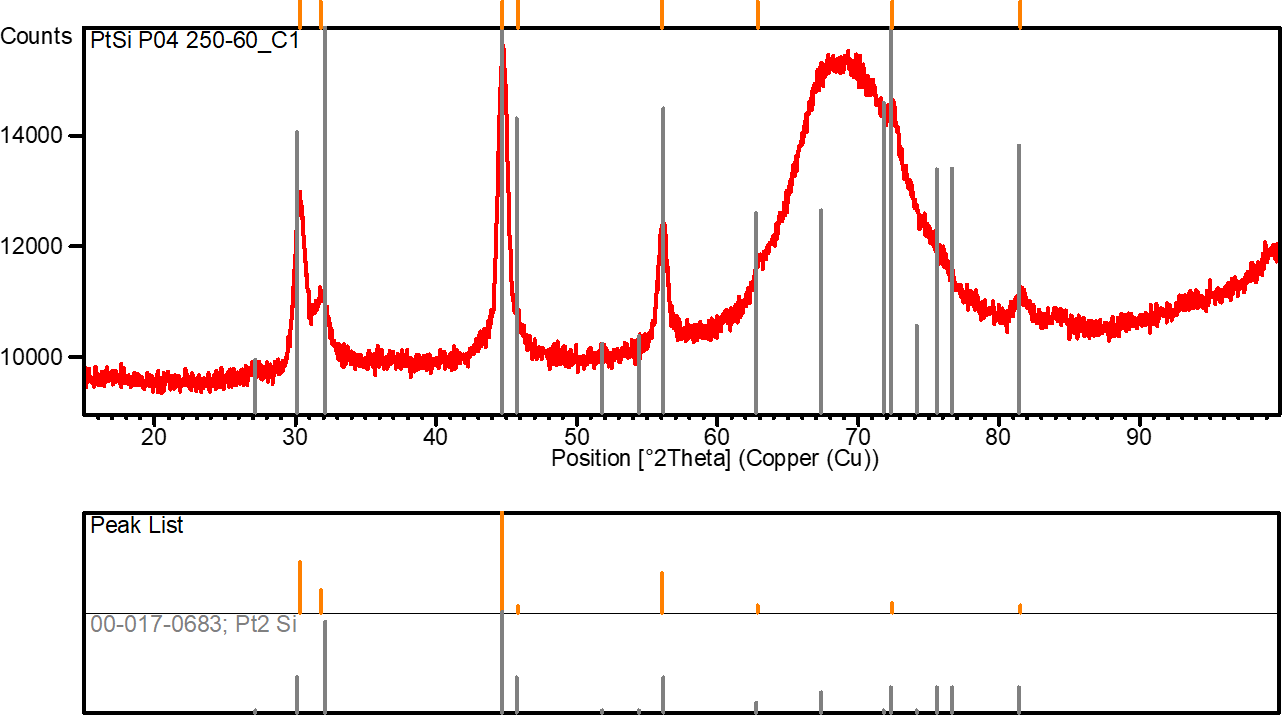}
			\vspace*{0.1\baselineskip}
			
			\includegraphics[width=0.8\textwidth]{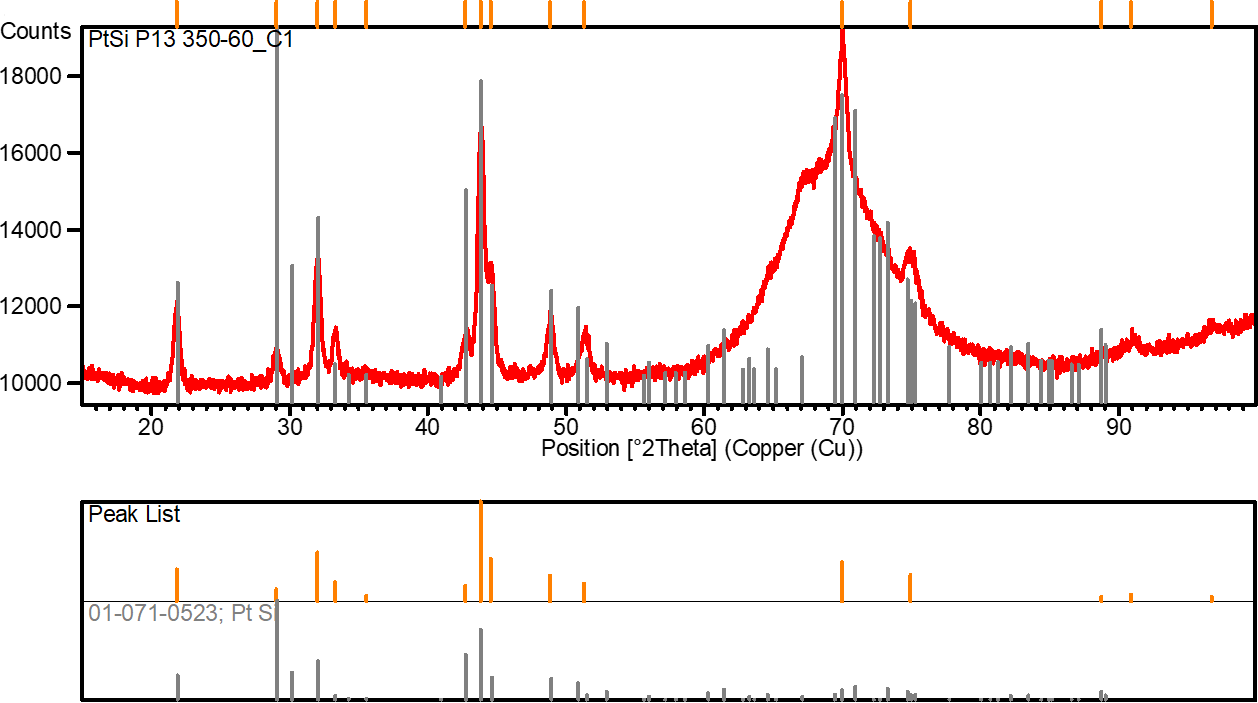}
			\caption{\label{fig:xrd_p01_p04_p13}Out-of-plane $\theta/2\theta$ XRD analyses performed on wafers P01, P04 and P13 of lot~3 (see Table~\ref{tab:lot3}). \B{(Top)} The as-deposited wafer shows strongly textured (111)-oriented Pt. Additional reciprocal space mapping confirmed the presence of (220)-oriented grains in-plane. \B{(Middle)} Annealing at \SI{250}{\celsius} leads to the formation of almost entirely untextured \ce{Pt2Si}. \B{(Bottom)} Annealing at \SI{350}{\celsius} results in untextured \ce{PtSi}.}
		\end{figure}

\begin{figure}
			\centering
			\begin{tikzpicture}[x=1cm,y=1cm]
				\node[anchor=center,inner sep=0] at (0,0) {
					\includegraphics[width=\textwidth]{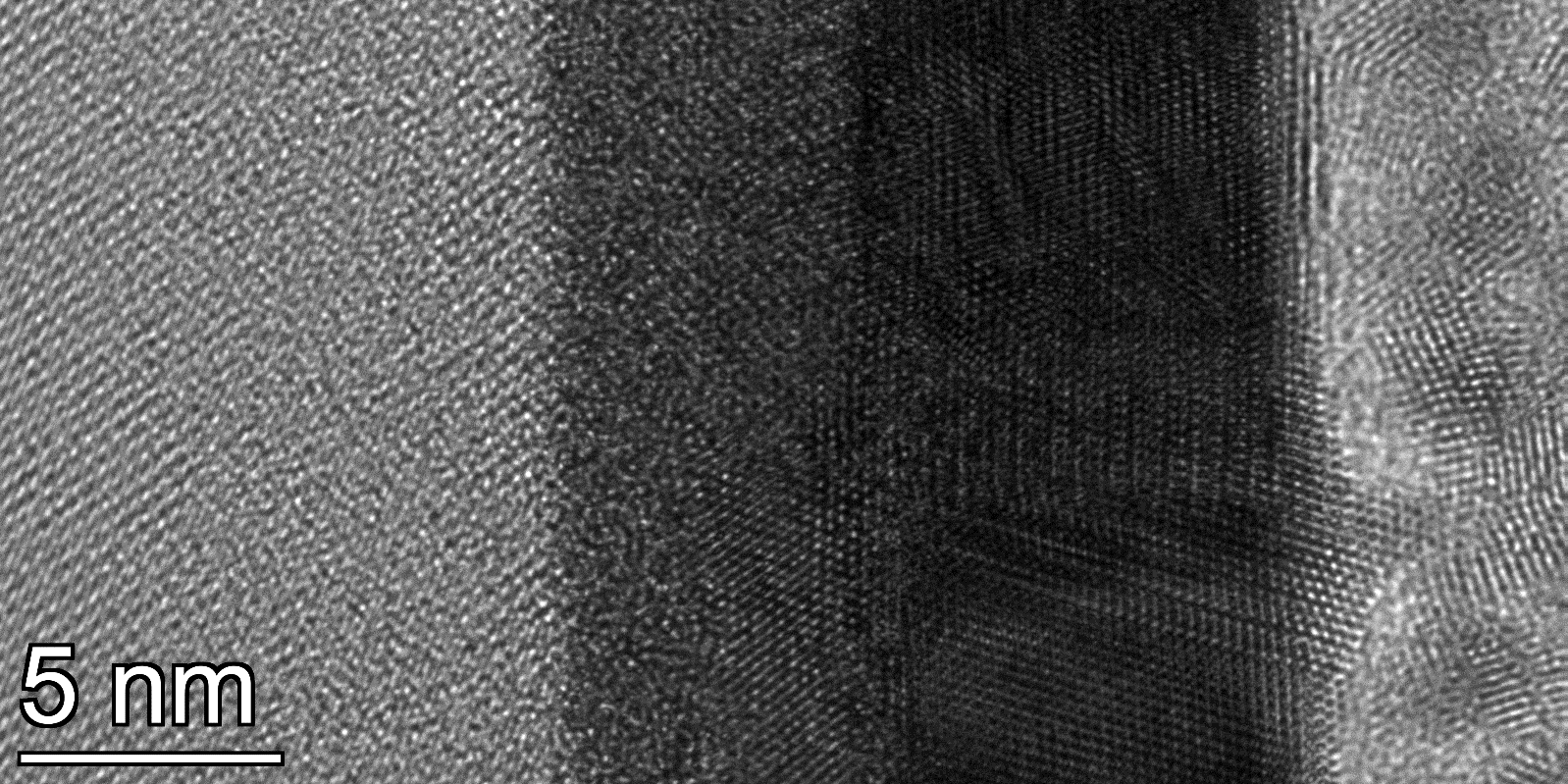}
					};
				
				\draw[red,ultra thick] (-2.25,3) -- (-2.20,1);
				\draw[red,ultra thick] (-0.75,3) -- (-0.70,1);
				\draw[red,ultra thick] (0.65,3) -- (0.70,1);
				\draw[red,ultra thick] (4.60,3) -- (4.65,1);
				
				\node at (-4.98,-0.02) {\Large \B{A}};
				\node[white] at (-5,0) {\Large \B{A}};
				\node at (-1.43,-0.02) {\Large \B{B}};
				\node[white] at (-1.45,0) {\Large \B{B}};
				\node at (-0.03,-0.02) {\Large \B{C}};
				\node[white] at (-0.05,0) {\Large \B{C}};
				\node at (2.62,-0.02) {\Large \B{D}};
				\node[white] at (2.60,0) {\Large \B{D}};
				\node at (6.22,-0.02) {\Large \B{E}};
				\node[white] at (6.20,0) {\Large \B{E}};
			\end{tikzpicture}
			\caption{\label{fig:p01_tem}Transmission electron microscope (TEM) image of a FIB-milled lamella of P01, with five zones indicated. EDS analysis indicated that zone \B{A} is silicon with progressive concentrations of up to 10\% of Pt dissolved close to the interface with zone \B{B}, which has a rather homogeneous concentration of around 50\% Pt. Zones \B{C} and \B{D} contain 60--70 and nearly 100\% of Pt, respectively. These results are consistent with the plasmon depth profile shown in Fig.~\ref{fig:plasmon_depth_p01}, which is likely broadened due to finite depth resolution.}
		\end{figure}

		A combination of XPS (X-ray Photo-electron Spectroscopy), AES (Auger Electron Spectroscopy) and REELS (Reflected Electron Energy Loss Spectroscopy) analyses performed in Budapest by the team of Labar Janos L\'aszl\'o and Miklos Menyhard provided further insight into the reactions that occurred in these films.
		Shown in Fig.~\ref{fig:plasmon_depth} are simulated depth profiles of the different phases of the Pt-Si system, which match the measured spectra of these measurements.
		Most importantly, we see that both silicide phases, \ce{Pt2Si} and \ce{PtSi}, are already present after deposition (P01, Fig.~\ref{fig:plasmon_depth}a), which was later confirmed by TEM analysis (see Fig.~\ref{fig:p01_tem}).
		Not all of these phases are simultaneously crystalline, however.
		XRD analysis (shown in Fig.\ref{fig:xrd_p01_p04_p13}) indicates that only a single phase appears in crystalline form at the same time: (111)-oriented Pt after deposition (P01), then \ce{Pt2Si} after RTP~1 at \SI{250}{\celsius} (P04), and untextured \ce{PtSi} after annealing at \SI{350}{\celsius}.
		Nor does the disappearance of a phase's crystalline form necessarily imply that it has disappeared entirely.
		This can be seen in Fig.~\ref{fig:plasmon_depth}a, where \ce{Pt2Si} was in fact still present in samples annealed at temperatures as high as \SI{350}{\celsius}.
		
		It is also notable that the PtSi that formed in P13, after annealing at \SI{350}{\celsius}, shows no signs of any texture; all peaks from different (hkl) planes are present with roughly the same intensities as they are in reference powder samples (shown as gray vertical lines in the bottom bar).
		The fact that the PtSi does not have any texture in this last sample is an interesting result in itself, since ``columnar'' growth has been reported for all film thicknesses when PtSi forms immediately~\cite{das1994thickness}, as confirmed by XRD analysis performed on samples annealed only once at \SI{500}{\celsius} (not shown).
		It is known that the preferred orientation depends on thickness and thermal history~\cite{sinha1972thermal}, and this disappearance of texture may be related to the increase in $T_\text{c}$ relative to the single-step annealing process as observed in Fig.~\ref{fig:D16S0667BCombinedPlot}, since axiotaxic PtSi with thin vertical columns would likely have a very small lateral grain size.
		One possible interpretation is that the intermediate appearance of untextured \ce{Pt2Si} prevents the inheritance of texture from the deposited Pt to the final PtSi.
		Since the interface between Si and epitaxial PtSi is often roughened by undulations and atomic steps due to the large lattice mismatch~\cite{kawarada1984structural}, removing texture could have the added benefit of smoothing the S/Sm interface to the channel.
		
		Perhaps the most surprising result is that all phases already co-exist after deposition, as can be seen in the TEM image in Fig.~\ref{fig:p01_tem}.
		Intermixing during room-T deposition has been shown before~\cite{das1994thickness,jin1999microstructural}, as has an initial reaction between Pt and Si at room temperature leading to amorphous products~\cite{ley1995initial,sinclair1994reactions}.
		Previous reports~\cite{donaton1997formation} have also observed the same full stack of Pt, \ce{Pt2Si}, PtSi and Si after annealing at low temperatures.
		It remains unexplained however, how the crystallinity of one phase (such as \ce{Pt2Si}) can disappear during the formation of the next (e.g. PtSi), while its chemical bonds remain.
	
	\FloatBarrier
	\subsection{Conclusion}
		
		PtSi has several advantages over \ce{V3Si};
		it is the thermodynamically favored phase within its binary system (which ensures its stability at higher annealing temperatures and allows for longer annealing times), has established process flows for its integration in MOSFET devices, and is compatible with the SALICIDE process.
		Furthermore, since it is formed after pure metal deposition, and thus consumes Si during its formation, its interface to the silicon channel can usefully be moved underneath the gate electrode.
		By introducing an additional annealing step at lower temperatures, around or above \SI{350}{\celsius}, the quality of the PtSi can be improved, while also controlling the lateral encroachment into the channel.
		As we will see in the next chapter, such a reduction of the channel length, paired with the placement of the Schottky barrier within the gate electrostatic field, gives rise to a favorable Schottky-barrier MOSFET (SBMOSFET) configuration.
		In such a device it is possible, even at cryogenic temperatures after dopants have frozen out, to tune the effective width of the Schottky barrier, and thus modulate the proximity effect induced from the PtSi contacts.
		
\printbibliography

\end{refsection}

\begin{refsection}
	\graphicspath{{img/ch4/}}
	\setcounter{chapter}{3}
\chapter{\label{sec:jofets}Josephson Field Effect Transistors}
	
	\section{Introduction}
	
		\Quote{``If you wish to make an apple pie from scratch, you must first invent the universe.''}{Carl Sagan}
		
		We are aiming to fabricate a transmon using only CMOS-compatible materials and processes.
		Before we can begin designing the actual LC circuit itself, it is necessary to thoroughly understand each of its components, not least the Josephson Field Effect Transistor (JoFET) that makes up the nonlinear inductance.
		In turn, we need to have some control over such properties as the superconducting critical temperature and coherence length in the source and drain, the transparency and contact resistance of the interface with the semiconductor, and the carrier mobility and scattering rates inside the channel, before it makes sense to design this JoFET.
		It is most efficient to extract each of these from simplified test structures, or even just blanket thin films, rather than spending time fabricating transistors immediately.
		To this end, both an industrial mask set, TASP (\underline{T}LM for \underline{As} and \underline{P}-based wafers\footnote{Any similarity to actual French words, \emph{verlan} or otherwise, is purely coincidental.}) and an academic mask set for test structures were used.
	
	\FloatBarrier
	\section{Test structures}
		
		\subsection{Academic mask set: Python code for adjustable patterns}
			
			A mask set was designed with Hall bars, Greek crosses, meanders and Josephson junctions~\cite{vethaak2021superge}, for wafers where a superconducting layer (typically Al, Nb, TiN, NbTiN, ...) is deposited on top of a semiconductor (Si, SiGe, Ge, ...).
			This set can be used both on wafers where the superconductor has already been deposited, and on wafers where this layer will be defined by lift-off instead.
			The structures are fabricated with three lithography steps:
			\begin{enumerate}
				\item Contact pads with extensions towards the center of the cell are defined with optical wavelength laser lithography.
				\item A mesa is drawn, removing both the superconductor and semiconductor layers everywhere else.
				\item Only the superconductor is selectively etched on part of the mesa, leaving the semiconductor exposed.
			\end{enumerate}
			Although we did not have the opportunity to apply this process to silicon on insulator (SOI) wafers with \ce{V3Si} or PtSi films, the publicly available Python code may be useful to future studies.
			The main advantage of such ``academic'' fabrication processes are the speed with which samples can be produced, and the flexibility in adjusting the mask sets.
			One of these types of structure, for transmission line measurements (TLM), was studied also with the industrial mask set, as we will see in the following section.
			
			\begin{figure}
				\centering
				\includegraphics[width=\textwidth]{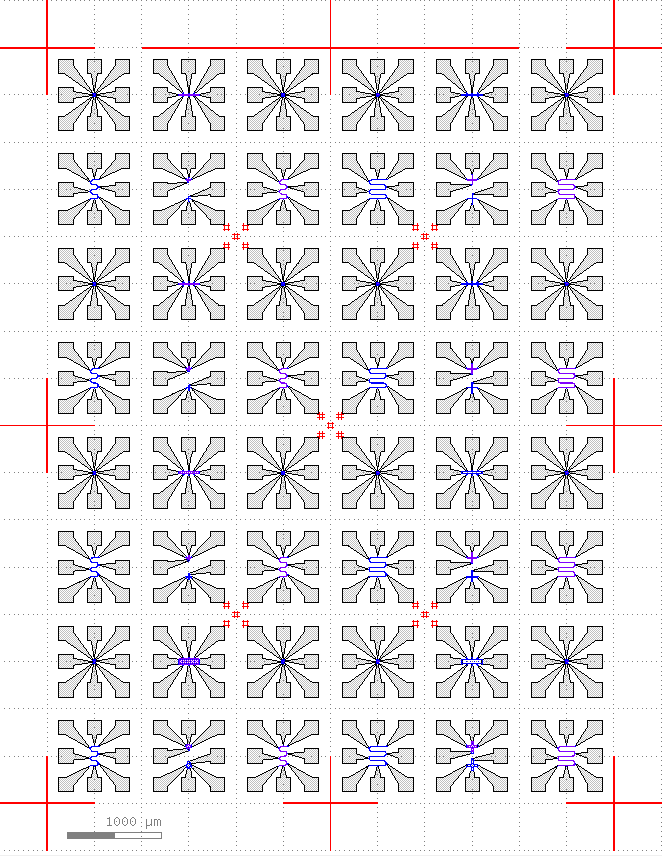}
				\caption{The mask set designed for a combination of laser and e-beam lithography. In this example, $6\times8$ cells are prepared, with 16 different TLM structures, 16 meanders (8 of Al/Ge, 8 Ge only), 8 Hall bars (4 of Al/Ge, 4 Ge only) and 16 Greek crosses (8 of Al/Ge, 8 Ge only). Each cell has different device parameters.}
			\end{figure}
			
			\begin{figure}
				\centering
				\begin{subfigure}[b]{0.48\textwidth}
					\centering
					\includegraphics[width=\textwidth]{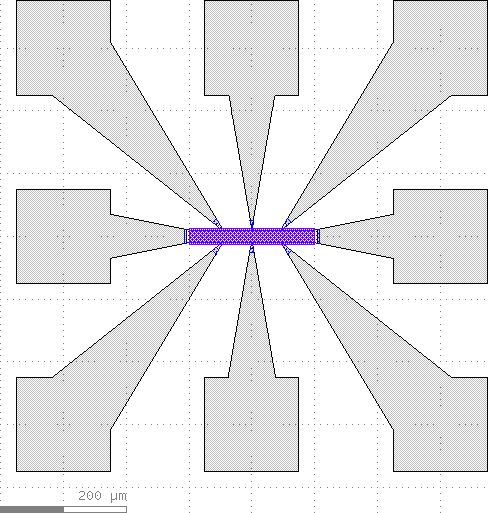}
					\caption{Hall bar.}
				\end{subfigure}\hfill\begin{subfigure}[b]{0.48\textwidth}
					\centering
					\includegraphics[width=\textwidth]{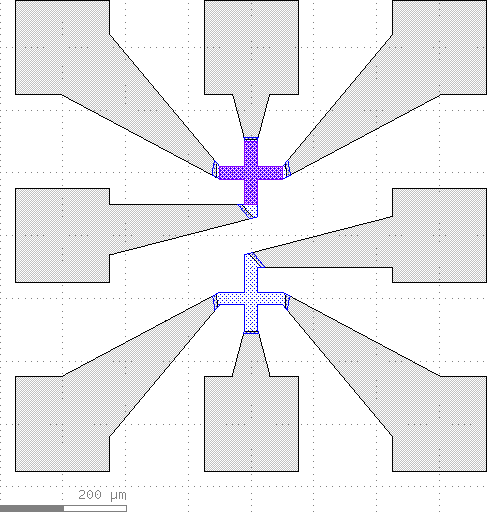}
					\caption{Greek crosses.}
				\end{subfigure}
				\vspace*{\baselineskip}
				
				\begin{subfigure}[b]{0.48\textwidth}
					\centering
					\includegraphics[width=\textwidth]{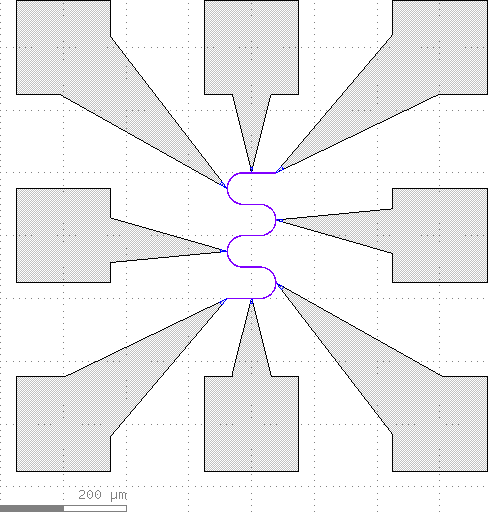}
					\caption{Meander.}
				\end{subfigure}\hfill\begin{subfigure}[b]{0.48\textwidth}
					\centering
					\includegraphics[width=\textwidth]{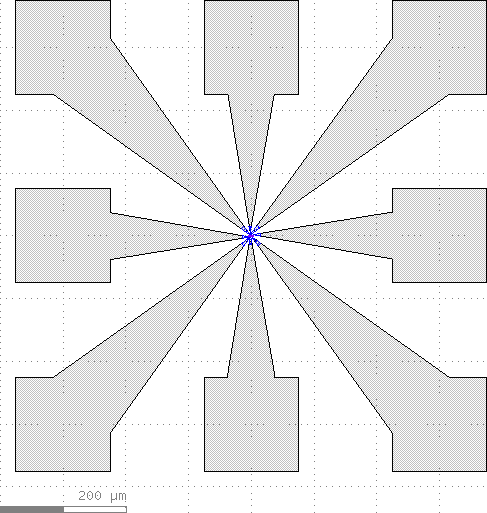}
					\caption{TLM/Josephson junctions.}
				\end{subfigure}
				\caption{The four types of structure included in the mask set. \B{(a)} Hall bars can be used to determine both the sign and the density of carriers. \B{(b)} Greek crosses are convenient structures to accurately determine the sheet resistance. \B{(c)} Long and thin meanders to check homogeneity. \B{(d)} TLM/Josephson structures to characterize the superconductor/semiconductor interface and to study the Josephson effect in junctions of different lengths.}
			\end{figure}

			\begin{figure}
				\centering
				\begin{subfigure}[b]{0.48\textwidth}
					\centering
					\includegraphics[width=\textwidth]{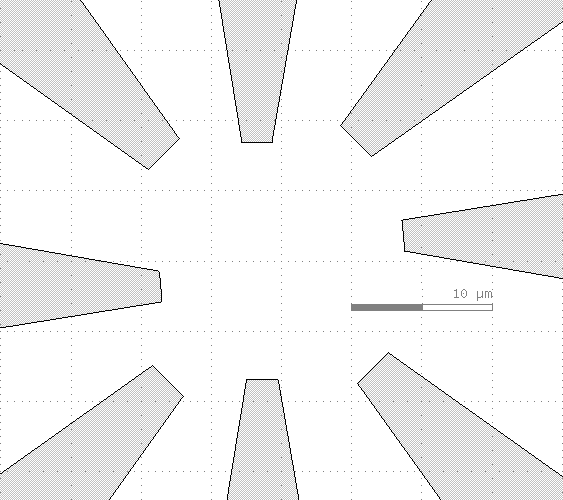}
					\caption{Step 1: Contacts (laser).}
				\end{subfigure}\hfill\begin{subfigure}[b]{0.48\textwidth}
					\centering
					\includegraphics[width=\textwidth]{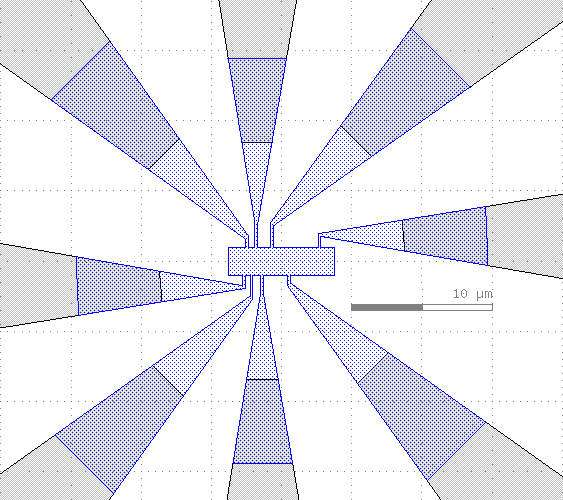}
					\caption{Step 2: Al/Ge mesa (e-beam).}
				\end{subfigure}
				\vspace*{\baselineskip}
				
				\begin{subfigure}[b]{0.48\textwidth}
					\centering
					\includegraphics[width=\textwidth]{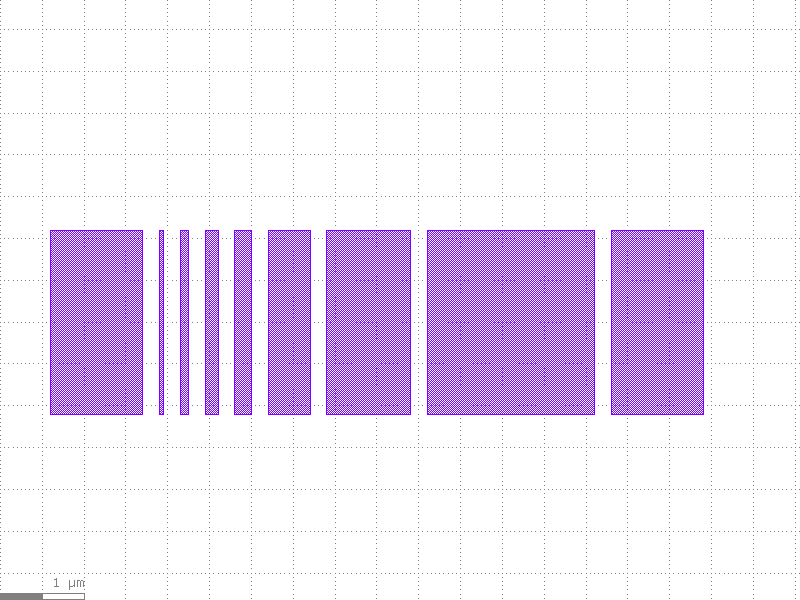}
					\caption{Step 3: Remove Al (e-beam).}
				\end{subfigure}\hfill\begin{subfigure}[b]{0.48\textwidth}
					\centering
					\includegraphics[width=\textwidth]{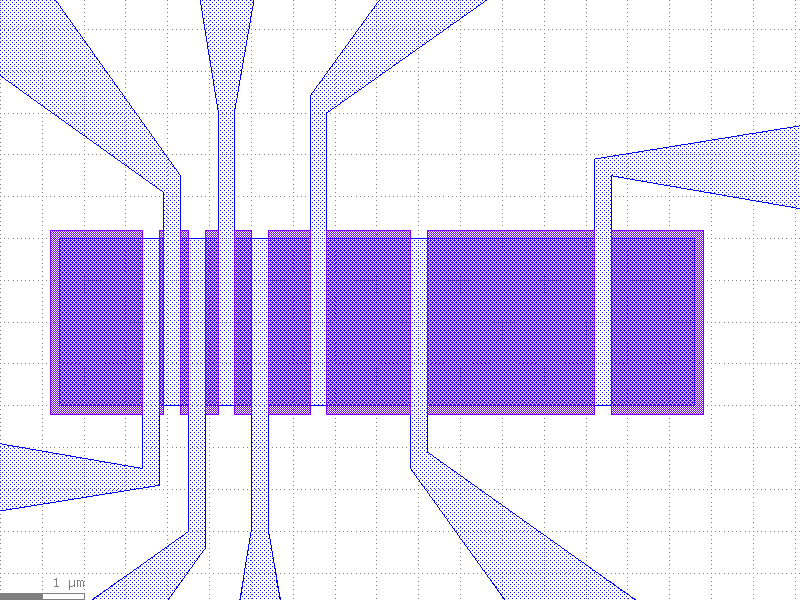}
					\caption{Final structure.}
				\end{subfigure}
				\caption{The TLM structure can be made with three lithography steps. \B{(a)} Contact pads with extensions towards the center of the cell are defined, either in a separate step by lifting off metal, or by choosing a resist where this laser lithography can be combined with the e-beam in the next step. \B{(b)} With some (adjustable) overlap, an Al/Ge mesa is drawn, removing both Al and Ge everywhere else. \B{(c)} Only the aluminum is selectively etched on part of the mesa, leaving Ge. \B{(d)} With 8 contacts, 7 different junction lengths can be measured on a single device.}
			\end{figure}
		
		\FloatBarrier
		\subsection{\label{sec:tasp}Industrial mask set: TASP}

			\begin{table}
				\centering
				\caption{\label{tab:lot4}Lot 4 (D19S1723): A set of patterned wafers are prepared, with deposited Pt thicknesses varying from 5 to \SI{25}{\nano\metre}. Same as the first half of Table~\ref{tab:lot1} (without any doubles).}
				\rowcolors{2}{white}{gray!15}
				\begin{tabular}{l c c c c c}
					\rowcolor{gray!30}\hline
					\B{Lot 4}				& \multicolumn{5}{c}{Wafer \#}\\
					\rowcolor{gray!30}
															& 7		& 9		& 10	& 11	& 12	 \\\hline\hline
					Patterned SOI wafer						& \DOT	& \DOT	& \DOT	& \DOT	& \DOT	\\
					Deposit $<>$ \si{\nano\metre} Pt + TiN	& 5		& 10	& 15	& 20	& 25	\\
					RTA	\SI{500}{\celsius} 120s				& \DOT	& \DOT	& \DOT	& \DOT	& \DOT	\\
					Etch TiN								& \DOT	& \DOT	& \DOT	& \DOT	& \DOT	\\
					Etch Pt									& \DOT	& \DOT	& \DOT	& \DOT	& \DOT	\\
					XRD										& \DOT	& \DOT	& \DOT	& \DOT	& \DOT	\\
					Complete back-end						& \DOT	& \DOT	& \DOT	& \DOT	& \DOT	\\
				\end{tabular}
			\end{table}
			
\begin{figure}
				\centering
				\begin{tikzpicture}
				\node[anchor=center,inner sep=0] at (0,0) {%
					\includegraphics[width=0.8\textwidth]{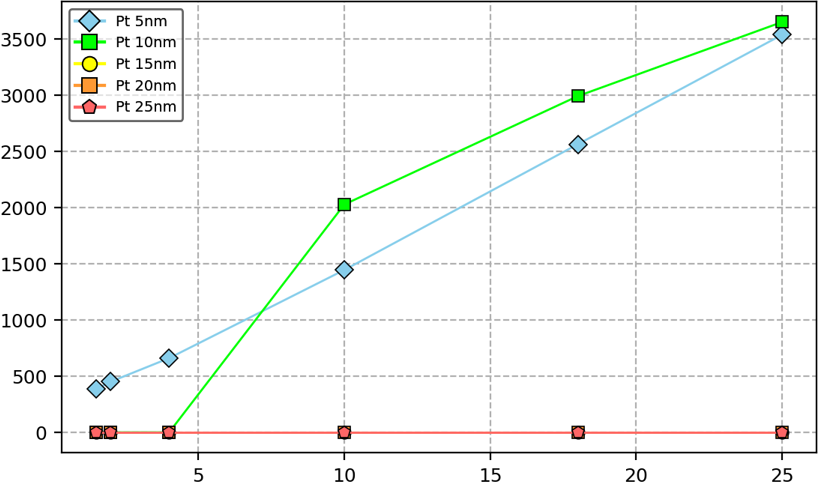}%
					};
				\node at (0.5,-3.9) {Contact spacing (\si{\micro\metre})};
				\node[rotate=90] at (-6.3,0) {$2R_\text{total}$ (\si{\ohm})};
				\end{tikzpicture}
				\caption{\label{fig:rac01_graph}The resistances of the six TLM structures (see Fig.~\ref{fig:DS16S0667AP20FIBSEM}) on scribe RA0C1 were measured, which each have a contact width of $W=\SI{1}{\micro\metre}$ and a contact length of $L=\SI{5}{\micro\metre}$, but have varying channel lengths between these two contacts from $L_\text{channel}=\SI{1.5}{\micro\metre}$ to \SI{25}{\micro\metre}.
				This scribe was tested on seven dice of each wafer, each square represents the average value obtained for a single channel length on a given wafer (\SI{5}{\nano\metre} = P07, \SI{10}{\nano\metre} = P09, etc).
				The $y$-axis intersect for P07 corresponds to $2R_\text{contact}=\SI{145\pm5}{\ohm}$, and an extrapolation of the linear regression gives a transfer length of \SI{0.8\pm0.05}{\micro\metre}.}
			\end{figure}

			In section~\ref{sec:split1} we discussed PtSi lot~1, a split in the thickness of deposited Pt.
			Unfortunately, we did not obtain any working devices on the patterned wafers in the first iteration of this lot, and a second lot (D19S1723) was launched that included only the patterned wafers, shown in Table~\ref{tab:lot4}.
			In Fig.~\ref{fig:rac01_graph} are shown the total resistances of the TLM structures in scribe RA0C1,
			\begin{equation}R_\text{total}=2R_\text{contact}+\dfrac{L_\text{channel}}{W_\text{channel}}\,R_{\square,\text{Si}},\end{equation}
			for each wafer in this lot, which shows a linear dependence on the channel length for only a single wafer (P07, \SI{5}{\nano\metre} of Pt).
			For wafers P10--P12, with deposited Pt thicknesses of 15--\SI{25}{\nano\metre}, the automatic probe station returned zero for each TLM structure on nearly every die, indicating either a short or no contact at all.
			We can understand what happened by looking instead at wafer P09 (\textcolor{green!60!black}{green curve}), where a reasonable resistance is obtained only for the longer channels: it is likely that the structures with $L_\text{channel}\leq\SI{4}{\micro\metre}$ were shorted by lateral encroachment of PtSi through the entire channel.
			As discussed in section~\ref{sec:ptsi}, this behavior is to be expected after all the silicon in the contacts has been consumed.
			Alternatively, the absence of any signal on wafers where thicker Pt layers were deposited could be attributed to the lifting of PtSi due to poor adhesion to \ce{SiO2}~\cite{wang1998sub}.

			A few scribes from wafer P07 were then cooled down by master student Axel Leblanc.
			Shown in Fig.~\ref{fig:rac01_graph} (left) is a similar plot, this time repeated at both room temperature and \SI{4}{\kelvin} for structures with slightly wider, but shorter contacts: RB0C1, RB0C2, RB0C1RE and RB0C2RE, with $W_\text{contact}=\SI{1.25}{\micro\metre}$ and $L_\text{contact}=\SI{0.35}{\micro\metre}$.
			Although the silicon channel has been doped with boron only up to $10^{15}\si{\per\centi\metre\cubed}$, we see that the channel sheet resistance (the slope in this graph) in fact drops from \SI{870}{\ohm} to \SI{635}{\ohm} after cooling to \SI{4}{\kelvin}.
			Such a drop suggests degenerate doping ($\gtrapprox10^{18}\si{\per\centi\metre\cubed}$); the resistance values are still at least an order of magnitude above that expected for real metals, so we can be confident that we are not just measuring a bar of PtSi.
			More consistent with expectation is that the contact resistance increases from $2R_\text{c}=\SI{300}{\ohm}$ at \SI{300}{\kelvin} to $2R_\text{c}=\SI{470}{\ohm}$ after cooling, as thermionic emission over the Schottky barriers at the PtSi/Si interfaces becomes suppressed.
			Note that this contact resistance is not simply proportional to the surface area of the contact: since the transfer length is less than one micron, only the edge of the larger contacts measured for Fig.~\ref{fig:rac01_graph} actually contribute to the transmission.
			
			\begin{figure}
				\centering
				\begin{subfigure}[b]{0.49\textwidth}
					\centering
					\includegraphics[width=\textwidth]{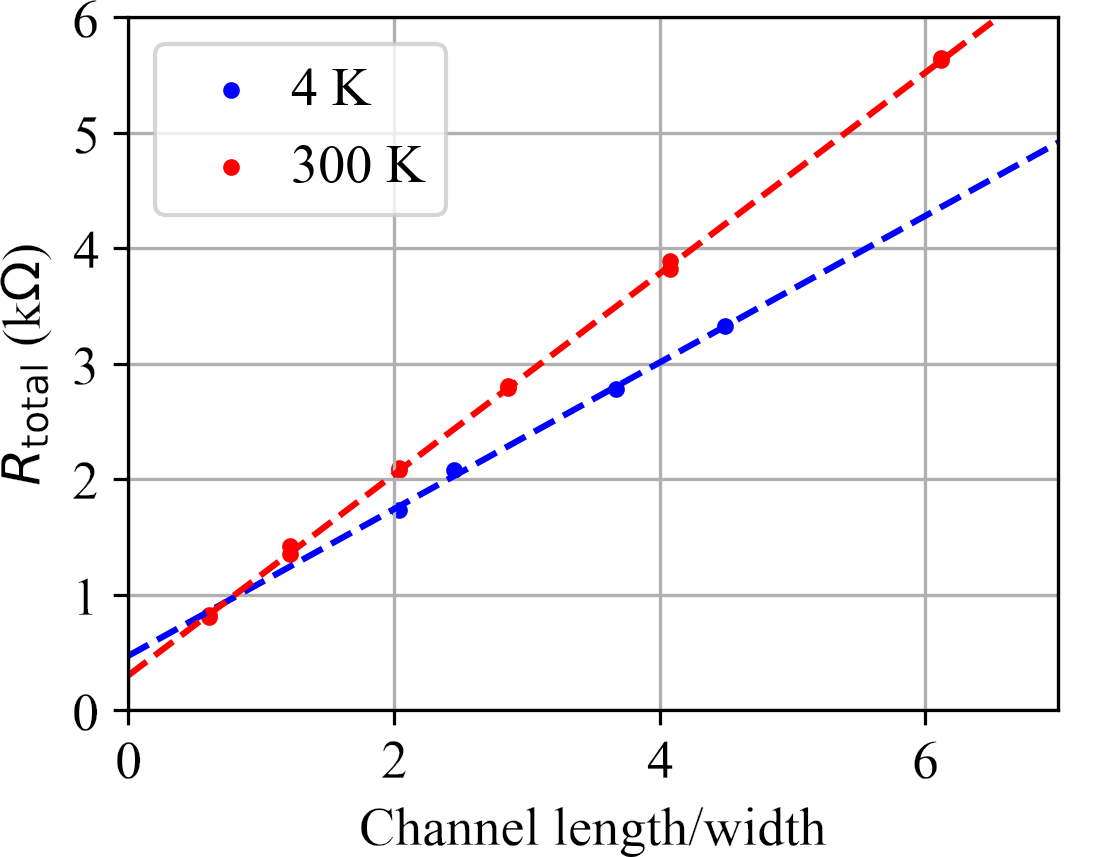}
				\end{subfigure}\hfill\begin{subfigure}[b]{0.49\textwidth}
					\centering
					\includegraphics[width=\textwidth]{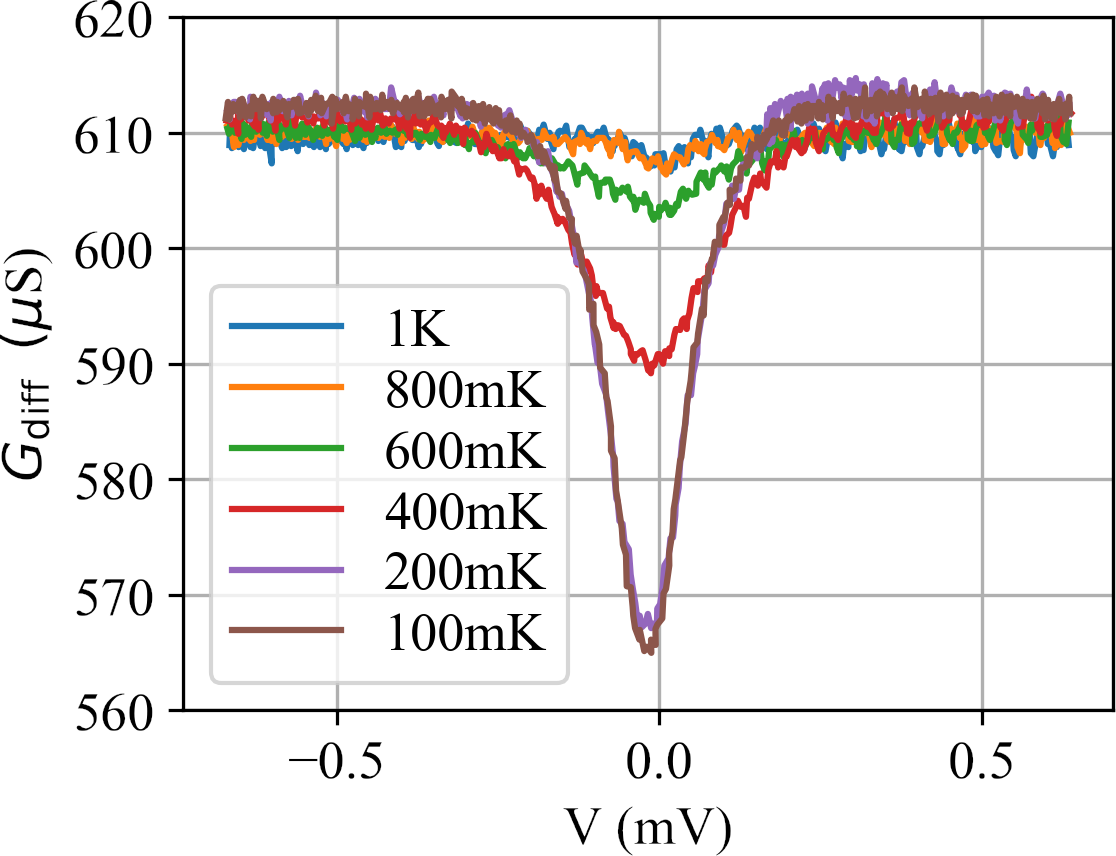}
				\end{subfigure}
				\caption{\label{fig:tasp_axel}\B{(Left)} Scribes RB0C1, RB0C2, RB0C1RE and RB0C2RE, all with contacts \SI{1.25}{\micro\metre} wide and \SI{0.35}{\micro\metre} long, were measured on wafer P07, which had \SI{5}{\nano\metre} Pt deposited (see Table~\ref{tab:lot4}). At room temperature (\Bred{red curve}) the slope indicates a silicon sheet resistance of \SI{870}{\ohm}, which drops to \SI{635}{\ohm} at \SI{4}{\kelvin} (\Bblue{blue curve}). The opposite trend is seen for the contact resistance, which increases from $2R_\text{c}=\SI{300}{\ohm}$ at \SI{300}{\kelvin} to $2R_\text{c}=\SI{470}{\ohm}$ at \SI{4}{\kelvin}. \B{(Right)} One of these structures, RB0C2RE5 (the fifth structure of scribe RB0C2RE, $L_\text{channel}=\SI{25}{\micro\metre}$) was further cooled to below the superconducting critical temperature of PtSi. Shown is the differential conductance versus the voltage bias across the junction, for different temperatures from \SI{100}{\milli\kelvin} to \SI{1}{\kelvin}.
				It is clear that the transmission across the interface is partially suppressed at $E<\Delta$, although no clear feature of the superconducting gap can be observed.
				This is likely a combination of strong inelastic scattering ($\Gamma\approx0.5\Delta\approx\SI{16}{\micro\electronvolt}$) and a high electronic temperature inside the channel (note that there is no difference between the curves for $100$ and \SI{200}{\milli\kelvin}).}
			\end{figure}
			
			More insight into the PtSi/Si interface was gained by measuring the differential conductance of one of these junctions, RB0C2RE5 ($L_\text{channel}=\SI{25}{\micro\metre}$), at different temperatures between \SI{100}{\milli\kelvin} and \SI{1}{\kelvin}, shown in the right of Fig.~\ref{fig:tasp_axel}.
			Although a clear reduction in transmission can be seen around $|eV|<2\Delta=\SI{31}{\milli\volt}$, indicative of a low-transparency S/Sm interface, no sharp coherence peaks are observed.
			This suggests that the inelastic scattering length is shorter than \SI{25}{\micro\metre}, which is not unreasonable for a dirty material~\cite{black2000influence}.		
	
	\clearpage
	\section{Proximity effect in a PtSi SBMOSFET}
			
		\subsection{\label{sec:laurie_sample_description}Description of the samples}
			
\begin{figure}
				\centering
				\begin{subfigure}[b]{0.48\textwidth}
					\centering
					\includegraphics[width=\textwidth]{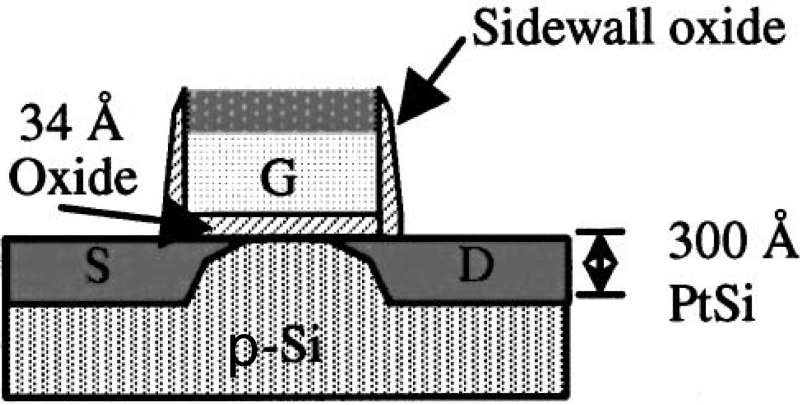}
				\end{subfigure}\hfill\begin{subfigure}[b]{0.48\textwidth}
					\centering
					\includegraphics[width=\textwidth]{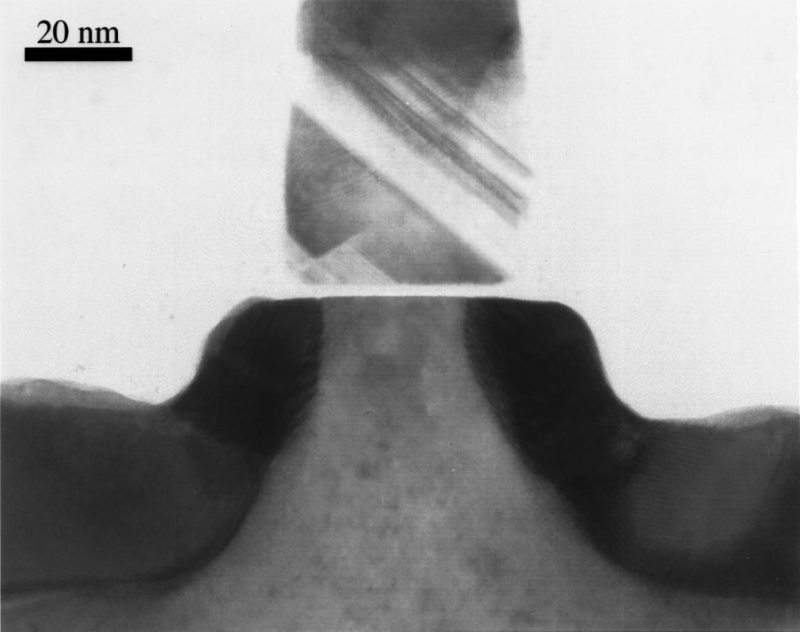}
				\end{subfigure}
				\caption{\label{fig:laurie_illustration_tem}\B{(Left)} A schematic of the SB-MOSFETs that were measured. Adapted from Ref.~\citenum{calvet2002suppression}. \B{(Right)} An XTEM of a device with a \SI{27}{\nano\metre} channel length, reproduced from Ref.~\citenum{calvet2001electrical}. Lateral encroachment occurred underneath the spacers during silicidation, after which a sidewall etch cut into the PtSi vertically, resulting in the upward ``bend'' in the silicide contacts.}
			\end{figure}

			The measurements in the coming sections were performed on transistors from a wafer provided by Laurie Calvet, manufactured in 1997--1998 at National Semiconductor in Santa Clara, California by Chinlee Wang and John Snyder~\cite{calvet2001electrical,calvet2002suppression}, with further details provided in Appendix~\ref{sec:app_low_t_devices}.
			A schematic drawing and a cross-section TEM micrograph of a representative device with a channel length of \SI{27}{\nano\metre} is shown in Fig.~\ref{fig:laurie_illustration_tem}.
			These Schottky-barrier MOSFETs were originally designed to improve the scaling of transistors~\cite{wang1998sub}; introducing a Schottky barrier further away from the gate electrode in devices without encroachment prevents short-channel effects, while moving the contact edges and associated Schottky barrier into the channel allows for barrier modulation in addition to charge accumulation, and thus a steeper sub-threshold slope.
			These devices find a new application in JoFETs, with the advantage of short channels on the order of a few tens of nanometers thanks to the encroachment, and direct gate control of the superconductor/channel interface transparency.
			
			The bulk silicon was implanted with boron to reach a p-doping 
			that ranges from $\SI{5E15}{\per\centi\metre\cubed}$ in the bulk to $10^{19}\si{\per\centi\metre\cubed}$ right below the \SI{3.5}{\nano\metre} gate oxide~\cite{calvet2001electrical}.
			After gate and spacer definition, \SI{29\pm3}{\nano\metre} of Pt was DC sputter deposited onto boron-implanted polycrystalline silicon~\cite{wang1998sub}.
			An estimated \SI{55\pm6}{\nano\metre} layer of PtSi was then formed by a one-hour thermal processing at \SI{450}{\celsius} under a \ce{N2} atmosphere, after which unreacted Pt was removed with aqua regia (4:3:1 \ce{H2O}:HCl:\ce{HNO3}) at \SI{85}{\celsius} for \SI{10}{\minute}.
			Note that there was no buried oxide underneath the silicon in the contacts and channel, such that silicidation occurred isotropically, with no concern of a lateral speed-up underneath the gate as discussed in section~\ref{sec:ptsi}.
			Nonetheless, a large variation in channel length was observed, with a reduction of up to \SI{330}{\nano\metre} in channel length from the target~\cite{calvet2001electrical}, and a standard deviation of \SI{3}{\nano\metre} between devices on the same quad~\cite{wang1998sub}.
			Additional variations across the wafer are due to nonuniform Pt deposition, leading to a range in channel lengths from 30 to \SI{70}{\nano\metre} on the shortest devices.
			In the discussions below, we will assume an average length for these transistors of \SI{50}{\nano\metre}.
			The resulting PtSi layer had a resistivity of \SI{40}{\micro\ohm\centi\metre}~\cite{wang1998sub}, a factor $1.6\times$ higher than what would be expected by extrapolating from our own experiments on single-anneal thick PtSi layers (see Fig.~\ref{fig:D16S0667A}).
			This had no negative impact on the superconductivity, as a relatively high critical temperature of \SI{1.03}{\kelvin} was observed on these devices (see Fig.~\ref{fig:laurie_bcs_fit}).

			Transistors usually have counterdoping with opposite sign in the channel to suppress current in the OFF state.
			In these devices however, both the contacts and the channel were implanted with boron, leading to a high $I_\text{off}$ at \SI{300}{\kelvin} that would make them unsuitable for room-temperature logic operation.
			Such an OFF current disappears when the device is cooled down, as thermionic emission is suppressed and dopants in the channel freeze out.
			At the operational temperature of a transmon qubit (preferably below \SI{50}{\milli\kelvin}), any transport across the interface will occur by quantum tunneling, at which point it becomes crucial to have a low potential barrier.
			Assuming a triangular barrier shape, the tunneling probability due to field emission will be suppressed super-exponentially as $\exp(-C\phi^{3/2}/E)$ (where $E$ is the electric field)~\cite{fowler1928electron}\footnote{Thermionic emission, which scales as $\exp(-\phi/k_\text{B}T)$, is irrelevant at these temperatures.}.
			The low Schottky barrier of \SI{0.16}{\electronvolt}~\cite{wang1998sub} due to the all-p doping~\footnote{The cited thesis mentions that the low value of \SI{0.16}{\electronvolt}, as compared to 0.19--\SI{0.25}{\electronvolt} in the literature, may be due to inaccuracies in the temperature measurement. However, a later report~\cite{dubois2004measurement} addressed a similar mismatch with the same ``commonly accepted numbers'' by fabricating a range of devices to remove short-channel effects and account for lateral transport and tunneling, and still found that only values as low as $0.14$--\SI{0.145}{\electronvolt} matched the data.} is thus essential for proper JoFET operation.
			For comparison, an n-type channel would have led to a Schottky barrier of $0.81$--\SI{0.88}{\electronvolt}~\cite{zhu1999beem}.
			A downside of the intentional channel doping is that reduced mobility was observed due to the high impurity concentration~\cite{wang1998sub}.
			
			Measurements of the gate capacitance with gate voltage indicated a high density of interface states,
			which were attributed to contamination during gate oxidation and silicon deposition~\cite{wang1998sub}.
			This is of special concern for the integration of these devices in superconducting qubits; although circuits in the transmon limit ($E_\text{J}\gg E_\text{C}$) are less prone to decoherence through the coupling of the qubit degree of freedom to these two-level systems~\cite{koch2007charge} than other designs~\cite{martinis2005decoherence}, fluctuations in Josephson coupling strength due to their charging will complicate qubit operation.
			
		\subsection{Room-temperature tests}
			
\begin{figure}
				\centering
				\begin{subfigure}[b]{0.48\textwidth}
					\centering
					\includegraphics[width=\textwidth]{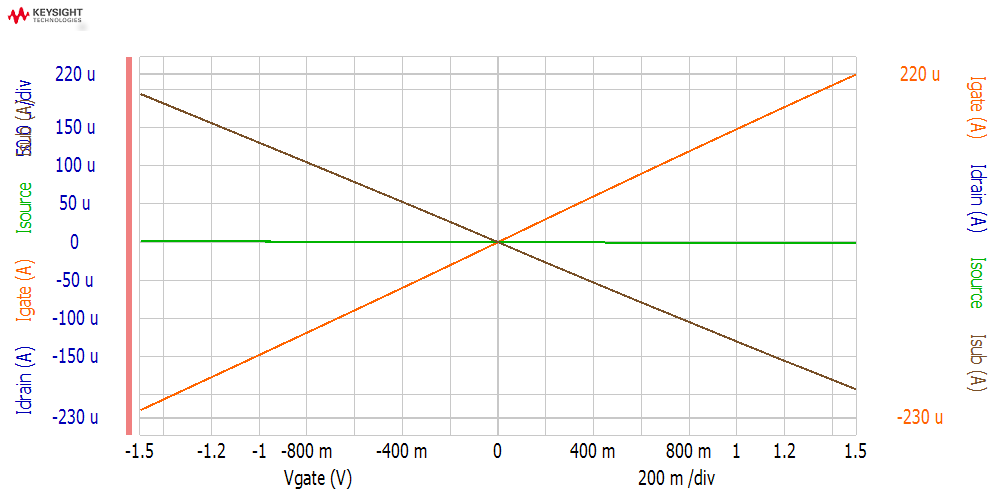}
					\caption{}
				\end{subfigure}\hfill\begin{subfigure}[b]{0.48\textwidth}
					\centering
					\includegraphics[width=\textwidth]{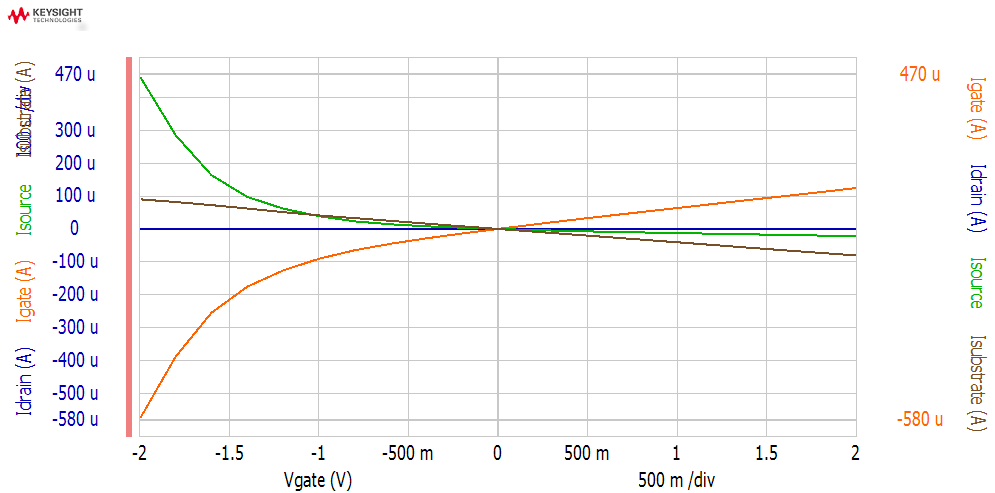}
					\caption{}
				\end{subfigure}
				
				\begin{subfigure}[b]{0.48\textwidth}
					\centering
					\includegraphics[width=\textwidth]{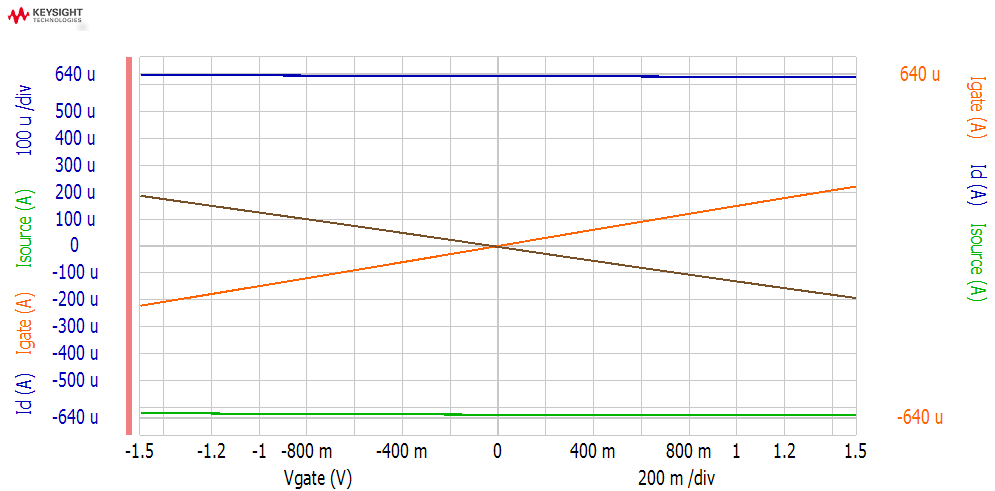}
					\caption{}
				\end{subfigure}\hfill\begin{subfigure}[b]{0.48\textwidth}
					\centering
					\includegraphics[width=\textwidth]{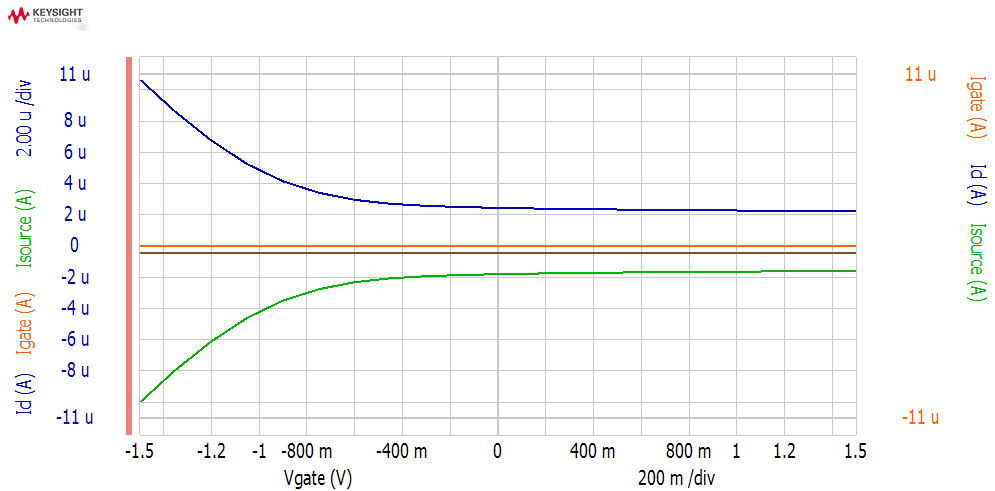}
					\caption{}
				\end{subfigure}
				\caption{\label{fig:probe_station_vg}Currents through the \B{\textcolor{orange!90!black}{gate}}, \B{\textcolor{brown!80!black}{substrate}}, \B{\textcolor{blue!80!black}{drain}} and \B{\textcolor{green!70!black}{source}} at $V_\text{d}=\SI{1}{\milli\volt}$ as $V_\text{g}$ is varied, for different devices. \B{(a)} Leakage from gate to substrate, no field effect. \B{(b)} Field-enhanced leakage from gate to source. \B{(c)} Shorted from source to drain. \B{(d)} Field effect, part of the current injected at the drain flows through the substrate (as expected).}
			\end{figure}

\begin{figure}
				\centering
				\begin{subfigure}[b]{0.48\textwidth}
					\centering
					\includegraphics[width=\textwidth]{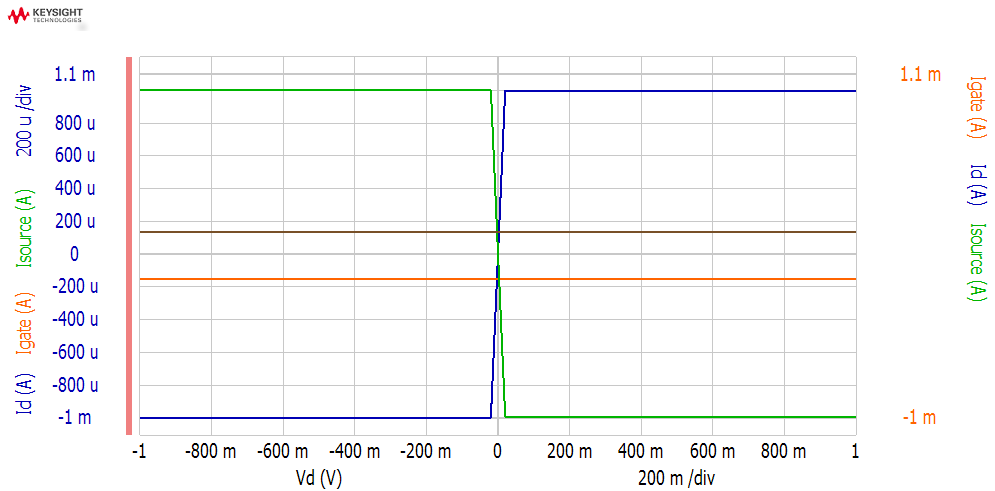}
					\caption{}
				\end{subfigure}\hfill\begin{subfigure}[b]{0.48\textwidth}
					\centering
					\includegraphics[width=\textwidth]{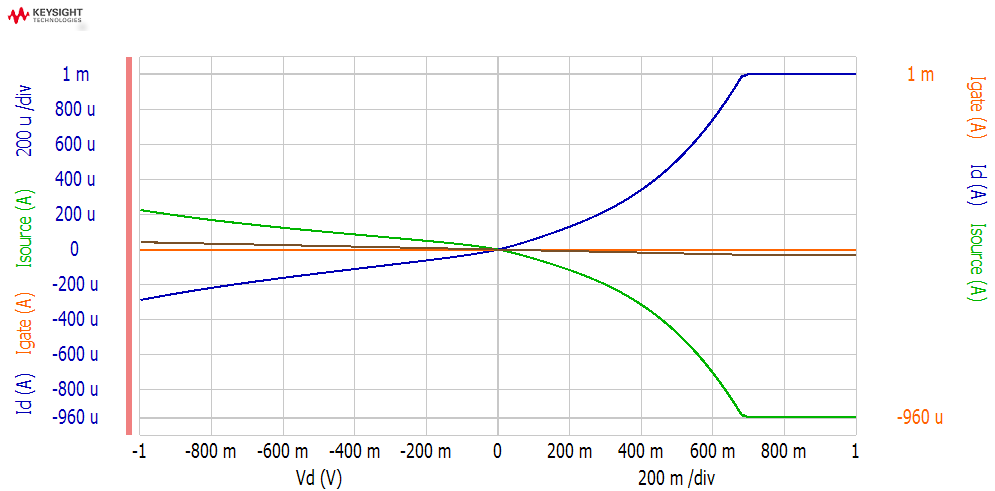}
					\caption{}
				\end{subfigure}
				\caption{\label{fig:probe_station_vd}Currents through the \B{\textcolor{orange!90!black}{gate}}, \B{\textcolor{brown!80!black}{substrate}}, \B{\textcolor{blue!80!black}{drain}} and \B{\textcolor{green!70!black}{source}} at $V_\text{g}=-\SI{1}{\volt}$ as $V_\text{d}$ is varied, for different devices. \B{(a)} Short from source to drain. \B{(b)} Nonlinear source-drain current due to the presence of a Schottky barrier (saturation of the instrument at $I_\text{d}=\SI{1}{\milli\ampere}$).}
			\end{figure}

			An automatic probe station was used to perform rapid tests on a series of devices at room temperature.
			The station switched between devices by moving the wafer underneath the probes, allowing for a large number of transistors to be characterized simply by providing a spreadsheet of coordinates.
			Shown in Figs.~\ref{fig:probe_station_vg} and~\ref{fig:probe_station_vd} are measured currents through gate, substrate, drain and source probes for the four types of device most often observed.
			As described in the figure captions, such automatically generated graphs are useful to identify working or broken devices, and helped select transistors that would then be measured at low temperature.
	
			It should be noted that a current from gate to substrate on the connected devices does not per se imply a broken oxide, since this path was shunted by a parallel diode (see Fig.~\ref{fig:separate_connected_pmos} in the appendices).
			This current will disappear at low temperatures as the substrate freezes out~\cite{clark1980feasibility,simoen1989freeze}.
			Nor does a large source-drain current linear in $V_\text{d}$ necessarily indicate that the channel has been fully consumed by silicidation; such large currents can be expected for the shortest devices due to the low Schottky barrier, and a good field effect is often observed at cryogenic temperatures as soon as dopants have been frozen out and thermionic emission is reduced.
			As detailed graphically in appendix~\ref{sec:app_low_t_devices}, working devices were cut from the wafer, glued onto a sample holder and wire bonded for cryogenic measurements.
			
		\subsection{Cryogenic measurement setup}
			
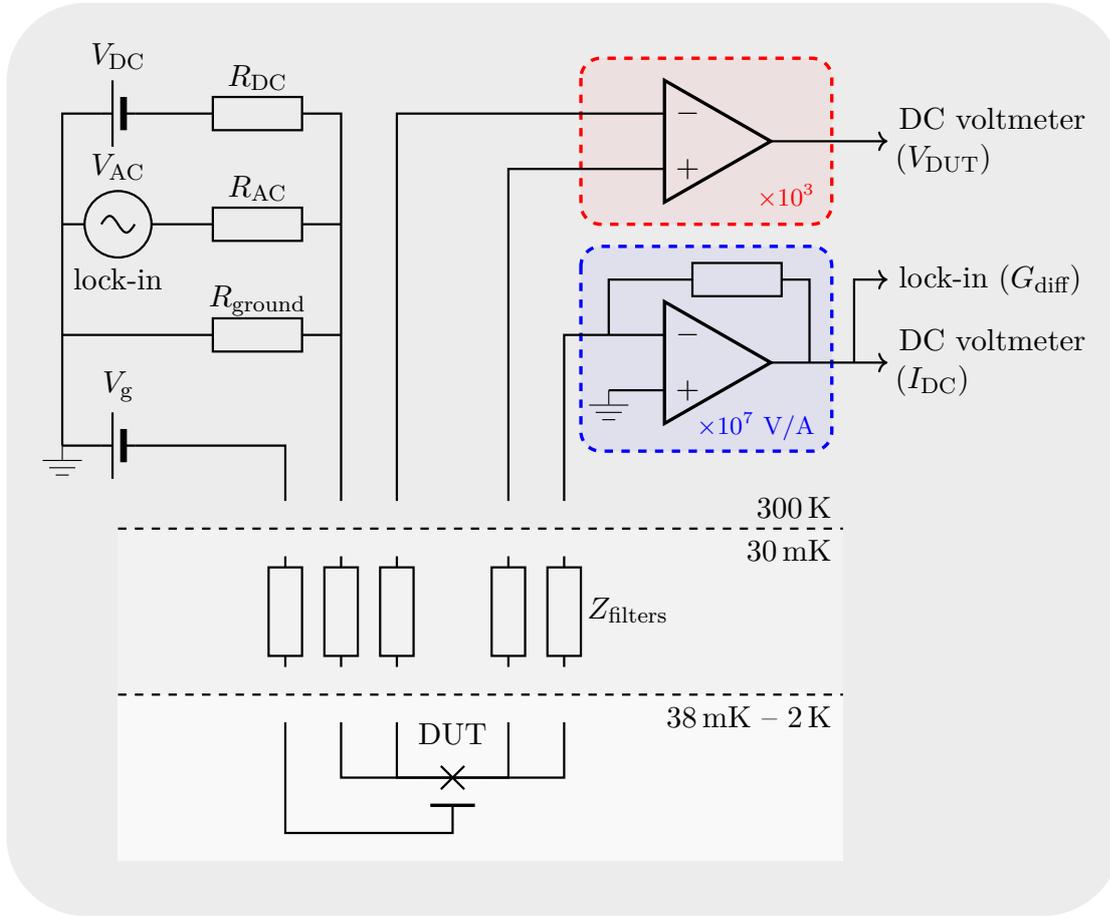
\begin{figure}
				\centering
				\resizebox{\columnwidth}{!}{%
					\begin{circuitikz}[circuit ee IEC,european,x=1.4cm,y=1.4cm,thick]
						\fill[rounded corners=1cm,gray!15!white] (-0.5,1) rectangle (9.5,-7.25);
					
						\draw (0.5,0) node[battery2shape,rotate=180] (Vdc) {} (Vdc.left) -- (1,0);
						\draw (Vdc.below) node[anchor=south] {$V_\text{DC}$};
						\draw (0.5,-1) node[vsourcesinshape,rotate=90] (Vac){} (Vac.below) -- (1,-1);
						\draw (Vac.right) node[anchor=south] {$V_\text{AC}$};
						\draw (Vac.left) node[anchor=north] {lock-in};
						\draw (0.5,-3) node[battery2shape,rotate=180] (Vg) {} (Vg.left) -- (2,-3) -- (2,-3.5);
						\draw (Vg.below) node[anchor=south] {$V_\text{g}$};
						
						\draw[thin] (0,-3) to[ground={pos=1}] (0,-3.2);
						\draw (0,-3) -- (Vg.right);
						\draw (0,-3) -- (0,0) -- (Vdc.right);
						\draw (0,-1) -- (Vac.above);
						\draw (1,0) 			to[R=$R_\text{DC}$] (2.5,0) -- (2.5,-3.5);
						\draw (1,-1) 			to[R=$R_\text{AC}$] (2.5,-1);
						\draw (0,-2) -- (1,-2) 	to[R=$R_\text{ground}$] (2.5,-2);
						
						\fill[gray!10] (0.5,-3.75) -- (7,-3.75) -- (7,-5.25) -- (0.5,-5.25) -- cycle;
						\fill[gray!5] (0.5,-6.75) -- (7,-6.75) -- (7,-5.25) -- (0.5,-5.25) -- cycle;
						
						\draw[dashed] (0.5,-3.75) -- (7,-3.75) node[anchor=south east] {\SI{300}{\kelvin}} node[anchor=north east] {\SI{30}{\milli\kelvin}};
						
						\draw (2,-4) 	to[R] (2,-5);
						\draw (2.5,-4) 	to[R] (2.5,-5);
						\draw (3,-4) 	to[R] (3,-5);
						\draw (4,-4) 	to[R] (4,-5);
						\draw (4.5,-4) 	to[R=$Z_\text{filters}$] (4.5,-5);
						\draw[dashed] (0.5,-5.25) -- (7,-5.25) node[anchor=north east] {\SI{38}{\milli\kelvin} -- \SI{2}{\kelvin}};
						
						\draw (2.5,-5.5) -- (2.5,-6) to[barrier] (4.5,-6) -- (4.5,-5.5);
						\node[anchor=south] at (3.5,-5.8) {DUT};
						\draw (3,-5.5) -- (3,-6) -- (4,-6) -- (4,-5.5);
						\draw (2,-5.5) -- (2,-6.5) -- (3.5,-6.5) -- (3.5,-6.25);
						\draw[very thick] (3.3,-6.25) -- (3.7,-6.25);
						
						\draw[rounded corners=0.25cm,dashed,red,very thick,fill=gray!15!white!95!red] (4.65,0.5) rectangle (6.9,-1);
						\node[anchor=south east, red] at (6.85,-0.95) {\footnotesize$\times10^3$};
						\draw[very thick] (5.4,0.3) -- (5.4+0.953,-0.25) -- (5.4,-0.8) -- cycle;
						\draw (3,-3.5) -- (3,0) -- (5.4,0) node[anchor=west] {$-$};
						\draw (4,-3.5) -- (4,-0.5) -- (5.4,-0.5) node[anchor=west] {$+$};
						\draw[->] (5.4+0.953,-0.25) -- (7.4,-0.25) node[anchor=west,text width=7em] {DC voltmeter ($V_\text{DUT}$)};
						
						\draw[rounded corners=0.25cm,dashed,blue,very thick,fill=gray!15!white!95!blue] (4.65,-1.2) rectangle (6.9,-3.05);
						\node[anchor=south east, blue] at (6.85,-3.05) {\footnotesize$\times10^7$ V/A};
						\draw[very thick] (5.4,0.3-2) -- (5.4+0.953,-0.25-2) -- (5.4,-0.8-2) -- cycle;
						\draw (4.9,-2) -- (4.9,-1.5) -- (5.4,-1.5) to[R] (6.7,-1.5) -- (6.7,-2.25);
						\draw (4.5,-3.5) -- (4.5,-2) -- (5.4,-2) node[anchor=west] {$-$};
						\draw[thin] (4.9,-2.5) to[ground={pos=1}] (4.9,-2.7);
						\draw (4.9,-2.5) -- (5.4,-2.5) node[anchor=west] {$+$};
						\draw[->] (7.1,-2.25) -- (7.1,-1.5) -- (7.4,-1.5) node[anchor=west] {lock-in ($G_\text{diff}$)};
						\draw[->] (5.4+0.953,-2.25) -- (7.4,-2.25) node[anchor=west,text width=7em] {DC voltmeter ($I_\text{DC}$)};
					\end{circuitikz}}
				\caption{\label{fig:setup}Schematic of the measurement setup.}
			\end{figure}

			Shown in Fig.~\ref{fig:setup} is a schematic of the measurement setup used in the experiments discussed below.
			The channel of the transistor was biased with a DC voltage ($V_\text{DC}$), together with an AC oscillation from a lock-in amplifier ($V_\text{AC}$).
			These were connected using three resistors, $R_\text{DC}=\SI{10}{\kilo\ohm}$, $R_\text{AC}=\SI{100}{\kilo\ohm}$ and a shunt to the ground $R_\text{ground}=\SI{10}{\ohm}$, effectively dividing the DC and AC voltages by $10^3$ and $10^4$, respectively, to be able to use voltage steps smaller than the resolution of the voltage source.
			The resulting current was amplified using a trans-impedance amplifier that delivered $10^7\si{\volt/\ampere}$, while a differential amplifier was used to measure the voltage across the device.
			
			Low-pass RC filters were anchored to the mixing chamber at 30 mK, ensuring that their impedance did not vary during the measurement.
			The sample holder, on the other hand, had only a weak thermal coupling to the mixing chamber, such that it could be heated from \SI{38}{\milli\kelvin} to \SI{2}{\kelvin}.
			
			Since the resistance of the filters is much greater than that of the shunt resistance to the ground ($2R_\text{filter}\approx\SI{40}{\kilo\ohm}\gg R_\text{ground}=\SI{10}{\ohm}$), we can safely assume that the effective applied voltage at the top of the fridge does not depend on the resistance of the device under test (DUT).
			In our case, that means that the AC and DC voltages at the second line entering the fridge in Fig.~\ref{fig:setup} are
			\begin{equation}\begin{array}{r@{\;}c@{\;}l}
				V_\text{applied,AC}	& =	& \alpha\,\dfrac{R_\text{ground}}{R_\text{AC}}\,V_\text{AC,source}\approx10^{-4}\,V_\text{AC,source}\quad,\text{and}\\\\
				V_\text{applied,DC}	& =	& \beta\,\dfrac{R_\text{ground}}{R_\text{DC}}\,V_\text{DC,source}\approx10^{-3}\,V_\text{DC,source},
			\end{array}\end{equation}
			where $\alpha=0.88$ and $\beta=1.00$ are fitting parameters to take into account any instrumental inaccuracies.
			However, this does not equal the actual bias across the device itself, especially when $R_\text{DUT}$ becomes smaller than $|Z_\text{filters}|$, which will be the case when the device becomes passing at large negative gate voltages.
			To take this into account, the voltage drop $R_\text{filter}I_\text{DC}$ on each side of the device is subtracted,
			\begin{equation}\label{eq:vdc_dut}V_\text{DC,DUT}=V_\text{DC,applied}-2R_\text{filter}I_\text{DC}.\end{equation}
			Similarly, the filter impedance will limit the measured differential conductance, which is corrected for by calculating~\cite{bethe1942theory4311}
			\begin{equation}\label{eq:gdiff_dut}G_\text{diff,DUT}=\left|\left(\dfrac{\alpha}{G_\text{diff}}-2Z_\text{filter}\right)^{-1}\right|,\quad\text{where}\quad Z_\text{filter}=\dfrac{R_\text{filter}}{1+iR_\text{filter}C_\text{filter}\omega_\text{lock-in}}.\end{equation}
			The two lines to the top operational amplifier in Fig.~\ref{fig:setup} that measured $V_\text{DUT}$ directly were disconnected once the resistance and capacitance of the filters were extracted, to avoid introducing additional noise.
			In all the figures that follow, the drain voltage $V_\text{d}$ will be bias across the device as defined in eq.~\eqref{eq:vdc_dut}, and the differential conductance $G_\text{diff}$ will be that of the device only, as in eq~\eqref{eq:gdiff_dut}.
			
			\begin{table}
				\centering
				\caption{\label{tab:laurie_devices}An overview of the devices where a gate effect was detected at low temperature, indicated in green those that showed a ZBCP.}
				\resizebox{\columnwidth}{!}{%
				\rowcolors{2}{white}{gray!15}
				\begin{tabular}{l l l | l l | l l}
					\rowcolor{gray!30}\hline
					\multicolumn{3}{c|}{}													& \multicolumn{2}{c|}{Scan in $V_\text{g}$}			& \multicolumn{2}{c}{Scan at fixed $V_\text{g}$}\\\rowcolor{gray!30}
					Device 					& W (\si{\micro\metre})	& L (\si{\nano\metre})	& $H=0$							& $H>H_0$			& $H$, $V_\text{g}=$	& $T$, $V_\text{g}=$\\\hline\hline
					D61D1\fns{1}			& ?						& ?						& $[-4.0,-1.45]$				& \Cross			& \Cross	& -3.5, -3, -2.5, -2.25	\\
					D61D2\fns{1}			& ?						& ?						& $[-3.5,-1.56]$				& $[-1.5,-1.4926]$	& -1.5		& \Cross\\
					\Bgreen{D61D4}\fns{1}	& 2.5					& 50						& $[-4.15,-2.0]$				& \Cross			& \Cross	& \Cross\\
					D61D5\fns{1}			& ?						& ?						& $[-3.0,-1.87]$				& \Cross			& \Cross	& \Cross\\
					D63D1\fns{1}			& ?						& ?						& $[-3.5,-1.6]$					& \Cross			& \Cross	& -3.5, -2\\
					\Bgreen{D63D3}\fns{1}	& 2.5					& 50						& $[-4.9,-1.0]$					& $[-4.5,-1]$		& -4.5, -2.7& -4.5, -4.0, -2.7\\
					D63D4\fns{1}			& ?						& ?						& $\Bred{[-4.0,-1.0]}$\fns{3}	& \Cross			& -3.9		& \Cross\\
					\Bgreen{D73D4}\fns{1}	& 2.5					& 50					& $[-4.4,-2.4]$					& $[-2.5,-2.494]$	& \Cross	& \Cross \\
					D84D1\fns{2}			& ?						& ?						& $[-4.5,-2.1]$					& \Cross			& \Cross	& \Cross\\
					D84D3\fns{2}			& ?						& ?						& $[-3.1,-2.05]$				& \Cross			& \Cross	& \Cross$[-3,0]$ ($V_\text{d}=\SI{1}{\milli\volt}$)\\
					D84D4\fns{2}			& ?						& ?						& $[-5,-4.7]$					& \Cross			& \Cross	& \Cross\\
					D84D5\fns{2}			& ?						& ?						& $[-3.5,0]$					& \Cross			& \Cross	& \Cross\\\hline
				\end{tabular}}
				\begin{minipage}{\textwidth-2\parindent}\footnotesize
					\noindent\fns{1}Measured in \emph{Christophe's fridge}.\\
					\fns{2}Measured in the fridge \emph{Bigoudène}.\\
					\fns{3}Offset in magnetic field detected, $H\neq0$.
				\end{minipage}
			\end{table}
		
		\subsection{Demonstration of the field effect at low temperature}
			
			As its name suggests, thermionic emission is activated by heat.
			This contribution to the current, due to the hot tail of the electron energy distribution that extends above the potential barrier, scales as~\cite{bethe1942theory4312,crowell1965richardson,rideout1970effects}
			\begin{equation}\label{eq:thermionic}I_\text{th}\propto T^2\,e^{-\phi_\text{Schottky}/k_\text{B}T},\end{equation}
			where we will use $\phi_\text{Schottky}=\SI{0.16}{\electronvolt}$ for our devices~\cite{wang1998sub}.
			When a bias $V_\text{d}$ is applied, the net current across a single barrier will then be~\cite{crowell1966current,dubois2004measurement}
			\begin{equation}\label{eq:thermionic_v}I_\text{th}\propto T^2\,e^{-\phi_\text{Schottky}/k_\text{B}T}\left(e^{-qV_\text{d}/k_\text{B}T}-1\right).\end{equation}
			In the case of pure thermionic emission ($qV_\text{d}\ll k_\text{B}T$, no field emission), the voltage drop is independent of the bias direction and (using a normalized $V'=qV/k_\text{B}T$),
			\begin{equation}I(\raisebox{-0.4ex}{\begin{circuitikz}[scale=0.5,transform shape]\draw (1,0)to[empty diode](0,0);\draw (1,0)to[empty diode](2,0);\end{circuitikz}})=\left(e^{-V'/2}-1\right)-\left(e^{+V'/2}-1\right)\stackrel{V'\ll1}{\approx}e^{-V'}-1=I(\raisebox{-0.4ex}{\begin{circuitikz}[scale=0.5,transform shape]\draw (1,0)to[empty diode](0,0);\end{circuitikz}}),\end{equation}
			such that we can approximate the SBMOS channel as a single diode.
			The key thing to note here is that the effective resistance of the two Schottky barriers in the channel sharply increases at low temperatures.

			\begin{figure}
				\centering
				\begin{subfigure}[t]{0.47\textwidth}
					\centering
					\includegraphics[width=\textwidth]{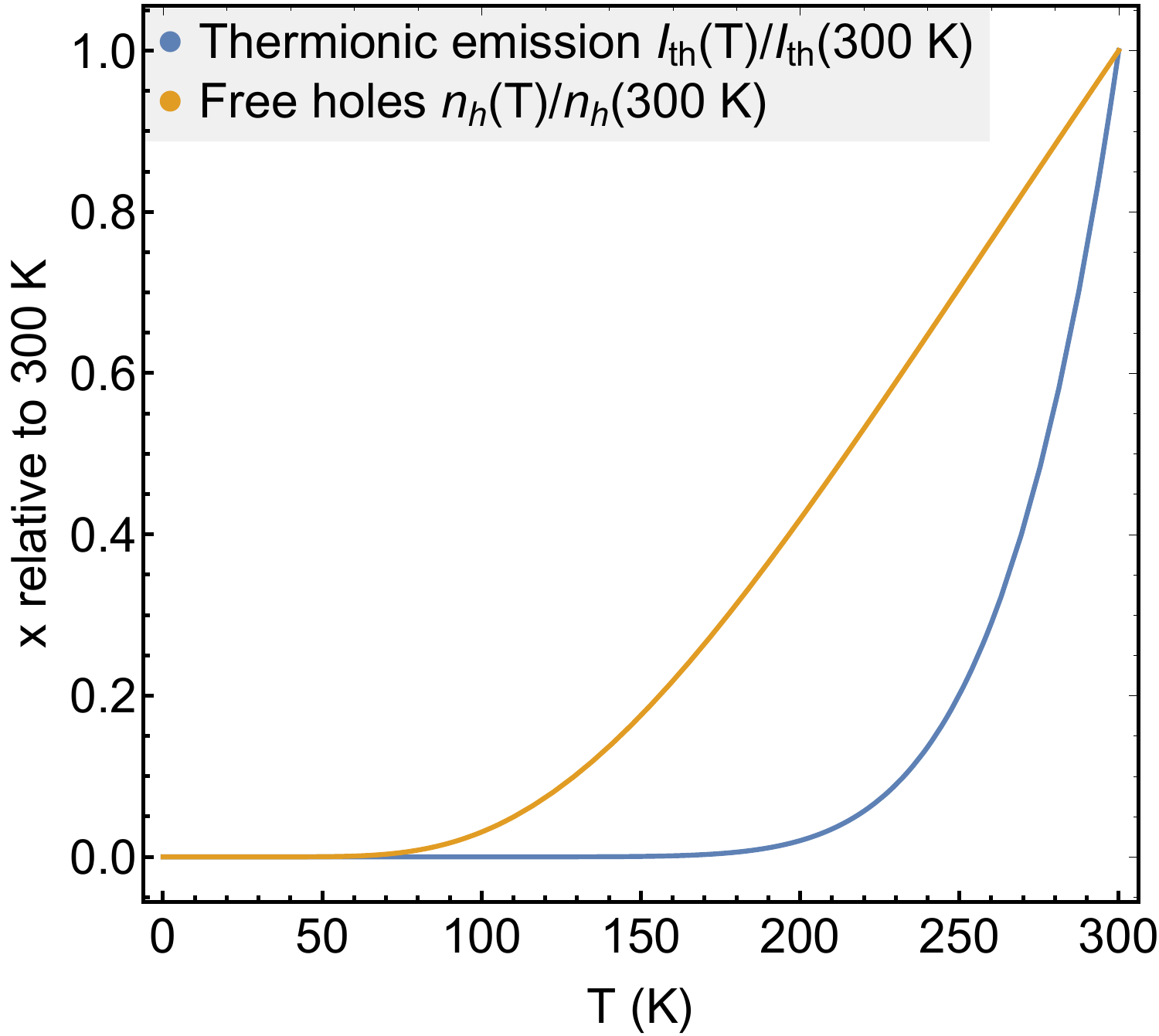}
					\caption{\label{fig:ThermalSuppressionPlot}}
				\end{subfigure}\hfill\begin{subfigure}[t]{0.49\textwidth}
					\centering
					\includegraphics[width=\textwidth]{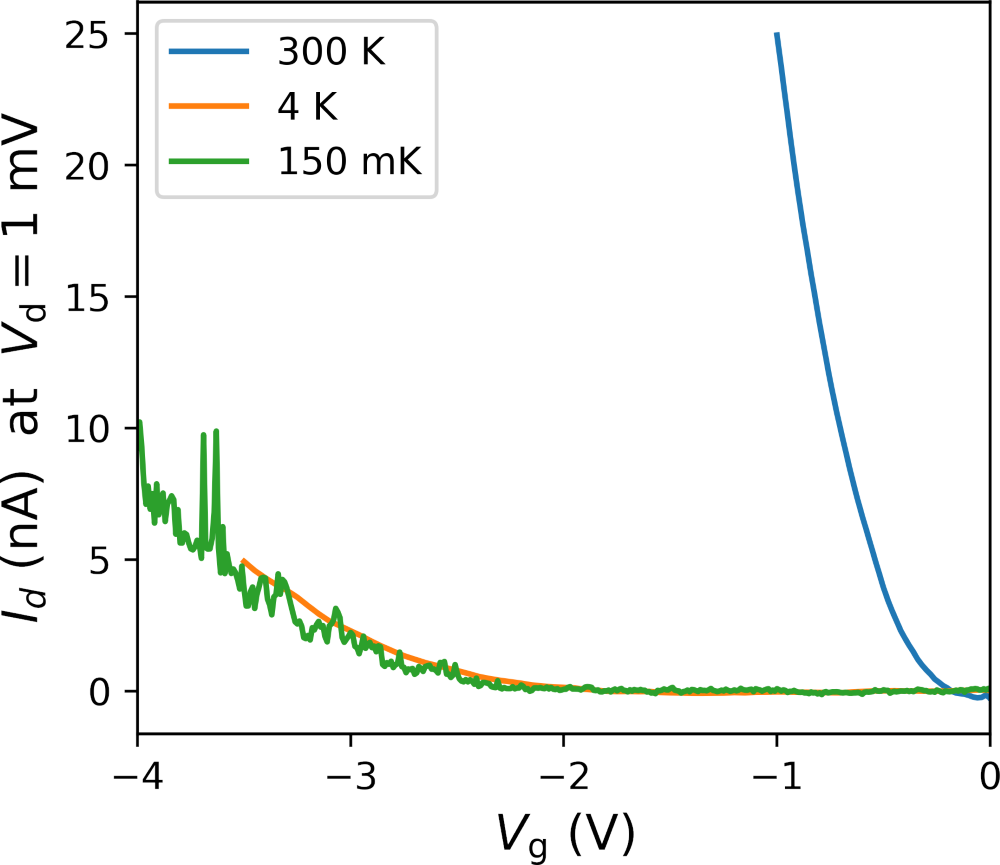}
					\caption{\label{fig:D84D4_Id_Vg_300K_4K_150mK}}
				\end{subfigure}
				\caption{\B{(a)} Theoretical estimates of the thermal suppression of the thermionic emission (\B{\textcolor{blue!80!black}{blue}}) and free carrier concentration (\B{\textcolor{orange!90!black}{orange}}), after equations~\eqref{eq:thermionic} and~\eqref{eq:free_holes}. \B{(b)} Device D84D4 conducts at negative gate voltages. Cooling the device from 300 to \SI{4}{\kelvin} suppresses thermionic emission and freezes out dopant carriers, shifting the threshold voltage down. At lower temperatures of \SI{150}{\milli\kelvin}, reproducible fluctuations appear due to charging effects.}
			\end{figure}
			
			The channel itself is made of boron-doped\footnote{Boron, a column to the left of silicon in the periodic table, has only 3 electrons in the outer shell.} Si, where the product of the free electrons and holes remains equal to the square of the intrinsic carrier density of pure silicon, as discussed in section~\ref{sec:field_effect}.
			This means that in the absence of an accumulating electrostatic field, the number of majority carriers (holes) will be~\cite{chrzanowska1989bilow}
			\begin{equation}\label{eq:free_holes} p\propto\,e^{(E_\text{V}-E_\text{F})/k_\text{B}T},\end{equation}
			where the negative $E_\text{V}$, measured from the gate-dependent Fermi level $E_\text{F}$, is the relative energy of the valence band edge.
			By introducing easily excited acceptors with energy levels at $E_\text{A}$, the boron pins the Fermi level at a distance of $E_\text{A}-E_\text{V}$ from the valence band, such that we can equivalently write
			\begin{equation}\label{eq:free_holes_ea} p\propto\,e^{(E_\text{V}-E_\text{A})/k_\text{B}T}.\end{equation}
			The intrinsic (i.e. field-independent) acceptor energy $E_\text{A}-E_\text{V}$ of boron depends on temperature, and decreases from around 62 to \SI{45}{\milli\electronvolt} when cooled from room temperature to anywhere between \SI{100}{\kelvin} and \SI{0}{\kelvin}~\cite{pires1990carrier}.
			At low temperature, holes introduced by the missing bond of boron atoms in the channel will be frozen out~\cite{balestra1987influence,dickstein1989carrier}, and accumulation can only be achieved by either ionizing these acceptors with the gate electrostatic field~\cite{frenkel1938pre,saks1979time}, or by exciting carriers from metallic atoms  in the contacts across the Schottky barrier.

			\begin{figure}
				\centering
				\includegraphics[width=0.8\textwidth]{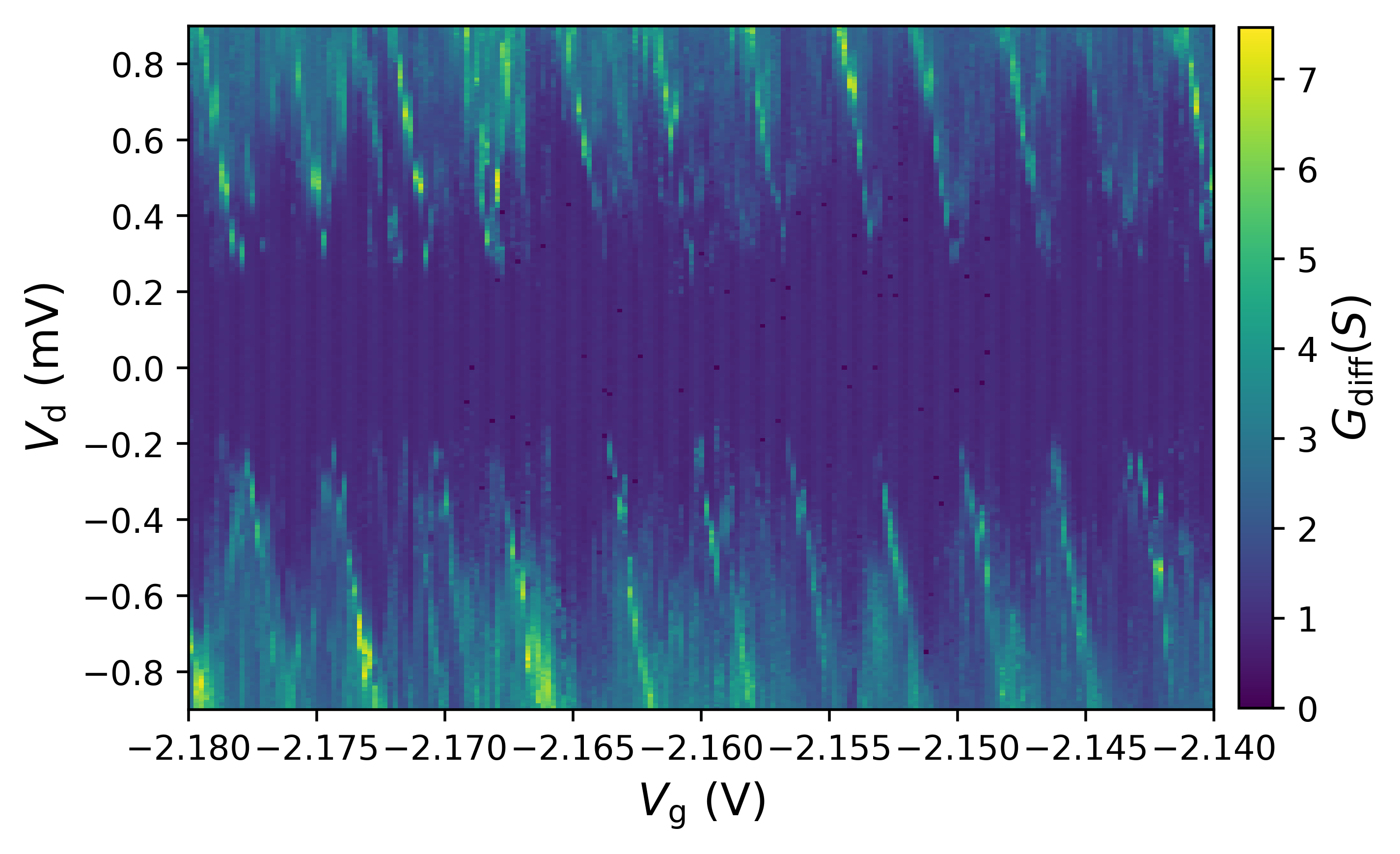}
				\caption{\label{fig:D63D4S02_DUT}Data from device D63D4. Coulomb diamonds are most clearly visible in the differential conductance at small absolute gate voltages, when the only mode of transport is by resonantly tunneling through the charging centers.}
			\end{figure}
			
			As the device is cooled down, thermionic emission and dopant excitation will both be suppressed, as shown in Fig.~\ref{fig:ThermalSuppressionPlot}, leading to larger threshold voltages at lower temperatures, illustrated in Fig.~\ref{fig:D84D4_Id_Vg_300K_4K_150mK}.
			In this last figure, an additional feature can be observed at \SI{150}{\milli\kelvin}: reproducible peaks in conductance at fixed values of $V_\text{g}$.
			These can be better understood by scanning also the source-drain bias, as is done for a different device in Fig.~\ref{fig:D63D4S02_DUT}.
			These Coulomb diamonds, spaced unevenly by around 4 or \SI{5}{\milli\electronvolt}, are likely\footnote{Unless something went wrong during fabrication and the wide channel is effectively broken up into many smaller parts due to variations in gate length.} not due to the charging of the channel itself.			
			Since this device has a channel of \SI{2.5}{\micro\metre} by \SI{50}{\nano\metre} and a \SI{3.5}{\nano\metre} gate oxide, we would expect its charging energy to be two orders of magnitude smaller,
			\begin{equation}E_\text{C,channel}=\varepsilon_0\,\kappa_\text{\ce{SiO2}}\,\dfrac{A_\text{channel}}{d_\text{oxide}}\approx\SI{65}{\micro\electronvolt}.\end{equation}
			Instead, these should probably be attributed to the charging of surface states in the oxide detected earlier by capacitance measurements~\cite{wang1998sub} (see section~\ref{sec:laurie_sample_description}).
		
		\subsection{Evidence of superconductivity}

			\begin{figure}
				\centering
				\includegraphics[width=\textwidth]{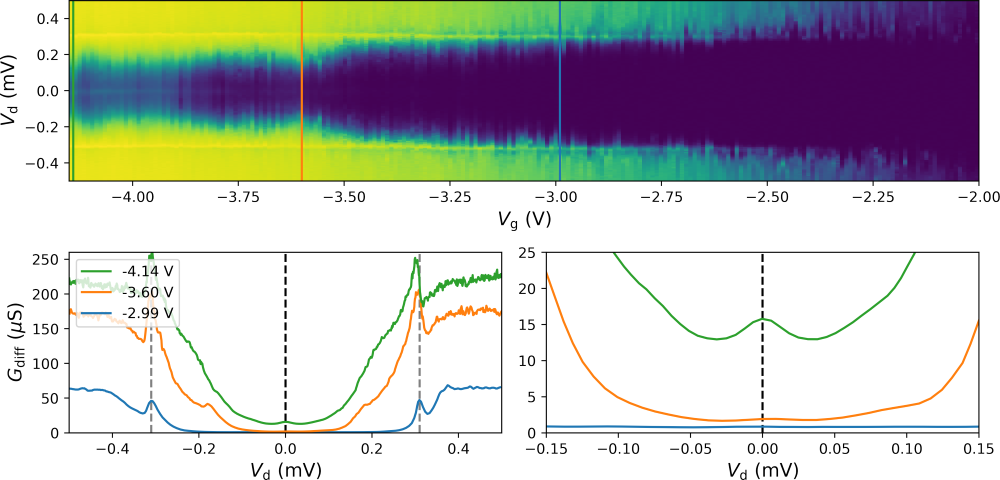}
				\caption{\label{fig:D61D4Gdiff_Vd_combo_plot3}Data from device D61D4. Both source-drain and gate voltage were scanned while the differential conductance was measured. \B{(Top)} As larger negative voltages are applied to the gate, we see first the appearance of coherence peaks at $V_\text{d}=\pm\SI{0.31}{\milli\volt}$ around $V_\text{g}\approx\SI{-2.5}{\volt}$, then large sub-gap conductance from $V_\text{g}\approx\SI{-3.5}{\volt}$, and finally a clear zero-bias conductance peak at $V_\text{g}\lesssim\SI{-3.75}{\volt}$. \B{(Bottom, left)} Selected line curves from the density plot shown above. \B{(Bottom, right)} The same curves, in a smaller $V_\text{d}$ range.}
			\end{figure}
			
\begin{figure}
				\centering
				\begin{subfigure}[b]{0.48\textwidth}
					\includegraphics[width=\textwidth]{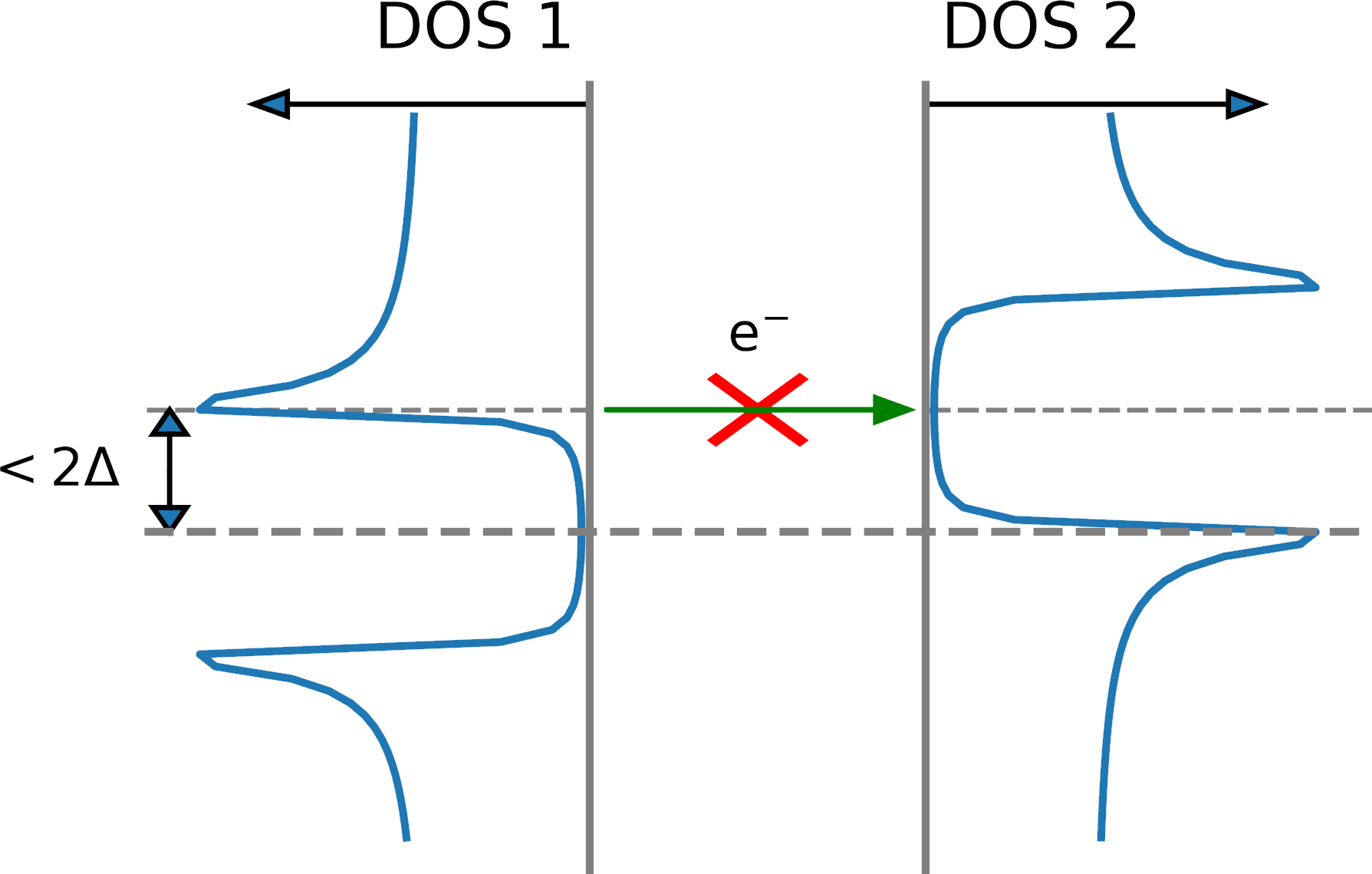}
				\end{subfigure}\hfill\begin{subfigure}[b]{0.48\textwidth}
					\includegraphics[width=\textwidth]{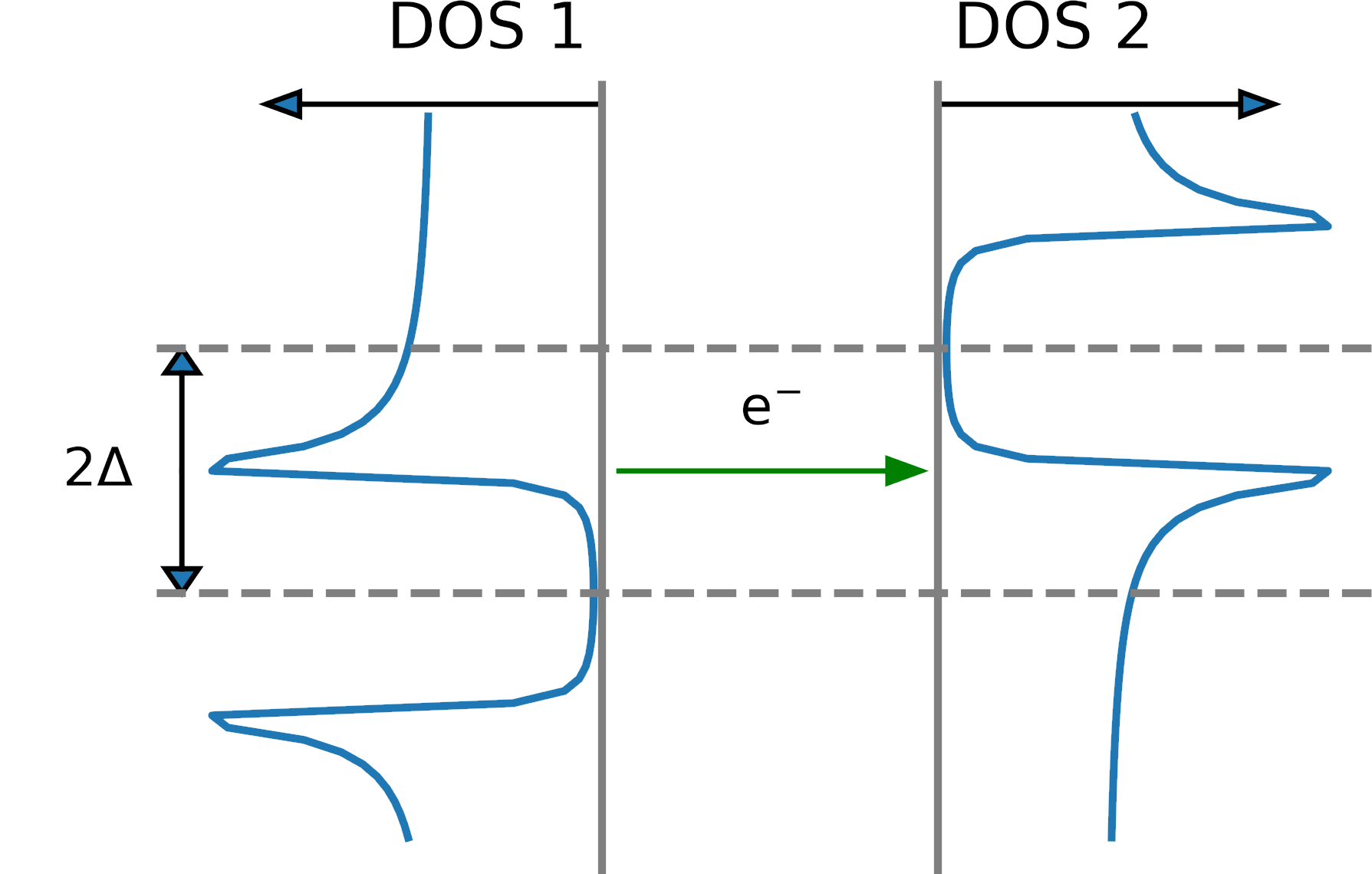}
				\end{subfigure}
				\caption{\label{fig:Superconducting_DOS}\B{(Left)} If the densities of states in the superconducting source and drain are biased by less than twice the gap, transport at the Fermi level will be prevented by the absence of available states. Since the electronic temperature in the device is far below $\Delta/k_\text{B}$, no thermionic emission to higher-energy states will occur either. \B{(Right)} Single quasiparticles will be able to enter the superconductor when $eV_\text{d}\geq2\Delta$, where they either contribute to the supercurrent immediately when $eV_\text{d}\approx2\Delta$, or relax into the condensate within $\sim10^{-9}\si{\second}$~\cite{clarke1972experimental,tinkham1972theory}.}
			\end{figure}

			An additional feature can be noticed in fig.~\ref{fig:D63D4S02_DUT}: the diamonds are spaced vertically by $\SI{0.62}{\milli\volt}$.
			This gap becomes clearer at more negative gate voltages, as can be seen in Fig.~\ref{fig:D61D4Gdiff_Vd_combo_plot3}, and has characteristic coherence peaks.
			If this is due to superconductivity in the source and drain, then we expect the observed value to equal four times the superconducting gap, as explained graphically in Fig.~\ref{fig:Superconducting_DOS}.
			Since its magnitude corresponds to~\cite{bardeen1957theory,johnston2013elaboration}
			\begin{equation}\SI{0.62}{\milli\electronvolt}=4\Delta_\text{0}=4\,\pi\,e^{-\gamma}\,k_\text{B}T_\text{c},\quad\text{and thus}\quad T_\text{c}\approx\SI{1.02}{\kelvin},\end{equation}
			where $\gamma\approx0.5772$ is Euler's constant~\cite{euler1740progressionibus}, we can be confident that this is due to superconductivity in the PtSi source and drain.
			To our knowledge, no other superconductors were present in the contacts or vias.

\begin{figure}
				\centering
				\begin{subfigure}[b]{0.48\textwidth}
					\includegraphics[width=\textwidth]{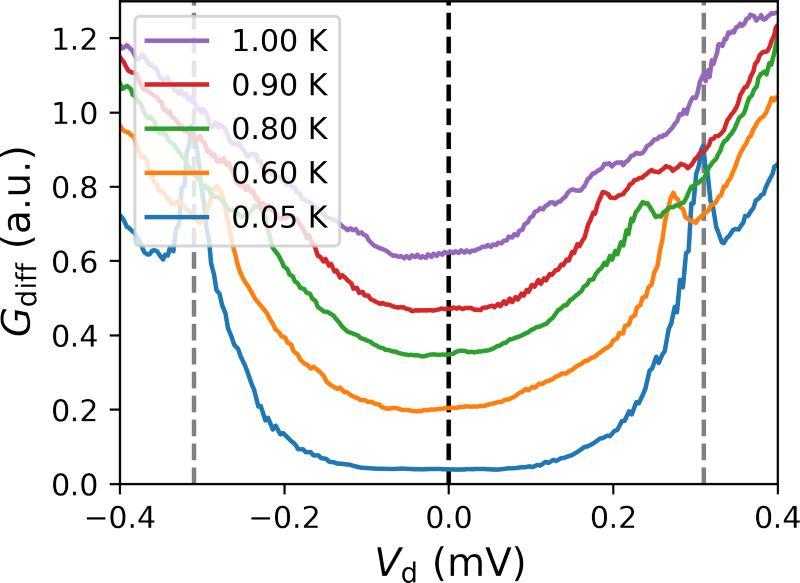}
				\end{subfigure}\hfill\begin{subfigure}[b]{0.48\textwidth}
					\includegraphics[width=\textwidth]{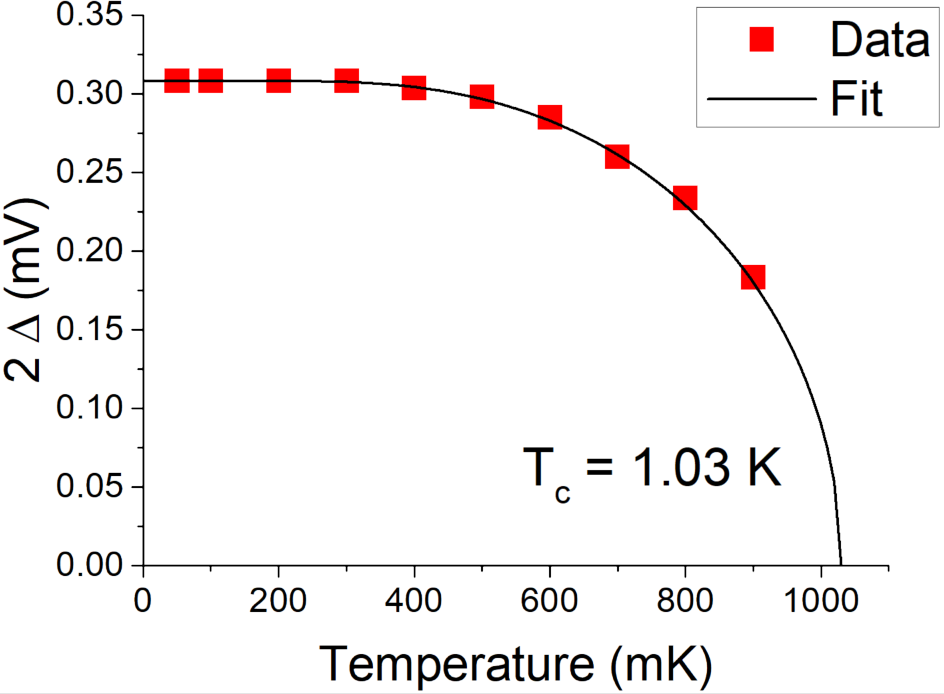}
				\end{subfigure}
				\caption{\label{fig:laurie_bcs_fit}Data from device D63D3. \B{(Left)} Differential conductance at $V_\text{g}=\SI{-2.7}{\volt}$, at selected temperatures between 0.05 and \SI{1.00}{\kelvin}. The coherence peaks (indicated by vertical dashed lines for $T=\SI{0.05}{\kelvin}$) move towards smaller values as the device is warmed up. \B{(Right)} A fit of the peak positions from the graph on the left, to the expected BCS effective gap $\Delta_\text{BCS}(T)$.}
			\end{figure}

			A fit of the peak positions versus temperature to that predicted by BCS theory, shown in Fig.~\ref{fig:laurie_bcs_fit} for a measurement performed on D63D3, confirms that PtSi is a weakly coupled superconductor (such that $\Delta\approx1.76\,k_\text{B}T_\text{c}$ is valid), with a critical temperature of $\SI{1.03}{\kelvin}$.
			This is slightly higher than what we found in section~\ref{sec:ptsi} for films of similar thickness, and implies a better quality of the PtSi obtained at National Semiconductor.

\begin{figure}
				\centering
				\begin{subfigure}[c]{0.50\textwidth}
					\centering
					\begin{tikzpicture}[x=1.55cm,y=1.55cm]
						\draw[fill={gray!15}] (5,2) to[out=0,in=110] (6,1.2) -- (6,0) -- (5,0) -- cycle;
						\draw[gray!15,fill={gray!15}] (6,0.8) to[out=-50,in=175] (8,0.1) -- (8,0) -- (6,0) -- cycle;
						\draw (6,0.8) to[out=-50,in=175] (8,0.1);
						
						\draw[thick] (6,0) -- (6,2);
						\draw (5.5,0.75) node[anchor=south]{S};
						\draw (7,0.75) node[anchor=south]{N};
						
						\draw[style=thick] (6.35,2.5) -- (6.35,2) -- (7.65,2) -- (7.65,2.5);
						\draw (7,1.95) node[anchor=south] {$V_\text{g}<0$};
						\draw (7,2.05) node[anchor=north] {$-$};
						
						\draw[->] (5,0) -- (5,2.3) node[anchor=south] {$|\psi(x)|$};
						\draw[->] (4.8,0) -- (8.5,0) node[anchor=west] {$x$};
					\end{tikzpicture}
				\end{subfigure}\hfill\begin{subfigure}[c]{0.44\textwidth}
					\centering
					\includegraphics[width=\textwidth]{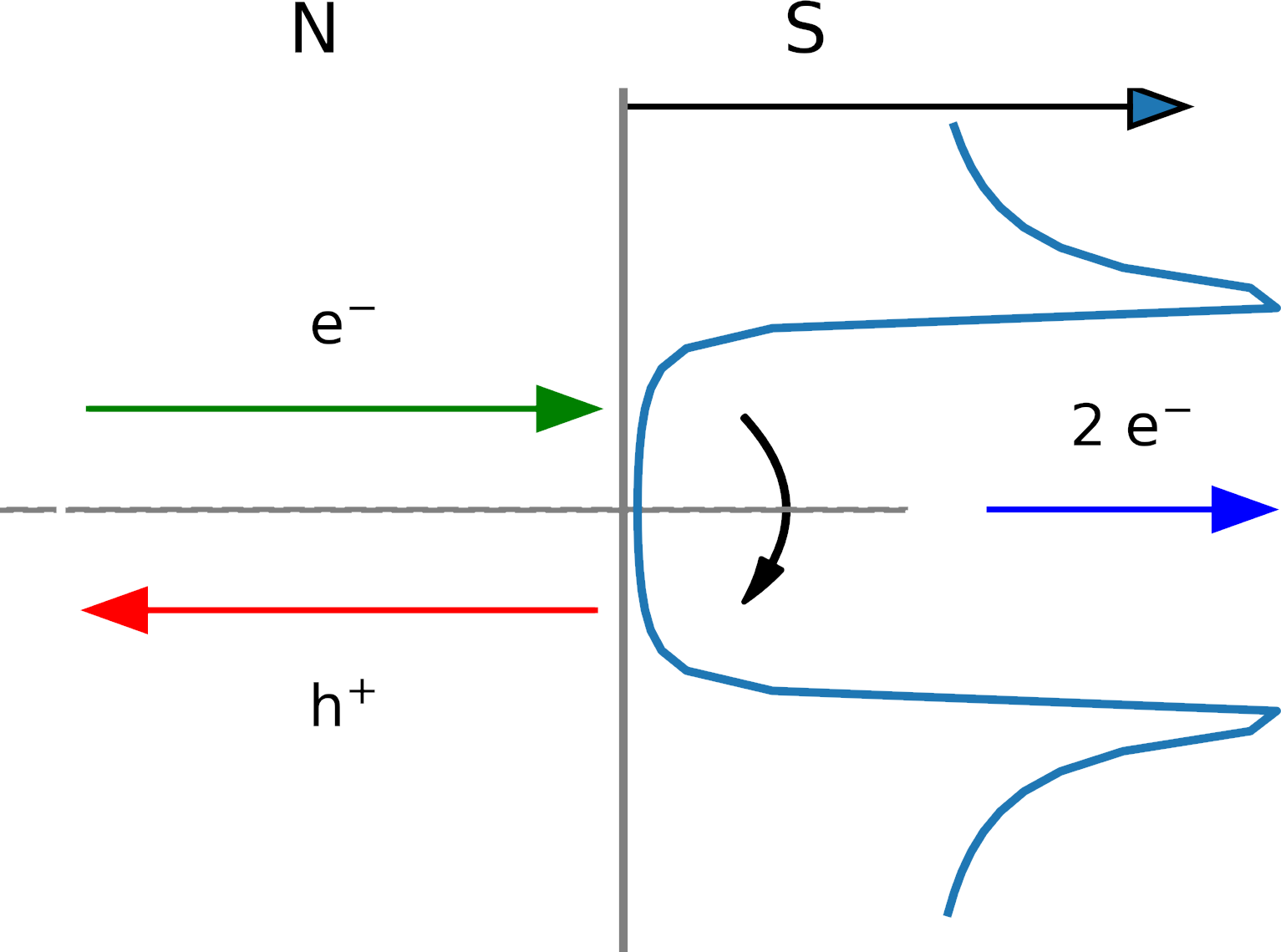}
				\end{subfigure}
				\caption{\label{fig:andreev_4}\B{(Left)} The sub-gap conductance that appears at negative gate voltages can be explained by the proximity effect: the ``leaking'' of superconductivity from the superconductor (S) into the normal material (N), here represented by the Ginzburg-Landau order parameter $\psi(x)$~\cite{landau1950k,landau2009theory,fossheim2004superconductivity}. \B{(Right)} Cartoon illustration of Andreev reflection: an incoming electron is reflected as a hole, introducing a Cooper pair inside the superconductor. The blue curve on the right represents the superconducting DOS.}
			\end{figure}
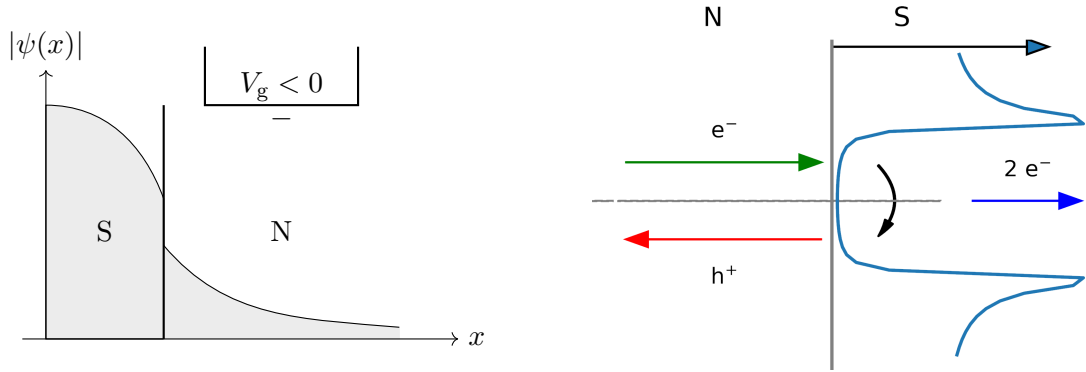

			Additionally, at larger negative gate voltages in Fig.~\ref{fig:D61D4Gdiff_Vd_combo_plot3}, we see the appearance of sub-gap conductance.
			Since no quasiparticle states are available in that energy range on either side of the channel, this must be due to a different mode of transport.
			When Andreev~\cite{andreev1964thermal} solved the Gor'kov equations~\cite{gor1958energy}, he found that an incident particle with an energy below the gap could be reflected as a hole, and vice versa, resulting in the effective transmission of regular quasiparticles as superconducting Cooper pairs, thus contributing to current within the gap.
			We now refer to this process as Andreev reflection, which has the additional feature that the paired electron and hole inside the normal material briefly maintain their coherence.
			This is associated with a minigap in the density of states (DOS) inside the normal material~\cite{belzig1996local} that decays over a distance that depends on the energy scale $E$ of the decohering processes (see section~\ref{sec:energy_length_time}),
			\begin{equation}\xi_\text{ballistic}=\dfrac{\hbar\,\vF}{2\pi\,E},\qquad \xi_\text{diffusive}=\sqrt{\dfrac{\hbar\,D}{2\pi\,E}}.\end{equation}
			This opening of a gap close to the boundary, which in the case of a perfectly transparent interface smoothly connects to the gap inside the superconductor, is commonly referred to as the proximity effect~\cite{de1999superconductivity}.
			The process is illustrated in Fig.~\ref{fig:andreev_4}, and explains the sub-gap conductance observed in Fig.~\ref{fig:D61D4Gdiff_Vd_combo_plot3}.
			
\begin{figure}
				\centering
				\begin{subfigure}[b]{0.46\textwidth}
					\includegraphics[width=\textwidth]{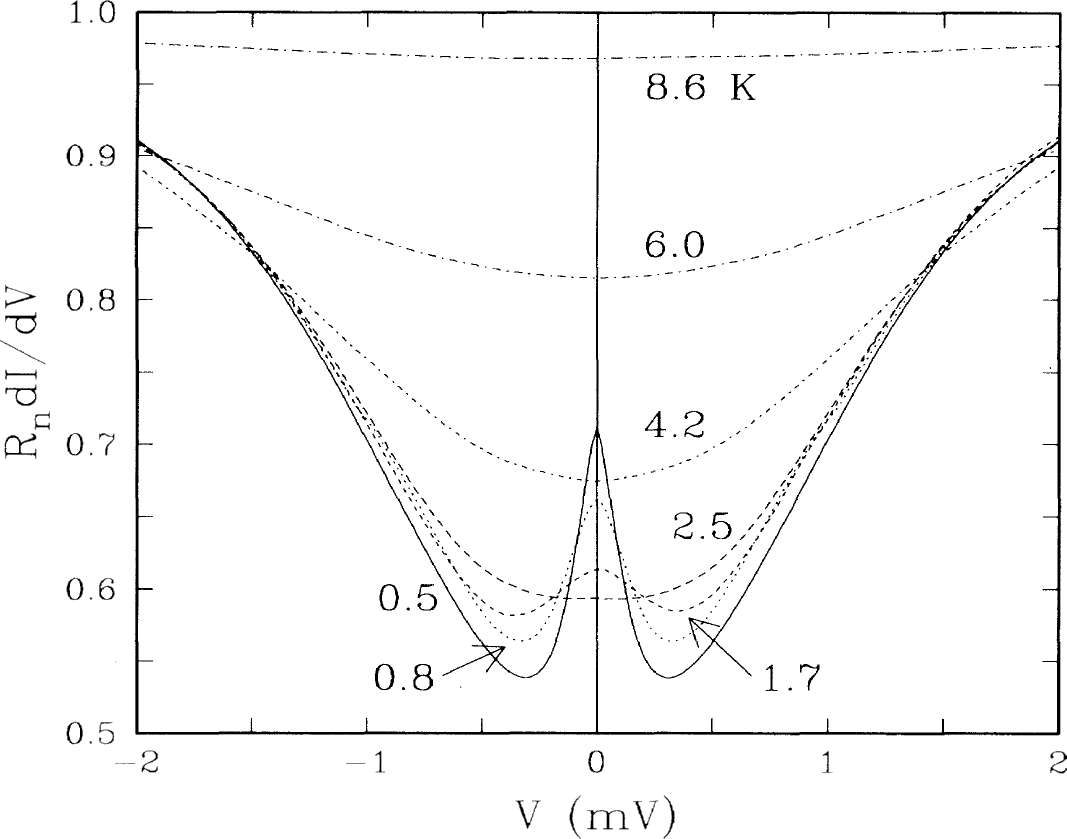}
				\end{subfigure}\hfill\begin{subfigure}[b]{0.50\textwidth}
					\includegraphics[width=\textwidth]{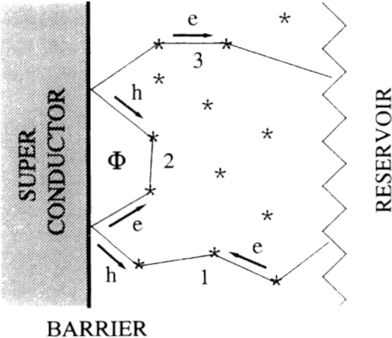}
				\end{subfigure}
				\caption{\label{fig:reflectionless_tunneling}\B{(Left)} Kastalsky et al observed a zero-bias peak in an SN junction --- i.e. a junction \emph{without} a second superconducting lead, and attributed it to a pair current~\cite{kastalsky1991observation}. \B{(Right)} Four months later, a team in Groningen submitted their explanation to the same journal~\cite{van1992excess}: the conductance could be enhanced by interacting with the interface multiple times before coherence is lost, by scattering many times inside the dirty normal material. We now refer to this process as ``reflectionless tunneling''.}
			\end{figure}

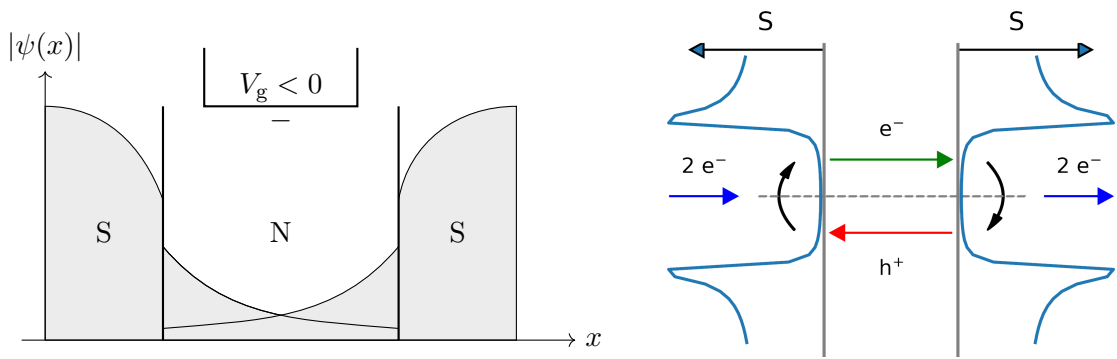
\begin{figure}
				\centering
				\begin{subfigure}[b]{0.56\textwidth}
					\begin{tikzpicture}[x=1.55cm,y=1.55cm]
						\draw[fill={gray!15}] (5,2) to[out=0,in=110] (6,1.2) -- (6,0) -- (5,0) -- cycle;
						\draw[fill={gray!15}] (6,0.8) to[out=-50,in=175] (8,0.1) -- (8,0) -- (6,0) -- cycle;
						\draw[fill={gray!15}] (8,0.8) to[out=-130,in=5] (6,0.1) -- (6,0) -- (8,0) -- cycle;
						\draw (6,0.8) to[out=-50,in=175] (8,0.1);
						\draw[fill={gray!15}] (9,2) to[out=180,in=80] (8,1.2) -- (8,0) -- (9,0) -- cycle;
						
						\draw[thick] (6,0) -- (6,2);
						\draw[thick] (8,0) -- (8,2);
						\draw (5.5,0.75) node[anchor=south]{S};
						\draw (7,0.75) node[anchor=south]{N};
						\draw (8.5,0.75) node[anchor=south]{S};
						
						\draw[style=thick] (6.35,2.5) -- (6.35,2) -- (7.65,2) -- (7.65,2.5);
						\draw (7,1.95) node[anchor=south] {$V_\text{g}<0$};
						\draw (7,2.05) node[anchor=north] {$-$};
						
						\draw[->] (5,0) -- (5,2.3) node[anchor=south] {$|\psi(x)|$};
						\draw[->] (4.8,0) -- (9.5,0) node[anchor=west] {$x$};
					\end{tikzpicture}
				\end{subfigure}\hfill\begin{subfigure}[b]{0.4\textwidth}
					\includegraphics[width=\textwidth]{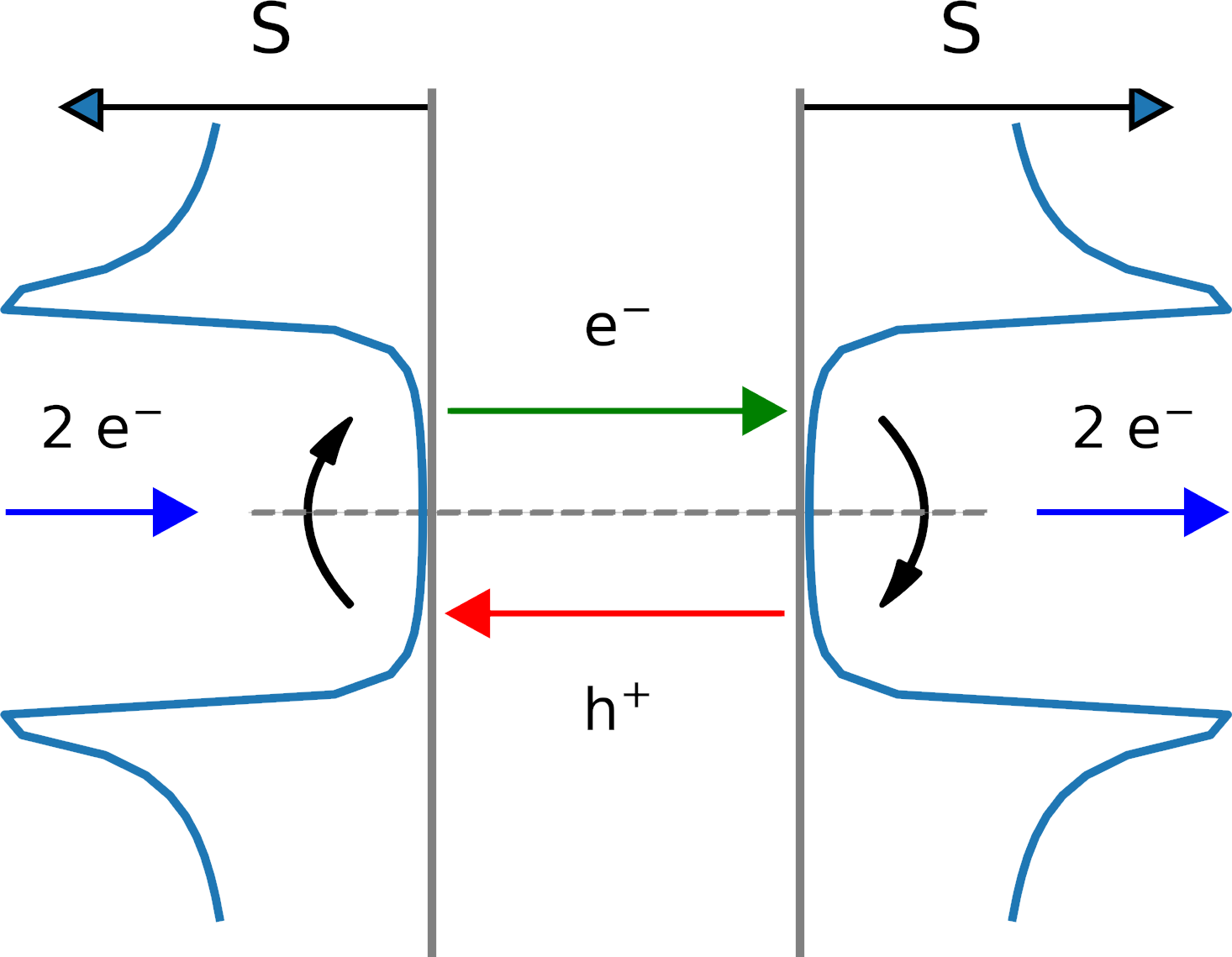}
				\end{subfigure}
				\caption{\label{fig:coherent_transport}\B{(Left)} As explained by de Gennes~\cite{de1964boundary}, a Josephson coupling can be established when the induced pair potentials from the two superconductors overlap inside the normal material. \B{(Right)} A Cooper pair entering the normal metal is transferred as an electron-hole pair, before forming a new Cooper pair on the other side. Note that the hole is moving in the opposite direction: the hole component of the pair travels as an electron, but ``backward in time''~\cite{de1994andreev,hoevers1988determination}. To be clear: in terms of a single reflection, the hole traces back the electron path \emph{after} the electron has been absorbed by the superconductor~\cite{de1994andreevsup}.}
			\end{figure}

			Since the differential conductance of a single S/N interface in the absence of any scattering is exponentially suppressed at $V=0$~\cite{tinkham2004introduction}, the zero-bias conductance peak (ZBCP) cannot be explained by Andreev reflection alone.
			Two different mechanisms that contribute to zero-bias conductance are relevant in the current situation.
			Either transmission is enhanced across each interface individually, which can be done by successive reflections inside the normal material~\cite{kastalsky1991observation,van1992excess}, or a zero-resistance channel is created by coherent transport between the two interfaces~\cite{de1964boundary,kleinsasser1991critical,dubos2001josephson}.
			These two explanations for the ZBCP are shown in Figs.~\ref{fig:reflectionless_tunneling} and~\ref{fig:coherent_transport}.

\begin{figure}
				\centering
				\begin{subfigure}[b]{0.48\textwidth}
					\includegraphics[width=\textwidth]{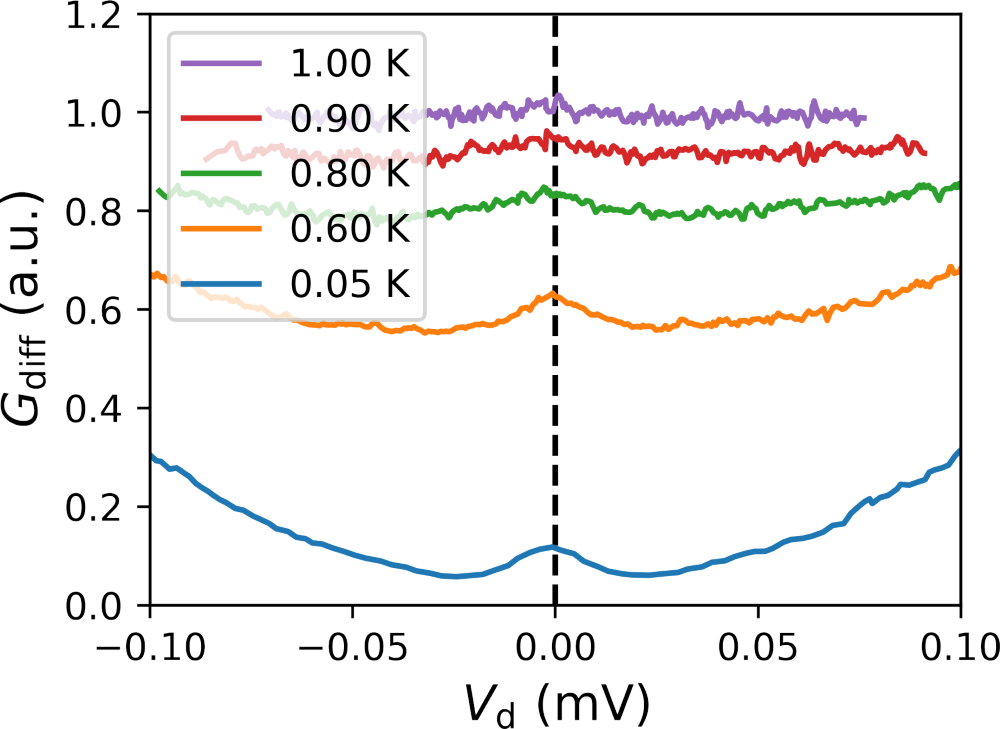}
				\end{subfigure}\hfill\begin{subfigure}[b]{0.48\textwidth}
					\includegraphics[width=\textwidth]{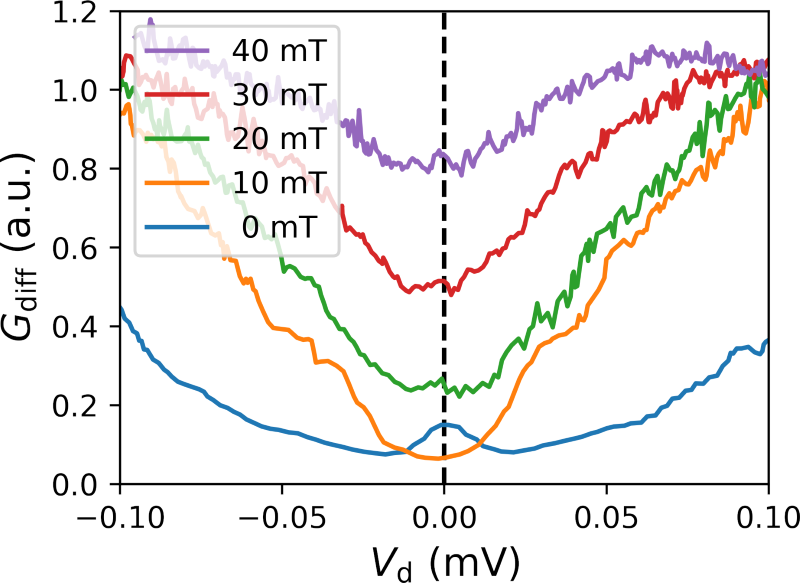}
				\end{subfigure}
				\vspace*{0.5\baselineskip}
				
				\begin{subfigure}{\textwidth}
					\centering
					\includegraphics[width=0.98\textwidth]{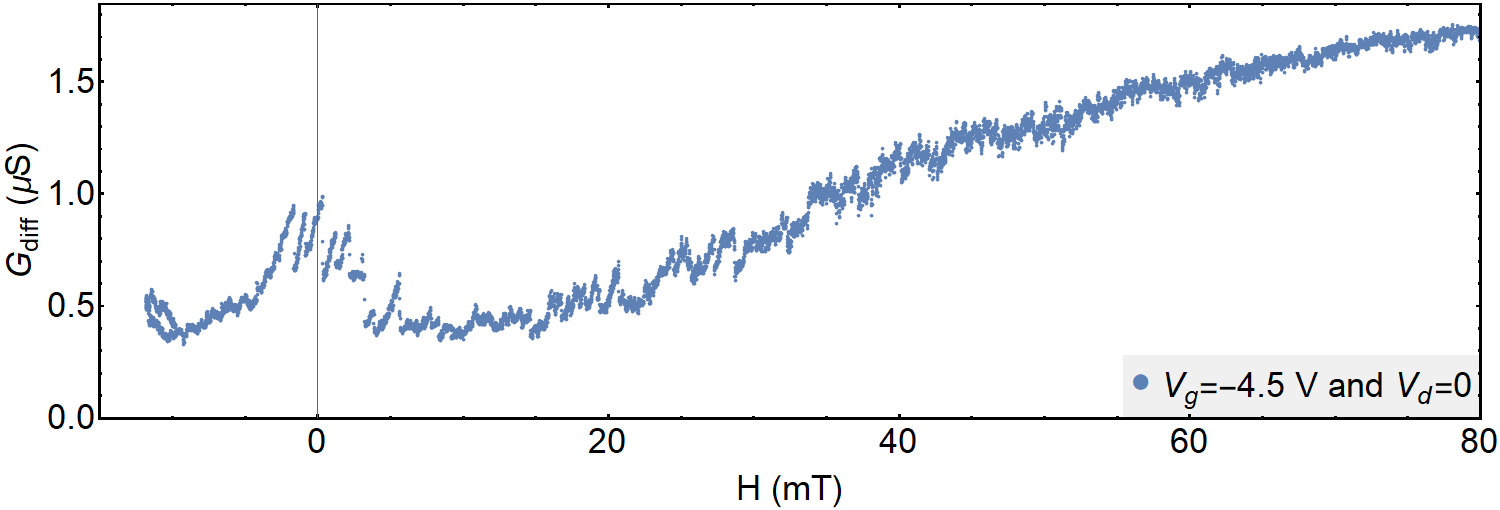}$\;$
				\end{subfigure}
				\caption{\label{fig:D63D3Gdiff_T_H}Data from device D63D3. \B{(Top left)} The differential conductance is plotted for a small $V_\text{d}$ range within the gap, at $V_\text{g}=\SI{-4.0}{\volt}$. The zero-bias conduction peak disappears as the gap closes. \B{(Top right)} The differential conductance is shown for a range of applied magnetic field strengths, this time at $V_\text{g}=\SI{-4.5}{\volt}$. The critical field for thin films of PtSi (a type-II superconductor) was estimated to be \SI{84}{\milli\tesla}~\cite{oto1994superconductivity}. \B{(Bottom)} The differential conductance, again at $V_\text{g}=\SI{-4.5}{\volt}$, but now at fixed $V_\text{d}=\SI{0}{\milli\volt}$ versus the applied magnetic field. The zero-bias conductance peak is suppressed at around $H\approx\SI{10}{\milli\tesla}$.}
			\end{figure}

			Scans in temperature and field were also used to investigate the nature of the ZBCP observed below $V_\text{g}=\SI{-3.75}{\volt}$.
			As can be seen in Fig.~\ref{fig:D63D3Gdiff_T_H}, this peak diminishes but survives up to $T=T_\text{c}$, while it is already suppressed with magnetic fields far below $H=H_\text{c,2}$.
			This is consistent with both reflectionless tunneling and coherent transport.
			While small magnetic fields are not expected to directly impact the proximity effect at each individual S/N interface~\cite{andreev1964thermal}, the total phase difference acquired by an electron-hole pair can be multiplied by repeated reflections~\cite{kastalsky1991observation,van1992excess}.
			In the case of coherent transport, a single flux quantum would suffice to suppress an inhomogeneous supercurrent across the channel.
			Taking $W=\SI{2.5}{\micro\metre}$ and $L=\SI{50}{\nano\metre}$ for the device considered in Fig.~\ref{fig:D63D3Gdiff_T_H}, we arrive at an equivalent field of
			\begin{equation}\dfrac{\Phi_0}{A_\text{juction}}=\dfrac{h/2e}{\SI{2.5}{\micro\metre}\times\SI{50}{\nano\metre}}=\SI{17}{\milli\tesla},\end{equation}
			in the right ballpark for suppression above \SI{10}{\milli\tesla}.
			Note that the channel length is not exactly known as discussed in section~\ref{sec:laurie_sample_description}, and that flux focusing would increase the effective field in the channel.
			However, direct evidence of a supercurrent through the channel has not yet been observed.
			A back-of-the-envelope calculation, multiplying width (\si{\volt}) and height (\si{\siemens}=\si{\ampere}/\si{\volt}) of the ZBCP, suggests that such a measurement needs to be sensitive to $10^{-10}\si{\ampere}$ if performed with the same setup, while improving noise filtering may further increase the critical current itself.
			
		\subsection{Gate modulation of the proximity effect}
			
			It is clear from Fig.~\ref{fig:D61D4Gdiff_Vd_combo_plot3} that the behavior of the device changes with gate voltage.
			We would like to make this more precise; find out whether the transport is limited by the channel or the transparency of the interfaces, how the Schottky barrier impacts that transparency, and then get an idea of what would need to be changed in the fabrication to make these transistors suitable for integration in transmons.
			Central to this analysis is the characterization of the interface, for which we will rely on the methodology introduced by Blonder, Tinkham and Klapwijk (BTK).
			
\begin{figure}
				\centering
				\includegraphics[width=0.8\textwidth]{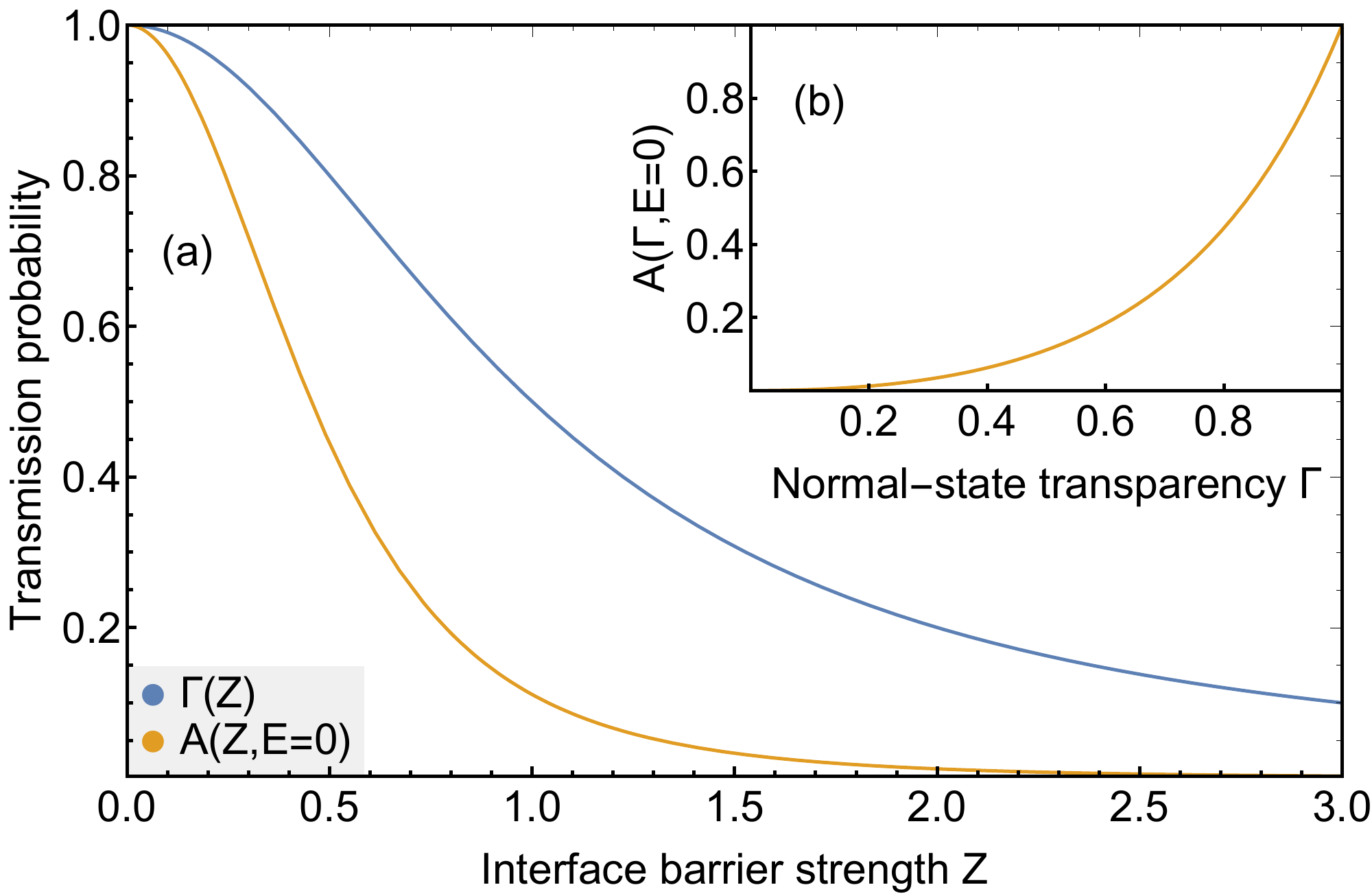}
				\caption{\label{fig:TransmissionProbabilityPlotBTK}\B{(a)} The parameters $\Gamma$ and $A$ are both relevant in describing the interface. The normal-state transparency $\Gamma$ gives the probability of regular electron transmission in the absence of superconductivity. Transfer of Cooper pairs is better described by the Andreev reflection probability $A$. \B{(b)} The parameter $A(E=0)$ at the Fermi level can be expressed directly in terms of $\Gamma$.}
			\end{figure}

			In their seminal 1982 paper~\cite{blonder1982transition}, BTK provided, among many other things, a simple method for extracting the interface barrier strength\footnote{The model assumes a delta function potential $H\delta(x)$ at the interface, that collects all possible causes of normal (i.e. non-Andreev) reflection: Schottky barrier, mismatch in effective electron mass or lattice parameter, grain boundaries, oxides etc.} $Z$ from the $I$--$V$ characteristics of an S/N junction.
			To extend this to our case, we will assume that our S/N/S junction is symmetric, such that exactly half the voltage drop occurs on either side, and treat the system as having a single interface with twice the resistance and half the bias.
			We refer to the probability of transmitting a particle as the \emph{transparency} of such a barrier, which at an S/N interface of course depends strongly on its energy --- recall the blocked quasiparticles at $eV<\Delta$ in Fig.~\ref{fig:Superconducting_DOS}.
			We therefore cannot just put a single number on the transparency for superconducting transport, and instead we customarily state its value in the normal state, which we call $\Gamma$ ($C$ in BTK notation),
			\begin{equation}\Gamma=\dfrac{1}{Z^2+1}:\qquad \lim_{Z\rightarrow0}\Gamma=1,\qquad\lim_{Z\rightarrow\infty}\Gamma=0.\end{equation}
			The Josephson effect is of course due to the transfer of Cooper pairs around the Fermi level ($E\ll\Delta$), and not to the transfer of normal electrons outside the gap, so we may gain some insight by discussing also the probability that Andreev reflection occurs.
			This, in turn, is given by BTK's coefficient $A$,
			\begin{equation}\begin{array}{r@{\;}c@{\;}l}
				\displaystyle\lim_{E\rightarrow0}A(Z)		& =	& \displaystyle\lim_{E\rightarrow0}\dfrac{\Delta^2}{E^2+(\Delta^2-E^2)(1+2Z^2)^2}=\dfrac{1}{(1+2Z^2)^2},\\\\
				\displaystyle\lim_{E\rightarrow0}A(\Gamma)	& =	& \left(1+\left|\dfrac{2}{\Gamma}-2\right|\right)^{-2}.
			\end{array}\end{equation}
			Shown in Fig.~\ref{fig:TransmissionProbabilityPlotBTK} are the relationships between $A$ at $V_\text{d}=0$ and $\Gamma$, and the barrier strength $Z$.
			
\begin{figure}
				\centering
				\includegraphics[height=0.8\textheight]{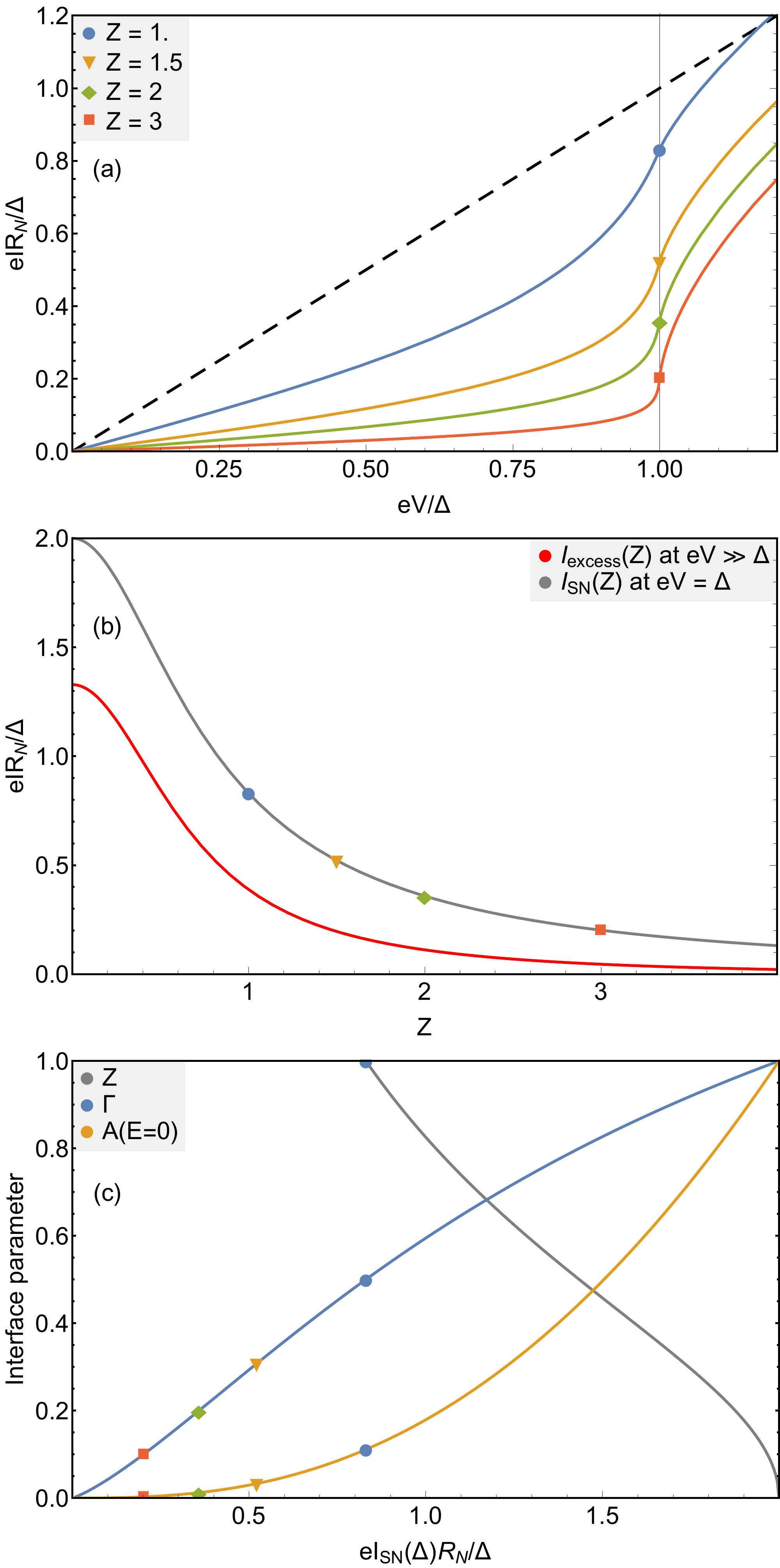}
				\caption{\label{fig:BTKExtractZCombinedPlot}\B{(a)} Normalized current $eI_\text{SN}R_\text{N}/\Delta$ versus applied bias voltage for selected values of $Z$. Markers indicate the current at $eV=\Delta$. \B{(b)} The \Bred{excess current} at $eV\gg\Delta$ as described by BTK, as well as the \B{\textcolor{gray}{current at $eV=\Delta$}}. The same markers are indicated. \B{(c)} Swapping the axes of the previous plot, we can find the interface parameters $Z$, $\Gamma$ and $A$ by solving $I_\text{SN}(Z)$ for $Z$. Note that for opaque interfaces, the Andreev reflection rate $A$ grows much more slowly with $I_\text{SN}(\Delta)$ than the normal-state interface transparency $\Gamma$. E.g., a moderate $Z=1.5$ gives $\Gamma=0.31$, while $A(Z=1.5,E=0)=0.03$.}
			\end{figure}

			In the simpler case (not ours) where the normal-state density of states has no additional features, we can at zero temperature express the current through an interface as
			\begin{equation}\label{eq:btk_eq17}I_\text{NS}(T=0,eV)=\dfrac{1+Z^2}{eR_\text{N}}\Int_0^{eV}\big[1+A(E,Z)-B(E,Z)\big]dE,\end{equation}
			which is plotted in the range of $0\leq eV\leq1.2\Delta$ for a selection of $Z$ values in Fig.~\ref{fig:BTKExtractZCombinedPlot}a.
			This current $I_\text{NS}(V)$ has a slope that always tends to $1/R_\text{N}$ at large voltages, and has an offset from $V/R_\text{N}$ that depends only on $Z$, allowing one to quickly extract the transparency of the interface either by drawing a tangent that intersects with the $I$ axis, or by subtracting $V/R_\text{N}$ to get the excess current,		
			\begin{equation}\begin{array}{r@{\;}c@{\;}l}
				I_\text{excess}	& =	& \left(I_\text{NS}-I_\text{NN}\right)\big|_{eV\gg\Delta}\\\\
											& =	& \dfrac{1+Z^2}{eR_\text{N}}\,\Int_0^{eV}\left[A(E,Z)-B(E,Z)+\dfrac{Z^2}{1+Z^2}\right]\,dE.
			\end{array}\end{equation}
			This excess current is plotted as a function of $Z$ in \Bred{red} in Fig.~\ref{fig:BTKExtractZCombinedPlot}b.
			
			\begin{figure}
				\centering
				\includegraphics[width=\textwidth]{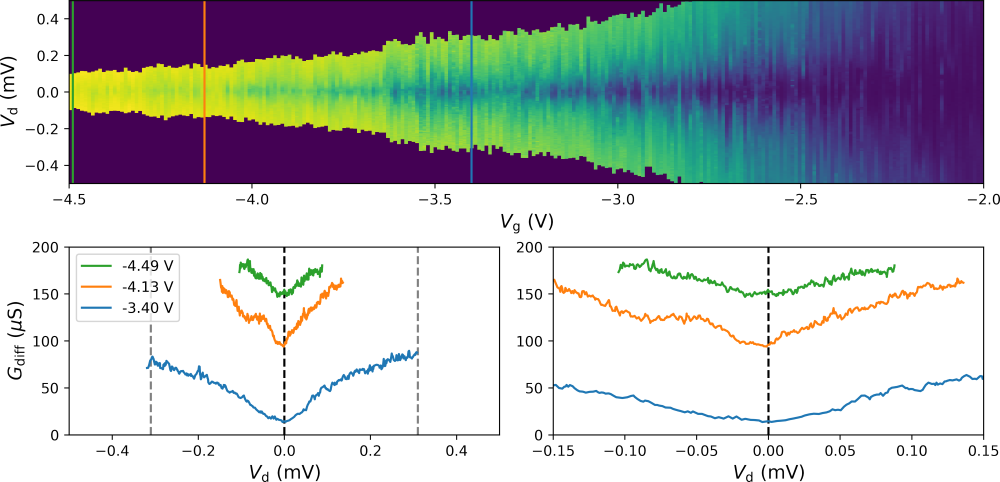}
				\caption{\label{fig:D63D3S11Gdiff_Vd_combo_plot}Data from device D63D3, $H=\SI{100}{\milli\tesla}>H_0$. A clear non-linearity can be observed in the differential conductance even when the superconductors in the source and drain are in the normal state.}
			\end{figure}
			
			This method of extracting $Z$ has a number of complications.
			In short, three conditions need to be met in order to reliably extract the interface transparency from the excess current:
			\begin{enumerate}
				\item The integral in eq.\eqref{eq:btk_eq17} needs to be valid, for which the normal-state conductance needs to be independent of $V_\text{d}$ around $E_\text{F}$;
				\item The error in $I_\text{SN}/I_\text{NN}$ should be much smaller than the desired accuracy in $Z$ (e.g., by a factor $6$ when $Z\approx2$)\footnote{See Fig.~\ref{fig:BTKExtractZCombinedPlot}b: since $I_\text{excess}$ falls off quickly with $Z$, a small error in the estimate of the normal-state resistance will lead to large uncertainty in $Z$, especially when the interface has a low transparency.}; and
				\item Both $I_\text{SN}$ and $I_\text{NN}$ should be known at $eV_\text{d}\gg\Delta$.
			\end{enumerate}
			Unfortunately, none of these were met in our experiments.
			Fig.~\ref{fig:D63D3S11Gdiff_Vd_combo_plot} shows the normal-state conductance for a similar device, which features a strong dip around zero bias, invalidating the BTK assumption of flat normal-state conduction.
			Second, as is clearest in Fig.~\ref{fig:D63D4S02_DUT}, the conductance is highly sensitive to small variations in gate voltage, while hysteresis (likely due to trapped charges in the oxide) has been observed in the gate field.
			This means that we can never directly compare superconducting and normal-state scans at $H=0$ and $H>H_\text{c}$, nor is it practical to fix a gate voltage and then change the field (you can only do this once, and then you need to heat up the coil recover $H=0$ exactly).
			Third, we never did scans wide enough to reach $eV>50\times2\Delta$~\cite{blonder1982transition}, at which point $I_\text{excess}/I_\text{total}$ would be smaller than our signal-to-noise ratio.

			Instead, a qualitative method was developed that, like the excess current in this situation, does not per se give an accurate estimate for $Z$ for a single given gate voltage, but that provides a way of comparing the $Z$ values for different $V_\text{g}$'s with limited data.
			This can then give us an idea of whether we are able to tune the transparency with the gate, by estimating the barrier strength $Z$ from only a single measurement in the superconducting state.

			\begin{figure}
				\centering
				\begin{subfigure}[b]{0.48\textwidth}
					\centering
					\includegraphics[width=\textwidth]{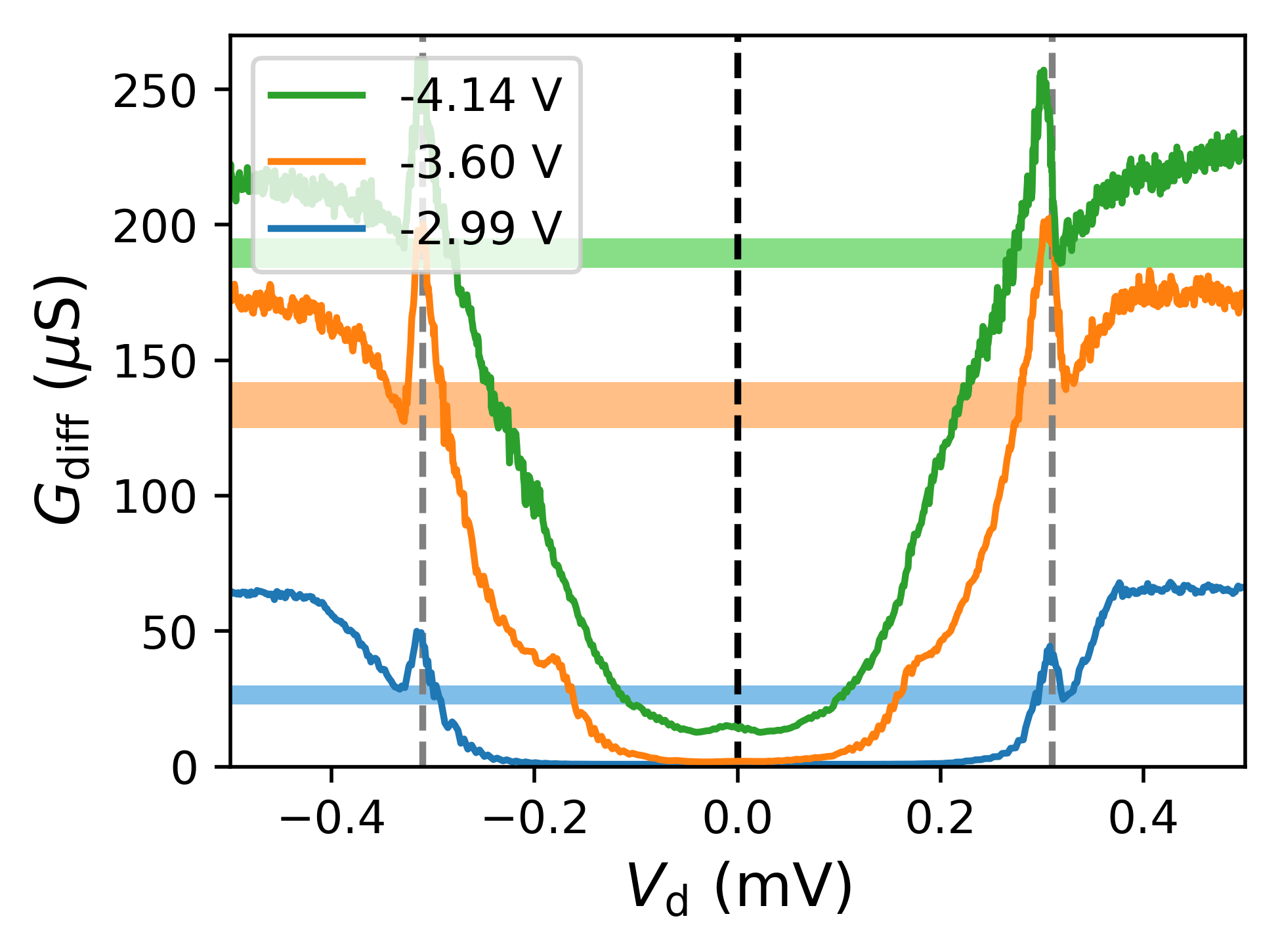}
				\end{subfigure}\hfill\begin{subfigure}[b]{0.48\textwidth}
					\centering
					\includegraphics[width=\textwidth]{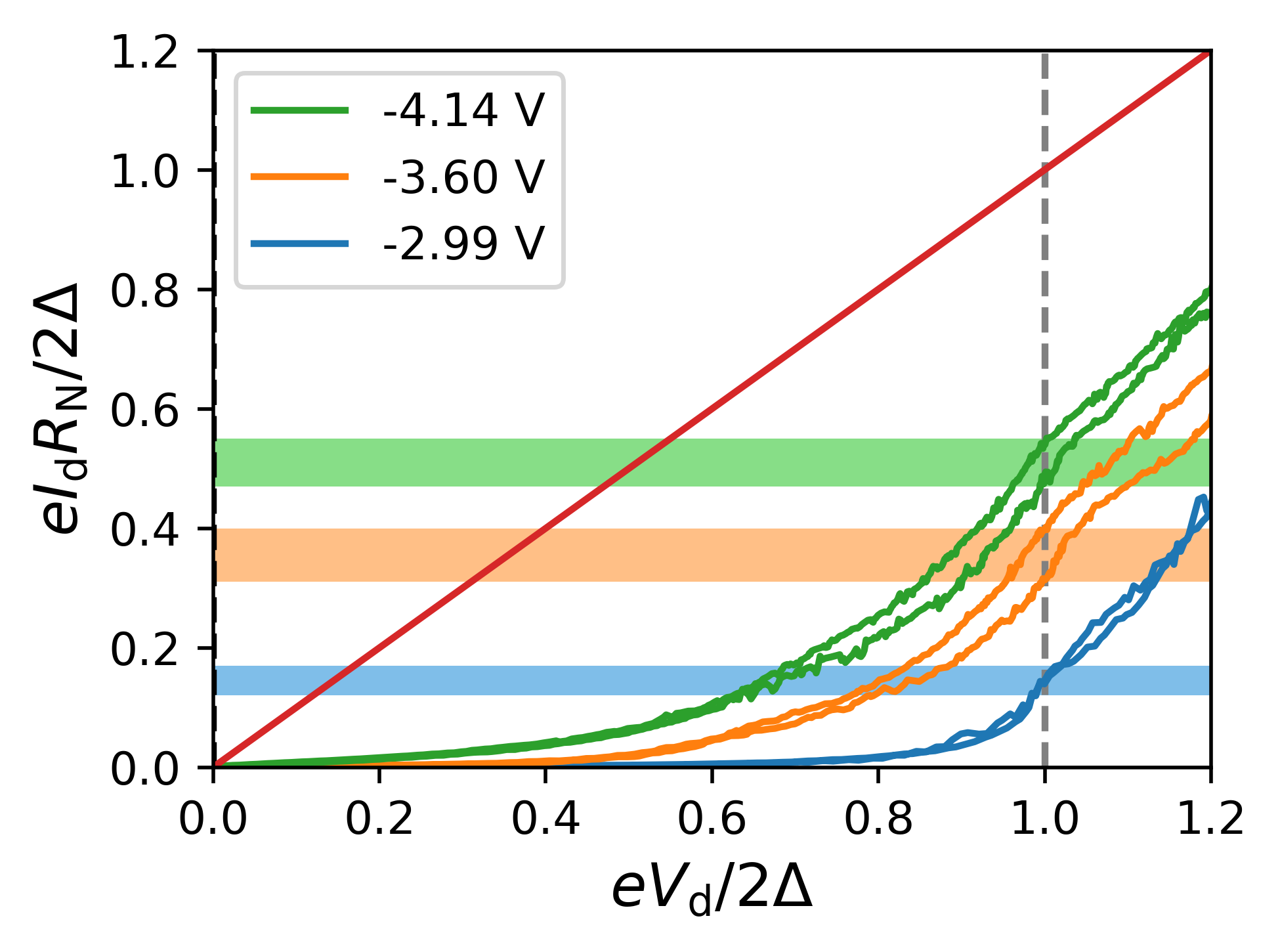}
				\end{subfigure}
				\caption{\label{fig:D61D4S01_Gdiff_I}\B{(Left)} The normal-state conductance $G_\text{diff,N}$ at $2\Delta$ can be approximated by taking the lowest value of $G_\text{diff}$ outside the coherence peaks, from which $eR_\text{N}(2\Delta)=2\Delta/G_\text{diff}$. Indicated in horizontal bars are the estimated ranges for $G_\text{diff,N}$. \B{(Right)} Using this estimate of $R_\text{N}$, we can now plot the normalized current $I_\text{SN}$ (data for negative bias voltages is mirrored and added). In this figure, horizontal bars represent the estimates for the current at $eV=2\Delta$.}
			\end{figure}
			
\begin{figure}
				\centering
				\includegraphics[width=\textwidth]{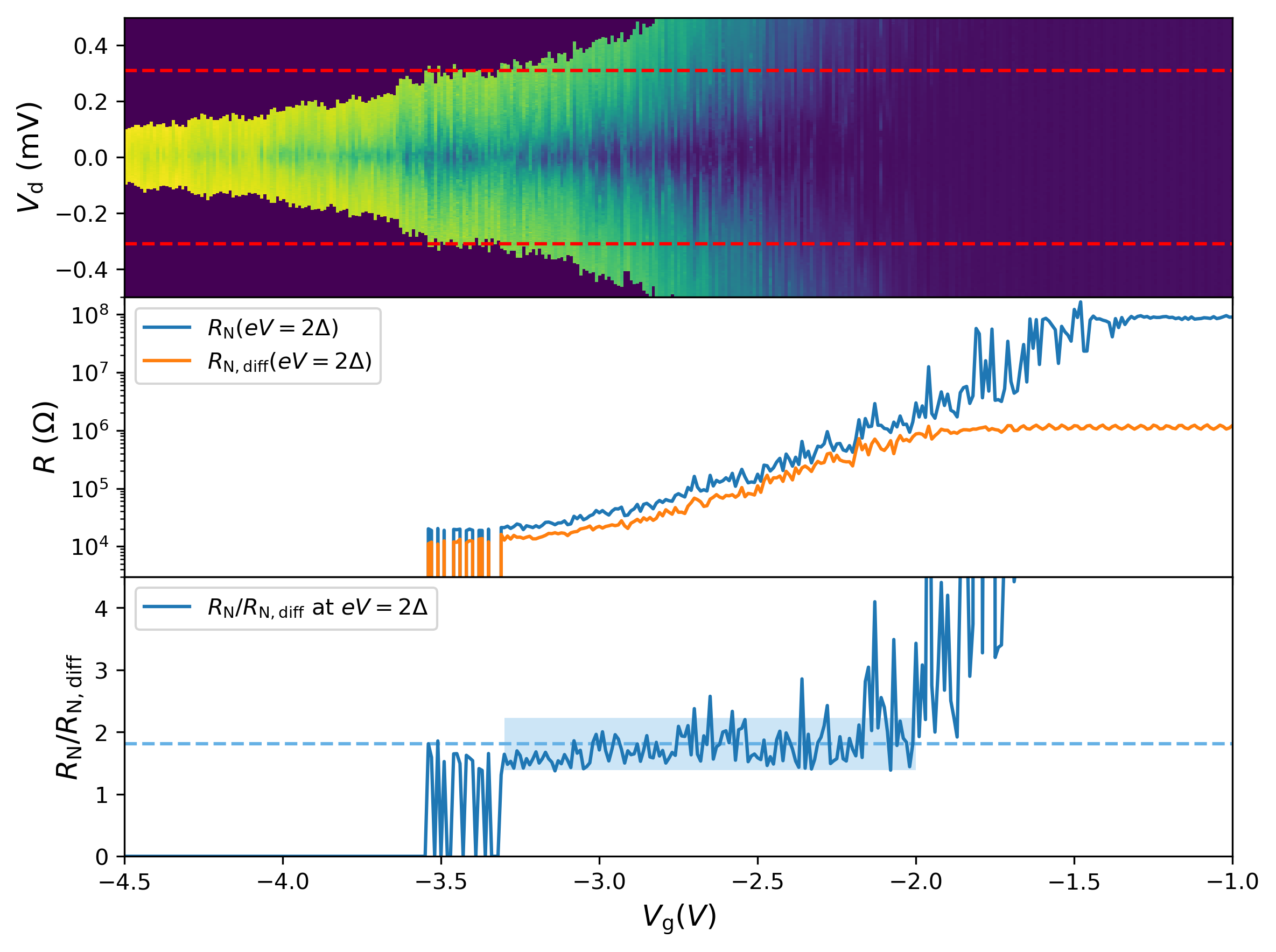}
				\caption{\label{fig:D63D3S11_Combo_Rdiff_RN_Vg_ratio}Data shown for D63D3, same measurement as shown in Fig.~\ref{fig:D63D3S11Gdiff_Vd_combo_plot}. \B{(Top)} The differential conductance versus $V_\text{g}$ and $V_\text{d}$, red dashed lines indicate $V=\pm2\Delta$. \B{(Middle)} The normal-state resistance at $eV=2\Delta$ is calculated from the current, $R_\text{N}=\Delta/eI$, while the differential resistance is simply $R_\text{N,diff}=1/G_\text{N,diff}$. \B{(Bottom)} In the range where we have data ($V_\text{g}\gtrsim\SI{-3.3}{\volt}$), and a decent signal-to-noise ratio ($V_\text{g}\lesssim \SI{-2.0}{\volt}$) (indicated in light blue), $R_\text{N}$ and $R_\text{N,diff}$ are off by about a factor $1.8(4)$.}
			\end{figure}

			Shown in in Fig.~\ref{fig:D61D4S01_Gdiff_I} are selected curves of the differential conductance and normalized current in device D61D4, from the same measurement as shown in Fig.~\ref{fig:D61D4Gdiff_Vd_combo_plot3}.
			Sub-gap conductance increases with applied gate voltage, as expected for an increasingly transparent interface.
			Now, we know that expression~\eqref{eq:btk_eq17} for the current at the gap (or twice the gap in our SNS junction) is not valid, since the current in both the superconducting and normal state is suppressed by the process\footnote{At the moment, we suspect that this dip is due to a dynamical Coulomb blockade.} that gives rise to the dip around $V=0$ in Fig.~\ref{fig:D63D3S11Gdiff_Vd_combo_plot}.
			But what's important, is that we can be reasonably sure that it remains a smoothly varying one-to-one relationship on larger $V_\text{g}$ scales\footnote{See Figs.~\ref{fig:D63D3S11Gdiff_Vd_combo_plot} and~\ref{fig:D63D4S02_DUT}: charging effects on the order of $\Delta V_\text{g}\approx\SI{5}{\milli\volt}$ will affect our estimates of $R_\text{N}$ and $I_\text{NS}$ differently. This doesn't matter when we just want to know the trend over a wider range in $V_\text{g}$.}, even when we replace $R_\text{N}\rightarrow R_\text{N,diff}$,
			\begin{equation}\label{eq:btk_eq17_diff}I_\text{NS}(T=0,eV)\approx\dfrac{1+Z^2}{eR_\text{N,diff}}\Int_0^{eV}\big[1+A(E,Z)-B(E,Z)\big]dE.\end{equation}
			We know that $R_\text{N}\neq R_\text{N,diff}$, shown in Fig.~\ref{fig:D63D3S11_Combo_Rdiff_RN_Vg_ratio}, but we will here make the approximation that $R_\text{N,diff}$ and $I_\text{NS}$ are suppressed by about the same amount within the gap.
			Using $R_\text{N,diff}$ essentially allows us to approximately compensate for the depression in $R_\text{N}$.
			This will be an overcompensation, since $R_\text{N}G_\text{diff,S}$ increases between $eV=0$ and $eV=2\Delta$ for any value of $Z$~\cite{blonder1982transition}, such that conductance near the edge of the gap makes up a larger relative share of total current in the superconducting state than it does in the normal state.
			This means that the normalized $eIR_\text{N,diff}/2\Delta$ that we will calculate at $eV=2\Delta$ will be a lower bound.
			
			To estimate $R_\text{N,diff}$ from data in the superconducting state, we make one further approximation, justified by earlier data on PtSi transistors~\cite{schwarz2021physics}, that
			\begin{equation}G_\text{diff,N}(eV=2\Delta)\approx G_\text{diff,S}(eV=2\Delta+\delta),\end{equation}
			where $\delta$ is a small shift in voltage from the coherence peak, such that we take a $G_\text{diff,S}(V)$ value just outside the gap, at the point where it is smallest.
			For clarity, this is drawn in Fig.~\ref{fig:D61D4S01_Gdiff_I} as shaded horizontal bars that intersect the minimum of $G_\text{diff,S}(V)$ outside the gap.
			Once we have this estimate for $R_\text{N,diff}$, we can proceed to normalize the measured current at $eV=2\Delta$. 
			
\begin{figure}
				\centering
				\includegraphics[width=\textwidth]{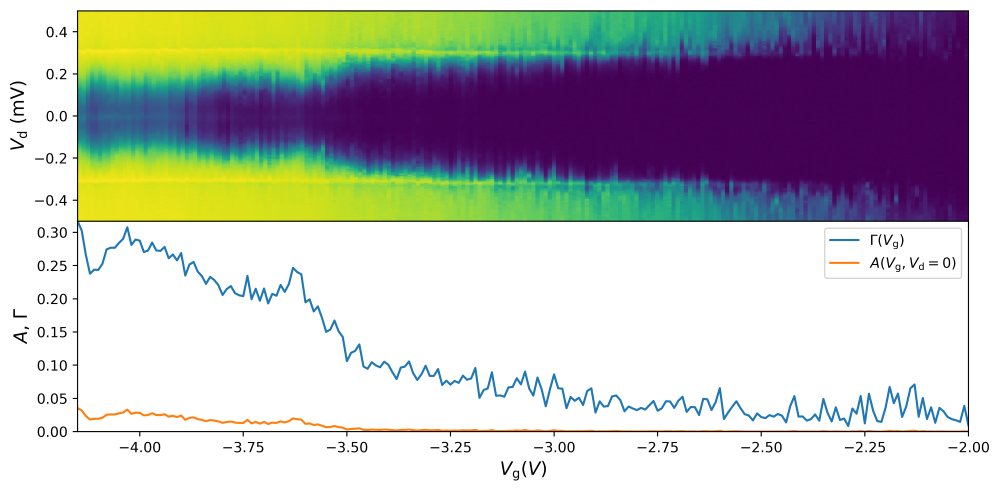}
				\caption{\label{fig:D61D4S01_Gdiff_interface}Data from device D61D4. \B{(Top)} The differential conductance of device D61D04 versus gate and drain voltage. On the left, around $V_\text{d}=0$, a zero-bias conductance peak (ZBCP) can be observed. \B{(Bottom)} Around the same range in $V_\text{g}$, the extracted $A(E=0)$ increases, consistent with this peak being due to the proximity effect.}
			\end{figure}

			Integrating the modified expression in eq.~\eqref{eq:btk_eq17_diff} to $eV=\Delta$ (not $eV=2\Delta$, BTK's formula is for a single interface) yields the following,
			\begin{equation}\label{eq:theoretical_isn}I_\text{SN}(T=0,eV=\Delta)=\dfrac{\Delta}{eR_\text{N}}\,\dfrac{\sqrt{1+Z^2}}{Z(1+2Z^2)}\,\atanh\left(\dfrac{2Z\sqrt{1+Z^2}}{1+2Z^2}\right),\end{equation}
			which is plotted in Fig.~\ref{fig:BTKExtractZCombinedPlot}b together with the excess current.
			Given an estimate for $I_\text{SN}$ at $eV=2\Delta$, we can then solve for $Z$, as is detailed in algorithm~\ref{alg:estimatez}.
			The last step, solving the equality, is done by minimizing the error using Brent's algorithm~\cite{brent1971algorithm} with the built-in function \texttt{minimize\_scalar()} from the Python library scipy, bounded between $Z=0$ and $Z=100$.
			Plotted in Fig.~\ref{fig:D61D4S01_Gdiff_interface} are the resulting estimates for the normal-state transparency $\Gamma$ and the probability of Andreev reflection at $eV=0$, given by BTK's $A(eV=0)$.
			
			\noindent\begin{minipage}{\textwidth}\begin{algorithm}[gobble=12,tabsize=4,caption={Estimate $Z$ from $G_\text{diff}(V)$ and $I(V)$.},label={alg:estimatez},mathescape]
			function Estimate_RN(Gdiff):
				GdiffWindowLeft 	$\leftarrow$ select(Gdiff,$-(1+\delta)\Delta<$eV$<-\Delta$)
				GdiffWindowRight 	$\leftarrow$ select(Gdiff,$\Delta$<eV<$(1+\delta)\Delta$)
				{RNLeft,RNRight}	$\leftarrow$ {1/min(GdiffWindowLeft),1/min(GdiffWindowLeft)}
				return mean(RNLeft,RNRight)
			
			function Estimate_Isn_at_$\Delta$(Isn):
				IsnLeft 			$\leftarrow$ -select(Isn,eV$=-\Delta$)
				IsnRight 			$\leftarrow$ select(Isn,eV$=+\Delta$)
				return mean(IsnLeft,IsnRight)
			
			function Theoretical_Isn(Z):
				return sqrt(1+Z^2)/(Z(1+2Z^2))atanh(2Z*sqrt(1+Z^2)/(1+2Z^2))
			
			RN 						$\leftarrow$ Estimate_RN(Gdiff(Vg))
			Isn						$\leftarrow$ Estimate_Isn_at\_$\Delta$(Isn)
			solve Theoretical_Isn(Z) = Isn$\times$RN/$\Delta$ for Z
			\end{algorithm}\end{minipage}
			
		\subsection{Estimation of the Schottky barrier height}
			
			An additional way that we can analyze the dependence of the interface transparency on variations in the applied gate voltage is by extracting the Schottky barrier height.
			As long as the temperature is high enough, transmission at the interface will mostly occur by thermionic emission, as opposed to regular or field-assisted tunneling\footnote{Thermionic emission of course has no role to play in the JoFET operation itself. Both the temperature and the superconducting gap are much smaller than the Schottky barrier, and Josephson coupling depends on transport at $E=eV_\text{d}=0$.}.
			Using eq.~\eqref{eq:thermionic_v} for the thermionic emission current, we can then relate the Schottky barrier height $\phi_\text{Schottky}$ to the increase in current with temperature,
			\begin{equation}\label{eq:arrhenius_sbh}\phi_\text{Schottky}=-k_\text{B}T\,\ln\left(\dfrac{I_\text{th}}{T^2}\right)+C,\quad\text{or}\quad\ln\left(\dfrac{I_\text{th}}{T^2}\right)=-\dfrac{\phi_\text{Schottky}}{k_\text{B}T}\,+C',\end{equation}
			such that $\phi_\text{Schottky}$ can be extracted as the slope in an Arrhenius plot of $\ln\left(I_\text{th}/T^2\right)$ versus $1/k_\text{B}T$.
			We will not attempt the more comprehensive analysis that takes into account the change in shape of the Schottky barrier with gate voltage~\cite{rideout1970effects}.
			
\begin{figure}
				\centering
				\includegraphics[width=0.8\textwidth]{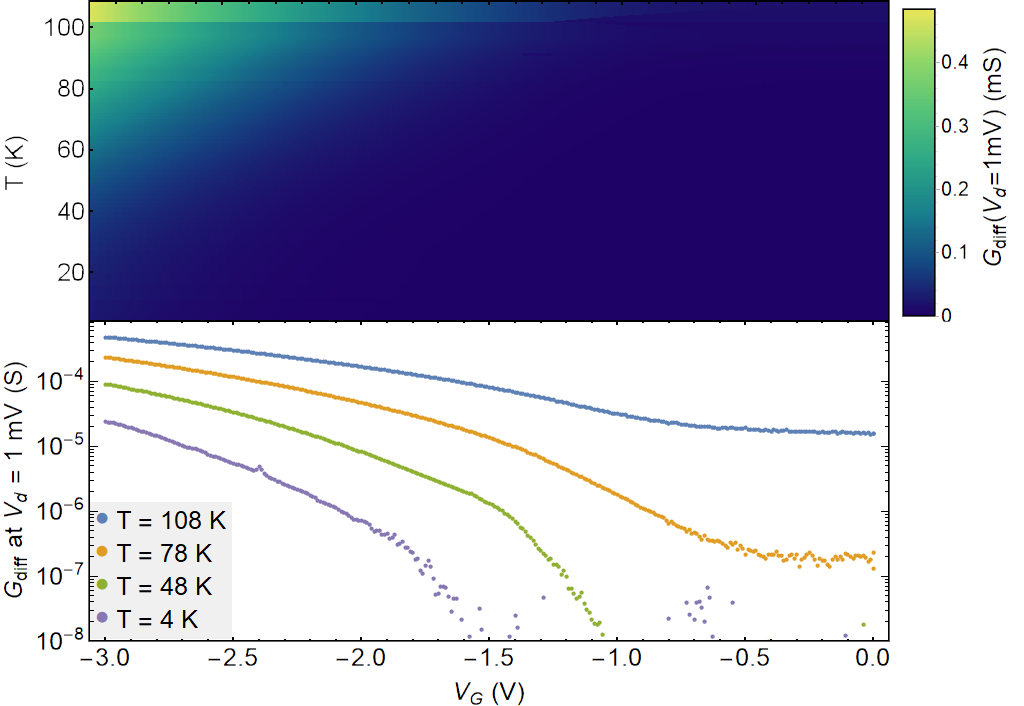}
				\caption{\label{fig:D84D3TSweepCombinedPlot}Data shown for device D84D3. The measurement was performed over a period of 1 week,
				and also gives some insight into the quality of the thermal insulation of the cryostat. \B{(Top)} The differential conductance at $V_\text{d}=\SI{1}{\milli\volt}$ for $\SI{-3}{\volt}\leq V_\text{g}\leq\SI{0}{\volt}$ and temperatures from \SI{4}{\kelvin} to \SI{108}{\kelvin}. \B{(Bottom)} Selected curves $G_\text{diff}(V_\text{g},V_\text{d}=+\SI{1}{\milli\volt})$.}
			\end{figure}

			The device for which data is shown in Fig.~\ref{fig:D61D4S01_Gdiff_interface} was characterized only at \SI{300}{\kelvin}, \SI{4}{\kelvin} and base temperature.
			We will therefore take data from a similar device, for which a scan in $V_\text{g}$ at fixed $V_\text{d}=\SI{1}{\milli\volt}$ was repeated continuously, after all the \ce{^4He} had evaporated from the cryostat.
			These data are shown in Fig.~\ref{fig:D84D3TSweepCombinedPlot}, and clearly indicate an increase in conductance with temperature.
			
			\begin{figure}
				\centering
				\includegraphics[width=\textwidth]{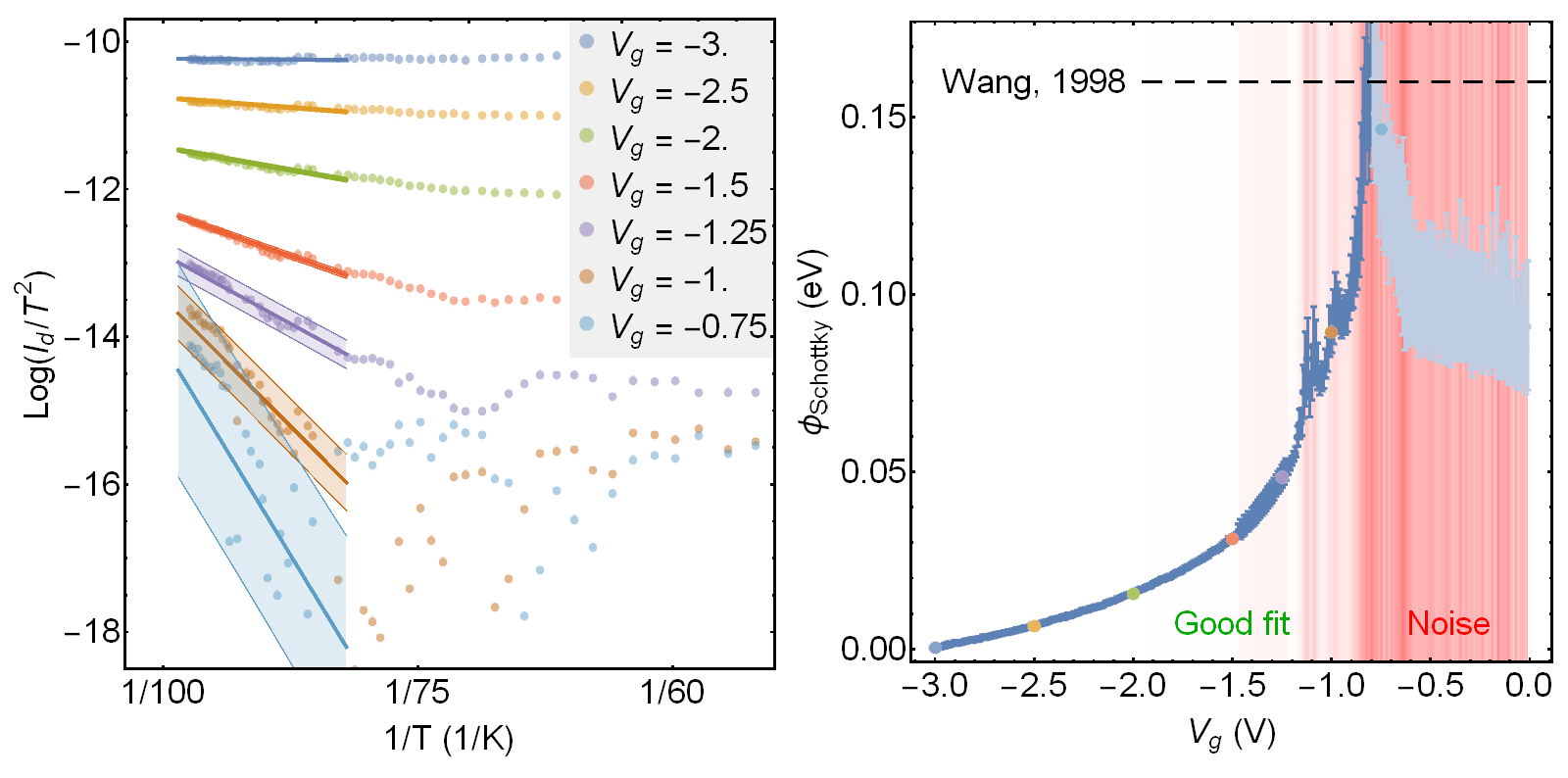}
				\caption{Data shown for device D84D3, same measurement as Fig.~\ref{fig:D84D3TSweepCombinedPlot}. \B{(Left)} By plotting $\log(I_\text{d}/T^2)$ at fixed source-drain bias $V_\text{d}$ and gate voltage $V_\text{g}$ versus the reciprocal temperature $1/T$, it is possible to extract the Schottky barrier height $\phi_\text{Schottky}$ using eq.~\eqref{eq:thermionic}. The linear regression fails when the detected current is on the same order as the noise, as is the case for e.g. $V_\text{g}=\SI{-0.75}{\volt}$ (bottom curve). \B{(Right)} Extracted SBH versus $V_\text{g}$, colored dots for the curves shown on the left superposed. The red background shading is proportional to the error bar on the slope estimate. For $V_\text{g}>\SI{-0.8}{\volt}$ (shown in lighter blue), the signal to noise ratio became too poor to perform good fits. The dashed horizontal line at $\phi_\text{SBH}=\SI{0.16}{\electronvolt}$ was extracted from a reverse-bias PtSi Schottky diode on a different wafer with the same doping~\cite{wang1998sub}.}
			\end{figure}
			
			The regime where thermionic emission dominates over tunneling depends on the width of the barrier~\cite{hugunin1995superconductor}, which in turn in the case of an SBMOSFET depends on the applied gate voltage.
			Heuristically, we can guess from linearity in the Arrhenius plot that thermionic emission dominates above $\sim\SI{80}{\kelvin}$, similar to that used in other sources~\cite{wang1998sub,dubois2004measurement}.
			At higher temperatures, perhaps $T>\SI{150}{\kelvin}$ for a typical PtSi SBMOSFET~\cite{dubois2004measurement}, resistance may be limited by the channel instead of the Schottky barrier.
			The magnitude of the tunnel/field-emission current depends only on the gate voltage and not the temperature, and so does not need to be taken into account.
			Nor do we have to worry about non-linearities due to the dependence of $I_\text{th}$ on the source-drain bias $V_\text{d}$ in eq.~\eqref{eq:thermionic_v}, since $eV_\text{d}=\SI{1}{\milli\electronvolt}\approx k_\text{B}\times\SI{12}{\kelvin}$, far below the temperatures at which we perform the fit.

			As can be seen in Fig.~\ref{fig:D84D3TSweepCombinedPlot}, the Schottky barrier is suppressed by the gate from \SI{0.16}{\electronvolt} around the threshold voltage (equal to the value measured in a diode~\cite{wang1998sub}, see section~\ref{sec:laurie_sample_description}), to nearly zero at $V_\text{g}=\SI{-3}{\volt}$.
			Though we cannot directly compare these results to the improved interface transparency observed in Fig.~\ref{fig:D61D4S01_Gdiff_interface} (the data are from different devices), we can be confident that the suppression of the Schottky barrier is responsible for at least part of the improvement in transparency.
			Transport then occurs by tunneling through this reduced barrier, though it is not yet clear what role is played by gate-stimulated emission.
		
		\FloatBarrier
		\subsection{Conclusion}
			
			Towards the end of this PhD we have begun measuring PtSi SBMOSFET devices with channel lengths on the order of \SI{50}{\nano\metre} and doping levels that varied from $\SI{5E15}{\per\centi\metre\cubed}$ deeper in the bulk to $10^{19}\,\si{\per\centi\metre\cubed}$ underneath the oxide.
			We have found that the source and drain are superconducting, and that this superconductivity can be induced by proximity effect inside the channel.
			More importantly, we saw that the extent of this proximity effect can be tuned with an electrostatic field, likely by tuning the Schottky barrier width even after dopants have frozen out, which suggests that the shielding charges that reduce the barrier width~\cite{aslamazov1981temperature} are instead ejected by field emission from the PtSi contacts~\cite{fowler1928electron}.
			The appearance of a zero-bias conductance peak is consistent both with reflectionless tunneling and coherent transport across the junction, as is its suppression by small magnetic fields on the order of \SI{10}{\milli\tesla}.
			It remains to be seen if a supercurrent can be detected by moving to a cryogenic setup with better noise filtering.
			
			As discussed in section~\ref{sec:gatemons}, a Josephson junction for transmon applications only needs a critical current on the order of 1--\SI{10}{\nano\ampere}, which corresponds to a current density on the order of 1--\SI{10}{\ampere/\centi\metre\squared} for the devices of the dimensions that we have studied, still 3 to 4 orders of magnitude below what can be obtained on degenerately doped (\SI{7E19}{}--\SI{1.8E20}{\per\centi\metre\cubed}) silicon~\cite{huang1974josephson}.
			Gate modulation has been demonstrated in devices with lower doping of \SI{5E18}{\per\centi\metre\cubed} in an experiment where a gate electrode was introduced by etching a hole from the back of the wafer~\cite{nishino1985three,nishino1986carrier}.
			The channel lengths in these experiments ranged from 40 to \SI{200}{\nano\metre}, well within the reach of CMOS fabrication techniques.
			Such structures seem to fall within the intermediate regime, where critical currents are reported to be limited by both the interface~\cite{kleinsasser1991critical,becker1995transport}, and decoherence in the channel~\cite{okamoto1992gate}, a crossover between which occurs as the Schottky barriers are suppressed by the gate~\cite{kleinsasser1989superconducting,kleinsasser1990crossover}.
			
			Now that a two-step annealing process has been found for PtSi formation, whereby the encroachment underneath the gate can be controlled, and undulations in the PtSi/Si interface can be reduced (see section~\ref{sec:lot2}), we have begun to fabricate transistors with pre-encroachment gate lengths down to \SI{40}{\nano\metre}.
			Since both the doping level and the channel length can be well controlled in CMOS technology, it should be possible to optimize these designs to obtain a gate-controllable supercurrent on the order of a few nano-amps, and then move on to designing a transmon circuit around it.
			With this, I wish the next student the best of luck.
			
\printbibliography

\end{refsection}

\begin{refsection}
	\graphicspath{{img/epilogue/}}
	\chapter*{Conclusion}
	
	Quantum computers are more powerful than their classical counterparts~\cite{bernstein1997quantum}, in the sense that they can solve certain classes of tasks more efficiently.
	For example, there exist $n$-bit problems that would take $\mathcal{O}(\exp(n))$ time on a classical machine, but that can be solved to arbitrary accuracy within $\mathcal{O}(n^k)$ (with $k\in\mathbb{N}$) on a quantum computer.
	Other problems benefit from more modest reductions by polynomial factors, or can be solved on a quantum computer while the classical computation cost is not yet clear.
	It is important to point out that this feat is achieved not by ``performing exponentially many calculations in parallel'', but by creating interference patterns in the collective wave function of the entangled qubits, with the probability density in the final state concentrated near solutions.
	This gives an advantage only in the asymptotic scaling of the computation time; performing a single fault-tolerant two-qubit operation can actually take up to $10^7$ times longer than the simple switch of a transistor~\cite{babbush2020q2b20}.
	For most problems, a quantum advantage can therefore only be achieved when large numbers of qubits are involved, at which point logical qubits need to be encoded by error correction in collections of perhaps thousands of physical ones~\cite{fowler2012surface}.
	
	To control entanglement of a few million qubits, it is necessary that both their operation and fabrication become scalable.
	Much of the progress in the past one and a half decade has been achieved with the transmon design~\cite{koch2007charge,arute2019quantum}, which as a lithographically defined ``artificial atom'' has the advantage that nearly any aspect of it can be engineered.
	This has allowed scalable operation by moving from flux-controlled SQUID-based transmons to circuits made with gate-tunable Josephson junctions~\cite{larsen2015semiconductor,de2015realization}, while mass production may be facilitated by developing a fabrication process that is fully compatible with the technologies of the semiconductor industry.
	Our approach is to modify existing transistor designs by using superconducting materials for the source and drain, such that the gate voltage can be used to tune the magnitude of a small Josephson current on the order of a few nano-Amperes.
	The resulting JoFET is then connected to a planar superconducting capacitor to form a transmon, and can be directly integrated alongside cryo-CMOS control electronics~\cite{le2020low,le2020FDSOI}.
	
	Silicon is however not the most natural candidate material for the weak link in a Josephson junction, and the main challenge is to achieve a sufficiently large supercurrent while maintaining a degree of gate control.
	The first concern is the transparency of the superconductor/semiconductor interface, since the relevant mechanism, Andreev reflection, is a second-order process that scales with the square of this parameter.
	Depending on the length of the channel $L$ and the diffusion constant $D$ inside it, the supercurrent is then limited by either the superconducting gap $\Delta$ of the contacts, or the Thouless energy $E=\hbar D/L^2$ of the channel (whichever is smaller).
	Both the transparency of the interfaces and the diffusion between them depend on the number of carriers that are attracted by the gate~\cite{volkov1996effect,kleinsasser1990crossover}, such that the limiting factor for a given device can change as the gate voltage is varied.
	To maximize supercurrent, we thus start by selecting contact materials with high transparency to silicon.
	Luckily, similar concerns for regular transistors have already driven the microelectronics industry to develop materials with low contact resistance, which has led to a thorough understanding of the fabrication and behavior of silicides.
	We can take advantage of this by selecting those that are superconducting, and then further optimize the fabrication to increase the superconducting gap and Thouless energy.
	It is not yet clear which of these two energy scales will typically be smallest in devices that can be fabricated with conventional CMOS technology, so we hedge our bets by studying both \ce{V3Si}, the silicide with the highest known critical temperature, and PtSi, which during its formation offers good control over the channel length.
	
	The integration of \ce{V3Si} in transistors comes with the challenge that it is neither the first, nor the final stable phase to form in the presence of a silicon reservoir.
	One way to overcome this challenge is to abandon the SALICIDE procedure in favor of direct sputtering of the silicide in the right stoichiometry, followed by crystallization annealing.
	This thermal processing in turn presents the challenge of \ce{VSi2} formation at the interface with the underlying silicon, with the risk of either consuming the entire channel, or all available superconducting \ce{V3Si}.
	Before either occurs, it could however also bring down the Thouless energy due to faster carrier diffusion in this metallic material, and move the Schottky barrier within range of the electrostatic gate field.
	This encroachment underneath the spacers, or even the gate itself, is also a central part of PtSi-based JoFETs.
	At the temperature where this final phase in the Pt-Si system obtains its desired properties, both species of atoms diffuse so fast that it becomes impossible to control the location of the silicide-silicon interface by timing of the thermal processing alone.
	Since in a silicon-on-insulator (SOI) design the planar contact openings have a vastly larger surface area than the vertical sides of the channel edges, it is also challenging to moderate the encroachment by tuning the metal deposition.
	A two-step annealing process provides a solution, in which the slower growth of either \ce{Pt2Si} or PtSi during a first heating step at lower temperatures is used to define the channel, and a selective etch of Pt and \ce{Pt2Si} before the final anneal prevents further silicon consumption.
	
	Measurements on a PtSi Schottky-barrier MOSFET confirmed that when the silicide extends underneath the spacers, the interface transparency can be modulated with the gate voltage.
	This allows for control of the proximity effect, the first step towards a semiconductor-based tunable Josephson junction.
	Further reduction of the Schottky barrier width and enhancement of the carrier diffusion could be achieved with higher levels of doping, where a second boron implantation after gate stack definition would minimize the barrier height while ensuring semiconducting behavior below the gate.
	Once a critical current on the order of \SI{10}{\nano\ampere} is obtained, and can be modulated with the gate, a surrounding planar capacitor can be lithographically defined to form a gatemon.
	A fully CMOS-compatible design may then be achieved by defining both this capacitor and the surrounding control and readout circuits in a superconducting silicide thin film, possibly on the same level as the JoFET contacts.
	The main strength of this approach is that by leveraging decades of industrial experience, a high degree of reproducibility can be achieved, with little variation in device performance within a single die, or even across the wafer.
	Overcoming this bottleneck~\cite{kreikebaum2020improving} in scalable fabrication may enable the development of larger NISQ circuits, and could be a stepping stone towards hardware suitable for error correction, a necessary component of fault-tolerant quantum computing.

\printbibliography
\end{refsection}

\pagenumbering{Roman}
\begin{appendices}
	\begin{refsection}
		\graphicspath{{img/appendices/}}
		\chapter{Samples for material studies}
	
	\section{\label{sec:samples_v}Samples prepared by pure V sputtering}
	
		\begin{table}[H]
			\caption{\label{tab:samples_vdep}Samples prepared for the study of \ce{V3Si} formation by V deposition.}
			\resizebox{\columnwidth}{!}{%
			\rowcolors{2}{gray!15}{white}
			\begin{tabular}{llllllll}
				\rowcolor{gray!30}\hline
				\# & \text{Sample} & Substrate/cap (TiN = cap) & V deposition (nm) & RTA (\si{\celsius}) & atmosphere & $R_\square$ & \text{$E(R_\square)$} \\\hline\hline
				1 & \text{A32A} & \text{15nm TiN} & 80 & 840. & \text{N2} & 4.78 & 0.046366 \\
				2 & \text{A32B} & \text{15nm TiN} & 80 & 0. & \text{N2} & 8.94 & 0.286974 \\
				3 & \text{A32C} & \text{15nm TiN} & 80 & 840. & \text{N2} & 5.16 & 0.087204 \\
				4 & \text{A32D} & \text{15nm TiN} & 80 & 0. & \text{N2} & 8.78 & 0.15365 \\
				5 & \text{A33B1} & \text{15nm TiN} & 40 & 800. & \text{N2} & 11.1724 & 0.08196 \\
				6 & \text{A33B2} & \text{15nm TiN} & 40 & 800. & \text{UHV} & 8.61458 & 0.0130839 \\
				7 & \text{A33C1} & \text{15nm TiN} & 40 & 900. & \text{N2} & 8.4348 & 0.0613132 \\
				8 & \text{A33C2} & \text{15nm TiN} & 40 & 900. & \text{UHV} & 7.99213 & 0.00692336 \\
				9 & \text{A33D1} & \text{15nm TiN} & 40 & 1000. & \text{N2} & 8.2248 & 0.0114063 \\
				10 & \text{A33D2} & \text{15nm TiN} & 40 & 1000. & \text{UHV} & 8.33357 & 0.00523356 \\
				11 & \text{A35B} & \text{15nm TiN} & 20 & 0. & \text{N2} & 20.1782 & 0.0197563 \\
				12 & \text{A35A} & \text{15nm TiN} & 20 & 500. & \text{N2} & 55.8833 & 0.670025 \\
				13 & \text{A35C} & \text{15nm TiN} & 20 & 600. & \text{N2} & 30.0367 & 0.070946 \\
				14 & \text{A36A} & \text{15nm TiN} & 20 & 700. & \text{N2} & 23.8267 & 0.0929157 \\
				15 & \text{A36C} & \text{15nm TiN} & 20 & 800. & \text{N2} & 21.8767 & 0.292973 \\
				16 & \text{A35D} & \text{15nm TiN} & 20 & 900. & \text{N2} & 16.7983 & 0.186969 \\
				17 & \text{A36B} & \text{15nm TiN} & 20 & 1000. & \text{N2} & 16.084 & 0.242474 \\
				18 & \text{A42A} & \text{HF dip + 15nm TiN} & 80 & 840. & \text{N2} & 4.11 & 0.02055 \\
				19 & \text{A42B} & \text{HF dip + 15nm TiN} & 80 & 0. & \text{N2} & 6.04 & 0.281464 \\
				20 & \text{A42C} & \text{HF dip + 15nm TiN} & 80 & 840. & \text{N2} & 3.95 & 0.00237 \\
				21 & \text{A42D} & \text{HF dip + 15nm TiN} & 80 & 0. & \text{N2} & 6.07 & 0.226411 \\
				22 & \text{A66A} & \text{} & 100 & 300. & \text{N2} & 14.9569 & 0.402849 \\
				23 & \text{A66B} & \text{} & 100 & 400. & \text{N2} & 20.5318 & 0.197563 \\
				24 & \text{A66C} & \text{} & 100 & 500. & \text{N2} & 21.6951 & 0.566701 \\
				25 & \text{A66D} & \text{} & 100 & 600. & \text{N2} & 13.5972 & 0.197563 \\
				26 & \text{A66E} & \text{} & 100 & 700. & \text{N2} & 10.2885 & 0.119916 \\
				27 & \text{A66F} & \text{} & 100 & 800. & \text{N2} & 6.58709 & 0.145696 \\
				28 & \text{A66G} & \text{} & 100 & 900. & \text{N2} & 5.30895 & 0.0911377 \\
				29 & \text{A66H} & \text{} & 100 & 1000. & \text{N2} & 4.42211 & 0.287882 \\
				30 & \text{A66I} & \text{} & 100 & 0. & \text{N2} & 11.3612 & 0.232585 \\
				31 & \text{B34B} & \text{20nm \ce{SiO2} + 15nm TiN} & 40 & 0. & \text{N2} & 12.7436 & 0.175792 \\
				32 & \text{B34A} & \text{20nm \ce{SiO2} + 15nm TiN} & 40 & 500. & \text{N2} & 37.0733 & 0.275379 \\
				33 & \text{B34C} & \text{20nm \ce{SiO2} + 15nm TiN} & 40 & 600. & \text{N2} & 35.62 & 0.155885 \\
				34 & \text{B23B} & \text{20nm \ce{SiO2} + 15nm TiN} & 40 & 700. & \text{N2} & 33.28 & 0.18735 \\
				35 & \text{B23D} & \text{20nm \ce{SiO2} + 15nm TiN} & 40 & 800. & \text{N2} & 29.8967 & 0.023094 \\\hline
			\end{tabular}}
		\end{table}

		\begin{table}
			\ContinuedFloat
			\caption{\B{(Continued)} Samples prepared for the study of \ce{V3Si} formation by V deposition.}
			\resizebox{\columnwidth}{!}{%
			\rowcolors{2}{gray!15}{white}
			\begin{tabular}{llllllll}
				\rowcolor{gray!30}\hline
				\# & \text{Sample} & Substrate/cap (TiN = cap) & V deposition (nm) & RTA (\si{\celsius}) & atmosphere & $R_\square$ & \text{$E(R_\square)$} \\\hline\hline
				36 & \text{B34D} & \text{20nm \ce{SiO2} + 15nm TiN} & 40 & 900. & \text{N2} & 27.771 & 0.514728 \\
				37 & \text{B23C} & \text{20nm \ce{SiO2} + 15nm TiN} & 40 & 1000. & \text{N2} & 29.158 & 0.841264 \\
				38 & \text{B54A} & \text{20nm \ce{SiO2} + 15nm TiN} & 80 & 840. & \text{N2} & 11.44 & 0.088088 \\
				39 & \text{B54B} & \text{20nm \ce{SiO2} + 15nm TiN} & 80 & 0. & \text{N2} & 8.12 & 0.1624 \\
				40 & \text{B54C} & \text{20nm \ce{SiO2} + 15nm TiN} & 80 & 840. & \text{N2} & 11.08 & 0.212736 \\
				41 & \text{B54D} & \text{20nm \ce{SiO2} + 15nm TiN} & 80 & 0. & \text{N2} & 8. & 0.1472 \\
				42 & \text{B55A} & \text{20nm \ce{SiO2} + 10nm HfO2 + 15nm TiN} & 100 & 840. & \text{N2} & 20.9 & 3.98354 \\
				43 & \text{B55B} & \text{20nm \ce{SiO2} + 10nm HfO2 + 15nm TiN} & 100 & 0. & \text{N2} & 11.88 & 0.105732 \\
				44 & \text{B55C} & \text{20nm \ce{SiO2} + 10nm HfO2 + 15nm TiN} & 100 & 840. & \text{N2} & 15.15 & 0.67266 \\
				45 & \text{B55D} & \text{20nm \ce{SiO2} + 10nm HfO2 + 15nm TiN} & 100 & 0. & \text{N2} & 11.85 & 0.103095 \\
				46 & \text{B56A} & \text{20nm \ce{SiO2}} & 100 & 300. & \text{N2} & 17.8879 & 0.291393 \\
				47 & \text{B56B} & \text{20nm \ce{SiO2}} & 100 & 400. & \text{N2} & 23.6138 & 0.522702 \\
				48 & \text{B56C} & \text{20nm \ce{SiO2}} & 100 & 500. & \text{N2} & 24.4901 & 0.228126 \\
				49 & \text{B56D} & \text{20nm \ce{SiO2}} & 100 & 600. & \text{N2} & 18.885 & 0.228126 \\
				50 & \text{B56E} & \text{20nm \ce{SiO2}} & 100 & 700. & \text{N2} & 19.6253 & 0.119916 \\
				51 & \text{B56F} & \text{20nm \ce{SiO2}} & 100 & 800. & \text{N2} & 11.6181 & 0.228126 \\
				52 & \text{B56G} & \text{20nm \ce{SiO2}} & 100 & 900. & \text{N2} & 11.2857 & 0.252353 \\
				53 & \text{B56H} & \text{20nm \ce{SiO2}} & 100 & 1000. & \text{N2} & 14.1109 & 0.188699 \\
				54 & \text{B56I} & \text{20nm \ce{SiO2}} & 100 & 0. & \text{N2} & -- & -- \\
				55 & \text{B73A} & \text{20nm \ce{SiO2} + 15nm TiN} & 80 & 800. & \text{N2} & 16.7291 & 0.448868 \\
				56 & \text{B73B} & \text{20nm \ce{SiO2} + 15nm TiN} & 80 & 900. & \text{N2} & 12.8161 & 0.359815 \\
				57 & \text{B73C} & \text{20nm \ce{SiO2} + 15nm TiN} & 80 & 1000. & \text{N2} & 13.3812 & 1.43864 \\
				58 & \text{B73D} & \text{20nm \ce{SiO2} + 15nm TiN} & 80 & 0. & \text{N2} & 8.05906 & 0.173023 \\
				59 & \text{B74A} & \text{20nm \ce{SiO2} + 15nm TiN} & 80 & 800. & \text{UHV} & 723.371 & 36.1826\\
				60 & \text{B74D} & \text{20nm \ce{SiO2} + 15nm TiN} & 80 & 900. & \text{UHV} & 116.029 & 14.2681 \\
				61 & \text{C26A} & \text{aSi (Ge) + 15nm TiN} & 80 & 800. & \text{N2} & 11.5168 & 0.216467 \\
				62 & \text{C26B} & \text{aSi (Ge) + 15nm TiN} & 80 & 900. & \text{N2} & 8.36832 & 0.1638 \\
				63 & \text{C26C} & \text{aSi (Ge) + 15nm TiN} & 80 & 1000. & \text{N2} & 8.54811 & 0.447562 \\
				64 & \text{C26D} & \text{aSi (Ge) + 15nm TiN} & 80 & 0. & \text{N2} & 7.42377 & 0.194495 \\
				65 & \text{C33A} & \text{aSi (Ge) + 0.1$\%$ O + 15nm TiN} & 80 & 800. & \text{N2} & 11.0243 & 0.322353 \\
				66 & \text{C33B} & \text{aSi (Ge) + 0.1$\%$ O + 15nm TiN} & 80 & 900. & \text{N2} & 8.73242 & 0.275111 \\
				67 & \text{C33C} & \text{aSi (Ge) + 0.1$\%$ O + 15nm TiN} & 80 & 1000. & \text{N2} & 7.31983 & 0.404604 \\
				68 & \text{C33D} & \text{aSi (Ge) + 0.1$\%$ O + 15nm TiN} & 80 & 0. & \text{N2} & 7.3168 & 0.099541 \\
				69 & \text{C53A} & \text{aSi (Ge) + 1$\%$ O + 15nm TiN} & 80 & 800. & \text{N2} & 11.5425 & 0.193164 \\
				70 & \text{C53B} & \text{aSi (Ge) + 1$\%$ O + 15nm TiN} & 80 & 900. & \text{N2} & 9.04818 & 0.274101 \\
				71 & \text{C53C} & \text{aSi (Ge) + 1$\%$ O + 15nm TiN} & 80 & 1000. & \text{N2} & 7.63105 & 0.136073 \\
				72 & \text{C53D} & \text{aSi (Ge) + 1$\%$ O + 15nm TiN} & 80 & 0. & \text{N2} & 7.06752 & 0.140894 \\
				73 & \text{C65A} & \text{aSi (Ge) + 5$\%$ O + 15nm TiN} & 80 & 800. & \text{N2} & 11.3642 & 0.156308 \\
				74 & \text{C65B} & \text{aSi (Ge) + 5$\%$ O + 15nm TiN} & 80 & 900. & \text{N2} & 8.36137 & 0.136723 \\
				75 & \text{C65C} & \text{aSi (Ge) + 5$\%$ O + 15nm TiN} & 80 & 1000. & \text{N2} & 7.69148 & 0.249364 \\
				76 & \text{C65D} & \text{aSi (Ge) + 5$\%$ O + 15nm TiN} & 80 & 0. & \text{N2} & 7.68091 & 0.203218 \\
				77 & \text{D32A} & \text{40nm SiN + 15nm TiN} & 100 & 840. & \text{N2} & 11.58 & 0.26055 \\
				78 & \text{D32B} & \text{40nm SiN + 15nm TiN} & 100 & 0. & \text{N2} & 10.69 & 0.132556 \\
				79 & \text{D32C} & \text{40nm SiN + 15nm TiN} & 100 & 840. & \text{N2} & 11.46 & 0.121476 \\
				80 & \text{D32D} & \text{40nm SiN + 15nm TiN} & 100 & 0. & \text{N2} & 10.62 & 0.174168 \\
				81 & \text{D33A} & \text{40nm SiN + 15nm TiN} & 100 & 800. & \text{N2} & 14.6638 & 0.304189 \\
				82 & \text{D33B} & \text{40nm SiN + 15nm TiN} & 100 & 900. & \text{N2} & 10.9896 & 0.300657 \\
				83 & \text{D33C} & \text{40nm SiN + 15nm TiN} & 100 & 1000. & \text{N2} & 10.355 & 0.208507 \\
				84 & \text{D33D} & \text{40nm SiN + 15nm TiN} & 100 & 0. & \text{N2} & 10.0272 & 0.130288 \\
				85 & \text{E55A} & \text{10$\%$ O + 15nm TiN} & 40 & 800. & \text{N2} & 19.8444 & 0.177999 \\
				86 & \text{E55B} & \text{10$\%$ O + 15nm TiN} & 40 & 900. & \text{N2} & 21.5002 & 1.07972 \\
				87 & \text{E55C} & \text{10$\%$ O + 15nm TiN} & 40 & 1000. & \text{N2} & 19.654 & 0.838755 \\
				88 & \text{E55D} & \text{10$\%$ O + 15nm TiN} & 40 & 0. & \text{N2} & 20.0045 & 0.65509\\\hline
			\end{tabular}}
		\end{table}

		\begin{table}
			\caption{Superconducting critical temperature ($T_\text{c}$), residual resistance ratio (RRR) and low-temperature resistance ($R(\SI{20}{\kelvin})$) of some of the samples listed in table~\ref{tab:samples_vdep}.}
			\resizebox{\columnwidth}{!}{%
			\rowcolors{1}{gray!15}{white}
			\begin{tabular}{llllllll}
				\rowcolor{gray!30}\hline
				\# & \text{Sample} & $T_\text{c}$ & $E(T_\text{c})$ (10,90$\%$) & \text{RRR} & $E$(RRR) & Normalized R(20K) & Normalized $E$(R(20K)) \\\hline\hline
				-- & \text{Ti10nm} & 0 & \text{$\{$0, 0$\}$} & 1.41449 & 0.00838401 & 30.1146 & 0.0431351 \\
				31 & \text{B34B} & 3.64205 & \text{$\{$-0.222872, 0.0725929$\}$} & 1.63019 & 0.0441452 & 7.81723 & 0.239143 \\
				34 & \text{B23B1A} & 0 & \text{$\{$0, 0$\}$} & 1.04629 & 0.000958445 & 31.8075 & 0.181416 \\
				35 & \text{B23D1A} & 0 & \text{$\{$0, 0$\}$} & 1.04629 & 0.000958445 & 28.5739 & 0.0342406 \\
				38 & \text{B54A} & 4.68876 & \text{$\{$-0.154002, 0.2653$\}$} & 0.998304 & 0.0530377 & 11.4594 & 0.623714 \\
				39 & \text{B54B} & 3.74038 & \text{$\{$-0.0409306, 0.0395684$\}$} & 2.09872 & 0.279515 & 3.86903 & 0.56646 \\
				55 & \text{B73A} & 0 & \text{$\{$0, 0$\}$} & 1.03719 & 0.0484716 & 16.1292 & 0.880549 \\
				56 & \text{B73B} & 2.68131 & \text{$\{$-0.188784, 0.347914$\}$} & 0.952515 & 0.0711176 & 13.455 & 1.0919 \\
				57 & \text{B73C} & 3.99827 & \text{$\{$-0.135969, 0.772515$\}$} & 0.918061 & 0.0437025 & 14.5755 & 1.71775 \\
				61 & \text{C26A} & 0 & \text{$\{$0, 0$\}$} & 1.42372 & 0.148817 & 8.08926 & 0.911885 \\
				62 & \text{C26B} & 5.85612 & \text{$\{$-0.16484, 0.145038$\}$} & 1.11062 & 0.0904084 & 7.53485 & 0.645378 \\
				62 & \text{C26B} & 5.86252 & \text{$\{$-0.169211, 0.151172$\}$} & 1.11062 & 0.0904084 & 7.53485 & 0.645378 \\
				63 & \text{C26C} & 7.80634 & \text{$\{$-0.0813111, 0.872293$\}$} & 1.18734 & 0.0844744 & 7.19937 & 0.646353 \\
				74 & \text{C65B} & 7.45921 & \text{$\{$-0.418516, 0.283958$\}$} & 1.15235 & 0.107453 & 7.25593 & 0.711175 \\
				75 & \text{C65C} & 8.02589 & \text{$\{$-1.04534, -0.125081$\}$} & 1.23567 & 0.139476 & 6.22453 & 0.7722 \\
				75 & \text{C65C} & 8.16673 & \text{$\{$-0.181696, 0.197914$\}$} & 1.23567 & 0.139476 & 6.22453 & 0.7722 \\
				86 & \text{E55B} & 5.60622 & \text{$\{$-0.203318, 0.261955$\}$} & 1.14946 & 0.0365311 & 18.7047 & 1.11395 \\
				87 & \text{E55C} & 7.06924 & \text{$\{$-0.173071, 0.189863$\}$} & 1.17895 & 0.0516317 & 16.6708 & 1.02769 \\\hline
			\end{tabular}}
		\end{table}
	
	\FloatBarrier
	\section{\label{sec:samples_v3si}Samples prepared by compound \ce{V3Si} sputtering}
	
		\begin{table}[H]
			\centering
			\caption{\label{tab:v3si_test}Before the processing of lots D19S2345--7, sputtering tests were performed on wafers A through L. Pieces were annealed, listed below.}
			{\scriptsize
			\rowcolors{1}{gray!15}{white}
			\begin{tabular}{llllllll}
				\rowcolor{gray!30}\hline
				\# & \text{Sample} & \text{Substrate} & \ce{V3Si} deposition (nm) & \text{RTA} & $R_\square$ & $E(R_\square)$ \\\hline
				1 & \text{C01} & \SI{300}{\nano\meter} \ce{SiO2} & 59. & 600. & 30.0196 & 0.811203 \\
				2 & \text{C02} & \SI{300}{\nano\meter} \ce{SiO2} & 59. & 700. & 17.48 & 0.159173 \\
				3 & \text{C03} & \SI{300}{\nano\meter} \ce{SiO2} & 59. & 800. & 17.1023 & 0.228126 \\
				4 & \text{C04} & \SI{300}{\nano\meter} \ce{SiO2} & 59. & 900. & 23.9764 & 3.33185 \\
				5 & \text{C06} & \SI{300}{\nano\meter} \ce{SiO2} & 59. & 0. & 34.0685 & 0.488855 \\
				6 & \text{D01} & \SI{300}{\nano\meter} \ce{SiO2} & 17. & 600. & 149.569 & 4.5324 \\
				7 & \text{D02} & \SI{300}{\nano\meter} \ce{SiO2} & 17. & 700. & 134.461 & 2.61678 \\
				8 & \text{D03} & \SI{300}{\nano\meter} \ce{SiO2} & 17. & 800. & 1042.45 & 181.296 \\
				9 & \text{D04} & \SI{300}{\nano\meter} \ce{SiO2} & 17. & 900. & 27606.8 & 15840.9 \\
				10 & \text{D06} & \SI{300}{\nano\meter} \ce{SiO2} & 17. & 0. & 133.479 & 2.47296 \\
				11 & \text{E01} & \SI{300}{\nano\meter} \ce{SiO2} & 63.5 & 600. & 34.7786 & 0.340182 \\
				12 & \text{E02} & \SI{300}{\nano\meter} \ce{SiO2} & 63.5 & 700. & 18.5526 & 0.145696 \\
				13 & \text{E03} & \SI{300}{\nano\meter} \ce{SiO2} & 63.5 & 800. & 16.8454 & 0.369143 \\
				14 & \text{E04} & \SI{300}{\nano\meter} \ce{SiO2} & 63.5 & 900. & 17.7519 & 0.321556 \\
				15 & \text{E06} & \SI{300}{\nano\meter} \ce{SiO2} & 63.5 & 0. & 37.4678 & 0.41126 \\
				16 & \text{F01} & \SI{300}{\nano\meter} \ce{SiO2} & 30. & 600. & 83.0336 & 0.891628 \\
				17 & \text{F02} & \SI{300}{\nano\meter} \ce{SiO2} & 30. & 700. & 41.8038 & 0.0943494 \\
				18 & \text{F03} & \SI{300}{\nano\meter} \ce{SiO2} & 30. & 800. & 46.5629 & 1.71594 \\
				19 & \text{F04} & \SI{300}{\nano\meter} \ce{SiO2} & 30. & 900. & 221.937 & 21.4959 \\
				20 & \text{F06} & \SI{300}{\nano\meter} \ce{SiO2} & 30. & 0. & 67.6989 & 0.900036 \\
				21 & \text{G01} & \text{} & 200. & 600. & 5.52953 & 0.119916 \\
				22 & \text{G02} & \text{} & 200. & 700. & 1.58634 & 0.0207701 \\
				23 & \text{G03} & \text{} & 200. & 800. & 2.49131 & 0.0301782 \\
				24 & \text{G04} & \text{} & 200. & 900. & 2.66203 & 0.0228126 \\
				25 & \text{G06} & \text{} & 200. & 0. & 9.06631 & 0.0228126 \\
				26 & \text{H01} & \text{} & 200. & 600. & 3.11527 & 0.0346168 \\
				27 & \text{H02} & \text{} & 200. & 700. & 601.298 & 49.925 \\
				28 & \text{H06} & \text{} & 200. & 0. & 4.62456 & 0.0525315 \\
				29 & \text{I02} & \text{} & 56. & 700. & 20.4109 & 0.265574 \\
				30 & \text{I03} & \text{} & 56. & 800. & 17.8425 & 0.171594 \\
				31 & \text{I05} & \text{} & 56. & 0. & 36.7427 & 0.308514 \\
				32 & \text{J02} & \text{} & 111. & 700. & 8.36983 & 0.214193 \\
				33 & \text{J05} & \text{} & 111. & 0. & 18.4469 & 0.353992 \\
				34 & \text{K02} & \text{} & 223. & 700. & 4.36319 & 0.123072 \\
				35 & \text{K03} & \text{} & 223. & 800. & 4.0429 & 0.0437089 \\
				36 & \text{K05} & \text{} & 223. & 0. & 9.0784 & 0.233452 \\
				37 & \text{K10} & \text{} & 223. & 800. & -- & -- \\
				38 & \text{K11} & \text{} & 223. & 800, \SI{5}{\minute} & -- & -- \\
				39 & \text{L01} & \text{} & 54. & 0. & 34.5 & -- \\
				40 & \text{L02} & \text{} & 54. & 0. & 34.5 & -- \\\hline
			\end{tabular}}
		\end{table}

		\begin{table}[H]
			\caption{Superconducting critical temperature ($T_\text{c}$), residual resistance ratio (RRR) and low-temperature resistance ($R(\SI{20}{\kelvin})$) of some of the samples listed in table~\ref{tab:v3si_test}.}
			\resizebox{\columnwidth}{!}{%
			\rowcolors{1}{gray!15}{white}
			\begin{tabular}{llllllll}
				\rowcolor{gray!30}\hline
				\# & \text{Sample} & $T_\text{c}$ & $E(T_\text{c})$ (10,90$\%$) & \text{RRR} & $E$(RRR) & Normalized R(20K) & Normalized $E$(R(20K))\\\hline\hline
				2 & \text{C02B} & 10.9133 & \text{$\{$-0.176991, 0.355359$\}$} & 3.10273 & 0.0417793 & 2.30049 & 0.0267165 \\
				3 & \text{C03B} & 11.7983 & \text{$\{$-0.194216, 0.363596$\}$} & 4.31187 & 0.0846629 & 1.48415 & 0.027792 \\
				12 & \text{E02} & 9.20809 & \text{$\{$0.120734, -0.734057$\}$} & 3.23786 & 0.0430004 & 2.32326 & 0.0278059 \\
				13 & \text{E03} & 8.51099 & \text{$\{$0.205673, -0.0168385$\}$} & 4.76806 & 0.0831607 & 1.63769 & 0.0274138 \\
				17 & \text{F02} & 10.0402 & \text{$\{$-0.210082, 0.40793$\}$} & 3.3369 & 0.0265089 & 4.54692 & 0.0270185 \\
				18 & \text{F03} & 10.9782 & \text{$\{$-0.19664, 0.39647$\}$} & 5.12474 & 0.0570779 & 3.0604 & 0.0296731 \\
				29 & \text{I02} & 10.6875 & \text{$\{$-0.229926, 0.492383$\}$} & 3.16905 & 0.0338262 & 2.58731 & 0.0249651 \\
				30 & \text{I03} & 11.7113 & \text{$\{$-0.203699, 0.382762$\}$} & 4.46587 & 0.0819486 & 1.57074 & 0.0260369 \\
				34 & \text{K02} & 7.84126 & \text{$\{$-0.121149, 0.270682$\}$} & 3.03805 & 0.125333 & 0.609604 & 0.0242081 \\
				37 & \text{K10} & 12.7378 & \text{$\{$-0.274433, 0.374017$\}$} & 4.98353 & 0.0520608 & 0.867358 & 0.00652876 \\
				38 & \text{K11} & 13.1118 & \text{$\{$-0.473417, 0.475283$\}$} & 5.08479 & 0.0680129 & 1.10151 & 0.0130036 \\
				39 & \text{L01} & 0.918545 & \text{$\{$-0.00185573, 0.00262192$\}$} & 0.965418 & 0.000770539 & 33.5135 & 0.0267525 \\
				40 & \text{L02} & 0.918545 & \text{$\{$-0.00185573, 0.00262192$\}$} & 0.965418 & 0.000770539 & 33.5135 & 0.0267525 \\\hline
			\end{tabular}}
		\end{table}

		\begin{table}
			\centering
			\caption{\label{tab:v3si_dep_rs}V3Si compound deposition.}
			\resizebox{\columnwidth}{!}{%
			\rowcolors{1}{gray!15}{white}
			\begin{tabular}{llllllll}
				\rowcolor{gray!30}\hline
				\# & Sample & Substrate & \ce{V3Si} deposition  (nm) & RTA (\si{\celsius}) & $R_\square$ & $E(R_\square)$ \\\hline
				1 & \text{B2P03A} & \text{Si + HF} & 20 & 500 & 103.43 & 10.0436 \\
				2 & \text{B2P03B} & \text{Si + HF} & 20 & 550 & 42.5054 & 2.12527 \\
				3 & \text{B2P03C} & \text{Si + HF} & 20 & 600 & 23.973 & 0.153255 \\
				4 & \text{B2P03D} & \text{Si + HF} & 20 & 650 & 24.3273 & 0.259712 \\
				5 & \text{B2P03E} & \text{Si + HF} & 20 & 700 & 23.0238 & 0.259712 \\
				6 & \text{B2P03F} & \text{Si + HF} & 20 & 750 & 22.2728 & 0.431383 \\
				7 & \text{B2P04A} & \text{Si + HF} & 50 & 500 & 35.1378 & 1.0697 \\
				8 & \text{B2P04A} & \text{Si + HF} & 50 & 500 & 35.1378 & 1.0697 \\
				9 & \text{B2P04B} & \text{Si + HF} & 50 & 550 & 23.9447 & 0.649281 \\
				10 & \text{B2P04B} & \text{Si + HF} & 50 & 550 & 23.9447 & 0.649281 \\
				11 & \text{B2P04C} & \text{Si + HF} & 50 & 600 & 10.6972 & 0.534848 \\
				12 & \text{B2P04C} & \text{Si + HF} & 50 & 600 & 10.6972 & 0.534848 \\
				13 & \text{B2P04D} & \text{Si + HF} & 50 & 650 & 8.26021 & 0.088482 \\
				14 & \text{B2P04E} & \text{Si + HF} & 50 & 700 & 7.80682 & 0.098162 \\
				15 & \text{B2P04F} & \text{Si + HF} & 50 & 750 & 7.43844 & 0.112459 \\
				16 & \text{B2P05B} & \text{Si + HF} & 200 & 600 & 269.909 & 114.647 \\
				17 & \text{B2P05C} & \text{Si + HF} & 200 & 650 & 5.67447 & 3.25166 \\
				18 & \text{B2P05C} & \text{Si + HF} & 200 & 650 & 5.67447 & 3.25166 \\
				19 & \text{B2P05C} & \text{Si + HF} & 200 & 650 & 5.67447 & 3.25166 \\
				20 & \text{B2P05D} & \text{Si + HF} & 200 & 700 & 1.56703 & 0.0171784 \\
				21 & \text{B2P05E} & \text{Si + HF} & 200 & 750 & 1.99634 & 0.00649281 \\
				22 & \text{B2P05F} & \text{Si + HF} & 200 & 800 & 2.1196 & 0.0129856 \\
				23 & \text{B2P05G} & \text{Si + HF} & 200 & 850 & 2.24003 & 0.0542673 \\
				24 & \text{B2P05H} & \text{Si + HF} & 200 & 900 & 2.48798 & 0.0249059 \\
				25 & \text{B2P05J} & \text{Si + HF} & 200 & 500 & \text{--} & \text{--} \\
				26 & \text{B2P05K} & \text{Si + HF} & 200 & 550 & \text{--} & \text{--} \\
				27 & \text{B2P05L} & \text{Si + HF} & 200 & 500, \SI{5}{\minute} & \text{--} & \text{--} \\
				28 & \text{B2P05Q} & \text{Si + HF} & 200 & 500, \SI{5}{\minute} & \text{--} & \text{--} \\
				29 & \text{B2P05R} & \text{Si + HF} & 200 & 500, \SI{1}{\second} & \text{--} & \text{--} \\
				30 & \text{B2P06A} & \text{Si + HF} & 100 & 500 & 17.0447 & 0.490196 \\
				31 & \text{B2P06B} & \text{Si + HF} & 100 & 550 & 13.3892 & 0.212527 \\
				32 & \text{B2P06C} & \text{Si + HF} & 100 & 600 & 5.5257 & 2.97538 \\
				33 & \text{B2P06D} & \text{Si + HF} & 100 & 650 & 3.99551 & 0.224917 \\
				34 & \text{B2P06E} & \text{Si + HF} & 100 & 700 & 3.47127 & 0.0245405 \\
				35 & \text{B2P06F} & \text{Si + HF} & 100 & 750 & 3.27292 & 0.0425054 \\
				36 & \text{B2P08B} & \text{Si + 20nm SiO2} & 200 & 600 & 205.726 & 68.6275 \\
				37 & \text{B2P08C} & \text{Si + 20nm SiO2} & 200 & 650 & 4.05218 & 0.578656 \\
				38 & \text{B2P08D} & \text{Si + 20nm SiO2} & 200 & 700 & 4.16553 & 0.112459 \\
				39 & \text{B2P08D} & \text{Si + 20nm SiO2} & 200 & 700 & 4.16553 & 0.112459 \\
				40 & \text{B2P08E} & \text{Si + 20nm SiO2} & 200 & 750 & 4.00968 & 0.088482 \\
				41 & \text{B2P08F} & \text{Si + 20nm SiO2} & 200 & 800 & 3.85382 & 0.088482 \\
				42 & \text{B2P08G} & \text{Si + 20nm SiO2} & 200 & 850 & 3.85382 & 0.088482 \\
				43 & \text{B2P08H} & \text{Si + 20nm SiO2} & 200 & 900 & 3.90908 & 0.0446475 \\
				44 & \text{B2P08I} & \text{Si + 20nm SiO2} & 200 & 900 & \text{--} & \text{--} \\
				45 & \text{B2P08P} & \text{Si + 20nm SiO2} & 200 & 800 & \text{--} & \text{--} \\\hline
			\end{tabular}}
		\end{table}

		\begin{table}
			\ContinuedFloat
			\centering
			\caption{\B{(Continued)} V3Si compound deposition.}
			\resizebox{\columnwidth}{!}{%
			\rowcolors{1}{gray!15}{white}
			\begin{tabular}{llllllll}
				\rowcolor{gray!30}\hline
				\# & Sample & Substrate & \ce{V3Si} deposition (nm) & RTA (\si{\celsius}) & $R_\square$ & $E(R_\square)$ \\\hline
				46 & \text{B2P09A} & \text{Si + 20nm SiO2} & 50 + \SI{10}{\nano\meter} Si & 500 & \text{--} & \text{--} \\
				47 & \text{B2P09B} & \text{Si + 20nm SiO2} & 50 + \SI{10}{\nano\meter} Si & 550 & \text{--} & \text{--} \\
				48 & \text{B2P09C} & \text{Si + 20nm SiO2} & 50 + \SI{10}{\nano\meter} Si & 600 & \text{--} & \text{--} \\
				49 & \text{B2P09D} & \text{Si + 20nm SiO2} & 50 + \SI{10}{\nano\meter} Si & 650 & \text{--} & \text{--} \\
				50 & \text{B2P09E} & \text{Si + 20nm SiO2} & 50 + \SI{10}{\nano\meter} Si & 700 & \text{--} & \text{--} \\
				51 & \text{B2P09F} & \text{Si + 20nm SiO2} & 50 + \SI{10}{\nano\meter} Si & 750 & \text{--} & \text{--} \\
				52 & \text{B2P09G} & \text{Si + 20nm SiO2} & 50 + \SI{10}{\nano\meter} Si & 800 & \text{--} & \text{--} \\
				53 & \text{B2P09H} & \text{Si + 20nm SiO2} & 50 + \SI{10}{\nano\meter} Si & 850 & \text{--} & \text{--} \\
				54 & B2P09I & \text{Si + 20nm SiO2} & 50 + \SI{10}{\nano\meter} Si & \text{--} & \text{--} & \text{--} \\
				55 & \text{B2P17B} & \text{Si} & 200 & 600 & 280.111 & 136.952 \\
				56 & \text{B2P17C} & \text{Si} & 200 & 650 & 2.81952 & 0.789884 \\
				57 & \text{B2P17D} & \text{Si} & 200 & 700 & 1.58545 & 0.0297538 \\
				58 & \text{B2P17E} & \text{Si} & 200 & 750 & 2.2117 & 0.0371365 \\
				59 & \text{B2P17F} & \text{Si} & 200 & 800 & 2.39447 & 0.0507104 \\
				60 & \text{B2P17G} & \text{Si} & 200 & 850 & 2.5149 & 0.0401745 \\
				61 & \text{B2P17H} & \text{Si} & 200 & 900 & 2.71184 & 0.0449835 \\
				62 & \text{B3P11A} & \text{Si + Amorphisation + HF} & 200 & 600 & 8.50108 & \text{--} \\
				63 & \text{B3P11B} & \text{Si + Amorphisation + HF} & 200 & 650 & \text{--} & \text{--} \\
				64 & \text{B3P11C} & \text{Si + Amorphisation + HF} & 200 & 700 & 20.686 & 5.14181 \\
				65 & \text{B4PP02} & \text{Si + PS5} & 200 & 700 & 3.91758 & 0.26573 \\
				66 & \text{B4PP03} & \text{Si + PS5} & 200 & 800 & 3.84107 & 0.0538215 \\
				67 & \text{B5PSA} & \text{Sapphire} & 200 & 600 & 5.6381 & \text{--} \\
				68 & \text{B5PSB} & \text{Sapphire} & 200 & 650 & \text{--} & \text{--} \\
				69 & \text{B5PSC} & \text{Sapphire} & 200 & 700 & \text{--} & \text{--} \\
				70 & \text{B5PSD} & \text{Sapphire} & 200 & 750 & \text{--} & \text{--} \\
				71 & \text{B5PSE} & \text{Sapphire} & 200 & 800 & \text{--} & \text{--} \\
				72 & \text{B5PSE} & \text{Sapphire} & 200 & 800 & \text{--} & \text{--} \\
				73 & \text{B5PSE} & \text{Sapphire} & 200 & 800 & \text{--} & \text{--} \\
				74 & \text{B5PSF} & \text{Sapphire} & 200 & 850 & \text{--} & \text{--} \\
				75 & \text{B5PSG} & \text{Sapphire} & 200 & 900 & 3.90908 & \text{--} \\
				76 & \text{B5PSJ} & \text{Sapphire} & 200 & \text{--} & 9.2095 & \text{--} \\\hline
			\end{tabular}}
		\end{table}

		\begin{table}
			\caption{Superconducting critical temperature ($T_\text{c}$), residual resistance ratio (RRR) and low-temperature resistance ($R(\SI{20}{\kelvin})$) of some of the samples listed in table~\ref{tab:v3si_dep_rs}.}
			\resizebox{\columnwidth}{!}{%
			\rowcolors{1}{gray!15}{white}
			\begin{tabular}{llllllll}
				\rowcolor{gray!30}\hline
				\# & Sample & $T_\text{c}$ & $E(T_\text{c})$ (10,90$\%$) & RRR & $E$(RRR) & Normalized R(20K) & Normalized $E$(R(20K)) \\\hline\hline
				16 & \text{B2P05B} & 9.39975 & \text{$\{$-0.225125, 0.3252$\}$} & 2.06268 & 0.00234758 & 130.854 & 55.5819 \\
				17 & \text{B2P05C} & 10.6252 & \text{$\{$-0.300275, 0.37477$\}$} & 2.7949 & 0.0111599 & 2.03029 & 1.16346 \\
				18 & \text{B2P05C} & 3.21457 & \text{$\{$-0.149301, -0.374715$\}$} & 2.79629 & 0.0070044 & 2.02929 & 1.16286 \\
				19 & \text{B2P05C} & 8.42608 & \text{$\{$-1.12277, 0.744664$\}$} & \text{--} & \text{--} & \text{--} & \text{--} \\
				25 & \text{B2P05J} & 2.64123 & \text{$\{$-0.773577, 1.77471$\}$} & 0.976846 & 0.000912291 & 14.4724 & 0.011331 \\
				26 & \text{B2P05K} & 7.81542 & \text{$\{$-0.1994, 0.226192$\}$} & 1.51343 & 0.00666554 & 5.72973 & 0.0247937 \\
				27 & \text{B2P05L} & 7.06625 & \text{$\{$-0.201275, 0.323417$\}$} & 1.19422 & 0.000981507 & 9.37858 & 0.00750012 \\
				28 & \text{B2P05Q} & 8.0154 & \text{$\{$-0.1737, 0.250225$\}$} & 1.62802 & 0.00389435 & 0.00563819 & \text{7.126726579471347$\grave{ }$*${}^{\wedge}$-6} \\
				31 & \text{B2P06B} & 6.99141 & \text{$\{$-0.650431, 0.574157$\}$} & 1.31487 & 0.000501889 & 10.1829 & 0.16168 \\
				32 & \text{B2P06C} & 9.36222 & \text{$\{$-0.202775, 0.353863$\}$} & 2.27716 & 0.0102915 & 2.42658 & 1.30667 \\
				36 & \text{B2P08B} & 9.17528 & \text{$\{$-0.250425, 0.299725$\}$} & 1.85604 & 0.00841305 & 110.841 & 36.9786 \\
				37 & \text{B2P08C} & 10.6248 & \text{$\{$-0.25025, 0.350187$\}$} & 2.26837 & 0.0145972 & 1.78639 & 0.255358 \\
				38 & \text{B2P08D} & 10.7321 & \text{$\{$-0.238435, 0.529402$\}$} & 2.73546 & 0.0176014 & 1.52279 & 0.0422633 \\
				39 & \text{B2P08D} & 8.33437 & \text{$\{$-0.460382, 0.461381$\}$} & \text{--} & \text{--} & \text{--} & \text{--} \\
				40 & \text{B2P08E} & 11.8377 & \text{$\{$-0.324225, 0.374675$\}$} & 3.58144 & 0.0238104 & 1.11957 & 0.0258034 \\
				41 & \text{B2P08F} & 12.4874 & \text{$\{$-0.300166, 0.350533$\}$} & 4.52791 & 0.0320953 & 0.851126 & 0.0204526 \\
				42 & \text{B2P08G} & 13.1619 & \text{$\{$-0.474246, 0.425721$\}$} & 5.80453 & 0.0228682 & 0.663934 & 0.0154681 \\
				43 & \text{B2P08H} & 12.9878 & \text{$\{$-0.40045, 0.475787$\}$} & 7.97082 & 0.0562111 & 0.490424 & 0.006596 \\
				44 & \text{B2P08I} & 12.6124 & \text{$\{$-0.2502, 0.4626$\}$} & 6.5462 & 0.0676341 & 1.01451 & 0.00928348 \\
				45 & \text{B2P08P} & 3.0011 & \text{$\{$-0.00183321, 0.149609$\}$} & 4.50756 & 0.0197293 & 1.68563 & 0.00723219 \\
				47 & \text{B2P09B} & 7.91395 & \text{$\{$-0.201025, 0.301025$\}$} & 1.60018 & 0.00911362 & 19.233 & 0.108694 \\
				48 & \text{B2P09C} & 8.76345 & \text{$\{$-0.2242, 0.293588$\}$} & 1.97441 & 0.0173875 & 14.1185 & 0.123979 \\
				49 & \text{B2P09D} & 9.61453 & \text{$\{$-0.200125, 0.352287$\}$} & 2.2364 & 0.0267886 & 12.6645 & 0.152503 \\
				50 & \text{B2P09E} & 11.8152 & \text{$\{$-0.301651, 0.400862$\}$} & 3.09353 & 0.061513 & 10.0559 & 0.202928 \\
				51 & \text{B2P09F} & 12.3402 & \text{$\{$-0.350916, 0.449673$\}$} & 3.76488 & 0.0987009 & 8.19155 & 0.219781 \\
				52 & \text{B2P09G} & 11.941 & \text{$\{$-0.324925, 0.475233$\}$} & 4.75268 & 0.0394714 & 6.20588 & 0.0419948 \\
				53 & \text{B2P09H} & 12.241 & \text{$\{$-0.274217, 0.450363$\}$} & 5.53965 & 0.0538192 & 4.34384 & 0.0359899 \\
				55 & \text{B2P17B} & 9.61425 & \text{$\{$-0.201925, 0.325233$\}$} & 2.03642 & 0.00230517 & 137.551 & 67.2518 \\
				56 & \text{B2P17C} & 10.6754 & \text{$\{$-0.30025, 0.424525$\}$} & 2.88818 & 0.0234438 & 0.97623 & 0.273604 \\
				62 & \text{B3P11A} & 9.2303 & \text{$\{$-0.26265, 0.286975$\}$} & 1.85695 & 0.00508018 & 4.57798 & 0.0125477 \\
				63 & \text{B3P11B} & 10.9663 & \text{$\{$-0.30055, 0.27475$\}$} & 2.6311 & 0.00418009 & 1.90483 & 0.0028325 \\
				65 & \text{B4PP02} & 11.0002 & \text{$\{$-0.2753, 0.49984$\}$} & 2.76095 & 0.0130342 & 1.41893 & 0.0964786 \\
				66 & \text{B4PP03} & 12.6254 & \text{$\{$-0.375513, 0.474538$\}$} & 4.66919 & 0.0372196 & 0.822642 & 0.0132642 \\
				67 & \text{B5PSA} & 11.0747 & \text{$\{$-0.200125, 0.250125$\}$} & 2.04521 & 0.0167994 & 2.75673 & 0.0227421 \\
				68 & \text{B5PSB} & 11.9254 & \text{$\{$-0.200475, 0.24955$\}$} & 2.45945 & 0.0168331 & 2.10274 & 0.00982798 \\
				69 & \text{B5PSC} & 12.6747 & \text{$\{$-0.199525, 0.1997$\}$} & 3.04876 & 0.0317685 & 1.58516 & 0.0140517 \\
				70 & \text{B5PSD} & 13.7998 & \text{$\{$-0.175025, 0.175225$\}$} & 3.71176 & 0.0416517 & 0.00171005 & 0.0000161791 \\
				71 & \text{B5PSE} & 14.4515 & \text{$\{$-0.1525, 0.173175$\}$} & 4.64903 & 0.0672126 & 0.00138091 & 0.0000182585 \\
				74 & \text{B5PSF} & 14.9495 & \text{$\{$-0.1992, 0.17575$\}$} & 6.21999 & 0.0615928 & 0.786249 & 0.00536839 \\
				75 & \text{B5PSG} & 15.3249 & \text{$\{$-0.199625, 0.224892$\}$} & 8.39993 & 0.30363 & 0.46537 & 0.0174079 \\
				76 & \text{B5PSJ} & 1.19738 & \text{$\{$-0.243652, 0.431044$\}$} & 0.955918 & 0.000396282 & 9.63419 & 0.00399455 \\\hline
			\end{tabular}}
		\end{table}
	
\chapter{\label{sec:app_low_t_devices}Devices measured at low temperature}
		
		Some tips for future students:
		\begin{itemize}
			\item Don't go above $|V_\text{g}|=\SI{2}{\volt}$ at room temperature.
			\item Devices that leak through the gate at room temperature, will still do so when cooled down. However, devices that seem shorted from source to drain often come back to life at \SI{4}{\kelvin} (remember: we are \emph{supposed} to have enormous OFF current in a device with all-p doping).
			\item Use silver paste to glue the sample.
			\item Ground yourself, the bonding machine and the sample holder to a common ground before bonding anything.
			\item Always bond the substrate contact first, and the gate last. Bond all substrate pins to the back plate, so that all devices can be grounded together without the RC delay of the filters.
			\item Keep the substrate grounded while cooling down. If no LED is used, the substrate will be frozen out at low temperatures, and any accumulated charges can no longer be evacuated.
			\item Only apply a magnetic field once all the zero-field measurements are finished. It is absolutely impossible to get the coil back to $H=0$ without warming it up, and the measurements are extremely sensitive to the field.
		\end{itemize}
	
		\clearpage
		\newgeometry{top=1cm,bottom=1cm,left=0.7cm,right=0.7cm}
		\pagestyle{empty}
			\begin{figure}
				\centering
				\begin{tikzpicture}[x=2.162cm,y=2.312cm]
					\node[anchor=south west,inner sep=0] at (0,0) {\includegraphics[width=\textwidth]{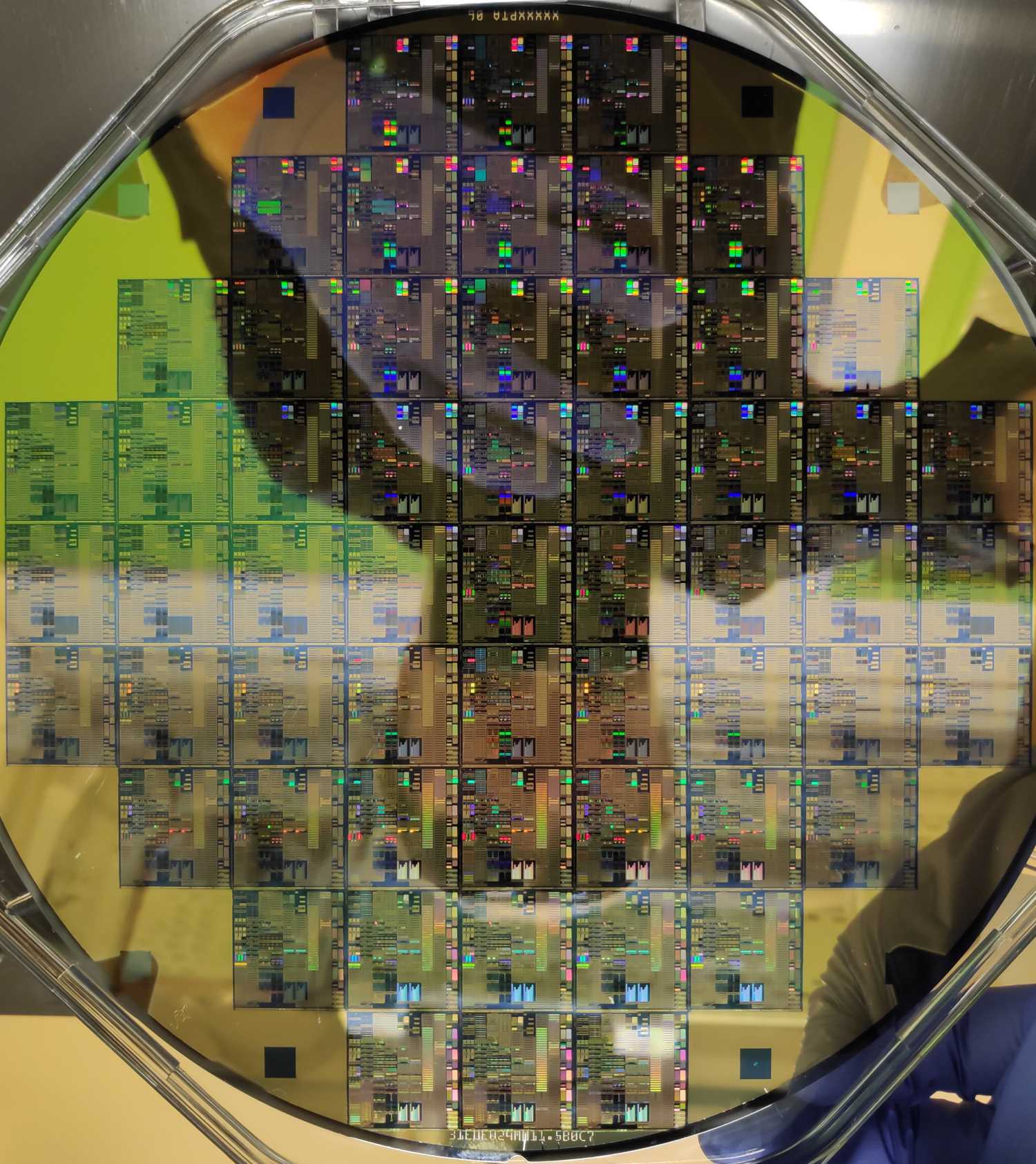}};
					
					\def\BadDieList{72,74,75,76,77,78,83,84,86,95};
					\def\GoodDieList{61,63,73,85};
					
					\foreach \x in {0,...,9}{
						\draw[red] (\x+0.04,0.25) -- (\x+0.04,9.25);
						}
					\foreach \y in {0,...,9}{
						\draw[red] (0.04,\y+0.25) -- (9.04,\y+0.25);
						}
					\foreach \x in {0,...,8}{
						\foreach \y in {1,...,9}{
							\pgfmathtruncatemacro{\labelx}{10-\y}
							\pgfmathtruncatemacro{\labely}{\x+1}
							\IfStringInList{\labelx\labely}{\BadDieList}{\node[baddie={\labelx}{\labely}]{}}{};
							\IfStringInList{\labelx\labely}{\GoodDieList}{\node[gooddie={\labelx}{\labely}]{}}{};
							\node[black,anchor=north west] at (\x+0.05,\y+0.24) {\large \B{D\labelx\labely}};
							\node[white,anchor=north west] at (\x+0.04,\y+0.25) {\large \B{D\labelx\labely}};
							}
						}
				\end{tikzpicture}
				\caption{Wafer PTA06 provided by Laurie Calvet, manufactured by Chinlee Wang and John Snyder at National Semiconductor. Dies with successful measurements at low temperature indicated in green, dies with only shorted/leaking devices in red.}
			\end{figure}
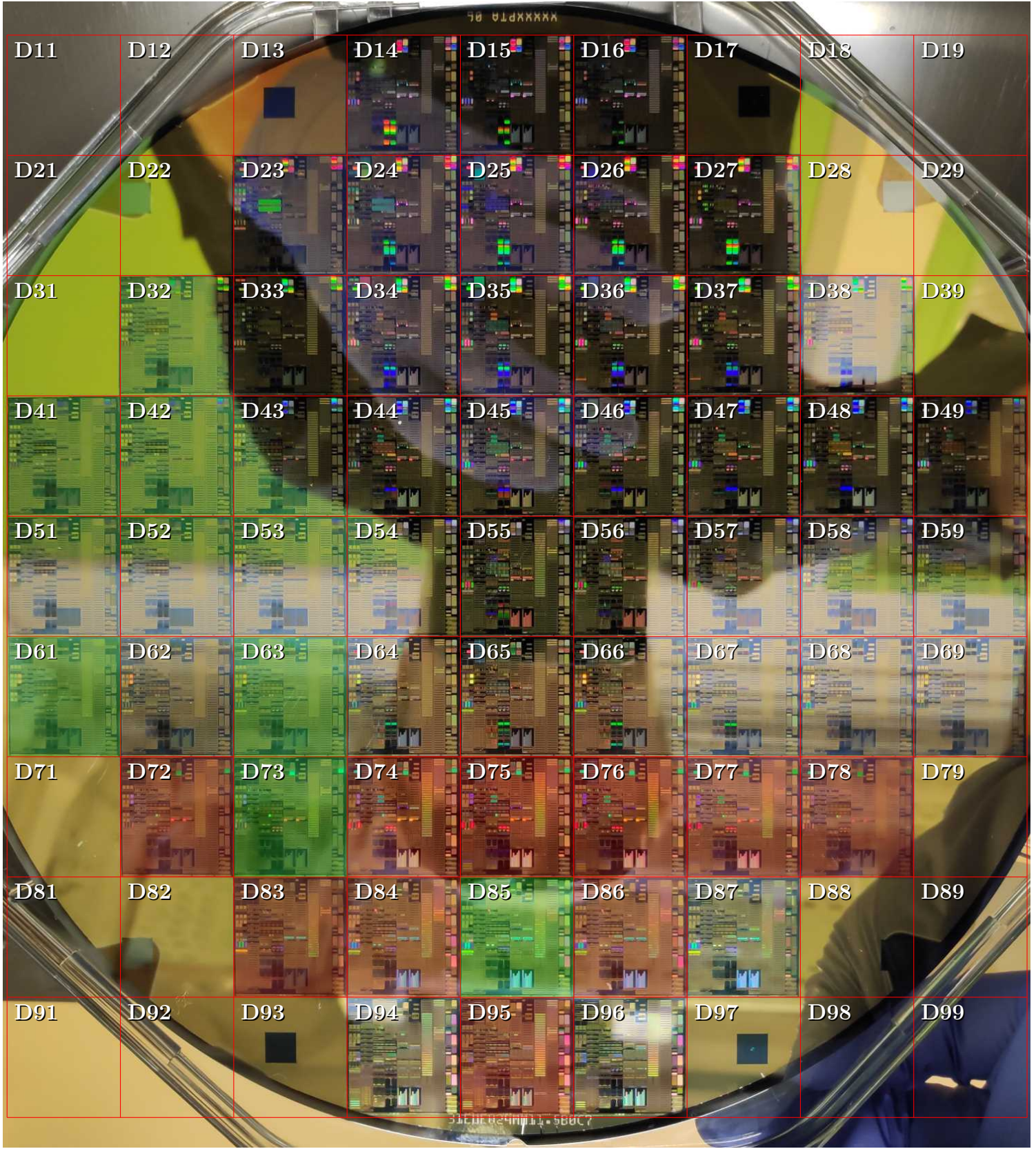
			
			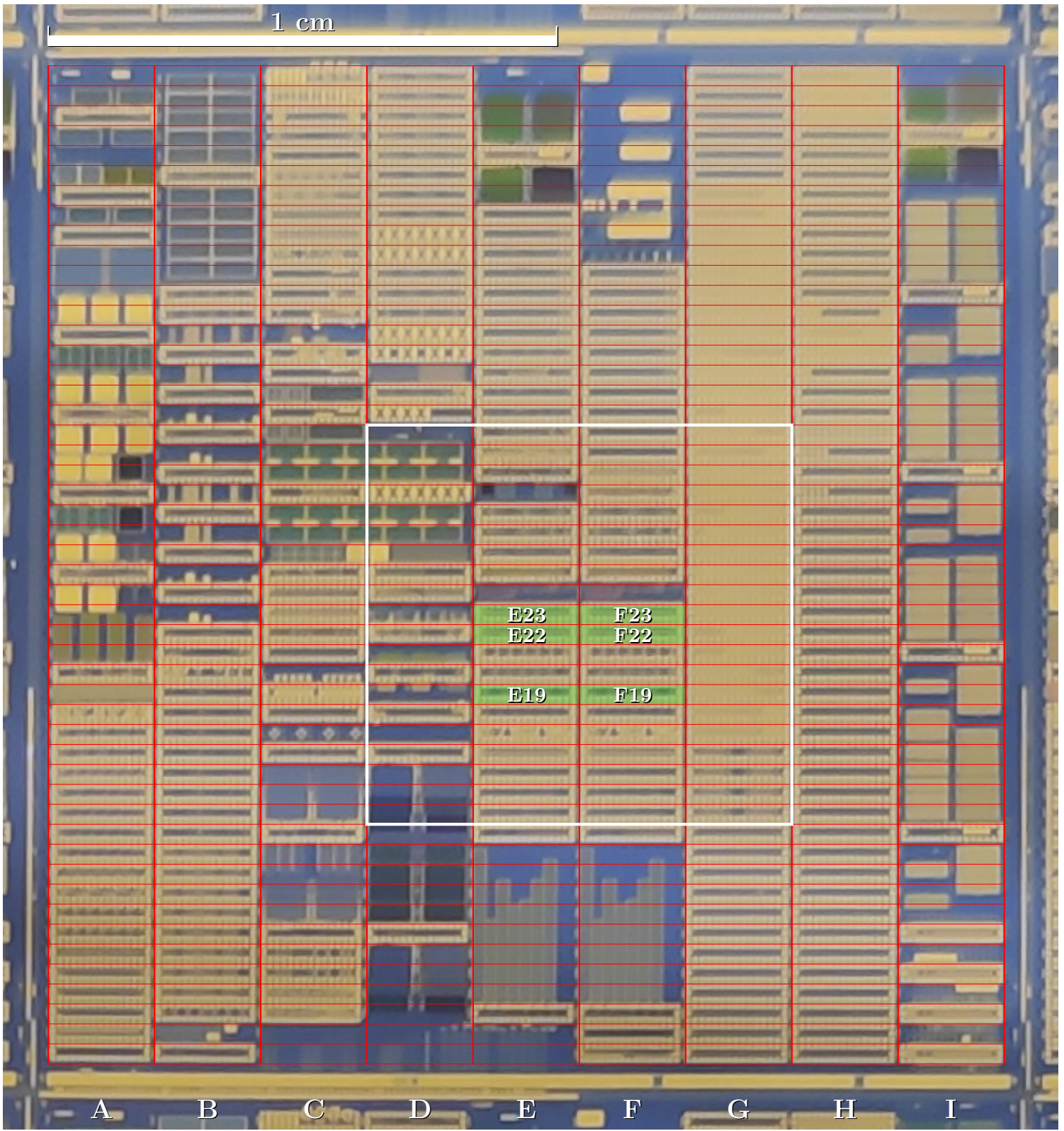
\begin{figure}
				\centering
				\begin{tikzpicture}[x=1.973cm,y=0.3735cm]
					\node[anchor=south west,inner sep=0] at (0,0) {\includegraphics[width=\textwidth]{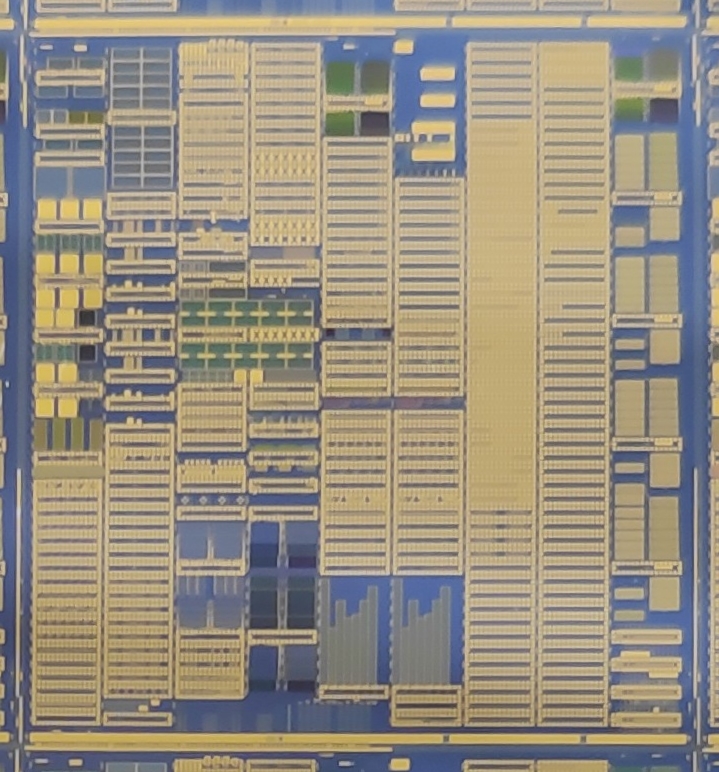}};
					
					\goodscribe{5}{19}{E19}
					\goodscribe{5}{22}{E22}
					\goodscribe{5}{23}{E23}
					\goodscribe{6}{19}{F19}	
					\goodscribe{6}{22}{F22}
					\goodscribe{6}{23}{F23}
					
					\def\Alphabet{A,B,C,D,E,F,G,H,I};		
					\foreach \x in {0,...,9}{
						\draw[red, thick] (\x+0.43,3.3) -- (\x+0.43,53.3);
						}
					\foreach \x [count = \xi] in \Alphabet{
						\node[black,anchor=center] at (\xi-0.06,1.00) {\Large \B{\x}};
						\node[white,anchor=center] at (\xi-0.07,1.05) {\Large \B{\x}};
						}
					\foreach \y in {0,...,50}{
						\draw[red] (0.43,\y+3.3) -- (9.43,\y+3.3);
						}
					
					\fill[black] (0.44,54.25) -- (5.225,54.25) -- (5.225,54.75) -- (0.44,54.75) -- cycle;
					\draw[black, thick] (0.44,54.25) -- (0.44,55.25);
					\draw[black, thick] (5.225,54.25) -- (5.225,55.25);
					\fill[white] (0.43,54.3) -- (5.215,54.3) -- (5.215,54.8) -- (0.43,54.8) -- cycle;
					\draw[white, thick] (0.43,54.3) -- (0.43,55.3);
					\draw[white, thick] (5.215,54.3) -- (5.215,55.3);
					\node[black] at (2.832,55.45) {\Large\B{1 cm}};
					\node[white] at (2.822,55.5) {\Large\B{1 cm}};
					
					\draw[white,ultra thick] (3+0.43,12+3.3) -- (7+0.43,12+3.3) -- (7+0.43,32+3.3) -- (3+0.43,32+3.3) -- cycle;
				\end{tikzpicture}
				\caption{Each die has 9 columns (letters) and 50 rows (numbers). Scribes E,F:19,22,23 contain the SBMOSFETs that were measured. The white square was cut out with a dicing saw and glued to a KYOCERA sample holder with silver paste. A bonding template is provided on the next page.}
			\end{figure}
		
		\clearpage
		\restoregeometry
		\newgeometry{top=0.2cm,bottom=0.2cm,left=0.2cm,right=0.2cm}
		\def\ImageScale{10.50}
		\begin{figure}
			\centering
			\begin{tikzpicture}[x=\ImageScale mm,y=\ImageScale mm]
				\draw (0,0) -- (15,0) -- (15,30) -- (0,30) -- cycle;
				\draw[fill=Gold] (0.2,5) -- (9.5,5) -- (9.5,0) -- (10.5,0) -- (10.5,5) -- (14.8,5) -- (14.8,25) -- (0.2,25) -- cycle;
				\draw (1.2,6.2) -- (13.8,6.2) -- (13.8,23.8) -- (1.2,23.8) -- cycle;
				\draw (2.1,7.1) -- (12.9,7.1) -- (12.9,22.9) -- (2.1,22.9) -- cycle;
				
				\foreach \y in {0,1}{
					\foreach \x in {0,...,4}{
						\draw[black,fill=white] (2.05*\x+2.8,6.2+16.7*\y) -- (2.05*\x+4,6.2+16.7*\y) -- (2.05*\x+4,7.1+16.7*\y) -- (2.05*\x+2.8,7.1+16.7*\y) -- cycle;
						}
					}
				\foreach \x in {0,1}{
					\foreach \y in {0,...,6}{
						\draw[black,fill=white] (1.2+11.7*\x,8.4+2*\y) -- (2.1+11.7*\x,8.4+2*\y) -- (2.1+11.7*\x,9.6+2*\y) -- (1.2+11.7*\x,9.6+2*\y) -- cycle;
						}
					}
					
				\foreach \x in {0,...,2}{	
					\pgfmathtruncatemacro{\labelx}{3-\x}
					\node at (2.375+2.05*\x,5.6) {\large \labelx};
					}
				\foreach \x in {3,...,5}{	
					\pgfmathtruncatemacro{\labelx}{27-\x}
					\node at (2.375+2.05*\x,5.6) {\large \labelx};
					}
				\foreach \x in {0,...,5}{	
					\pgfmathtruncatemacro{\labelx}{10+\x}
					\node at (2.375+2.05*\x,24.4) {\large \labelx};
					}
				\foreach \y in {0,...,5}{	
					\pgfmathtruncatemacro{\labely}{4+\y}
					\node at (0.7,10+2*\y) {\large \labely};
					}
				\foreach \y in {0,...,5}{	
					\pgfmathtruncatemacro{\labely}{21-+\y}
					\node at (14.3,10+2*\y) {\large \labely};
					}
				
				\def\chipwidth{8.1};
				\def\chiplength{8.6};
				
				\draw[black,fill=white] (7.5-0.5*\chipwidth,15-0.5*\chiplength) -- (7.5+0.5*\chipwidth,15-0.5*\chiplength) -- (7.5+0.5*\chipwidth,15+0.5*\chiplength) -- (7.5-0.5*\chipwidth,15+0.5*\chiplength) -- cycle;
				
				\pgfmathtruncatemacro{\imageheight}{\ImageScale*\chiplength}
				
				\node[anchor=center,inner sep=0] at (7.5,15) {\includegraphics[trim = 0 0 0 0, clip, height=\imageheight mm,angle=0]{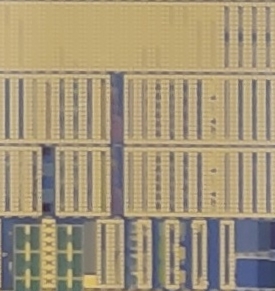}};
				
				\Tempgoodscribe{5}{19}{E19}
				\Tempgoodscribe{5}{22}{E22}
				\Tempgoodscribe{5}{23}{E23}
				\Tempgoodscribe{6}{19}{F19}
				\Tempgoodscribe{6}{22}{F22}
				\Tempgoodscribe{6}{23}{F23}
			
				\def\Alphabet{D,E,F,G};		
				\foreach \y in {0,...,4}{
					\draw[red] (7.5-0.5*\chipwidth,15-0.5*\chiplength+\y*0.25*\chiplength) -- (7.5+0.5*\chipwidth,15-0.5*\chiplength+\y*0.25*\chiplength);
					}
				\foreach \x in {0,...,20}{
					\draw[red] (7.5-0.5*\chipwidth+\x*\chipwidth/20,15-0.5*\chiplength) -- (7.5-0.5*\chipwidth+\x*\chipwidth/20,15+0.5*\chiplength);
					}
			\end{tikzpicture}
		\end{figure}
		\pagestyle{plain}
		\restoregeometry

		\newgeometry{left=0.5cm,right=0.5cm}
		\begin{figure}
			\centering
			\begin{subfigure}[b]{0.32\textwidth}
				\centering
				\includegraphics[width=\textwidth]{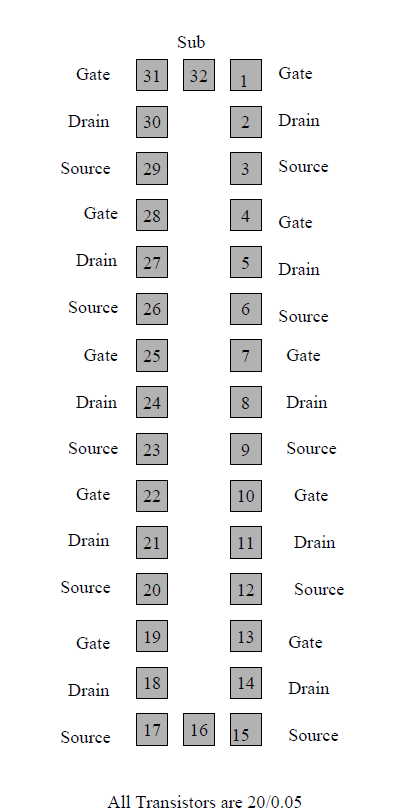}
				\vspace*{0.5\baselineskip}
				
				\caption{Quads E,F:19.}
			\end{subfigure}
			\begin{subfigure}[b]{0.32\textwidth}
				\centering
				\includegraphics[width=\textwidth]{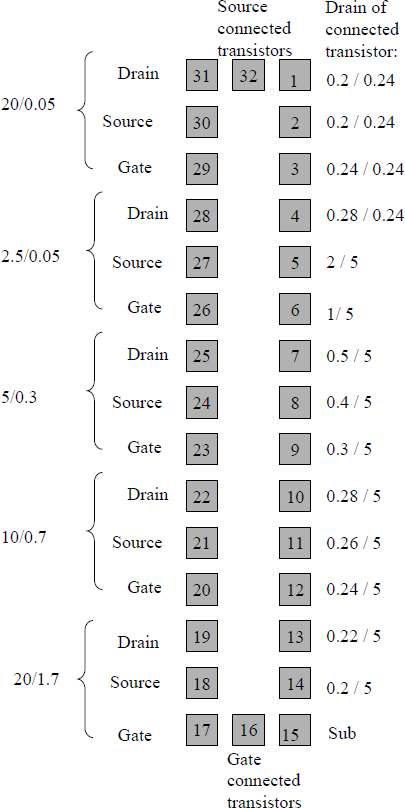}
				\vspace*{0.5\baselineskip}
				
				\caption{Quads E,F:22.}
			\end{subfigure}
			\begin{subfigure}[b]{0.32\textwidth}
				\centering
				\includegraphics[width=\textwidth]{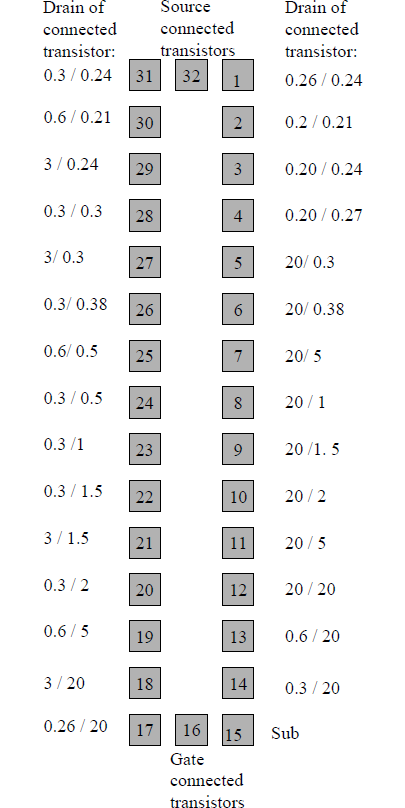}
				\vspace*{0.5\baselineskip}
				
				\caption{Quads E,F:23.}
			\end{subfigure}
			\caption{Layouts of quads E19 through F23, adapted from Ref.~\citenum{calvet2001electrical} (note that on wafer PTA06, \B{all} quads have PMOS devices).}
		\end{figure}	
		\restoregeometry
		
		\begin{figure}[H]
			\centering
			\begin{circuitikz}[x=1.5cm,y=1.5cm]
				\fill[rounded corners=1cm,gray!15!white] (-2,2.5) rectangle (4.75,-1);
				\node[anchor=south west] at (-1.75,-0.75) {(a)};
				\draw (0,1) node[pnp,rotate=-90](T1){};
				\draw (T1.E) -- (T1.E |-,0) node[ground](Vs){};
				\draw (T1.C |-,0.5) node[anchor=north]{$V_\text{d,1}$} to[short,*-] (T1.C);			
				\draw (T1.B |-,1.75) node[anchor=south]{$V_\text{g,1}$} to[short,*-] (T1.B);
				
				\draw[dashed] (1,|- T1) -- (2,|- T1);
							
				\draw (3,1) node[pnp,rotate=-90](T1){};
				\draw (T1.E) -- (T1.E |-,0) node[ground](Vs){};
				\draw (T1.C |-,0.5) node[anchor=north]{$V_\text{d,n}$} to[short,*-] (T1.C);			
				\draw (T1.B |-,1.75) node[anchor=south]{$V_\text{g,n}$} to[short,*-] (T1.B);
				
				\fill[rounded corners=1cm,gray!15!white] (-2,-1.5) rectangle (4.75,-5);
				\node[anchor=south west] at (-1.75,-4.75) {(b)};
				\draw (1,-3) node[pnp,rotate=-90](T1){};
				\draw (T1.E) -- (T1.E |-,-4) node[ground](Vs){};
				\draw (T1.C |-,-3.5) node[anchor=north]{$V_\text{d,1}$} to[short,*-] (T1.C);
				
				\draw (3.5,-3) node[pnp,rotate=-90](T2){};
				\draw (T2.E) -- (T2.E |-,|- Vs) -- (3,|- Vs);
				\draw[dashed] (2,|- Vs) -- (3,|- Vs);
				\draw (2,|- Vs) -- (Vs);
				\draw (T2.C |-,-3.5) node[anchor=north]{$V_\text{d,n}$} to[short,*-] (T2.C);
	
				\draw (-0.5,-4) node[ground]{} to[diode] (-0.5,|- T1.B);
				
				\draw (-1,|- T1.B) node[anchor=east]{$V_\text{g}$} to[short,*-] (T1.B) -- (2,|- T2.B);
				\draw[dashed] (2,|- T2.B) -- (3,|- T2.B);
				\draw (3,|- T2.B) -- (T2.B);
			\end{circuitikz}
			\caption{\label{fig:separate_connected_pmos} \B{(a)} The separate PMOS transistors (all devices on E,F:19; drains 19--31 on E,F:22) each have their own source, drain and gate contact pads. A voltage was systematically applied to the drain, and the current measured from the source. \B{(b)} The connected PMOS transistors (drains 1--14 on E,F:22, all devices on E,F:23) share the source and the gate, each has its own drain contact. A parallel reverse biased ESD diode prevents large currents from flowing across the gate.}
		\end{figure}
		
\printbibliography

	\end{refsection}
\end{appendices}

\end{document}